%% file: spaper.tex
\documentstyle{sespart}
\begin{document}
\begin{frontmatter}
\title{Approximation Methods for Non--linear
Gravitational Clustering}
\author{Varun Sahni}
\address{IUCAA -- Inter--University
Centre for Astronomy \& Astrophysics,
\\ Post Bag 4,  Ganeshkind, Pune 411007, India}
\and
\author{Peter Coles}
\address{Astronomy Unit,
School of Mathematical Sciences,
Queen Mary
\& Westfield College, Mile End Road,
London E1 4NS, United Kingdom}

\begin{abstract}
We discuss various analytical approximation methods
for following the evolution of cosmological density perturbations
into the strong (i.e. nonlinear) clustering regime. We start
by giving a thorough treatment of linear gravitational instability
in cosmological models and discussing
the statistics of primordial density fluctuations
produced in various scenarios of structure formation,
and the role of non--baryonic dark matter.
We critically review various methods for dealing with the
non--linear evolution of density inhomogeneities, in the context
of theories of
the formation of galaxies and large--scale structure. These methods
can be classified into five types: (i) simple extrapolations from linear
theory, such as the high--peak model and the lognormal model; (ii)
{\em dynamical} approximations, including the Zel'dovich approximation
and its extensions; (iii) non--linear models based on purely geometric
considerations, of which the main example is the Voronoi model; (iv)
statistical solutions involving scaling arguments, such as the
hierarchical closure {\em ansatz} for BBGKY, fractal models and the
thermodynamic model of Saslaw; (v) numerical techniques based on particles
and/or hydrodynamics. We compare the
results of full dynamical evolution using particle codes and the various
other approximation schemes. To put the models we discuss into perspective,
we give a brief review of the observed properties of galaxy clustering
and the statistical methods
used to quantify it, such as
correlation functions, power spectra, topology and spanning trees.
\end{abstract}
\end{frontmatter}

\input{ssec1}
\input{ssec2}
\input{ssec3}
\input{ssec4}
\input{ssec5}

\input{ssec6}

\input{ssec7}
\input{ssec8}

\input{ssec9}

\begin{ack}
The authors acknowledge stimulating conversations with Robert Brandenberger,
John Dubinsky, Jim Fry, Nikolai Gnedin, Salman Habib,
Alan Heavens, Anatoli Klypin,
Lev Kofman, Bob Mann, Adrian Melott, Dipak Munshi, Dima Pogosyan, Dave Salopek,
B.S. Sathyaprakash, Sergei Shandarin, Tarun Souradeep and Alexei Starobinsky.
We also thank Ed Bertschinger for his detailed comments on an earlier
version of this paper.
Peter Coles receives a PPARC Advanced Research Fellowship and thanks PPARC
for travel support for the visit to IUCAA during which most of
this paper was written.
He also thanks IUCAA for hospitality.
Both authors thank S.S. Pawar for help in preparing the figures.
\end{ack}

\input{srefs}
\end{document}

%% file: ssec1.tex
\section{Introduction}
The last two decades have been an exciting period for astrophysicists
working on the problem of the origin of the large--scale structure
of the Universe. Over this period, as a result of a continual
interplay between theory and observation, a picture has emerged
within which most of the properties of galaxy clustering
can be understood.

In this picture the Universe is almost homogeneous in the
early stages, but has some small amplitude density inhomogeneities
with a characteristic spectrum.
The theory of inflation can offer explanations for both the
initial almost--smoothness and the small density perturbations.
As the Universe evolves, these perturbations grow by gravitational
instability, at first linearly (so that the evolution of the
different spatial Fourier modes are independent and the
perturbations have a small amplitude), but eventually
forming non--linear concentrations of mass (where the
Fourier modes are coupled together and the density perturbations
are large); these non--linear concentrations are the eventual
sites for galaxy and cluster formation. An
important ingredient in this picture is the (postulated) existence
of large quantities of non--baryonic dark matter which allow
fluctuations to begin to grow even in the early phases
of the evolution of the Universe when radiation pressure inhibits
the growth of structure in the baryonic material. This general
idea has some empirical confirmation from both galaxy clustering
data and, most recently, by the discovery by the COBE satellite
of small anisotropies in the Cosmic Microwave Background radiation,
which may be the signature of primordial density perturbations.

Successful though this basic picture has been, it has to be admitted
that at present there is no model of galaxy and structure formation
that can account for all the observational data. Models with
different kinds of dark matter, different initial fluctuation
spectra, different relationships between luminous material and
dark matter have all been suggested at various times but none
has been successful enough to deserve the status of a
``standard model'' of galaxy formation.

The most fundamental problem one must tackle in a theory
of the origin of large--scale structure is that structures
around us in the Universe today correspond to density
fluctuations many orders of magnitude greater than the
mean density of the Universe. A galaxy cluster, for example,
will have a density around a factor 1000 greater than
the mean density of cosmic material. This means that such
structures are non--linear objects. While the growth
of density fluctuations in an expanding Universe is
tractible analytically if the fluctuations are small
(using linear perturbation theory), there is no general
exact solution for the non--linear regime. Traditionally,
astrophysicists have therefore resorted to numerical
$N$--body methods to understand this latter regime.
Alternatively, one can try to look to scales large
enough so that density fluctuations are small enough that
linear theory is still applicable. The problem with the
former approach is that such methods are extremely
time--consuming and can only be used to look at a small
part of the parameter space of possible models. The
latter approach has two weaknesses. One is that
it is extremely hard to obtain direct empirical information
about galaxy clustering on such large scales
(and galaxy clustering need not give us direct information
about the mass distribution). The second problem is that
we actually want to understand the formation of non--linear
objects, not just the super--large structures which are
still evolving more--or--less linearly.

What has been missing is a good theoretical understanding
of the non--linear regime of gravitational clustering.
Being able to produce a particular clustering property
in a numerical simulation is not the same as understanding
it!

In recent years, however, there has been a great deal of
interest in analytical methods for studying this difficult
problem. Various approximation schemes have been suggested
which can be applied to study different aspects of the
growth of non--linearity in the cosmological mass
distribution. These range from straightforward
perturbative extensions of linear theory, to simplified
kinematical models, scaling solutions and arguments
based on statistical mechanics. The description of
galaxy clustering is essentially statistical, but its
origin is dynamical. Different aspects of the clustering
pattern are therefore reproduced with differing
degrees of accuracy by different approximation schemes.

Our aim in this review is to give a thorough treatment
of the most important of these approximation schemes,
explaining their physical motivation and, where
appropriate, suggesting situations where they
might be sufficiently accurate to be useful.
We have tried to make the paper as self--contained
as possible, so we have included a fairly thorough
grounding in linear perturbation theory before
introducing the non--linear approximations. To put
these methods in context, we have also included
a discussion of numerical simulation methods and
some aspects of the statistical analysis of galaxy
clustering.
We do, however, assume basic knowledge
of physical cosmology such as can be obtained
from \cite{kt90,p93,zn83}.

%% file: ssec2.tex
\section{Cosmological Perturbations and Linear Evolution}

\subsection{Field Equations for the Unperturbed Universe.}
The Universe is generally believed to be homogeneous and isotropic on
scales greater than several hundred Mpc. This assumption is largely
based upon the observed homogeneity in the large scale distribution of
galaxies, radio galaxies and QSO's. In addition, the extremely small degree
of anisotropy of the Cosmic Microwave Background Radiation (CMBR) (which
travels to us virtually unhindered from the surface of last scattering
located at a cosmological redshift of $\sim 1100$) indicates that the
Universe was extremely isotropic (to one part in $\sim 10^5$) at the time
when matter and radiation decoupled. This suggests that our basic
starting point\footnote{We should mention that, although considerable
effort has been directed towards the study of perturbations, rather
less effort has been expended on the subject of how to fit a background
model to an observed lumpy Universe in an optimal way \protect{\cite{es87}}.}
 for the study of the origin of inhomogeneity should be
an isotropic and homogeneous background model.


\subsubsection{FRW Models}
The Friedmann--Robertson--Walker line element describes
a homogeneous and isotropic Universe\footnote{We
set the velocity
of light $c = 1$ for simplicity.}
\begin{equation}
ds^2 = dt^2 - a^2(t)\left({dr^2\over 1 - \kappa r^2} + r^2 d\theta^2 +
r^2\sin^2\theta d\phi^2\right)  ~~~~\kappa = 0, ~\pm 1
\label{eq:l1a}
\end{equation}
where $a(t)$ is the {\em cosmic scale factor}.
The dynamics of a FRW Universe is determined by the Einstein field equations:
$G_\mu^\nu = 8\pi G T_\mu^\nu$ where $T_\mu^\nu$ is the energy--momentum
tensor describing the contents of the Universe.
For a perfect fluid $\rho = T_0^0$, P = $- T_\alpha^\alpha$
and the Einstein equations simplify considerably to
\begin{eqnarray}
3\left( \frac{\dot{a}}{a} \right)^{2}
& = & 8\pi G\rho - {3\kappa\over a^2} + \Lambda\nonumber\\
{\ddot{a}\over a} & =& - {4\pi G\over 3} (\rho + 3 P) + {\Lambda\over 3}
\nonumber\\
\dot{\rho}& =& - 3 {\dot{a}\over a}(\rho + P).
\label{eq:l1b}
\end{eqnarray}
The cosmological constant $\Lambda$ increases the acceleration of the Universe
compared to models containing only matter and/or radiation.
It has been introduced on several occasions into the theory in attempts to
make cosmological models more compatible with observations.
A central issue, presently unresolved in cosmology, is the current value of
the Hubble parameter and the matter density. Although the density of
the Cosmic Microwave Background
radiation is known to a high degree of accuracy to be
$\rho_r = 4.7\times 10^{-34}$ g cm$^{-3}$,
the density of matter is known only to
within an order of magnitude, the uncertainty in $\rho$ being usually
expressed in terms of the dimensionless density parameter
\begin{equation}
\Omega \equiv {\rho\over \rho_{cr}} = {8\pi G \rho\over 3 H^2}.
\label{eq:omega_def}
\end{equation}
Present observations limit $\Omega$ to lie within the range $0.02 \le
\Omega \le$ few. The value of the Hubble parameter is known
to within about a
a factor of two to be $0.4 \le h \le 1$ where $h$ is the
dimensionless Hubble parameter expressed in units of 100 km s$^{-1}$
Mpc$^{-1}$. The uncertainties in $H_0$ and $\Omega$ are of course
related, because of the definition of $\Omega$ (\ref{eq:omega_def}).
The value of the cosmological constant is equally uncertain, being loosely
constrained by the inequality
$\vert \Omega_{\Lambda}\vert  = |\Lambda|/ 3H^2 < 0.6$ with 90\%
confidence \cite{t90,mr93,mk94}.

{}From (\ref{eq:l1a}) \& (\ref{eq:l1b}) we can obtain analytical
forms for the scale factor for some
physically interesting cases.

\subsubsection{Spatially Flat Universe with One Matter Component}
In this case we have $\Omega=1$ and the scale factor evolves
according to
\begin{equation}
a(t) = \left({t\over t_0}\right)^p \equiv \left({\eta \over \eta_0}
\right)^{\half - \nu}
\label{eq:l2}
\end{equation}
for a perfect fluid with
\begin{eqnarray}
p  & = & {2\over 3(1+w)}\nonumber\\
w  & =  & {P\over \rho} \nonumber\\
\nu & = & {3p -1\over 2(p-1)}
\end{eqnarray}
and $\eta$ is the conformal
time coordinate
\begin{equation}
\eta = \int {dt\over a(t)}.\label{eq:eta_def}
\end{equation}
Some special instances of eq. (\ref{eq:l2}) are:
\begin{enumerate}
\item pressureless matter
 or dust ($w = 0$): $a(t) \propto t^{2/3},$
$a(\eta) \propto \eta^2$
\item
radiation ($w = {1\over 3}): a(t) \propto \sqrt{t}, $
$a(\eta) \propto \eta$.
\end{enumerate}

\subsubsection{Spatially Flat Models with Two Non--interacting Fluids}
Here the interesting solutions are:
\begin{enumerate}
\item dust + radiation: $a(\eta) \propto \eta~(\eta + \eta_0)$,
\item
radiation + stiff matter ($w = 1$): $a(\eta) \propto
\sqrt{\eta~(\eta + \eta_0)}$,
\item
dust + cosmological constant: $a(t) \propto \left(\sinh {{3\over 2}}
\sqrt{{\Lambda\over 3}}ct\right)^{2/3}$
\end{enumerate}

\subsubsection{Spatially Curved Models}
In this case we divide into {\em open} models
$(\Omega<1)$, where we have
\begin{enumerate}
\item Dust: $a(\eta)=A(\cosh\eta-1)$, $ct=A(\sinh\eta-\eta)$
\item Radiation: $a(\eta) = A'\sinh \eta$, $ct=A'(\cosh\eta-1)$,
\end{enumerate}
while for {\em closed} modes $(\Omega>1)$ we have
\begin{enumerate}
\item Dust: $a(\eta)=A(1-\cos\eta)$, $ct=A(\eta-\sin\eta)$
\item Radiation: $a(\eta)=A'\sin\eta$, $ct=A'(1-\cos\eta)$.
\end{enumerate}

\subsubsection{Other Useful Relations}
It is sometimes convenient to express the Hubble parameter and the
FRW equations in terms of the cosmological redshift parameter
\begin{equation}
1 + z = {a_0\over a(t)}.\label{eq:z_def}
\end{equation}
In terms of
this new parametrisation, the Hubble parameter
in a multicomponent Universe consisting of several non-interacting matter
species characterised by equations of state $P_{\alpha} =
w_{\alpha}\rho_{\alpha}$ is given in a very simple form:
\begin{equation}
H(z) = H_0(1 + z) \left[ 1 - \sum_{\alpha}\Omega_{\alpha} + \sum_{\alpha}
\Omega_{\alpha}(1 + z)^{1 + 3w_{\alpha}}\right]^{\half}. \label{eq:l3a}
\end{equation}
For a spatially flat Universe $1 - \sum_{\alpha}\Omega_{\alpha} = 0$ and
equation (\ref{eq:l3a}) reduces to
\begin{equation}
H(z) = H_0(1 + z) \left[\sum_{\alpha}\Omega_{\alpha}(1 + z)^{1 + 3w_{\alpha}}
\right]^{\half}
\label{eq:l3b}
\end{equation}
In a Universe consisting only of dust,
\begin{equation}
H(z) = H_0(1 + z)(1 + \Omega_0z)^{\half} \label{eq:l3c}
\end{equation}
In this case the expression for the density parameter turns
out to be particularly simple:
\begin{equation}
\Omega (z) = {\Omega_0 (1 + z)\over 1 + \Omega_0 z}
\label{eq:l4}
\end{equation}
for large values of $\Omega z >> 1$, $\Omega (z) \rightarrow 1$,
indicating that the density always
approaches the critical value at early times.

A useful relationship between the cosmological time parameter $t$ and the
cosmological redshift $z$ can be obtained by differentiating
$1 + z = a_0/a(t)$ with respect to time so that $dz/ dt = - H(z)(1 + z)$.
For a Universe consisting only of dust this leads to
\begin{equation}
H_0t(z)  = \int_z^\infty {dz'\over (1 + z')^2 (1 + \Omega_0 z')^{\half}}.
\label{eq:l5a}
\end{equation}
Evaluating the integral in eq. (\ref{eq:l5a})
we get \cite{zn83}:
\begin{eqnarray}
t(z)  & = & {H_0^{-1}\over 1 - \Omega_0}\left\{ {(1 + \Omega_0 z)^{\half}
\over 1 + z} +
{\Omega_0\over 2(1 - \Omega_0)^{\half}} \log \left[{(1 + \Omega_0 z)^{\half}
- (1 - \Omega_0)^{\half} \over (1 + \Omega_0z)^{\half} +
(1 - \Omega_0)^{\half}}\right] \right\}\nonumber\\
t(z) & = &
{H_0^{-1}\over \Omega_0 - 1}\left\{{\pi\over 2}{\Omega_0\over
(\Omega_0 - 1)^{\half}} -
{\Omega_0\over (\Omega_0 - 1)^{\half}}\arctan \left[({1 +
\Omega_0 z\over \Omega_0 - 1})^{\half}
\right] - {(1 + \Omega_0 z)^{\half}\over 1 + z}\right\}
\nonumber\\
t(z) & = &  {2H_0^{-1}\over 3(1 + z)^{3\over 2}}
\label{eq:l5b}
\end{eqnarray}
for the cases $\Omega_0<1$, $\Omega_0>1$ and $\Omega_0=1$,
respectively.

For a spatially flat Universe with a cosmological constant ($\Omega_{\Lambda}
= 1 - \Omega$) the present age of the Universe is given by \cite{kgb93}:
\begin{equation}
t_0 = 6.5 h^{-1}\bigg\lbrack\frac{1}{2\sqrt{1-\Omega}}
\log {\frac{1+\sqrt{1-\Omega}}{1-\sqrt{1-\Omega}}}\bigg\rbrack \times 10^9
{}~{\rm yr}.
\end{equation}
In the limit $\Omega \rightarrow 1$ we recover $t_0 = 6.5$ h$^{-1}$ Gyr.

\subsection{Gravitational Instability: Linear Theory}
Having briefly described some properties of the evolution equations
governing the expansion of a
homogeneous and isotropic Universe, let us now consider small departures from
homogeneity and isotropy which eventually grow to become galaxies and
other cosmic structures.
The line element which appropriately describes the perturbed FRW Universe
in the synchronous gauge, the most convenient gauge choice for
many purposes, has the form
\begin{equation}
ds^2 = ds_{FRW}^2 + h_{\alpha\beta} dx^{\alpha}dx^{\beta},
\label{eq:pert_FRW}
\end{equation}
where $\alpha, \beta = 1, 2, 3$, and $ds_{FRW}^2$ is the
FRW line element given
by eq. (\ref{eq:l1a}).
In the linearised approximation which we shall follow, perturbations
of the metric are assumed to be small $h_{\alpha\beta}h^{\alpha\beta} << 1$,
and the length scale of the perturbations is always much smaller than the
Horizon scale $\lambda_H \simeq 2ct$ so that a Newtonian treatment of the
subject is valid.
The relativistic generalisation of the
linearised treatment of gravitational instability presented
here
has been developed in several papers
\cite{l46,h66,ss84,sstar85,eb89,hv90,bde92}.

If  the mean free path of a particle is small, matter can
be treated as an ideal fluid and
the Newtonian equations governing the motion of gravitating
collisionless particles in an expanding Universe can be written in terms of
$\vec{x} = \vec{r} / a$ (the comoving spatial coordinate),
$\vec{v} = \dot {\vec{r}} - H\vec{r} = a\dot {\vec{x}}$ (the peculiar
velocity field), $\phi (\vec{x} , t)$ (the peculiar Newtonian gravitational
potential) and $\rho (\vec{x}, t)$ (the matter density). In this manner
 we obtain  the following set of equations \cite{p80}:

First, {\em the Euler equation},
\begin{equation}
{\partial (a\vec{v})\over \partial t} + (\vec{v}\cdot\vec{\nabla}_x)\vec{v} =
- {1\over \rho}\vec{\nabla}_x P - \vec{\nabla}_x\phi
\label{eq:Euler}
\end{equation}
The second term on the right hand side of eq. (\ref{eq:Euler}) is the
peculiar gravitational
force responsible for the acceleration of the fluid element,
which we shall denote  $\vec{g} = -\vec{\nabla_x}\phi/a$.
If the velocity flow is irrotational, $\vec{v}$ can be rewritten in terms
of a velocity potential $\phi_v$: $\vec{v} = - \vec{\nabla} \phi_v/a$.
For a polytropic fluid with equation of state $P \propto \rho^{\gamma}$
the Euler equation can be integrated to give the Bernoulli equation
\begin{equation}
{\partial \phi_v\over \partial t} - {1\over 2 a^2} (\nabla_x\phi_v)^2 =
\frac{\gamma}{\gamma-1}{P\over \rho} + \phi.
\label{eq:Berno}
\end{equation}
Next we have the continuity equation:
\begin{equation}
{\partial\rho\over \partial t} + 3H\rho + {1\over a} \vec{\nabla}_x
(\rho\vec{v}) = 0,
\label{eq:continuity}
\end{equation}
or, equivalently,
\begin{equation}
{\partial\delta\over \partial t} +
{1\over a} \vec{\nabla}_x\left[(1 + \delta)\vec{v}\right] = 0.
\label{eq:continuityb}
\end{equation}
And, finally, the Poisson equation:
\begin{equation}
\nabla_x^2\phi = 4\pi G a^2(\rho - \rho_0) = 4\pi Ga^2\rho_0\delta,
\label{eq:Poisson}
\end{equation}
where $\rho_0$ is the mean background density
and $\delta = \rho/\rho_0 - 1$ is the {\em density contrast}.

Expanding $\rho$, $\vec{v}$ and $\phi$ perturbatively
and keeping only the first--order terms in equations
(\ref{eq:Euler}), (\ref{eq:continuity}) and (\ref{eq:Poisson})
gives the linearised continuity equation:
\begin{equation}
{\partial\delta\over \partial t} = - {1\over a}\vec{\nabla}_x\cdot
\vec{v}
{}~~~\hbox{or, equivalently,}~~~~
\delta = - {1\over a H f}\left(\vec{\nabla}_x\cdot\vec{v}\right).
\label{eq:l9}
\end{equation}
The function $f$ is given by
\begin{equation}
f \equiv {d\log\delta\over d\log a}
\simeq \Omega_0^{0.6} + {\Omega_{\Lambda}\over 70}\left(1 + {\Omega_0\over 2}
\right),
\label{eq:9b}
\end{equation}
where $\Omega_{\Lambda} = \Lambda/ 3H_0^2$ is the effective
fraction of the
critical energy density contributed by a cosmological constant term
\cite{llpr91,p80,p84,m91}.

The linearised Euler and Poisson equations are as follows:
\begin{equation}
{\partial \vec{v}\over\partial t} + {\dot a\over a}\vec{v} = -
{1\over \rho a}\vec{\nabla}_x P -{1\over a}\vec{\nabla}_x\phi
\label{eq:l10}
\end{equation}
\begin{equation}
\nabla_x^2\phi = 4\pi G a^2\rho_0\delta
\label{eq:l11}
\end{equation}
($|v|, |\phi|, |\delta| << 1$ in equations (\ref{eq:l9})--
(\ref{eq:l11})). From these equations
we can obtain a second order differential equation
for the linearised density contrast $\delta$
\begin{equation}
{\partial^2\delta\over\partial t^2} + 2H{\partial\delta\over\partial t}
- 4\pi G\rho_0\delta = {\nabla_x^2 P\over \rho_0a^2}
\label{eq:l12}
\end{equation}
Ignoring pressure forces for the moment  we obtain
\begin{equation}
\ddot\delta + 2H\dot\delta - 4\pi G\rho_0\delta = 0,
\label{eq:l13a}
\end{equation}
which can also be rewritten as
\begin{equation}
\ddot\delta + 2H\dot\delta - {3\over 2}\Omega H^2\delta = 0.
\label{eq:l13b}
\end{equation}
For a spatially flat matter--dominated Universe
$\rho_0 = (6\pi Gt^2)^{-1}$, and
eq. (\ref{eq:l13a})
admits two linearly independent power law solutions $\delta(\vec{x},t)
= D_{\pm}(t)\delta(\vec{x})$ where
$D_+(t) \propto a(t) \propto t^{2\over 3}$  is the growing  mode
and $D_-(t) \propto t^{-1}$ is the decaying mode.
For initial conditions demanding $\delta = \delta_1$ and $\dot\delta_1 = 0$
at $t = t_1$ (i.e. the perturbation begins to grow from rest), the final
form of $\delta$ is a superposition of increasing and decreasing modes:
\begin{equation}
D(t) = {3\over 5}\delta_1 \left({t\over t_1}\right)^{2\over 3} +
{2\over 5}
\delta_1\left({t\over t_1}\right)^{-1},
\label{eq:l14}
\end{equation}
i.e. three fifths of the initial amplitude is in the
increasing mode and the
remaining two fifths in the decreasing mode.
Substituting $\delta(\vec{x}, t) \equiv D_+(t)\delta(\vec{x})$ in
eq. (\ref{eq:l11}) we obtain the linearised Poisson equation
\begin{equation}
\vec{\nabla_x}^2\phi = 4\pi G D_+(t)\delta(\vec{x})\rho_0 a^2
\label{eq:lzzzz}
\end{equation}
In a spatially flat, matter--dominated Universe:
$\rho_0 \propto a^{-3}, D_+(t) \propto a(t)$, and we find that
$\vec{\nabla_x}^2\phi \propto \delta(\vec{x})$ i.e. the linearised
gravitational
potential does not evolve with time.

{}From equations (\ref{eq:l9})--(\ref{eq:l11}),
ignoring pressure, we get
\begin{equation}
\vec{v} = - {2f\over 3\Omega H a}\vec{\nabla}_x\phi + {\mbox{ const}\over
a(t)}.
\label{eq:l15}
\end{equation}
Equation (\ref{eq:l15}) demonstrates that the velocity
flow associated with the growing
mode in the linear regime is potential: $\vec{v} = - \vec{\nabla}\phi_v/a$
where $\phi_v = \tilde{A}\phi, \tilde{A} = 2f/3\Omega H$,
a fact that is used to great advantage
in non-linear approximations of gravitational clustering such as the
Zeldovich approximation (\S 4.2)
and in reconstruction methods such as POTENT (\S 8.10).
In the absence of a gravitational force field $\vec{g} = - \vec{\nabla}_x\phi/a
= 0$, so
the peculiar velocity in an expanding Universe decays kinematically as
$\vec{v} \propto a(t)^{-1}$ \cite{p80}.

Since
\begin{equation}
\vec{g} = - {\vec{\nabla}\phi\over a} = G\rho_0 a(t)\int d^3\vec{x}'
{\delta(\vec{x}',t) (\vec{x}' - \vec{x})\over |\vec{x}' - \vec{x}|^3},
\label{eq:l16}
\end{equation}
we find that the peculiar gravitational acceleration associated with a
given mode is simply
\begin{equation}
\vec{g}_\pm \propto \rho_0 a D_\pm (t).
\label{eq:l17}
\end{equation}
{}From equations (\ref{eq:l15}) and (\ref{eq:l17})
we
find that the peculiar
velocity and the peculiar acceleration are parallel to one another, and
that in a flat matter--dominated Universe, the velocities associated
with the increasing and decreasing modes respectively are
$v_+ \propto t^{1\over 3}$ and $v_- \propto t^{-{4\over 3}}$.
An expression similar to equation (\ref{eq:l15}) can also be derived
for an isolated spherical overdensity (\S3.1).

The linearised equations of motion  provide an excellent description of
gravitational instability at very early times when density fluctuations
are still small ($\delta \ll 1$). Linear theory may also
provide a qualitatively good description of large scale clustering in
hierarchical scenarios (such as the cold dark matter model which will be
considered shortly) in which large scale power is suppressed relative to
small scale power in the fluctuation spectrum
leading to $\delta >> 1$ on scales $< 1$ Mpc, while
$\delta < 1$ on scales greater than $10 - 30$ Mpc.

For spatially open and closed FRW models it is convenient to find general
solutions of eq. (\ref{eq:l13b})
by noting that the decreasing mode is simply
proportional to the Hubble parameter \cite{h77,zn83}. To see this,
differentiate  the Einstein equation $\dot H + H^2 = - 4\pi G(\rho -
2\Lambda)/3$, and use the continuity condition $\dot\rho_0 = - 3H\rho_0$,
leading to
$\ddot H + 2H \dot H - 4\pi G\rho H = 0$. This shows that $D_-(t) \propto H$
is indeed a solution to eq. (\ref{eq:l13a}). The second solution of
eq. (\ref{eq:l13a}) can be constructed from $D_{-}(t)$ and the
Wronskian of eq. (\ref{eq:l13a}): $W(t) = \dot{D_+}D_- - \dot{D_-}D_+
\propto a^{-2}$.
This gives
\begin{equation}
D_+(t) = D_-(t)\int^t
W(t')D_-^{-2}(t')\d t' = H(t)\int^t{\d t'\over a^2(t')H^2(t')}
= H(a)\int^a{\d a\over (Ha)^3}.
\label{eq:l19}
\end{equation}
Substituting $a(z) = a_0(1 + z)^{-1}$ with $H(z)$ defined in
(\ref{eq:l3c})
we obtain
\begin{equation}
D_+(z) = (a_0H_0)^{-2} (1 + z)(1 + \Omega_0 z)^{\half}
\int_z^{\infty}
{\d z'\over (1 + z')^2(1 + \Omega_0 z')^{3/2}}.
\label{eq:l20}
\end{equation}
Equation (\ref{eq:l20}) can be integrated to give
\begin{equation}
D_{+}(z) = \frac{1+2\Omega_0 + 3\Omega_0 z}{|1-\Omega_0|^{2}}
+ 3\Omega_0
{(1 + z)(1 + \Omega_0 z)^\half\over |1 - \Omega_0|^{5\over 2}}f(\Omega_0, z),
\label{eq:l21}
\end{equation}
where
\begin{eqnarray}
f(\Omega_0, z) & = &
- \half\log\left[{(1 + \Omega_0 z)^\half + (1 - \Omega_0)^\half
\over (1 + \Omega_0 z)^\half - (1 - \Omega_0)^\half}\right]\nonumber\\
f(\Omega_0, z)  & = &
\arctan\left[{1 + \Omega_0 z\over \Omega_0 - 1}\right]^{\half},
\label{eq:l21a}
\end{eqnarray}
for the cases $\Omega_0<1$ and $\Omega_0>1$, respectively. In the
simpler case where $\Omega_0=1$, we have
\begin{equation}
D_+(z)  \propto  ~(1 + z)^{-1}
\label{eq:l21b}
\end{equation}
A good approximation to $D_+(z)$ for $\Omega_0 < 1$
can be provided by the fitting formula \cite{zn83}
\begin{equation}
D_+(z) \simeq \left(1 + {3\over 2}\Omega_0\right)\left(1 + {3\over 2}\Omega_0
+ {5\over 2}\Omega_0 z\right)^{-1}.\label{eq:l22}
\end{equation}
{}From equations (\ref{eq:l21}) \& (\ref{eq:l22}),
we find that, for $z >> \Omega_0^{-1}$, $D_+ \propto z^{-1}$
-- a consequence of the fact that the Universe at early times expands at the
critical rate $a \propto t^{2/3}$
regardless of the value of $\Omega_0$. For open models
we find that, at late times (corresponding to
$z << \Omega_0^{-1}$), the growth of
$D_+$ freezes to a constant value due to
the  rapid  expansion of the
Universe $a \propto t$, caused by the dominance of the curvature term
in the equations of motion (\ref{eq:l1b}).
The associated velocity perturbation
first grows according to $(1 + z)^{-1/2}$ and then
decreases as $1 + z$.

Although the decreasing mode can be important in some circumstances,
we shall hereafter mainly deal only with the increasing mode because this
is responsible for the formation of cosmic structure in the gravitational
instability picture.

In the preceding discussion we limited our attention to gravitational
instability in a single component Universe. We shall now extend it to more
general situations by considering the behavior of linearised perturbations
in a Universe consisting of, firstly, pressureless matter
(dust) and radiation, and,
secondly, dust and a cosmological constant. In both cases we shall
consider flat  FRW models.
The resulting equations in this case are
\begin{eqnarray}
\ddot{\delta} + 2\left({\dot{a}\over a}\right)\dot{\delta} - 4\pi G
\bar{\rho}_m
\delta  & =  & 0\nonumber\\
\left({\dot{a}\over a}\right)^2  =  {8\pi G\over 3} (\rho_m + \rho_X)
& = & {8\pi G\over 3} \rho_{tot}
\label{eq:l23}
\end{eqnarray}
where in the first case $\rho_X \equiv \rho_r \propto a^{-4}$, and in the
second $\rho_X \equiv \Lambda/ 8\pi G =$ constant.
We have here assumed that the radiation component is unperturbed which is
generally not exact, but this calculation serves to illustrate a
qualitative point. From eq. (\ref{eq:l23}) it is clear that the
presence of an additional smooth component
will, by increasing the rate of expansion of the Universe, inhibit the
growth of  $\delta$ \cite{gz70,m74,gp75,sv80}.
This effect can be quantitatively
demonstrated by solving eq. (\ref{eq:l23}) with $\rho_X \equiv \rho_r$
which after
a change of variables $\tau = \rho_m/ \rho_r \propto a(t)$ becomes
\cite{m74,gp75,e90}
\begin{equation}
{\d^2\delta\over \d\tau^2} + {(2 + 3\tau)\over 2\tau(1 + \tau)}{\d\delta\over
\d\tau}
- {3\over 2\tau(1 + \tau)}\delta = 0.
\label{eq:l24}
\end{equation}
Equation (\ref{eq:l24}) has the
growing solution $D_+ \propto 1 + 3\tau/2$,
which demonstrates that fluctuations remain frozen so long as the Universe
is radiation dominated ($\tau << 1$), and begin to grow only after
matter--domination ($\tau > 1$).

A similar effect (but in reverse chronological order) occurs for a Universe
with a cosmological constant.
The decaying mode in this case is given by
$D_{-}(a) \propto H(a)$, with $H(a)$ given by (\ref{eq:z_def}),
(\ref{eq:l3a}); $D_{+}(a)$ is given by (\ref{eq:l19}). For a spatially
flat Universe with $\Omega_{\Lambda} + \Omega_m = 1$:
\begin{eqnarray}
D_{-}(a)  & \propto &
a^{-{3\over 2}}\left(1+{\Omega_{\Lambda}\over \Omega_m}a^3\right)^{1\over 2}
\nonumber\\
D_+(a) & = & A\,{5\over 6}{\cal B}_x\left({5\over 6},{2\over 3}\right)\,
\left({\Omega_m\over \Omega_{\Lambda}}\right)^{1\over 3}\left[1+
{\Omega_m\over \Omega_{\Lambda}a^3}\right]^{1\over 2}
\label{eq:l25}
\end{eqnarray}
where
\begin{equation}
x = {\Omega_{\Lambda}a^3\over \Omega_m + \Omega_{\Lambda}a^3}\,\,\,\,\,\,\,
a = (1+z)^{-1},
\end{equation}
$A$ is an arbitrary constant, ${\cal B}_x(\alpha,\beta)$ is the incomplete Beta
function \cite{bbk92}.
As in open models, the growth of density perturbations slows down
considerably during later epochs
$z < z_* = \left(\Omega_{\Lambda}/\Omega_m\right)^{1/3} - 1$,
when the expansion of the Universe begins to be driven by the cosmological
constant \cite{ds90,w87}.
At very late times $D_+(a)$ approaches a constant value
\begin{equation}
\lim_{a\rightarrow\infty} D_+(a) = A\,{5\over 6}{\cal B}\left({5\over 6},
{2\over 3}\right)\,
\left({\Omega_m\over \Omega_{\Lambda}}\right)^{1\over 3}.
\end{equation}

In closed models of the Universe the presence of a cosmological constant
leads to additional interesting possibilities including
an intermediate {\em loitering} or {\em  coasting} stage
which arises because of the mutual cancellation of terms involving matter,
the curvature term and the cosmological constant in equation (\ref{eq:l1b})
leading to
an epoch when the scale factor does not change significantly with time:
$a(t) \simeq$ constant \cite{pss67,sfs92}.
Such models are interesting because during loitering density perturbations
grow very rapidly:
\begin{equation}
D_+(t) \propto {1\over a}\exp \left\{ \int\sqrt {4\pi G\rho} \d t\right\},
\label{eq:l26}
\end{equation}
approaching the exponential Jeans rate which arises in a nonexpanding Universe.
In addition to growing large amplitude inhomogeneities from small initial
conditions, a loitering Universe can also be very old,
much older than $H_0^{-1}$. This might ameliorate the age problem which
arises in a flat matter--dominated FRW Universe having $t_0 =
2H_0^{-1}/3$,
which for $H_0^{-1} = 1 - 2\times 10^{10}$ years is somewhat smaller than
the estimated age of globular clusters, $1.6\times 10^{10}$ years
\cite{jd83}.
As shown in \cite{sfs92},
loitering is not confined to
cosmological models with a cosmological constant, but in fact can arise in
all closed FRW models in which in addition to normal matter
there exists a form of matter whose equation of state violates the
{\em strong energy condition}
so that $w = P/ \rho < - 1/3$ ($w = - 1$ for a cosmological
constant). The loitering epoch $z_l$ can be related to the density parameter
of normal matter $\Omega_0$, and the equation of state $w$ by
\begin{equation}
\Omega_0 =  {\vert 1 + 3w\vert\over (1 + z_l)^{- 3w} + 3w~z_l - 1}.
\label{eq:l27}
\end{equation}
For instance if $\Omega = 0.1$ then loitering must occur between
$2 < z_l < 7.2$. Other features of a loitering Universe are discussed in
\cite{sfs92,dkov90,fe93,p80,swaga94}.

\subsection{Pressure Effects; the Jeans Length}
In our treatment of density inhomogeneities we have so far ignored the effect
of pressure terms, choosing to work with eq. (\ref{eq:l13a})
rather than (\ref{eq:l12}).
Including matter pressure into the analysis we recover
\begin{equation}
{\partial^2\delta\over\partial t^2} + 2H{\partial\delta\over\partial t}
- \left({v_s\over a}\right)^2 {\nabla_x}^2\delta - 4\pi G\rho_0\delta = 0.
\label{eq:l28}
\end{equation}
Expanding $\delta$ in a Fourier series $\delta (\vec{x},t) = \sum_k\delta_k
\exp (i\vec{k}\vec{x})$ we obtain
\begin{eqnarray}
\ddot\delta_k + 2 H\dot\delta_k + \omega_k^2\delta_k & = & 0\nonumber\\
\left({k v_s\over a}\right)^2 - 4\pi G\rho & = & \omega_k^{2},
\label{eq:l29}
\end{eqnarray}
where $k\equiv |\vec{k}|=2\pi a/\lambda$ is the comoving wavenumber and
$v_s=\sqrt{\d P/d\rho}$ is the speed of sound.

Equation (\ref{eq:l29}) is effectively a
damped oscillator equation, the
damping being caused  by the expansion of the Universe.
The condition $\omega_k^2 = 0$ defines the so-called {\em Jeans wavelength}:
\begin{equation}
\lambda_J = v_s\left({\pi\over G\rho}\right)^{\half}.
\label{eq:l30}
\end{equation}
The mass contained within a sphere
with diameter $\lambda_J$ is the associated Jeans mass.
Perturbations on scales $\lambda > \lambda_J$ remain unaffected by pressure
forces and continue to grow in accordance with the relations discussed in
\S 2.3.
On the other hand, small scale perturbations feel
the effects of pressure gradients and oscillate as acoustic waves with
a steadily decreasing amplitude.

Prior to recombination, Thomson scattering couples photons and electrons
together, as a result of which the two components matter and radiation behave
effectively as a single fluid in which the adiabatic speed of sound is
\cite{e90}
\begin{equation}
v_s = {c\over \sqrt 3}\left({3\rho_m\over 4\rho_\gamma} + 1\right)^{-1/2}.
\label{eq:l31}
\end{equation}
Substituting $\rho_r = 3/(32\pi Gt^2)$ and $v_s \simeq c/\sqrt 3$
in eq. (\ref{eq:l30}),  we obtain $\lambda_J = \alpha ct$ where
$\alpha = 4\sqrt{2}\pi/3$. As a result we find that the Jeans length
scales with the cosmological horizon during the radiation dominated epoch.
The associated Jeans mass at recombination has the value
\begin{equation}
M_J \simeq 9\times 10^{16}(\Omega h^2)^{-2} M_{\odot}
\label{eq:l32}
\end{equation}
which is similar to the estimated masses of superclusters
of galaxies.

After recombination is complete photons cease coupling to electrons (which
are no longer free)
so that the pressure support is now provided by neutral
hydrogen atoms instead of by radiation.
As a result the speed of sound drops abruptly to
$v_s \simeq (5k_BT/3m_p)^{1/2}$ ($m_p$ being the proton mass
and $k_B$ is Boltzmann's
constant),
the corresponding value of the Jeans
mass falls by over ten orders of magnitude to \cite{e90}
\begin{equation}
M_J \simeq 1.3\times 10^5 (\Omega h^2)^{-\half} M_\odot
\label{eq:l33}
\end{equation}
which is similar to the mass of  a globular cluster.

On scales smaller than the Jeans scale dissipative processes define yet another
length scale below which the amplitude of density perturbations is
exponentially damped. This scale (set by the so-called {\em  Silk length}
$\lambda_s$) arises because although Thomson scattering couples
photons and baryons together before recombination, this coupling weakens as
recombination is approached (due to the increasing mean free path of the
photon) leading photons to leak out from overdense regions and into
underdense regions
carrying the still tightly coupled electrons with them.
This leads to an exponential
damping of inhomogeneities in the photon-baryonic plasma on scales
$\lambda < \lambda_s$, where
$\lambda_s$ is the Silk length at recombination, the associated Silk mass is
\cite{e90}
\begin{equation}
M_s \simeq 1.3\times 10^{12}(\Omega h^2)^{-{3/2}} M_\odot
\label{eq:l35}
\end{equation}

\subsection{Statistics of Primordial Density Fluctuations}
So far we have talked only about the evolution of a single
Fourier mode of the density field $\delta(\vec x, t) = D_+(t)\delta(\vec{x})$.
In the
general case, however, the density field will consist of a stochastic
superposition of such modes with different amplitudes.
Writing the Fourier transform of $\delta(\vec{x})$ as
\begin{equation}
\hat{\delta}(\vec{k}) = \frac{1}{(2\pi)^{3}} \int \d^{3} \vec{x}e^{-
i\vec{k}\cdot\vec{x}} \delta(\vec{x}),
\label{eq:lad1}
\end{equation}
which has inverse
\begin{equation}
\delta(\vec{x}) = \int\d^{3} k e^{i\vec{k}\cdot\vec{x}} \hat{\delta}
(\vec{k}),
\label{eq:lad2}
\end{equation}
it is useful to specify the properties of $\delta$ in terms of
$\hat{\delta}$. Recall the Dirac  delta--function:
\begin{equation}
\delta^{D}(\vec{k}) = \frac{1}{(2\pi)^{3}} \int \d^{3}\vec{x}
e^{\pm i\vec{k}\cdot\vec{x}}.
\label{eq:lad3}
\end{equation}
In terms of this function, we can define the {\em power--spectrum}
of the field to be the autocovariance of $\hat{\delta}(\vec{k})$:
\begin{equation}
\langle \hat{\delta}(\vec{k}_1) \hat{\delta}(\vec{k}_2) \rangle
= P(k_1) \delta^{D} (\vec{k}_1+\vec{k}_2).
\label{eq:lad4}
\end{equation}
The presence of the delta--function in this definition takes
account of the symmetry and reality requirements for $P(k)$
\cite{b92}. The analogous quantity in real space is
called the two--point correlation function or, more correctly,
the autocovariance function:
\begin{equation}
\langle \delta(\vec{x}_1) \delta(\vec{x}_2) \rangle
=\xi (|\vec{x}_1-\vec{x}_2|) = \xi(r),
\label{eq:lad5}
\end{equation}
which is related to the power spectrum by the Wiener--Khintchin
relations which are discussed in \S 8.4. Both the power spectrum
and the covariance function of the present--day density field
can be estimated by calculating analogous quantities for
discrete galaxy counts, as described in \S 8.5.

The power spectrum or, equivalently, the autocovariance
function, of the density field is a particularly important
statistic because it provides a complete statistical
characterisation of the density field for a particular
kind of stochastic field: a {\em Gaussian Random Field}.
This class of field is the generic prediction of inflationary
models where the density fluctuations are generated
by quantum fluctuations in a scalar field during the
inflationary phase\footnote{Gaussian random processes also
occur generically in linear physics due to the
central limit theorem.}. We shall assume for virtually
all of this paper that the primordial $\delta(\vec{x})$
is a Gaussian random field. Formally, this means that
the Fourier modes have a Rayleigh distribution of amplitudes
and the phases are independent and uniformly random on
$[0,2\pi]$. Writing
\begin{equation}
\hat{\delta}(\vec{k})=|\delta_k|\exp(i\phi_k)
\end{equation}
gives
\begin{equation}
dP(|\delta_k|)= \frac{1}{\sigma^{2}} \exp\left( -\frac{|\delta_k|^{2}}
{2\sigma^{2}} \right) |\delta_k|\d|\delta_k|,
\end{equation}
where $\sigma^{2}\propto P(k)$. All the joint probability
distributions of $\delta(\vec{x}_1)\ldots\delta(\vec{x}_n)$
are $n$--variate Gaussian distributions for this type of
field and this property makes many properties
of the spatial distribution of density fluctuations
calculable analytically \cite{d70,a81,k84,ph85,bbks86}.

Many other physically--interesting properties can be
related to the power spectrum. For example, consider
the mean square density fluctuation\footnote{Both the spatial average and the
ensamble average are equivalent for a Gaussian random field, which is said to
be
ergodic.}
\begin{equation}
\langle \left(\frac{\delta\rho}{\rho}\right)^{2}\rangle
= \langle \delta(\vec{x})\delta(\vec{x})\rangle =
4\pi\int_{0}^{\infty} P(k) k^{2} \d k.
\label{eq:lad6}
\end{equation}
This definition of the mean--square fluctuation takes into
account contributions on all scales and may, depending on the
form of $P(k)$, actually diverge. A more physically realistic
quantity to use is the mean square fluctuation smoothed
on a given scale $R$, which can be shown to be
\begin{equation}
\sigma^{2}(R) = \int\d^{3}\vec{k} W^{2} (kR) P(k),
\label{eq:lad7}
\end{equation}
where $W(x)$ is the Fourier transform of the
filter function used to smooth the density field in real space.
for example, a spherical top--hat function in real space
gives
\begin{equation}
W(x) = \frac{3}{x^{3}} (\sin x-x\cos x);
\label{eq:lad8}
\end{equation}
whereas a Gaussian real--space filter function leads to
a Gaussian Fourier domain function (which is a useful
property).\footnote{Useful
mathematical properties of the top--hat window function are discussed in
\cite{b94b}.}
In the linear regime, the density field retains its initial
Gaussian character, so that
\begin{equation}
P(\delta_R)= \frac{1}{\sigma_R(t)\sqrt{2\pi}}
\exp \left[ - \frac{\delta_R^{2}}{2\sigma_R^{2}(t)}\right].
\label{eq:ladz1}
\end{equation}
For a Gaussian distribution all moments of order $n>2$
are either zero (odd) or can be expressed in terms
of $\sigma^{2}$ (even). The cumulants, or
connected moments are zero for $n>2$ for such a distribution,
\S 8.3.

Furthermore, assuming the linearised continuity equation and
Poisson equations are valid, one can calculate
such quantities as the mean
square velocity smoothed on a scale $R$ and the mean
square gravitational potential fluctuation. It is useful to
express these, and the mass fluctuation, in terms of the
contribution from a logarithmic interval in $k$-space:
\begin{eqnarray}
\frac{\d\sigma^{2}}{\d \log k} & = &
4\pi k^{3} P(k);\nonumber\\
\frac{\d \sigma_v^{2}}{\d \log k} & = &
4\pi(aHf)^{2} k P(k);
\nonumber\\
\frac{\d \sigma_{\phi}^{2}}{\d \log k} & = &
4\pi \left( \frac{3}{2} \Omega_0 H^{2} a^{2}\right)^{2} \frac{P(k)}{k}.
\label{eq:lad9}
\end{eqnarray}
We discuss the use of velocity statistics in \S 8.10.

The general picture that emerges is that, while the amplitude
of each Fourier mode remains small, i.e. $\delta(\vec{k})\ll 1$,
linear theory applies. In this regime, each Fourier mode
evolves independently and the power--spectrum therefore
just scales as $P(k,t)=P(k,t_1)D_{+}^{2}(k,t)/D_{+}^2(k,t_1)$.
Even if $\delta>1$ for large $k$, one can still
apply linear theory on large scales by applying
a filter as discussed above to remove the non--linear scales.
Generally, it is assumed that a scale evolves linearly as
long as $\sigma^{2}(R)<1$, though as we shall see this is not
always correct. For scales larger than the Jeans length, this
means that the shape of the power spectrum is preserved
during linear evolution. Only when scales go non--linear
does the shape of the spectrum begin to change. Of course,
scales below the Jeans length do not grow relative to
those above and so the shape of the spectrum is distorted
even in the linear phase. In the popular models with
non--baryonic dark matter, and models which have significant
Silk damping there are other effects suppressing the
growth of fluctuations in the linear regime.

To complete the picture we need to specify the shape of
some initial fluctuation spectrum which exists at some
arbitrary time before it can be distorted by damping processes.
The inflationary scenario often produces a power--law form
\begin{equation}
P(k)=Ak^{n},
\label{eq:lad10}
\end{equation}
with many models having the Harrison--Zel'dovich form
$n=1$. Even if inflation is not the origin of density
fluctuations, the form (\ref{eq:lad10}) is a useful
phenomenological model for the fluctuation spectrum.
Given the current plethora of Inflationary Universe models, the value of the
primordial amplitude $A$ is usually determined empirically \cite{kt90}.

First attempts to normalise $P(k)$ used counts
of galaxies. Because one wants to use linear theory to
normalise the spectrum -- and thus avoid any physical
process which distorts the spectrum's shape, rather than
its amplitude -- it is necessary to measure fluctuations
on rather large scales. A useful way to do this is
using $J_3$, which is defined by
\begin{equation}
J_3(R) \equiv \int_{0}^{R} \xi(r) r^{2} dr
=\frac{R^{3}}{3} \int \d^{3}\vec{k} W^{2}(kR) P(k)
\end{equation}
where $W(x)$ is defined by eq. (\ref{eq:lad8}). Alternatively,
one can measure the variance of galaxy counts--in--cells
on a scale $R$, using eq. (\ref{eq:lad7}). These methods
are discussed in \S 8.5. The problem with such attempts
to normalise the spectrum are two--fold. First, there is
the problem that redshift surveys probe only a comparatively
small volume of space and we need large scales for
linear theory to be reliable. Secondly, and more
fundamental, is the problem that we do not know how to
relate galaxy counts and mass fluctuations in an
unambiguous way. This ignorance is usually parametrised
by the phenomenological relationship
\begin{equation}
\left(\frac{\delta N}{N}\right) = b\left(\frac{\delta \rho}{\rho}\right),
\label{eq:lad12}
\end{equation}
but the bias factor\footnote{Although
we shall deal extensively with the origin of structure in this
paper, we shall not deal with the even harder problem of how
individual galaxies form. Quite apart from the need to accumulate
enough mass to form a galactic halo, one also needs to understand
star formation, hydrodynamical processes during the protogalactic
collapse and the possible role of environmental feedback processes
on the forming object. All this physics is encapsulated in the
bias parameter $b$. That an attempt should be made to embody
such complicated effects in a single parameter is ample
evidence that we are not close to a theory of galaxy formation.}
$b$ may well be
scale--dependent\cite{k84,dr87,k91,c93,bcfw93}.

The fact that the velocity flow induced by density perturbations is sensitive
to the total mass fluctuation, not just the part that is
in galaxies, suggests that one should use statistics of the
velocity field to normalise the spectrum. This idea
is discussed in \S 8.1.5 \& \S 8.10. The problem with this is that our
knowledge of the detailed statistics of the velocity field
is rather poor.

As we shall see in \S 2.7, the COBE satellite's detection of
CMBR anisotropy has allowed us to circumvent these difficulties
and obtain a relatively model--independent normalisation of the
power  spectrum appropriate for models of structure formation.
Before we discuss this, however, we need to look in more
detail at models of structure formation involving non--baryonic
dark matter.

\subsection{Non--Baryonic Dark Matter}
We would  now like to describe
currently popular cosmological models
in which the matter content of the Universe is
usually considered to be
in the form of two non-interacting fluids. The first
is presumed to be dissipationless and nonbaryonic
whereas the second is made up of
baryons (WIMPs such as massive neutrinos and SUSY particles, or axions,
are  possible
candidates for nonbaryonic
matter\footnote{WIMP means Weakly Interacting Massive
Particle.},
whereas the baryonic material is presumed   to be
composed predominantly of neutral hydrogen and helium at early times).
The need to incorporate  non--baryonic forms of matter into the standard
big bang model arises because of the necessity to generate large enough
perturbations to account for the presence of galaxies,
from sufficiently small initial values of the density contrast
$\delta\rho/\rho$,
without violating the CMBR constraints on small, i.e. arc minute angular
scales. This is a difficult task in baryonic models in which perturbation
growth takes place only after the cosmological recombination of hydrogen,
which occurs at redshifts $\sim 1100$.
Before recombination, radiation pressure caused by Thomson scattering
effectively prevents the growth of all perturbations having wavelenths
smaller than $\sim 180 h^{-1} $Mpc, the horizon size at recombination.
After recombination, density perturbations grow linearly with the scale factor
of the Universe, $a(t)$, until the Universe
becomes dominated by curvature or by a cosmological constant,
after which time perturbation growth is strongly
suppressed (if $\Omega_0<1$).

The maximum permitted growth factor for perturbations is therefore
$\delta/\delta_{rec} \leq a_0/ a_{rec} \simeq 1100$
(for $\Omega \le 1$ ).
Consequently, the requirement that $\delta  > 1$ today -- which is a
necessary (but not sufficient) condition for
the
formation of gravitationally bound systems such as galaxies -- leads to the
primordial amplitude: $\delta \approx 10^{-3}$ at recombination.
If the primordial perturbations are adiabatic
(i.e. fluctuations for which the
entropy per baryon is spatially constant), then at recombination
\begin{equation}
\delta_B =\frac{3}{4} \delta_{\gamma} = 3\frac{\delta T}{T}
\label{eq:l36}
\end{equation}
so that a fluctuation amplitude $\sim 10^{-3}$ in the baryon component would
invariably result in an anisotropy $\sim 10^{-4}$ in the CMBR temparature on
arc minute scales which has not been observed. On the other hand, perturbations
in a nonbaryonic component, can begin growing soon after matter--radiation
equality has been achieved (i.e. by $z \sim 10^4$ in the case of an $\Omega
= 1$ Universe)
with the result that large final perturbations can develop
from small initial values without violating the CMBR constraints on small
scales \cite{p82b,sdz83,sv90}.

The existence of non--baryonic matter is also strongly
suggested by the observed
density of matter in the Universe parametrised by
$\Omega_0$. Although the value of $\Omega_0$ is not precisely
known, most observations place it in the range bounded
by $0.1 \le \Omega \le 1$ with some indications that $\Omega$ could be
as large as unity, in agreement with predictions made by the Inflationary
Universe scenario \cite{k91,nd93,dbysdh93,ce94,dek94}.
A value of
$\Omega$ greater than $\sim 0.1$ would require the presence of some form
of non--baryonic matter since primordial nucleosynthesis imposes rather
stringent constraints on the contribution to $\Omega_0$ from baryons:
$\Omega_b \simeq 0.016 h^{-2}$ \cite{osw91}.

Gravitational instability in a two component medium
with $\Omega=1$ consisting of baryons
and dark matter is governed by the
following coupled system of equations (again to linear order):
\begin{eqnarray}
\ddot{\delta}_1 + 2\left(\frac{\dot{a}}{a}\right)
\dot{\delta}_1 + \left(\frac{kv_1}{a}\right)^{2}\delta_1
 & = &
 4\pi G(\rho_1\delta_1 + \rho_2\delta_2)\nonumber\\
\ddot{\delta}_2
+2 \left(\frac{\dot{a}}{a}\right) \dot{\delta_2}
+ \left(\frac{kv_2}{a}\right)^{2}\delta_2
& = & 4\pi G
(\rho_1\delta_1 + \rho_2\delta_2)\nonumber\\
\left({\dot{a}\over a}\right)^2 = {8\pi G\over 3}(\rho_1 + \rho_2).
\label{eq:l37}
\end{eqnarray}
If either $k = 2\pi a/\lambda = 0$, or $v_1 = v_2 = 0$ then equations
(\ref{eq:l37}) have the solutions \cite{w81}
\begin{eqnarray}
\delta_1 & = & B_1t^{2\over 3} + B_2 t^{-1} + B_3 + B_4 t^{-1/3}
\nonumber\\
\delta_2  & = & B_1t^{2\over 3} + B_2 t^{-1} + B_5 + B_6 t^{-1/3}
\nonumber\\
0 & = & \Omega_1B_3 + \Omega_2B_5 = \Omega_1B_4 + \Omega_2B_6.
\label{eq:l38}
\end{eqnarray}
It is interesting that, of the four linearly independent components in equation
(\ref{eq:l38}), only one corresponds to a growing mode.
An important subclass of equations (\ref{eq:l37}) \& (\ref{eq:l38}) arises if
$\rho_1\delta_1 >> \rho_2\delta_2$. This condition is satisfied in most
cosmological models since
the perturbation in the dark matter component
immediately after recombination is much larger than the corresponding baryonic
perturbation ($\rho_1\delta_1 \simeq 10 \rho_2\delta_2$),
consequently we can ignore the presence of $\rho_2\delta_2$ terms in the
right hand side of eqs. (\ref{eq:l37})
which leads to:
$\delta_1 \propto a(t), \delta_2 \propto \delta_1 (1 - \mbox{const}/a)$,
 indicating that perturbations in baryons rapidly fall into
potential wells created by dark matter, with the result
that the growth rate of both types of perturbations converges at late times.
The behavior of perturbations in both baryonic and nonbaryonic matter after
matter--radiation equality is illustrated in Figure (\ref{fig:2dm}).
\begin{figure}
\vspace{10cm}
\caption{The growth of perturbations in baryons (solid line)
and non-baryonic dark matter (dashed line -- `x')
is plotted against the cosmological redshift parameter $z$
for different values of $\Omega_b$ and $\Omega_x$.
Adapted, with permission, from \protect{\cite{sdz83}}.}
\label{fig:2dm}
\end{figure}

For other physically interesting solutions to the system of equations
(\ref{eq:l37}), see
\cite{gz81,f83,f85b,hb84,s80,sn85,sstar85}. For a general equation of state,
solutions to these equations can be expressed
in terms of the Meijer $G$--function
and  can therefore be studied by analysing the well documented
properties of this function \cite{be53}.
A relativistic generalisation of equations
(\ref{eq:l37}) can also be obtained\cite{sstar85,ss84,s84}.

Non--baryonic matter does not, in general, couple to
photons, so that perturbations
in it are not subject to collisional (Silk) damping. However depending upon
the momentum of the non--baryonic particle, collisionless phase mixing
caused primarily by the free streaming of relativistic particles as these
travel from regions of high density into regions of
low density (and {\em vice versa})
leads to a very effective mechanism whereby
fluctuations on scales smaller than the
free-streaming distance $\lambda_{fs} $ are wiped out;
$\lambda_{fs} $ is the mean distance traversed by a relativistic particle
species until its momentum becomes non-relativistic.
The cosmological properties of non-baryonic matter depend
mainly upon whether $\lambda_{fs} $ is large or small.
Particles for which the
free-streaming distance $\lambda_{fs} $ is large are said to constitute
{\em Hot Dark Matter} (HDM) (to underscore the relativistic nature of the
particle species). In the opposite case when $\lambda_{fs} $ is too small to
be cosmologically interesting, the particles are said to constitute
{\em Cold Dark Matter} (CDM) \cite{bonds83}.

For both HDM and CDM the shape of the processed {\em  final} spectrum
of fluctuations differs
from its primordial shape. In the case of HDM this arises because
fluctuations on scales smaller than $\lambda_{fs}$ are wiped out due to
free streaming so that the processed final spectrum has a well--defined
cutoff on scales smaller than $\lambda \sim \lambda_{fs}$. For HDM models
based on a light massive neutrino $\lambda_{fs}
\simeq 41 (30\mbox{ eV}/ m_\nu)$
Mpc, the associated mass scale for neutrinos of mass $m_\nu \simeq 30$ eV
is $M_{fs} \sim 10^{15}$($\Omega h^2)^{-2} M_\odot$ which is similar to
the observed mass of a rich cluster of galaxies \cite{bonds83,sdz83}.
Consequently, the first objects to collapse in such a model have
the mass of a cluster, smaller objects forming
later by the fragmentation of
primordial cluster-scale ``pancakes''. This scenario is commonly called the
``top-down'' scenario for galaxy formation and was originally suggested by
Zeldovich and co-workers in the 1970s in
connection with baryonic models of galaxy formation \cite{sdz83}.

If either the mass of the nonbaryonic particle is very large
(as in the case of SUSY particles such as the  photino and the gravitino),
or if its momentum is very small due to it being formed out of thermal
equilibrium (e.g. the axion), then the free streaming
distance $\lambda_{fs}$ is too small to be relevant and the processed
spectrum has no cutoff on any cosmological scale. However, even in such cases,
the final spectrum does show an appreciable departure from
its primordial form $P_0(k)$ on scales smaller than
$\lambda_{eq}$ (= $13/\Omega h^2$ Mpc) ($\lambda_{eq}$ is related to
the comoving horizon scale at
matter--radiation equality) approaching the asymptotic form
$P(k) \propto P_0(k)\times k^{-4}\log ^2 k$ for $k >> k_{eq}$
\cite{ss84,pb84}. This bending of the CDM spectrum occurs on scales
$\sim \lambda_{eq}$
and arises because small scale density fluctuations
(having wavelength $\lambda < \lambda_H < \lambda_{eq}$)
which enter the horizon prior to matter--radiation equality have
nothing to gravitate towards
(the free-streaming of photons having effectively
wiped out all fluctuations in the photon component on scales $< \lambda_{eq}$).
As a result their growth rate slows down considerably and approaches
the asymptotic form
$\delta \propto 1 + 3a(t)/2a_{eq}$ until matter dominance
$(a = a_{eq})$ \cite{m74,pb84}.
Perturbations on scales greater than the cosmological
horizon ($\lambda_H
< \lambda < \lambda_{eq}$), however, continue to grow at the usual rate
$\delta \propto a^2$, until they enter the horizon, resulting in a larger
amplitude for longer wavelength perturbations and, consequently, a bend in the
shape of $P(k)$. Very large wavelength perturbations
($\lambda > \lambda_{eq})$
are left virtually unscathed by the radiation dominated epoch, since they
enter the horizon after matter--radiation equality. As a result the
fluctuation spectrum preserves its primordial form on scales $ > \lambda_{eq}$
{}.

The combined effect of the various processes involved in changing the shape
of the original power spectrum can be summarised in a single quantity --
the transfer function $T(k)$ which relates the processed power spectrum $P(k)$
to its primordial form $P_0(k)$ via the transformation:
$P(k) = P_0(k)\times T^2(k)$. Inflationary models of the Universe
generically predict $P_0(k) \propto k^n$, $n = 1$ being the scale invariant
Harrison - Zeldovich spectrum \cite{h70,z72}. The results of full
numerical calculations of all the physical processes we have
discussed can be encoded in the transfer function of a particular
model. The main examples are listed here.

For {\em Cold Dark Matter}
\begin{equation}
T(k) = \left[1 + {(Ak)^2\over \log (1 + Bk)}\right]^{-1},
\label{eq:l39a}
\end{equation}
where $A = 3.08 \sqrt\kappa h^{-2}$ Mpc,
$B = 1.83 \sqrt\kappa h^{-2}$ Mpc, and
$\kappa = \Omega_{rel}/\Omega_{\gamma}$
is the ratio of the energy density in relativistic
particles to that in photons.
The value of $\kappa = 1.68$ corresponds to three
neutrino flavours plus photons contributing
to the total density in relativistic particles \cite{s84,ss84}.
An alternative fitting--formula is:
\begin{equation}
T(k) = {\log (1 + 2.34 q)\over (2.34 q)}\left[
1 + 3.89 q + (16.1 q)^2 + 5.46 q)^3 + (6.71 q)^4\right]^{- {1\over 4}},
\label{eq:l39b}
\end{equation}
where $q = k\theta^\half/(\Omega h^2)$
and $\theta = \kappa/1.68$ \cite{bbks86}.
In addition the following transfer function was given by Bond and Efstathiou
\cite{be84}
\begin{equation}
T(k) = \big\lbrack1 + (ak + (bk)^{3\over 2} +
(ck)^2)^\nu\big\rbrack^{- {1\over \nu}},
\label{eq:l39c}
\end{equation}
where $a = 6.4 (\Omega h^2)^{-1}$ Mpc, $b = 3.0 (\Omega h^2)^{-1}$ Mpc,
$c = 1.7 (\Omega h^2)^{-1}$ Mpc,  $\nu = 1.13$.
Approximations (\ref{eq:l39a},\ref{eq:l39b}) correctly reproduce the
small wavelength behaviour of the transfer function
$T(k) \propto k^{-2}\log k$ for $k >> k_{eq}$.
A comparison of different transfer functions is given in
\cite{ll93b}; see also \cite{p82,bfpr84,ks84,ks87,h89}.

For {\em Hot Dark Matter} we have
\begin{equation}
T_{HDM} \simeq 10^{- ({k/ k_\nu})^{1.5}},
\label{eq:l40a}
\end{equation}
by solving the Boltzmann equation for the collisionless
neutrinos. In eq. (\ref{eq:l40a}),
$\lambda_{\nu} = 2\pi/k_\nu = 13 (\Omega_\nu h^2)^{-1}$ Mpc
= $41 ~(m_\nu/30 \mbox{ eV})$ Mpc
is the damping length in neutrino fluctuations caused by free streaming.
For a single flavour neutrino having mass
$m_\nu$ and temperature $T_\nu = 1.9^\circ K$,
$\Omega_\nu h^2 = 0.31 (m_\nu/30 \mbox{ eV}.)$.
Several alternative fits to the HDM transfer function are given in Appendix G
of \cite{bbks86}, including:
\begin{equation}
T_{HDM} \simeq \exp[- (3.9 q + 2.1 q^2)],
\label{eq:l40b}
\end{equation}
where $q = k/(\Omega_\nu h^2$ Mpc$^{-1})$.

An alternative scenario is furnished if the primordial
density fluctuations are in the isocurvature mode, rather than
adiabatic. Isocurvature initial conditions are perturbations
to the equation of state, rather than true curvature perturbations:
the total energy density is constant, but the relative contribution
from each fluid component varies spatially; such perturbations do not
arise generically (as do adiabatic
initial conditions, which arise as fluctuations in the inflaton field
and are therefore predicted in most Inflationary Universe models)
but can be generated via specific mechanisms such as by an axionic field
in the early Universe and also in a certain category of Inflationary models
\cite{kt90}.

The evolution of isocurvature perturbations before and after matter--radiation
equality has been discussed
in the framework of the two fluid CDM model.
Starobinsky \& Sahni \cite{ss84,s84} give the following form for
the isocurvature transfer
function:
\begin{equation}
T(k) = (1 + Ak)^{-2},
\label{eq:l41a}
\end{equation}
where $A = 4.36\sqrt \kappa h^{-2}$ Mpc.
An alternative transfer function is given by Efstathiou \& Bond \cite{eb86}
\begin{equation}
T(k) = \big\lbrack1 + (ak + (bk)^{3\over 2} +
(ck)^2)^\nu\big\rbrack^{- {1\over\nu}},
\label{eq:l41b}
\end{equation}
where $a =  15.6 (\Omega h^2)^{-1}$ Mpc, $b = 0.9 (\Omega h^2)^{-1}$ Mpc,
$c = 5.8 (\Omega h^2)^{-1}$ Mpc,  $\nu = 1.24$.
An essential feature of the isocurvature transfer function is the absence of
the logarithmic growth factor that characterises the small scale behaviour
of the adiabatic CDM spectrum, indicating that density perturbations on
a range of scales  ($k >> k_{eq}$) grow to become
non-linear virtually instantaneously. Somewhat surprisingly, however,
isocurvature fluctuations generate larger anisotropies in the CMBR than
adiabatic fluctuations, for two reasons.
Firstly, initially isocurvature fluctuations
generate significant fluctuations in the
 gravitational potential when they enter the horizon; this
is due to the influence of pressure gradients. In addition,
isocurvature fluctuations  generate significant fluctuations
in the radiation density after
matter--domination, because the initial  perturbation in the equation
of state, which can be thought of as a perturbation in the entropy, is then
transferred into the  perturbation of the
radiation. The upshot of all this is that the net anisotropy seen is a factor
six larger for isocurvature fluctuations than for adiabatic
ones, making such fluctuations uninteresting from a cosmological
viewpoint \cite{ss84,eb86}, though
they have been discussed in the literature \cite{ll93b,ks87,e90}.

\subsection{Normalising the Power--Spectrum}
We have already mentioned that the power spectrum could in
principle be normalised using the observed clustering properties of galaxies.
In practice, such a normalisation runs into trouble
because of the absence of a
strict relationship between the clustering properties of dark and luminous
matter. Recent observations of the large scale anisotropy of the CMBR by the
COBE satellite remove this ambiguity by relating temperature fluctuations
$\delta T/T$ directly to the amplitude and shape of the underlying power
spectrum on scales that are still evolving according to linear theory
\cite{sea92}.
Furthermore, having normalised the power spectrum in such a manner,
we can
proceed to determine such quantities as
$\delta M/M(r)$ which measure the clustering of dark matter
and can therefore help predict the value of the
biasing factor $b = (\delta N/N)/(\delta M/M)$ for a given cosmological model.

In our discussion of the CMBR anisotropy we shall focus mainly on scales
greater than a few degrees, since the primordial
anisotropy on such scales has been unambiguously
measured. Observations on smaller scales have been used to constrain
the shape of the power spectrum on these scales and also to provide limits
on the gravity wave contribution \cite{cbdes93,bcdes94,bond94,wss94}.
The CMBR travels to us from the surface of last scattering whose comoving
distance from us practically coincides with the horizon scale
$d_H = 2H_0^{-1} \simeq 6000 h^{-1}$ Mpc.
The horizon at last scattering subtends an angle
$\theta \simeq 1.8^{\circ}
\Omega_0^{1/2} (1000/z_{rec})^{1/2} \simeq 1.8^{\circ}$
for $\Omega_0 \simeq 1$ and $z_{rec} \simeq 1000$.
As a result the CMBR anisotropy on angular scales greater than a few degrees
probes
length scales that were causally unconnected at the time of recombination,
thereby providing us with the cleanest possible probe of the primordial
spectrum prior to its distortion by astrophysical processes.

The dominant contribution to the CMBR anisotropy on large angular scales
($> 1^{\circ}$) arises because of the Sachs-Wolfe effect which
relates temperature fluctuations in the CMBR
to the integral over the variation of the metric
evaluated along the line of sight \cite{sw67}
\begin{equation}
{\delta T\over T} = - \frac{1}
{2}\int_{\eta_{rec}}^{\eta_0}\frac{\partial h_{\alpha\beta}}
{\partial\eta}e^{\alpha}e^{\beta}d\eta,
\label{eq:l55a}
\end{equation}
where $e^{\alpha}$ is the tangent vector of the photon
trajectory, $\eta = \int dt/a$
is the conformal time coordinate ($\eta_0$ corresponds to the present epoch)
and $\eta_{rec}$ to recombination;
an analysis with three massless neutrino species and $z_{rec} = 1100$
gives \cite{mss95} $\eta_0/\eta_{rec} = 49.6$.
In a flat, matter--dominated Universe, the gravitational potential does not
evolve with time and the above expression simplifies to
\begin{equation}
{\delta T\over T} \simeq {1\over 3}{\delta \phi\over c^2}.
\label{eq:l55}
\end{equation}
In other words, the CMBR anisotropy is directly related to fluctuations in the
gravitational potential on the surface of last scattering.

Perturbations in the primordial gravitational potential
(usually accompanied by gravitational waves) arise naturally
in models of the Universe based on the inflationary scenario which
usually predict
a primordial fluctuation spectrum $P_0(k) = Ak^n$ . The scale invariant
spectrum $n = 1$ is predicted by inflationary models with exponential
expansion,
models with power law inflation (PLI) in which the scale factor grows as a
power law $a(t) \propto t^p, ~p > 1$ predict a lower spectral index:
$n = (p - 3)/(p - 1) < 1.$

It is convenient to express the cosmic microwave temperature distribution as
\begin{equation}
T(\theta, \phi) = T_0\left[1 + \frac{\delta T}{T}
 (\theta, \phi) \right],
\label{eq:l56}
\end{equation}
where $T_0$ is the blackbody temperature $T_0 = 2.736 \pm 0.010^\circ K$
\cite{math90,math94}.
We can rewrite $\delta T/T$ as a multipole expansion
on the celestial sphere:
\begin{equation}
\frac{\delta T}{T} (\theta, \phi) =
\sum_{l=2}^\infty~\sum_{m=-l}^l a_{lm} Y_l^m (\theta, \phi),
\label{eq:l57}
\end{equation}
where $\theta,\phi$ are spherical angular cordinates associated with a
direction $\hat{n}$ in the sky, $Y_l^m(\theta,\phi)$ are spherical harmonics.
We have ignored the dipole contribution to $\delta T$ which
cannot be distinguished
from our local velocity\footnote{
The dipole anisotropy is usually interpreted as being due to the motion of
our local group of galaxies relative to the CMBR. For $v/c << 1$, $T(\theta)
\simeq T_0(1 + (v/c)\cos\theta)$, $\theta$ being the angle at which the
microwave temperature is being measured relative to the motion of the
observer. Observations indicate $\delta T/T\vert_{dipole}
\simeq 1.2\times 10^{-3}$
from which, after correcting for the motion of the
Solar system within the Milky
Way,  we can infer the value $V_{LG} \simeq 610 \pm 50$km/s for the
velocity of the mass center of the local group with respect to the CMBR.
The direction of $\vec V_{LG}$ is towards $\alpha = 11^h$,
$\delta = - 25^\circ$, which
is some $45^\circ$ away from the nearest large mass concentration --
the Virgo cluster, indicating that the Virgo cluster contributes only
partially to our overall peculiar velocity, and implying thereby that more
distant (and more massive) mass concentrations must exert a significant
influence on the motion of the local group.} with respect to the CMBR.
The Sachs-Wolfe effect allows us to relate the multipoles $a_{lm}$ to the
primordial spectrum
\begin{equation}
C_l^{(S)} \equiv \langle|a_{lm}|^2\rangle =
4\pi^{2} \left({H_0\over c}\right)^4\int_0^\infty
{dk\over k^2}P(k) j_l^2(kx),
\label{eq:l58}
\end{equation}
where $x = 2c/H_0$ is the present day
Horizon size and $j_l(kx)$ are spherical
Bessel functions.
For power law spectra $P(k) = Ak^n$ the
integral in eq. (\ref{eq:l58}) can be evaluated
analytically, with the result
\begin{equation}
C_l^{(S)} = {A\over 16} \left({H_0\over c}\right)^{n+3} f(n,l);
{}~~~ f(n,l) = {\Gamma(3-n)\over
\left[\Gamma ({4-n\over2})\right]^2}
{\Gamma[(2l + n - 1)/2]\over \Gamma[(2l + 5 - n)/2]}.
\label{eq:l59}
\end{equation}
Of considerable importance is the temperature auto-correlation function
defined by
\begin{equation}
C(\theta) = \langle \frac{\delta T}{T}
(\hat{n}_1)\frac{\delta T}{T} (\hat{n}_2)\rangle,
\label{eq:l60}
\end{equation}
where $\hat{n}_1\cdot\hat{n}_2 = \cos\theta$.
Substituting $\delta T/T$ from eq. (\ref{eq:l57})
and using the addition theorem for spherical harmonics we get
\begin{equation}
C(\theta) = {1\over 4\pi}\sum_{l\ge 2} (2l + 1)C_l P_l(\cos\theta).
\label{eq:l61}
\end{equation}
Equation (\ref{eq:l59})
allows us to relate the higher multipole
moments directly to the
quadrupole (for $n < 3$)
\begin{equation}
C_l^{(S)} = C_2^{(S)} {\Gamma\left[l + ({n - 1\over 2})\right]\over
\Gamma\left[l + ({5 - n\over 2})\right]}
{\Gamma\big ({9 - n\over 2}\big )\over \Gamma\big ({3 + n\over 2}\big )}.
\label{eq:l62}
\end{equation}
The quantity $Q_S^2 = (5/4\pi) C_2^{(S)}F_2$
is sometimes called the quadrupole
anisotropy of the CMBR; $F_2$ incorporates the finite beam width of the
detector, $F_2 \simeq 0.99$ for the COBE-DMR. We use the
subscript ``S''
for scalar perturbations to distinguish it from the quadrupole anisotropy
generated by  gravitational waves.
Relating $C_2^{(S)}$ to $A$ through eq. (\ref{eq:l59})
and substituting \cite{wr94}
$Q_S = Q_{COBE-DMR} \simeq 6.46\times 10^{-6}$ we obtain
\begin{equation}
Q_S^2 = Q_{COBE-DMR}^2 = {5A\over 512\pi^4} f(n,2)\left({H_0\over c}
\right)^{n + 3} F_2
\label{eq:l63a}
\end{equation}
and thus the COBE--normalised amplitude
\begin{equation}
A = {8\over 5\pi^{2} f(n,2)} \left({c\over H_0}\right)^{n+3}
{Q_{COBE-DMR}^2\over F_2}
\label{eq:l63b}
\end{equation}
($n = 1.1 \pm 0.32$ is indicated by COBE data \cite{wss94}.)
This normalisation however ignores the presence of
gravitational waves which are
generically
produced in most inflationary scenarios and also contribute to the
anisotropy in the CMBR by the Sachs--Wolfe effect
\cite{rsv82,aw84,sy85}. If the contribution of these modes to the
quadrupole is $Q_{T}^{2}$ then the total quadrupole seen
by COBE will be
\begin{equation}
Q^2 = Q_{S}^2 + Q_{T}^2.
\label{eq:l71}
\end{equation}
For particular inflationary models one can calculate the relative
contributions of scalar and tensor modes
\cite{sy85,dhsst92,sk92,ss92b,ss92a,lmm92,lc92,ll92}.
For power--law inflation this gives
\begin{equation}
4\pi A = \Big\lbrack{72.5\over 16\pi}~{(n - 1)\over(n - 3)}
{}~g(n,2) +
{5\pi\over 32} ~f(n,2)\Big\rbrack^{-1} \Big({c\over H_0}\Big)^{n + 3}
\frac{Q^2_{COBE-DMR}}{F_2},
\label{eq:l72}
\end{equation}
where $g(n,2) \simeq \exp{0.6(n-1)}$, $f(n,2)$ is defined in
equation(\ref{eq:l59}).
Instead of normalising to the quadrupole we could, instead, normalise
to the variance at $10^\circ$, this would lead to a slight change in the
value of $A$ \cite{ll93b}.
Having normalised the primordial spectrum  (at least
for one particular inflationary model)
we can now determine
statistical indicators of clustering, such as $\delta M/M$ and $\xi(r)$,
for spatially--flat matter--dominated cosmological models such as the standard
CDM scenario
\cite{ll92,dhsst92,lc92,lmm92,ss92b,s92,sss94,ll93b}.
Extensions of the model such as mixed dark matter (CHDM)
or CDM plus a cosmological constant were considered in
\cite{kstar85,gsv92,ebw92,ll93a,pstar93,khpr93,kgb93,phkc94}.

%% file: ssec3.tex
\section{Local extensions of Linear Theory}
In this section we shall discuss the simplest class of non-linear
approximations, those which involve extrapolations of the
linear properties of the density field into the non--linear
regime. This kind of model typically does not make
any attempt to account for the dynamical evolution of clustering, but
in some situations can give interesting insights into the
properties of the non--linear mass distribution.

\subsection{The Spherical Model}
One of the simplest and best studied models of nonlinear gravitational
instability is the spherical model. In this model
we ignore the tidal effects of neighbouring density
perturbations upon the evolution of an isolated, homogeneous,
spherical density perturbation. To justify this we can appeal
to Birkhoff's theorem in General Relativity,
or Gauss's law in Newtonian Gravity.
Under these simplifying assumptions an exact analytical treatment
is possible \cite{t34,b47,pp67,gg72}.

In order to understand the dynamics of non--linear spherical collapse, consider
a spherical density perturbation expanding in the background of a homogeneous
and isotropic background Universe. The density of the fluctuation is
characterised by $\Omega'$ whereas that of the background Universe by $\Omega$
($\Omega ' > \Omega$ will correspond to an overdensity and
$\Omega ' < \Omega$ to an underdensity).
The subsequent fate of the spherical density perturbation will depend
crucially upon the value of $\Omega'$. For $\Omega' > 1$ the perturbation
will behave just like
a part of a closed FRW Universe and will therefore expand to a maximum radius,
turn around at a time $t_{ta}$, and thereafter collapse to a point at
$t_{coll} \simeq 2t_{ta}$. A spherical density perturbation with $\Omega ' < 1$
on the other hand, will mimic an open Universe and never recollapse
(if $\Omega ' < \Omega$ then such an underdensity will correspond to a void).
In an idealised cosmological scenario spherical overdensities might be
thought of as progenitors of clusters of galaxies \cite{k84},
whereas underdensities would correspond
to voids.

In order to treat the collapse of a spherical overdensity quantitatively let us
consider a spherical shell of radius $R$ with an initial overdensity
$\delta_i$ and a constant mass $M = 4\pi R^3\bar\rho(1 + \delta_i)/3$, where
$\bar\rho$ is the density of the backround Universe.
Conservation of energy guarantees
\begin{equation}
\half\dot{R}^2 - {GM\over R} = E = \mbox{ const.}
\label{eq:lg1a}
\end{equation}
At early times the expansion of the shell is virtually indistinguishable from
that of the rest of the Universe so that $\dot R_i = H_iR_i$, $R_i$ being the
radius of the shell and $H_i$ is the Hubble parameter at an initial time
$t = t_i$. The kinetic energy of the shell is therefore $K_i = \half
H_i^2R_i^2$
and its potential energy is $U_i = -GM/ R_i = - K_i\Omega_i(1 + \delta_i)$
where $\Omega_i$ is the density parameter at $t_i$:
$3H_i^2\Omega_i/2 = 4\pi G\bar\rho_i$.
As a result we obtain
\begin{equation}
E = K_i + U_i = K_i\Omega_i\lbrack\Omega_i^{-1} - (1 + \delta_i)\rbrack.
\label{eq:lg1b}
\end{equation}
The requirement for collapse $E < 0$ leads to the condition
$1 + \delta_i > \Omega_i^{-1}$.
Substituting $\Omega_i \equiv \Omega (z) = \Omega_0 (1 + z)/(1 + \Omega_0 z)$,
$\delta_i \equiv \delta (z)$ we get
\begin{equation}
\delta(z) > \frac{1 - \Omega_0}{\Omega_0(1 + z)}
\label{eq:lg2}
\end{equation}
as a precondition for collapse to occur.
Equation (\ref{eq:lg2}) indicates that in
flat or closed cosmological models an
infinitesimal
initial density perturbation is sufficient to give rise to collapsed objects.
In open models on the other hand $\delta(z)$ must exceed a critical positive
value in order for collapse to occur.

It is relatively straightforward to relate the maximum expansion radius
reached by an overdensity at turnaround
$R_{ta}$ to its ``seed'' values $R_i$, and $\delta_i$ (equivalently
$R(z)$ and $\delta(z)$).
Since the mass of a perturbation is conserved, and $\dot R\big\vert_{ta} = 0$
we get
\begin{equation}
E = U_{ta} = -{GM\over R_{ta}} = - {R_i\over R_{ta}}K_i\Omega_i(1 + \delta_i).
\label{eq:lg3}
\end{equation}
Equating (\ref{eq:lg1b}) and (\ref{eq:lg3})
we get
\begin{equation}
{R_{ta}\over R_i} = {1 + \delta_i\over \delta_i - (\Omega_i^{-1} - 1)}
\equiv {1 + \delta(z)\over \delta(z) - {1 - \Omega_0\over \Omega_0(1 + z)}}.
\label{eq:lg4}
\end{equation}
The time evolution of a spherical mass shell is identical to that of a
spatially
open or closed FRW Universe.
The resulting equations of motion may be obtained by
integrating eq. (\ref{eq:lg1a}) giving \cite{p80,zn83}:
\begin{eqnarray}
R & = & A(1-\cos\theta)\nonumber\\
t & = & B(\theta-\sin\theta)\label{eq:lg5a}
\end{eqnarray}
for the case $E<0$, and
\begin{eqnarray}
R & = & A(\cosh\theta-1)\nonumber\\
t & = & B(\sinh\theta-\theta)\label{eq:lg5b}
\end{eqnarray}
for the case $E>0$. In eqs. (\ref{eq:lg5a}) and (\ref{eq:lg5b})
we have $A^{3}=GMB^{2}$. The behaviour of the background Universe
is described by similar equations.

Setting $\theta = \pi$ in  equation (\ref{eq:lg5a})
we can express the constants $A$ and $B$ in terms of the
turnaround radius $R_{ta}$ and the turnaround time $t_{ta}$:
$A = R_{ta}/2, B = t_{ta}/\pi$. Next using eq. (\ref{eq:lg4})
and the relationships
$B^2 = A^3/GM$, $M = 4\pi R^3\rho/3$ and
$8\pi G\rho = 3H^2\Omega$, we can re-express $A$ and $B$ in terms of $R_i$ and
$\delta_i$ \cite{p80,pad93}:
\begin{eqnarray}
A & = & \left({R_i\over 2} \right)
{1 + \delta_i\over \delta_i - (\Omega_i^{-1} - 1)}
\nonumber \\
B & =& {1 + \delta_i\over 2H\Omega_i^{1/2}\lbrack
\delta_i - (\Omega_i^{-1} - 1)\rbrack^{3/2}}.
\label{eq:lg6a}
\end{eqnarray}
In a spatially flat Universe, eq. (\ref{eq:lg6a}) becomes
\begin{equation}
A \simeq {R_i\over 2\delta_i},~~B \simeq {3\over 4}t_i\delta_i^{-3/2},
\label{eq:lg6b}
\end{equation}
where we assume $\delta_i << 1$.
It is now relatively straightforward to compute the
overdensity in each mass shell.
Since mass is conserved we get, using $M = 4\pi R^3\rho/3$ and
eq. (\ref{eq:lg5a}),
\begin{equation}
\rho(t) = {3M\over 4\pi A^3(1 - \cos\theta)^3}.
\label{eq:lg7}
\end{equation}
In a spatially flat matter dominated Universe the
background density scales
as
\begin{equation}
\bar\rho(t) = {1\over 6\pi Gt^2} =
{1\over 6\pi GB^2 (\theta - \sin\theta)^2}.
\label{eq:lg8}
\end{equation}
So, combining eqs. (\ref{eq:lg7}) and (\ref{eq:lg8}),
we get
\begin{equation}
\delta (\theta) \equiv {\rho (t) \over \bar\rho (t)} - 1 =
{9\over 2}{(\theta - \sin\theta)^2\over (1 - \cos\theta)^3} - 1,
\label{eq:lg9a}
\end{equation}
for positive density fluctuations, and
\begin{equation}
\delta (\theta) = {9\over 2}{(\theta - \sinh\theta)^2\over (\cosh\theta - 1)^3}
- 1
\label{eq:lg9b}
\end{equation}
for negative density fluctuations.

{}From equations (\ref{eq:lg9a}) and (\ref{eq:lg6b})
we  recover the linear limit for small $\theta, t$:
\begin{equation}
\lim_{\theta \rightarrow 0} \delta (\theta) \simeq
\frac{3\theta^{2}}{20}
\simeq {3\over 20}\left({6t\over B}\right)^{2/3} = {3\over 5}\delta_i
\left({t\over t_i}\right)^{2/3},
\label{eq:lg10}
\end{equation}
indicating that only $3/5$th of the initial amplitude is in the growing mode.
In view of eq. (\ref{eq:lg10})
the criticality condition (\ref{eq:lg2}) translates into
$\delta_i > 3(\Omega_i^{-1} - 1)/5$
or, equivalently \cite{zn83},
\begin{equation}
\delta(z) > {3\over 5}{1 - \Omega_0\over \Omega_0(1 + z)}.
\label{eq:lg11}
\end{equation}
{}From eq. (\ref{eq:lg9a})
we find $\delta (\theta = \pi) \simeq 4.6$ at the radius of
maximum expansion (``turnaround''),
and $\delta (2\pi) \rightarrow \infty$ at recollapse.
The corresponding extrapolated linear density contrast can be found from
equations (\ref{eq:lg10}), (\ref{eq:lg5a}) \& (\ref{eq:lg6b}):
\begin{equation}
\delta_L(\theta) \simeq {3\over 5} \left({3\over 4}\right)^{2\over 3}
(\theta - \sin\theta)^{2\over 3}.
\label{eq:lg12}
\end{equation}
We thus obtain $\delta_L(\pi) \simeq 1.063$ for the linear density
contrast at turnaround, and $\delta_L(2\pi) \simeq 1.686$ at recollapse.
These, and other, values of $\delta_L$ are compared with the exact
results $\delta$ in the Table.\footnote{It is interesting to note in this
context that in a low $\Omega$ Universe a spherical density
contrast can grow to
fairly large values $\delta \gg 1$, without ever recollapsing
\protect{\cite{p80}}.}

\begin{table}
\begin{center}
\caption{Linear theory and exact overdensities $\delta$ at various
stages of the collapse in the spherical top--hat model.}
\begin{tabular}{ccc}
$\theta$ & $\delta_{L}$ & $\delta$\\
$\theta\rightarrow 0$ & $\delta_L \propto \theta^{2}$ & $\delta
\propto \theta^{2}$ \\
$\frac{\pi}{2}$ & 0.341 & 0.466\\
$\frac{2\pi}{3}$ & 0.568 & 1.01\\
$\pi$ & 1.063 & 4.6\\
$2\pi$ & 1.686 &$\infty$\\
\end{tabular}
\end{center}
\end{table}

Knowing the linear density contrast corresponding to a given perturbation,
the redshift at which that perturbation ``turned around'' and ``collapsed''
can be found from
\begin{eqnarray}
1 + z_{ta} & \simeq  & {\delta_L\over 1.063}\nonumber\\
1 + z_{coll}&  \simeq  & {\delta_L\over 1.686}.
\label{eq:lg13}
\end{eqnarray}
In reality $\delta_{coll} \rightarrow \infty$
will never be achieved since exact spherical collapse is at best a rather
crude approximation, which will break down as the overdensity
begins to contract, dynamical relaxation and
shocks both ensuring that the system reaches virial equilibrium
at a finite density.
The maximum density at recollapse can be estimated using the virial theorem
and the fact that at $R = R_{ta}$ all the energy in the system
is potential:
\begin{equation}
 U(R = R_{vir}) = 2E = 2U(R = R_{ta}),
\end{equation}
since $U = -G M/ R$ we get $R_{vir} = R_{ta}/2$ and $\rho_{vir} =
8\rho_{ta}$ \cite{p67}.
The mean density of an object at turnaround is $\rho_{ta}/\bar\rho =
\delta_{ta} + 1 \simeq 5.6$ so that
$\rho_{ta} \simeq 5.6\bar\rho_{ta}$.  We therefore get
$\rho_{vir} \simeq 8\times 5.6 \bar\rho_{ta}$.
Since $\bar\rho = (6\pi Gt^2)^{-1}$ and setting
$t_{vir} \simeq t_{coll} \simeq 2 t_{ta}$
we finally get $\rho_{vir} \simeq 8\times 5.6\times 4\bar\rho_{vir}$ or
since $\bar\rho_{vir} = (1 + z)^3 \rho_0$
\begin{equation}
\rho_{vir} \simeq 179.2 (1 + z_{vir})^3\rho_0,
\label{eq:lg14}
\end{equation}
where $z_{vir}$ is the collapse redshift, and $\rho_0$ the present matter
density. Equation (\ref{eq:lg14})
permits us to relate the virialised density of a
collapsed object to the epoch of its formation \cite{zn83,p80,pad93}:
$z_{vir} \simeq 0.18(\rho/\rho_0)^{1/3} - 1$.
(Since the present overdensity
in clusters is $10^2 \le \rho/\rho_0 \le 10^4$, the above
arguments might indicate that clusters formed relatively recently at redshifts
$z \le 3$, provided they formed from spherical density enhancements.)

Generalisations of these arguments show that the addition of
a cosmological constant to the Einstein equations does not
significantly affect the dynamics of a spherical
overdensity \cite{llpr91,m91}.
The final (virial) radius of a spherical overdensity in this case
turns out to be
\begin{equation}
{R_{vir}\over R_{ta}} \simeq {1 - {\eta\over 2}\over 2 - {\eta\over 2}} <
\half,
\label{eq:lg15}
\end{equation}
where $\eta = \Lambda/ 8\pi G\bar\rho(t = t_{ta})$ is the ratio of the
cosmological constant to the background density at turnaround.
Equation (\ref{eq:lg15})
indicates that the presence of a positive cosmological constant
leads to a somewhat smaller final radius (and consequently higher density)
of a collapsed object. This effect is clearly larger for objects
that collapse later, when the turnaround density is lower.

As a final remark in this context, we should mention that studies of the
collapse of spherically--symmetric perturbations into the fully non--linear
regime have been discussed in \cite{gott75,gunn77,fg84b,b85b}.
These investigations showed that
spherical infall onto an initial overdensity was self-similar and
resulted in a density profile $\rho \propto r^{-2.25}$
around the accreting object
\cite{fg84b,b85b}. (Solutions with planar and cylindrical symmetry are
discussed in \cite{fg84b}.)

\subsection{Anisotropic Collapse}
Although very convenient from the point of view of an approximate analysis,
the spherical model
is nevertheless an idealisation and does not describe the collapse of realistic
astrophysical objects such as protogalaxies and protoclusters very accurately.
The reason
for this is that one does not expect a density perturbation to have started
out being exactly spherical. It is well known that once an overdensity
turns around all departures from spherical symmetry get amplified, a result
sometimes called the Lin--Mestel--Shu instability
\cite{lms65}. As a result the generic
shape of a collapsing obect at late times is likely to become spheroidal
even though
it might have been highly spherical to begin with\footnote{The reverse
is true for underdense regions which become progressively more
spherical as they expand \protect{\cite{i73,i84}}; see \S 5.}.
Studies of the motion of homogeneous spheroids have shown that
collapse usually occus along a preferred direction which coincides with the
smaller spheroidal axis, collapse to a point in this case
occuring much faster than it would have for an equivalent spherical density
enhancement \cite{l64,z64,lms65,nf72,mhp86}.
An interesting new feature of anisotropic collapse is that
the condition $\Omega' > 1$ essential for collapse in the spherical model
is no longer necessary for spheroids, for which collapse along one axis
can occur even if
the ratio's of the kinetic and potential energies marginally exceeds unity,
implying $\Omega' < 1$ for the fluctuation \cite{p80}.
These results have recently been both confirmed and extended in
\cite{hof86,hof89,zh93,vdb94,bj94} where it is shown that
(a) the presence of shear leads perturbations
to collapse faster than they would have in the spherical model, and
(b) shear can lead to the collapse of regions
which are initially underdense.
Some aspects of anisotropic collapse are also incorporated in the Zel'dovich
approximation which will be discussed in the next chapter.

\subsection{Peaks Theory}
A simple way to seek to extend linear theory into the non--linear
regime is to imagine that the density field evolves linearly until
high--density regions reach some critical `threshold' value $\delta
\sim 1$. Such regions then collapse in some manner assumed to be
similar to the top--hat collapse of an isolated spherical inhomogeneity.
This amounts to the assumption that the transition from linear
to non--linear evolution is `fast' and that peaks of the linear
density field are the sites for non--linear structure formation.
If this is true then we should be able to understand the clustering
of non--linear objects by looking at the clustering of appropriate
sites in the linear density field. An immediate problem in implementing
this prescription is posed by the need to smooth the density field
to select appropriate peaks. For power--spectra of the kind discussed
in \S 2.4, the density field has structure on all scales. This means
that $\delta$ actually has an infinite number of small scale peaks.
When we are looking at the properties of what we assume to be protogalaxies,
we do not want to take account of very high frequency modes which
correspond to structure on scales much smaller than galaxies. The
way to achieve this is to smooth the density field with some
kind of low--pass filter in the manner described in \S 2.4.
Implicit in this procedure is the assumption that high frequency
modes of the system do not affect the evolution of low frequency
modes, which therefore continue to evolve linearly; some attempt
to justify this is made in \cite{p80}.
As we shall see in \S 4, even with smoothing, this
turns out not to be a particularly good way of looking at the
formation of large--scale structure in most of the favoured models,
because non--linear structure formation does not in general occur
near peaks of the linear {\em density field}. Dynamical effects are
important even on large scales, and it appears that peaks in the final density
field are displaced significantly with respect to peaks in the
initial conditions\cite{kqg93}.
Nevertheless, this general approach has led to useful
results and has spawned a huge literature on the statistics
of primordial density maxima, so it is worth reviewing in some
detail here.

To proceed with this approach we need to construct a formalism to
describe the statistical properties of the linear density field.
In the usual models, primordial density fluctuations are a Gaussian
random field (see \S 2.4). Recall that for such a field, {\em all}
the statistical information required to specify its properties
is contained in the autocovariance function $\xi(r)=\langle
\delta(\vec{x})\delta(\vec{x}+\vec{r})\rangle_{\vec{x}}$, where
$\xi(0)=\sigma^{2}$, the variance of $\delta$,  and the subscript
means that the average is taken over all spatial configurations,
assuming that $\delta(\vec{x})$ is spatially ergodic, which is
actually true under very weak conditions for Gaussian fields
\cite{a81}.

The first attempt to investigate the properties of linear density
peaks was made by Doroshkevich in 1970 \cite{d70}, who presented his results
mainly in the framework of the baryonic pancake model of galaxy formation
\footnote{Doroshkevich
calculated a number of statistical quantities of great importance including:
the number density of peaks of a Gaussian random field in the high peak limit;
the shape of contours surrounding maxima; the average value of the Euler
characteristic defining the topology of contour surfaces of a given density.}.
More recently, this formalism was extended to hierarchical models
of structure formation, following the lead of Kaiser \cite{k84}.
Kaiser considered peaks to be
defined as regions where the density field $\delta(\vec{x})$ exceeds
some threshold value $\delta_c$; such regions are often called
{\em excursion regions} and the set comprising all
excursion regions is called the {\em excursion set}. We shall later
discuss the possibility that local maxima might be used rather than
excursion regions to indicate the positions of proto--objects.
With the spherical top-hat model
in mind, it perhaps makes sense to take $\delta_c=1.68$, which is
the (extrapolated) linear density field at recollapse in the spherical
collapse model.\footnote{Numerical simulations indicate a somewhat smaller
value of $\delta_c$: $\delta_c \sim 1.33 - 1.44$ \cite{cc89,er88}.}
It is straightforward to calculate the correlation
function of points exceeding $\delta_c$ using the Gaussian
prescription because the probability of finding two regions
separated by a distance $r$ both above the threshold will be just
\begin{equation}
 Q_2  =  \int_{\delta_c}^{\infty} \int_{\delta_c}^{\infty}
P_2 (\delta_1,\delta_2) \d \delta_1\d \delta_2. \label{eq:lg20}
\end{equation}
Now the $N$--variate joint distribution of a set of $\delta_i$
can be written as a multivariate Gaussian distribution:
\begin{equation}
P_N (\delta_1, ..\delta_N) = \frac{||{\bf M}||^{1/2}}{(2\pi)^{N/2}} \exp
\left( -\half \vec{V}^{T} {\bf M} \vec{V}\right),\label{eq:lg21}
\end{equation}
where ${\bf M}$ is the inverse of the correlation matrix ${\bf C}
=\langle \delta_{i} \delta_{j} \rangle$ and $\vec{V}$ is a column
vector made from the $\delta_i$. The ability to construct, not
only the $n$--dimensional joint distribution of values of $\delta$,
but also joint distributions of derivatives of arbitrary order
which are all of the form (\ref{eq:lg21}), is what makes
Gaussian random fields so useful from an analytical point of view;
see \cite{a81}.
The expression (\ref{eq:lg21}) is
considerably simplified
by the fact that $\langle \delta_i \rangle =0$ for all $i$.
Using the substitution $\delta_i=\nu_i\sigma$ and
$w(r)=\xi(r)/\sigma^{2}$ yields
\begin{equation}
P_2(\nu_1,\nu_2) = \frac{1}{2\pi}\frac{1}{\sqrt{1-w^{2}(r)}}
\exp \left\{ -\frac{\nu_1^{2}+\nu_2^{2}-2w(r)\nu_1\nu_2}{2[1-w^{2}(r)]}
\right\}.\label{eq:lg22}
\end{equation}
The two--point correlation function for points exceeding $\nu_c=
\delta_c/\sigma$ is then
\begin{equation}
1+\xi_{\nu_c}=\frac{Q_2}{Q_1^{2}},\label{eq:lg23}
\end{equation}
where $Q_1=\Pr(\nu>\nu_c)$. The exact calculation of the integrals
in this equation is difficult but various approximate relations have been
obtained. For large $\nu_c$ and small $w$ we have \cite{k84}
\begin{equation}
\xi_{\nu_c}\simeq \nu_c^{2} w(r),\label{eq:lg24}
\end{equation}
while another expression which is valid when $w$ is not necessarily small
is \cite{pw84}
\begin{equation}
\xi_{\nu_c} \simeq \exp[\nu_c^{2} w(r)] - 1.\label{eq:lg25}
\end{equation}
More accurate expansions are given in \cite{js86};
higher--order correlations in this model can be
expressed in the Kirkwood scaling form, see \S 8.3.
Kaiser \cite{k84}
initially introduced this model to explain the enhanced correlations
of Abell clusters compared to those of galaxies. Here the field
$\delta$ is initially smoothed with a filter of radius several
Mpc to pick out structure on the appropriate scale. If galaxies
trace the mass, and so have $\xi_{gg}(r)\simeq \xi(r)$ then
the simple relation (\ref{eq:lg24}) explains qualitatively why cluster
correlations might have the same slope, but a higher amplitude than
the galaxy correlations. This enhancement is natural because rich
clusters are defined as structures within which the density of matter
exceeds the average density by some fairly well--defined factor in
very much the way assumed in this calculation.

This simple argument spawned more detailed analyses of the statistics
of Gaussian random fields \cite{bbks86,cprw87,c89,c87a,c87b,cd93,lhp89,ph85}.
The interest
in most of these studies was the idea that galaxies themselves
might form only at peaks of the linear density field (this time
smoothed with a smaller filtering radius). If galaxies only form
from large upward fluctuations in the linear density field then
they too should display enhanced correlations with respect to the
matter. This seemed to be the kind of {\em bias} required to
reconcile the standard CDM model with observations of galaxy
peculiar motions \cite{defw85}. One should say, however, that there
is no reason {\em a priori} to believe that galaxy formation
should be restricted to peaks of particularly high initial density.
It is true that peaks collapsing later might produce
objects with a lower final density than peaks collapsing earlier,
but these could (and perhaps should) still correspond to galaxies.
Some astrophysical mechanism must be introduced which will
inhibit galaxy formation in the lower peaks. Many mechanisms
have been suggested, but none of these are particularly compelling.
Indeed, the COBE CMB fluctuations have recently given cause to
doubt the validity of this kind of biasing as a means for
reconciling CDM with observations.

Notwithstanding these reservations about the validity of peaks theory,
the idea has led to some interesting analyses of the properties
of Gaussian random fields and the possible effects of biasing.
One particular line of enquiry has been to look at {\em local
maxima} of the density field, rather than just points where
$\delta$ exceeds some particular value $\delta_c$. The
trouble with the original calculation by Kaiser \cite{k84}
is that it applies no constraint that the two field values
$\delta_1$ and $\delta_2$ are in different proto--objects.
The smoothing required to select objects
of the correct characteristic scale introduces fairly
large scale coherence into the density field. This means
that regions where $\delta$ exceeds some value may be quite large
compared with the mean spacing of clusters. Since each
disjoint region above the threshold is implicitly
assumed to form one cluster, the simple calculation given above
will lead to an overcounting of pairs of objects separated
by distances less than the typical size of {\em excursion regions}
\cite{c86}.

This suggests that a more accurate representation of reality
might be obtained by looking at local maxima of the field.
If the criterion for proto--object sites is that $\delta$
be a local maximum, it is impossible that two distinct
spatial positions satisfying this criterion can both
belong to the same proto--object. Dealing with local maxima
complicates the calculations considerably, but some progress
is possible. For illustrative purposes, we discuss here the
argument for fluctuation fields defined in one spatial dimension.
The difficulties with three dimensions will quickly become
obvious!

To handle local maxima, we need to consider not just the
value of $\delta$ at a given spatial location, but also
the first and second derivatives of $\delta$. Smoothing is
also required here, otherwise the field $\delta$ would not
be differentiable (in the mean--square sense: \cite{a81})
and local maxima would be consequently undefined.
We need first need to consider $P(\delta,\delta',\delta'')$,
where the primes denote derivatives with respect to the spatial
coordinate (which we denote $x$). This joint probability
is also a multivariate normal distribution (\ref{eq:lg21}), but
the column vector is written $\vec{V}=(\delta,\delta',\delta'')$
and the covariance matrix ${\bf C}$ has the form
\begin{equation}
{\bf C} = \left( \begin{array}{ccc}
                   \sigma^{2} & 0 & -\sigma_1^{2}\\
                           0  & \sigma_{1}^{2} & 0\\
                   -\sigma_1^{2} & 0 & \sigma_2^{2}\end{array}\right).
\label{eq:lg26}
\end{equation}
The $\sigma_i^{2}$ are moments of the power--spectrum of the
density fluctuations in one dimension:
\begin{equation}
\sigma_j^{2} = \int P_1(k) k^{2j} dk,\label{eq:lg27}
\end{equation}
where $P_1(k)$ is the one--dimensional Fourier transform
of $\xi(r)$ \cite{c89,lhp89}.
In this notation the variance
of $\delta$ is just $\sigma^{2}=\sigma_0^{2}$. To select
stationary points we need to specify $\delta'=0$. Since
such points form a set of measure zero, the appropriate probability
would seem to be identically zero. This problem can be avoided
by replacing the `volume' $\d \vec{V}=\d\delta\d\delta'\d\delta''$
by $\d\delta\d\delta'' |\delta''|\d x$: assuming that $\delta$
is spatially ergodic then means that the integral over $x$ can
be taken to be an integral over the probability distribution
and one avoids the measure problem. It helps to define,
in addition to $\nu$ (which is as above), the parameters
$\gamma=\sigma_1^{2}/\sigma\sigma_2$ (which is a measure
of the bandwidth of the spectrum $P_1$), and $R_{\ast}=\sigma_1/\sigma_2$,
a measure of the characteristic coherence length of the field.
Defining $\theta=-\delta''/\sigma_2$ gives the number--density
of stationary points as:
\begin{equation}
\d N_0=\frac{|\theta|\d \theta\d\nu}{(2\pi)^{3/2}
(1-\gamma^{2})^{1/2}R_{\ast}} \exp \left[ -\frac{1}{2}\left(
\frac{(\nu-\gamma \theta)^{2}}{1-\gamma^{2}} + \theta^{2} \right)\right].
\label{eq:lg28}
\end{equation}
To further specify local maxima, we require $\theta>0$:
$N_{pk}=\int_0^{\infty} \d N_0$ is then the number of local maxima
per unit length. By integrating over $\nu$ first, and then
over $\theta$ we find that $N_{pk}=1/2\pi R_{\ast}$. To find the
number--density of local maxima as a function of their
(dimensionless) height $\nu$ we need to integrate over
$\theta$. The result was found analytically in a classic
paper by Cartwright and Longuet--Higgins \cite{cl56} who give the probability:
\begin{equation}
P_{\rm max}(>\nu) = \frac{1}{2} \left[
{\rm erfc}\left( \frac{\nu}{\sqrt{2(1-\gamma^{2})}}\right)
+ \gamma e^{-\nu^{2}/2}
\left\{ 1+{\rm erf}\left(\frac{\gamma \nu}{\sqrt{2(1-\gamma^{2})}}\right)
\right\}\right].\label{eq:lg29}
\end{equation}
Notice that in the limit $\gamma\rightarrow 0$ the distribution
of heights at maxima is just a Gaussian. This is the limit of
infinitely broad spectrum (i.e. no filtering) and it simply
tells us that, without the low--pass filter, every point is
a local maximum! The limit for a $\delta$--function spectrum
$\gamma\rightarrow 1$ is a Rayleigh distribution. The fact that
a narrow spectrum produces peaks which have $\nu>0$ is not
surprising: a monochromatic plane wave has all its local maxima
at $\nu>0$.

It is conceptually straightforward to extend these results to three
spatial dimensions, but it is technically difficult.
The reason for the complexity is that, in three dimensions,
the local maximum constraint requires the three spatial derivatives
of $\delta(\vec{x})$ to be zero, and that the matrix of second
derivatives $\partial^{2}\delta/\partial x_i\partial x_j$ be
negative definite (having three negative eigenvalues). One therefore
needs a 10--dimensional probability density function
for $\delta$, its three first derivatives, and the
six independent second derivatives. One then
has to integrate over the region where the second
derivative matrix is negative definite \cite{a81}.
We shall not go into details here; the differential
number--density of local maxima has been obtained by \cite{bbks86}
as the function
\begin{equation}
\frac{\d N_{pk}}{\d \nu} = \frac{1}{(2\pi)^2 R_{\ast}^{3}}
e^{-\nu^{2}/2} \, G(\gamma,\gamma\nu),\label{eq:lg30}
\end{equation}
where $R_{\ast} = \sqrt{3}\sigma_1/\sigma_2$,
\begin{equation}
G(\gamma,y) = \int_{0}^{\infty} \frac{f(x)}{[2\pi(1-\gamma^{2})]^{1/2}}
\exp \left( -\frac{(x-y)^{2}}{2(1-\gamma^{2})} \right) \d x.\label{eq:lg31}
\end{equation}
The function $f(x)$ is given by \cite{bbks86}
\begin{eqnarray}
f(x) & = & \left(\frac{x^{3}-3x}{2}\right)\left\{
{\rm erf} \left[x\sqrt{\frac{5}{2}}\right] + {\rm erf}
\left[\frac{x}{2}\sqrt{\frac{5}{2}}\right]\right\}\nonumber\\
  & & +\sqrt{\frac{2}{5\pi}} \left[ \left(\frac{31x^{2}}{4}+\frac{8}{5}\right)
e^{-5x^{2}/8} +\left(\frac{x^{2}}{2}-\frac{8}{5}\right)e^{-5x^{2}/2}\right].
\end{eqnarray}
In one dimension, the corresponding result has the same form as
eq. (\ref{eq:lg31}), but with $f(x)=x$; in two dimensions the
result has $f(x)=x^{2}+e^{-x^{2}}-1$ \cite{p92}.
The number of local maxima per unit volume turns out to be $N_{pk} \simeq
0.016 R_{\ast}^{-3}$. For high peaks
\begin{equation}
N_{pk}(>\nu) = {1\over 4\pi^2}\left({\gamma\over R_{\ast}}\right)^3\nu^2
e^{-\nu^2/2}.
\end{equation}
As shown in \cite{bbks86} the number density of high peaks in the limit
$\gamma\nu \gg 1$ is related to
the density $N_{\chi}$ of the Euler characteristic of the surfaces
bounding the peaks by $N_{pk}(>\nu) = {1\over 2}N_{\chi}(>\nu)$.
\footnote{The relationship between the Euler characteristic $\chi$ and the
genus used in \S 8.8 is ${\rm genus} = 1 - \chi/2$.}

To calculate the two--point correlation function of local maxima is more
complicated still. In three dimensions one needs to construct
the 20--dimensional distribution of $\delta$ and its first two
spatial derivatives at two distinct spatial locations, and then
integrate imposing the peak constraint at each spatial location.
No exact analytic results are known for this calculation. In one
dimension (which we hope will give a good qualitative guide to
the behaviour of three--dimensional peaks), the problem boils
down to the construction and integration over a six--dimensional
probability distribution \cite{c89,lhp89}. Even here,
analytical results are difficult to obtain and one must
resort to some numerical integration. At very large
separations, however, the peak--peak correlation function
has a similar form to (\ref{eq:lg25}), but with $\nu$ defined
in a different way \cite{lhp89}. It is not possible, even in one
spatial dimension, to express the peak--peak correlation function
as a simple function of the underlying correlation function of the
density fluctuations. This is because the peak--peak correlation
function depends, not just on $\xi(r)$ but also upon its derivatives
with respect to $r$. For example, notice that the Politzer--Wise
formula, equation (\ref{eq:lg25}), gives a biased correlation
function which is zero whenever the underlying $w(r)$ is zero
which is not necessarily the case when one considers local
maxima \cite{c89,lhp89}. Higher--order
correlations of peaks in one--dimensional fields are discussed
in \cite{cd93}.

The problem with this entire approach is that it assumes that
non--linear evolution does not move proto--objects away from
the positions of local density maxima in the linear field.
This is unlikely to be the case, except for very large
separations (where the displacement of matter from its
linear position might be small compared to the
peak separation), or if the primordial spectrum has very
little large--scale power. A comparison between the `peaks'
of the density field evolved using a full numerical simulation,
with the peaks in the linear density field shows very little
correspondence between the two \cite{cms93,kqg93}. It seems
therefore that linear density maxima are probably not,
in themselves, very good candidates for the positions of galaxy formation;
the situation may be more favourable for cluster--scale peaks.

A closely related issue concerning the peaks picture of galaxy formation is the
influence of shear (both internal -- due to asphericity, and external --
due to the surrounding matter distribution) on the evolution of density peaks.
Recent studies conducted on galaxy scale peaks in a Gaussian random field
show that shear can occasionally break up a single density peak into two or
more halos \cite{vdb94,zh93,bm94a,bm94b,bm94c,bm94d}.
It can also, by boosting the accretion of
matter, occasionally promote the growth of smaller scale peaks into larger
ones.
Such studies demonstrate that the conventional peaks scenario in which halo's
of a characteristic scale and peaks in a Gaussian random field (filtered on
the same scale) are held in one-to-one correspondence, is somewhat simplistic
and needs to be considerably modified before it is taken seriously.\footnote{
Tidal effects are less important for high peaks which can be treated as
rare events. As demonstrated in \protect{\cite{b94}}
the non--linear evolution of a
high density peak follows spherical collapse if the spectral index $n < -1$.}
Some attempts have been made to
include dynamics in this kind of analysis, although it
is very difficult to do so convincingly \cite{bbks86}.
An impressive attempt at this was made by Bond \& Couchman
\cite{bc88}, who used the Zel'dovich approximation
(see later) to move linear density peaks away from their
initial positions; see also \cite{mhp93}.
The correlation function of the evolved
peaks can then be expressed in terms of the initial
correlations of the linear density peaks, modified by
corrections for the displacement caused by non--linear
evolution. Although this calculation leads to an exact
solution for $\xi$, it is not entirely satisfactory because
linear density peaks are not necesarily in one--to--one
correspondence with peaks in the evolved field
(some may disrupt, others may merge). Because clustering
evolves dynamically, the final density peaks
are formed by the action of particle displacements
generated by the gravitational potential, which is not
a simple local function of the density field.
A better procedure would therefore be to look for peaks
in the {\em final} density field, as is done in \cite{kqg93,bcm94}.

In a bold extension of the peak formalism, Bond and Myers
\cite{bm94a,bm94b,bm94c,bm94d} have developed the so--called ``peak--patch''
picture for identifying and tracing the dynamical evolution of
protostructure. In this picture, an entire hierarchy
of peaks on different scales is identified using different values of
the filter radius. An ellipsoidal collapse model is then used
to follow the evolution within each patch, including the
influence on its evolution of the
external tidal field around it. By following peak patches on different
scales simultaneously, one can allow for patches that overlap in the
evolved system. The Zel'dovich approximation (\S 4.2) is then used
to move the peak patches thus identified to their final
Eulerian position. Both the statistical and dynamic contributions
to the final correlation function are therefore included in this
treatment. The use of a ``catalogue'' of peaks as a function
of smoothing radius allows one to counteract the obvious limitation
of the standard peaks theory we have concentrated on so far, namely
that it assumes all peaks result from the application of a single
smoothing radius. The procedure is, however, technically complex
and it is not yet clear how useful will be the results it
generates. We shall also make some comments on the peak--patch
picture at the end of \S 3.4.

\subsection{The Press--Schechter Approach}
One of the problems involved in trying to identify structure
in density fields is that the power spectrum of the density field
may not possess a physical cutoff on small scales. In other
words, the filtered variance $\sigma (R)$ (\ref{eq:lad7}) diverges as the
filtering scale is reduced. Such fields always contain structure
on a small enough scales which is non--linear. One expects
structure thus to develop hierarchically, with large structures
forming by the merger of small--scale lumps. This seems
far too complicated a process to be dealt with in any variant
of linear theory, but there is a method, developed by
Press \& Schechter \cite{ps74} which attempts to explain some
aspects of hierarchical clustering in terms of properties
of the linear density field. We shall discuss this method
in some detail here, because of its widespread use,
and surprising effectiveness! We shall henceforward refer to this
method as the PS method.

The PS formalism rests on one critical assumption: that, even
if the density field is non--linear on some scale, the amplitude
of long--wave modes in the density field is close to that predicted
by linear theory. This effectively means that there must be enough
large--scale power in the density field that it is not swamped by
power generated by non--linear interaction of small--scale modes.
Consider a massive clump. As described in \S 3.1, the clump
will undergo collapse (assuming it is roughly spherical), if
its overdensity exceeds some critical value $\delta_c$ which
is of order unity. This suggests that one can identify the
positions of clumps which will be bound by smoothing the
linear density field on some scale $R_f$. On average,
a mass $M\sim \rho_0 R_f^{3}$ will be contained within a
filtering radius, so it is reasonable to assume that points
where $\delta$ smoothed on this scale exceeds $\delta_c$ will
be bound objects with a mass $M$. Nothing in this approach
determines the form of the filter function: it is conventional
to use a Gaussian for no other reason than analytical convenience.

Now, assuming a Gaussian distribution for $\delta$, the probability
that a randomly--selected point exceeds the threshold $\delta_c$
when smoothed on a scale $R_f$ is just
\begin{equation}
\Pr (\delta>\delta_c|R_f) = P_{\delta}=\frac{1}{2} \left[
1- {\rm erf} \left( \frac{\delta_c}{\sigma_f\sqrt{2}} \right)\right],
\label{eq:lg32}
\end{equation}
where $\sigma_f^{2}$ is the variance of the smoothed field (\ref{eq:lad7}).
We now
follow the PS argument. This probability, (\ref{eq:lg32}), is assumed
to be proportional to the probability that a given point {\em has ever
been contained in a collapsed object} on a scale $>R_f$. It is
not obvious that this assertion is correct: it amounts to assuming
that the only objects that exist at a given epoch are those which
have just collapsed. If an object has $\delta>\delta_c$ when
smoothed on some scale $R$, it will have $\delta=\delta_c$ after
$R$ is increased, and will therefore be counted as part of
a higher level in the hierarchy \cite{p92}.
One problem with this assumption
is immediately obvious, however: matter in underdense regions
is not correctly dealt with. As a result, {\em half} the mass
remains unaccounted for. PS `solved' this problem by simply
multiplying the probability by a factor of two and attributing
the additional mass to accretion. This has, however, long been
recognised as a fiddle. For the moment, we shall leave this
problem to one side and continue with the PS analysis.

Let us define the mass function of bound objects, $f(M)$, such that $f(M)\d M$
is the comoving number density of objects with masses in the range
$(M, M+\d M)$. This function is related to the probability
$P_{\delta}$ in equation (\ref{eq:lg32}) by
\begin{equation}
Mf(M) = \rho_0 |\d P_{\delta}/\d M |, \label{eq:lg33}
\end{equation}
where $\rho_0$ is the usual mean density of cosmic material.
Hence,
\begin{equation}
M^{2} f(M) = \frac{\delta_c \rho_0}{\sigma_f} \sqrt{\frac{2}{\pi}}
\exp \left(-\delta_c^{2}/2\sigma^{2}\right) |\d \log \sigma_f/
\d \log M|. \label{eq:lg34}
\end{equation}
It is useful to define a multiplicity function, $q(m)$, to be
the fraction of mass contained in objects
with mass in unit range of $\log M$: this is just $q(M)=M^{2} f(M)/\rho_0$.
This function has a simple form if the power--spectrum of the linear
field has a power--law shape ($P(k) \propto k^n$):
\begin{equation}
q(M)= \frac{(n+3)}{6} \sqrt{\frac{2}{\pi}} \nu \exp \left(-\frac{\nu^{2}}{2}
\right),\label{eq:lg35}
\end{equation}
where $\nu=\delta_c/\sigma_f$.

This calculation of the mass and multiplicity functions of
objects forming hierchically, using only the properties of the
linear density field, contains a number of assumptions which
are difficult to justify. Even if one accepts the basic assumption,
that there is a reasonable correspondence between the sites
where virialised structures form and regions of high density
in the linear density field, there is still the problem
of the factor of $2$ `fudge' in the PS analysis. One would be
inclined to consign the entire theory to the dustbin were it
not for the remarkable fact that the final result,
equation (\ref{eq:lg35}) is in remarkably good accord
with the results from $N$--body simulations
\cite{efwd88,er88}. This method has consequently been
used in a wide variety of different contexts to estimate
$f(M)$ for cosmological objects
\cite{ss85,cc89,ck88,ck89,efwd88,er88,ss88,bow91}.
The unreasonable effectiveness of the PS formalism has
spurred others to look at the problem from different
points of view to see why it works so well. Although the
issue is not fully resolved as yet, substantial progress
has been made using some of the techniques outlined
in \S 3.3. Since one might reasonably identify proto--objects
with peaks of the linear density field, adaptation of the
peaks formalism to the calculation of the mass function
would seem to be a fruitful line of enquiry
\cite{aj90,bbks86,bss88,bab90,bcek91,ph90}. The most
important outcome of these analyses has been a
clarification of the `cloud--in--cloud' problem. This is
simply that the PS analysis does not check
that objects
which are considered to be `bound' on a scale $M$ are
not subsumed into objects
on a larger scale as the hierarchy evolves. In fact, the Press--Schechter
formalism does not really deal with localised objects at all: it is
simply a scheme for labelling points in the linear density field.
It turns out that the failure
to deal correctly with substructure is at the root
of the infamous factor two.

Recall the one--dimensional analysis of local maxima
we referred to in the previous section. Here one looks
to find locations where the trajectory of $\delta(x)$
as a function of $x$ exceeds some value $\nu\sigma$, or
is a local maximum at some point $x$. For the mass function
analysis, it helps to consider the field $\delta$, not
as a function of spatial position but as a function
of smoothing radius $R$ (dropping the suffix $f$ for
brevity) at some fixed spatial location. Clearly $\delta(R)
\rightarrow 0$ as $R\rightarrow\infty$. As $R$ decreases,
$\delta(R)$ will develop fluctuations whose amplitude
continues to increase as $R$ gets smaller. If $\delta(R)<\delta_c$
for some value $R_1$, it may well exceed $\delta_c$ again for some
smaller value of $R$. (Indeed, if the small--scale power spectrum
diverges, this will happen with probability one.) Clearly
as one proceeds to filter the density field on a range of decreasing scales
$R_1,...,R_n$ the corresponding density contrast at a given location
$\delta(x)$ executes a random walk through the points
($\delta_1,...,\delta_n$).
The first upcrossing of the level $\delta_c$ will occur for the largest
value of $R_i$ for which $\delta_i \ge \delta_c$.

Here the analysis depends on the form of spatial filter one
uses. Consider the illustrative example of a sharp filter
in $k$--space, and consider what happens in the Fourier domain
as $R$ is decreased from infinity. Decreasing $R$ corresponds
to adding $k$--space shells which, for Gaussian fluctuations,
are independent of each other.
As a result the steps in the random walk of $\delta(R)$ become independent for
a sharp k-space filter and the solution follows very
easily\footnote{In the continuum limit the random walk is
described by a diffusion equation which can be solved analytically
\protect{\cite{bcek91}}.} \cite{ph90,bcek91,p92}.
Consider a point on the trajectory which is just sitting at the
threshold. The subsequent motion will be symmetric so that
it is equally likely to be above or below $\delta_c$ at any
smaller value of $R$. The probability that the trajectory
does not cross the threshold -- the {\em survival probability},
$P_s$ -- is then given by
\begin{equation}
\frac{\d P_s}{\d \delta} = \frac{1}{\sigma\sqrt{2\pi}}
\left[ \exp \left(-\frac{\delta^{2}}{2\sigma^{2}} \right)
- \exp \left( -\frac{(\delta-2\delta_c)^{2}}{2\sigma^{2}}\right)
\right].\label{eq:lg36}
\end{equation}
One can integrate this expression to give the probability of having
crossed the threshold at least once to be
\begin{equation}
P_{\geq 1} = 1-P_s = 1 - {\rm erf}\left(\frac{\delta_c}{\sigma\sqrt{2}}
\right), \label{eq:lg37}
\end{equation}
which is simply twice the unconditional probability of lying above the
threshold, given by eq. (\ref{eq:lg32}).
This therefore supplies exactly the missing factor of 2.
Unfortunately, however, this does not provide a full solution of
the problem because different filter functions behave differently
to this simple example. What happens generically, is that the
PS form, equation (\ref{eq:lg34}) is recovered for large $M$
but the shape of the function is changed at low $M$ to account
for the factor two in the total probability. Our problem to
figure out why the PS analysis works is, therefore, reduced to the
problem of knowing why to use a particular form for the filter
function. At present, we cannot say why the sharp $k$--space
filter best represents the process of hierarchical clustering
other than that that it is the choice that produces the PS
result which is obeyed empirically. It may also be
that the agreement between PS and numerical simulations
is essentially fortuitous. One reason for believing that this
may be the case is that, although the mass function of
objects may agree with the PS result, the spatial locations
of bound structures in simulations are very poorly
correlated with peaks in the linear density field
\cite{cms93,kqg93}. Although this does not in itself
rule out the PS analysis -- the mass function does not
specify the locations of the final, virialised structures
-- it certainly suggests that one should not take the
derivation too literally.
It also suggests that
attempts to calculate the spatial
correlations of virialised objects
of different masses
by mixing the PS formalism with the correlations of peaks
in the linear density field, e.g. \cite{ky87},
are likely to be of little
relevance to the real world.
On the other hand, in \S 4.5
we show how to estimate the mass function of collapsed objects
using adhesion theory, which at least attempts to treat
dynamical evolution in a reasonable way, and the results do
agree with the PS formalism reasonably well, so a method
may well be available for this kind of analysis.
Various alternative approaches have been explored
\cite{lc93,kw93,kwg93,kgw94,cafnz94}, but probably
the most promising is the peak--patch picture \cite{bm94a,bm94b,bm94c,bm94d}
which appears to have the most detailed treatment of the dynamics
of collapse and implicitly takes into account the hierarchical
nature of the evolution.
\footnote{An interesting formal connection
between the Press-Schechter formalism and that of Cayley trees has been
explored in \cite{cm94b}.}

\subsection{The Lognormal Model}
In \S 3.3 and \S 3.4, we dealt with attempts to identify
non--linear structures with structures in the initial linear
density field. As long as matter does not move too far away
from its primordial location as a result of non--linear evolution,
these methods might be fairly accurate. In the next
sections, we shall discuss ways of approximating the full
dynamical evolution of structure and thus improving upon
simple extrapolations of linear theory. Before we move onto
dynamical approximations, however, it is worth discussing
another way of extrapolating linear theory which, although
simple, seems to describe some aspects of the clustering
phenomenon reasonably well.

If the initial density fluctuations are Gaussian (\S 2.4), then
the form of the statistical distribution function for $\delta$
keeps the same form -- eq. (\ref{eq:ladz1}) --
as the fluctuations evolve in the linear
regime. To apply peaks theory and the Press--Schechter approach
one implicitly assumes that this distribution function is still
appropriate for the fluctuations when they are about to collapse.
But collapse occurs, in the simplest picture at any rate,
when $\delta$ is at least of order unity. At the onset of
structure formation at a given scale, we must have an
{\em rms} density contrast on that scale of order unity. But the
Gaussian distribution with $\sigma^{2}\sim 1$ assigns a sizeable
probability to regions with $\delta<-1$ which are clearly
unphysical ($\rho<0$). The assumption of Gaussian statistics
in the linear regime
is therefore physically unreasonable.

These thoughts suggest that one might achieve some measure
of improvement over linear theory by adopting a distribution
function for $\delta$ which does not allow $\delta<-1$. In the
spirit of the other approximations in this section, it is
reasonable to achieve this by a simple mapping of the initial
linear field $\delta_{LIN}\mapsto\delta_{NL}$ such that
the mapping is one--to--one but enforces the physical requirement
that there should not be regions with negative total density.
Of course, this {\em ansatz} does not include dynamics in a
physically reasonable way, but one might expect it be some kind
of improvement on linear theory. There are various ways to
impose the mapping \cite{cb87}, but the simplest and probably
most useful is with an exponential function:
\begin{equation}
{\rho\over \bar{\rho}} \equiv
1+\delta_{NL} \propto \exp(\delta_{LIN}).\label{eq:ln_1}
\end{equation}
We have implicitly assumed that the linear field is smoothed
on some scale, say $R$, which we have not specified. Obviously,
the mapping (\ref{eq:ln_1}) does not commute with smoothing
of $\delta$ so the actual mass distribution would not
have the same statistical form when smoothed on different
scales. The resulting distribution function is called a {\em lognormal}
distribution and its properties are discussed extensively in
\cite{cj91}. The constant of proportionality in eq. (\ref{eq:ln_1})
must be fixed by requiring that the mean value of $\delta_{NL}$
be zero; it will therefore depend upon the statistics of
$\delta_{LIN}$, i.e. upon the linear theory variance $\sigma^{2}_{L}$.
Of course, the variance of the non-linear $\delta_{NL}$ will not
be the same as the linearly--extrapolated field $\delta_{LIN}$.
In fact,
\begin{equation}
\sigma^{2}=\exp\left(\sigma^{2}_{LIN}\right)-1,\label{eq:ln_2}
\end{equation}
where $\sigma^{2}$ is now the variance of the non--linear field.
The probability distribution function of $\delta$ (dropping
the suffix $NL$) becomes
\begin{equation}
P(\delta) = \frac{(1+\delta)^{-1}}{\sqrt{2\pi\log (1+\sigma^{2})}}
\exp \left( -\frac{\log^{2} \left[ (1+\delta)\sqrt{1+\sigma^{2}}\right]}{
2\log(1+\sigma^{2})} \right).\label{eq:ln_dis}
\end{equation}
\cite{cj91,ccmm94}. The great advantage of this distribution is
that one can calculate analytically any property which can
be calculated for the Gaussian distribution, using the simple
local mapping, eq. (\ref{eq:ln_1}). The two--point correlation
function is given by
\begin{equation}
\xi(r) =\exp \left[ \xi_{L} (r) \right] -1,
\end{equation}
where $\xi_{L}$ is the linear theory correlation function.
Notice the similarity between this equation and eq. (\ref{eq:lg25}).
This similarity extends further:
the higher--order correlation functions for this model obey
the {\em Kirkwood} \cite{k35} scaling relationship,
equation (\ref{eq:kirkwood}),
{\em exactly} for all orders $n$. The lognormal field therefore
has correlation functions which are exactly like those of the
biased Gaussian density field computed by Politzer \& Wise \cite{pw84}.
Other statistical properties of the
$P(\rho)$ described by equation (\ref{eq:ln_dis}) are discussed
in \cite{ab69,cb87,c88,c89,cj91,ccmm94}. It is worth
mentioning just one of these properties here: the lognormal is an
example of a probability distribution that is not completely
specified by its moments. In other words, knowledge of all the
moments of the distribution of all orders $n$ is not sufficient
to specify the $P(\rho)$ completely. This is related to the
fact that the series expansion of the moment generating function
for this distribution does not converge \cite{cs88,cj91,szsz93}.
This apparently bizarre property is basically due to the
extremely long tail at large positive values in the lognormal
distribution. It serves as a warning that many of the statistical
tools one uses in the study of galaxy clustering rely on assumptions
that may not actually be realised in practice.

We should point out here that the lognormal distribution,
equation (\ref{eq:ln_dis}), is a model for the continuous
density field and cannot be directly compared with galaxy
counts, or even particle counts in an $N$--body simulation.
There is no unique way to obtain a discrete distribution from
a continuous one; the standard procedure in cosmology is
to assume that objects are a random (Poisson) sample of the
underlying density field with a mean rate proportional
to the local matter density. In other words the probability
of a small volume $\d V$ containing a galaxy is just
$\d P\propto\rho \d V$. One can then construct the
probability that a volume $V$ contains $N$ galaxies as
\begin{equation}
P(N) = \int_{0}^{\infty} P(\bar{\rho}) P(N|\bar{\rho}) \d
\bar{\rho},\label{eq:ln_disb}
\end{equation}
where $V\bar{\rho}=\int_V \rho \d V$, so that $\bar{\rho}$
is the density smoothed over the cell volume.
If the distribution is lognormal when smoothed on the scale
$V$, then $P(\rho)$ is given by eq. (\ref{eq:ln_dis}), and
$P(N\rho)$ is just a Poisson distribution. The resulting
compound distribution is called the Poisson--lognormal
distribution. Unfortunately, there is no simple closed--form
expression for $P(N)$ \cite{cj91,cs88}. This distribution
function seems to match the observed distribution function
of galaxies reasonably well \cite{sms93}. An alternative,
less transparent, route to the `discretised' form of
the pdf for a Poisson--sampled field starts from the moment
generating function \cite{f85,cj91}.

Although the lognormal model
density field must be thought of as mainly phenomenological,
and was introduced into the cosmological literature for this reason,
it is remarkable that numerical $N$--body experiments and other
analytical considerations (\S 4) show that the distribution function
of fluctuations in the non--linear regime is well--described by
the form (\ref{eq:ln_dis}). The reason why this is the case has been
the cause of much debate; various explanations are discussed
in \cite{bk94,kbgnd94}. There is some qualitative motivation for
the lognormal distribution in general terms because of the properties
of non--linear processes. The central limit theorem states that the
distribution of the sum of $n$ independent random variables
tends towards a Gaussian distribution for large $n$ regardless
of the form of the distribution function of the individual
random variables. If instead we consider the product of $n$
independent random variables, then it is clear that the logarithm
of this product should have a normal distribution by virtue of the
central limit theorem. Lognormal distributions occur widely
in non--linear physics, just as normal distributions do in linear
physics. This argument does not give us any motivation
for having taken a simple local extrapolation of the density field
like equation (\ref{eq:ln_1}); a more specific argument
is needed. Coles \& Jones (1991) took the continuity
equation (\ref{eq:continuity}) as the starting point
for a simple argument. If we
change variables, using $a\d \tau=\d t$ and $\varrho=\rho a^{3}$,
this equation can be written in the form
\begin{equation}
\frac{\partial \varrho}{\partial \tau} + (\vec{v}\cdot\vec{\nabla})
\varrho + \varrho(\vec{\nabla}\cdot\vec{v})=0. \label{eq:ln_3}
\end{equation}
Introducing the total time derivative with respect to conformal
time $\tau$, we can then write
\begin{equation}
\frac{1}{\varrho} \frac{\d \varrho}{\d \tau} = - (\vec{\nabla}\cdot\vec{v}).
\label{eq:ln_4}
\end{equation}
If the velocity field $\vec{v}$ is Gaussian, then $\vec{\nabla}\cdot\vec{v}$
will be Gaussian, too. By inspection, the solution of eq. (\ref{eq:ln_4})
will therefore be a lognormal distribution for $\varrho$. This does
not, however, prove that the real space distribution of $\varrho$ should
be accurately lognormal in the non--linear regime because (a)
the velocity field will not evolve exactly linearly in this regime and
(b) the total time derivative $\d/\d \tau$ follows the fluid elements.
At best, this argument can apply only in the very weakly non--linear
regime. Further confirmation that this is not the correct explanation
for the effectiveness of the lognormal distribution in the strongly
non--linear regime is afforded by the comparison between full
$N$--body evolution and the lognormal extrapolation via
equation (\ref{eq:ln_1}) of the same initial conditions. This
exercise was part of an extended study by \cite{cms93}, which
also looked at other approximations. The result shows that, although
the distribution function of fluctuations in the final $N$--body
simulations is fairly accurately described by a lognormal distribution,
the variance of this distribution is not that obtained by
the simple transformation of the linear variance via equation
(\ref{eq:ln_2}). This means that we need to look more deeply
into the problem before coming up with any real justification
for this model. We discuss more detailed dynamical justification
for the lognormal approximation in \S 4.2; see also \cite{bk94,kbgnd94,col94}.
Some other non-Gaussian distributions are discussed in \cite{mmlm91,wc92}.

%% file: ssec4.tex
\section{Non--linear Approximations based on Dynamics}
Although local extensions of linear theory do provide a qualitative first
step in comparing theory with observations, a deeper insight into gravitational
clustering is provided by the dynamical approximations we consider now.
Depending upon the degree of nonlinear clustering (characterised by the
{\rm rms} density fluctuation $\sigma$), a particular
dynamical approximation may be best suited for describing gravitational
clustering at that epoch. In the weakly non-linear regime when $\sigma
\ll 1$ (also known as the quasi-linear regime), Eulerian perturbation theory
(\S 4.1) provides us with an excellent description of gravitational
clustering and of the accompanying departures from Gaussianity
which result if the initial density field was Gaussian.
At later epochs ($\sigma \le 1$) the Zel'dovich approximation and its
perturbative extensions (\S 4.2, \S 4.4) provide us with
a quantitative understanding of gravitational dynamics until the
formation of pancakes ($\sigma \simeq 1$).
Several approximations have been suggested to model gravitational instability
during the strongly non-linear regime when $\sigma > 1$
(\S 4.3, \S 4.5 -- \S 4.7).
Perhaps the most promising of these is
the adhesion model (\S 4.5) which provides us with a deep understanding of
gravitational clustering and describes the growth of cellular structure in
the form of pancakes, filaments and clumps, as well as the dynamics of voids.
A comparison of these different approximations is outlined in \S 4.7.
We conclude this section with a brief discussion of approximations
inspired by a relativistic approach to perturbations.

\subsection{Higher--Order Eulerian Theory}
The linear regime of gravitational instability breaks down when $\delta$
becomes comparable to unity marking the commencement of the {\it quasi-linear}
(or weakly non-linear) regime. During this regime the density contrast remains
small ($\delta < 1$), but the phases of the fourier components $\delta_{\bf k}$
become substantially different from their initial values
resulting in the gradual development of a non--Gaussian distribution
function if the primordial
density field was Gaussian \cite{brmg91,ss91,sms91,sod92}.
The most obvious way to attempt to extend linear theory into the
fully non--linear regime is via a higher--order perturbative expansion.

In this approach \cite{p80,f84,gw87,bjcp92,buch92,b92b,mss94,clmm94}
we write the density contrast as a superposition
of terms:
\begin{eqnarray}
\delta(\vec{x},t) & = & \sum_{n=1}^{\infty} \delta^{(n)} (\vec{x},t)
\nonumber\\
 & = & \sum_{i=1}^{\infty} D_{+}^n (t) \Delta^{(n)}(\vec{x}).
\label{eq:het1}
\end{eqnarray}
For simplicity we shall assume the case of a flat universe which
is matter--dominated. Perturbative methods for open universe models
are discussed in \cite{mf91,bjcp92}. The first term in the above
series expansion corresponds to the linear growth law:
$\delta^{(1)}=D_{+}(t) \Delta^{(1)}(\vec{x})$. The second--order
contribution can be written:
\begin{equation}
\delta^{(2)} = \frac{10}{14} (\delta^{(1)})^{2} + \vec{\nabla}
\delta^{(1)}\cdot \vec{\nabla} \phi^{(1)} + \frac{4}{14}
\sum_{i,j} (\nabla_i\nabla_j \phi^{(1)})^{2},\label{eq:het2}
\end{equation}
where $\nabla$ denotes the derivative with respect to
the Eulerian (comoving) co-ordinate $\vec{x}$ and $\phi^{(1)}$
is the linear gravitational potential. Higher--order terms
become rapidly more complicated as $n$ increases.
For distributions
which are initially Gaussian, the weakly non--linear regime
witnesses the gradual emergence of non--Gaussian properties
which can be described by the following hierarchical
relationship for the connected moments of the density field,
as explained in \S 8.3:
\begin{equation}
\langle \delta^{n} \rangle_{c} \simeq S_n \langle \delta^{2} \rangle^{n-1}.
\end{equation}
The subscript $c$ refers to the connected part of the
moment, i.e. that part which cannot be expressed in terms of lower--order
moments of $\delta$; see \S 8.
The coefficent $S_3$ describes the skewness and
$S_4$ the kurtosis, as mentioned in \S 8.5. For a Gaussian distribution,
$S_n=0$ for all $n\geq 3$.
The values of $n$ can be calculated exactly by means of an elegant
formalism introduced by Bernardeau \cite{b92b}. In this approach
one constructs {\em vertex weights} for the various tree diagrams, according to
\begin{equation}
\nu_n = \langle \delta^{(n)} \rangle_{c}
= \frac{\int \langle \delta^{(n)}(\vec{x},a)\delta^{(1)}(\vec{x}_1,a)
\ldots \delta^{(1)}(\vec{x}_n,a) \rangle_{c}\d^{3}\vec{x} \d^{3}\vec{x}_1
\ldots \d^{3}\vec{x}_n }{ \left( \int \langle \delta^{(1)}(\vec{x},a)\,
\delta^{(1)}(\vec{x}',a) \rangle_{c} \d^{3}\vec{x} \d^{3}\vec{x}' \right)^{n}
}.
\label{eq:z48}
\end{equation}
It is helpful to obtain the weights from the generating function:
\begin{equation}
G_{\delta} = \sum_{i=1}^{\infty} \frac{\nu_{n}}{n!} \tau^{n}.
\label{eq:z49}
\end{equation}
The $S_n$ can then be straightforwardly obtained from the $\nu_n$:
\begin{eqnarray}
S_3 & = & 3\nu_2\nonumber\\
S_4 & = & 4\nu_3 + 12 \nu_2^{2}\nonumber\\
S_5 & = & 5\nu_4 + 60\nu_3\nu_2 + 60 \nu_2^{3}\nonumber\\
S_6 & = & 6\nu_5 + 120 \nu_4\nu_2 + 90 \nu_3^{2} + 720\nu_3\nu_2^{2}
+ 360\nu_2^{4}.
\label{eq:z50}
\end{eqnarray}
Remarkably, the generating function $G$ can be shown to evolve according
the same equation as a spherical top-hat solution:
\begin{equation}
G_{\delta} =\left( \frac{\rho}{\bar{\rho}} - 1\right)
= \frac{9}{2} \frac{ (\theta-\sin \theta)^{2} }{(1-\cos\theta)^{3}} - 1,
\label{eq:z51a}
\end{equation}
where the time--coordinate is reparameterised according to:
\begin{equation}
\tau = \delta_0 \frac{a}{a_{0}} = \frac{3}{5} \left[ \frac{3}{4} \left(
\theta-\sin\theta \right) \right]^{2/3}.
\label{eq:z51b}
\end{equation}
Expanding $G$ as a power--series in $\tau$ leads to
\begin{equation}
G_{\delta} = \tau + 0.810\tau^{2} + 0.601\tau^{3} + 0.426\tau^{4}
+0.293\tau^{5} + \ldots,
\label{eq:z51c}
\end{equation}
which leads to a straightforward determination of the $\nu_n$ and thence
the $S_n$. The results for $S_3$ are the same as those obtained by
a more straightforward analysis by \cite{p80}: we get
$S_3=34/7=4.857$, in this case. However, this result is only valid only
if the density field is not smoothed.
For smoothed final density fields the value of $S_n$
depends both on the form of the window function and upon the slope of
the power spectrum at values of $k$ corresponding to the smoothing length.
For scale invariant spectra $P(k) \propto k^n$ with spherical top-hat smoothing
$S_3 = 34/7 - (n + 3)$ \cite{jbc93,b94b}. For Gaussian smoothing the expression
for $S_3$ is more complicated \cite{ljwb94}.
(Redshift-space distortions to $S_3$ and other moments are
discussed  in \cite{hbcj94}, see also \cite{cm94}.)
We shall discuss these results later, in \S 4.7, where we compare
the full perturbative solution for the hiearchical coefficients with the
corresponding results obtained for various approximation schemes. (The
function $G_{\delta}$ always obeys the equation for spherical collapse
so all we need to do is to solve that problem in a given approximation
and then calculate the $S_n$ as above.) Skewness and kurtosis of
velocity fields are discussed in \cite{jbc93,b94b,mss94}, and also in \S 8.10.

For small values of $\sigma \ll 1$, the
moments $S_n$ can be used to reconstruct the probability density function
(PDF)\footnote{The
PDF can be thought as the volume fraction occupied by a given
value of the density.}
 in the quasi-linear regime using the Edgeworth expansion
\cite{ks77,bk94,jwacb93}
\begin{eqnarray}
P(\delta)d\delta &=& \frac{1}{(2\pi\sigma^2)^{1\over 2}} \exp{(-\nu^2/2)}
\bigg\lbrack1 + \sigma{S_3\over 6}  H_3(\nu) + \sigma^2 \bigg ({S_4\over 24}
H_4(\nu)
+ {S_3^2\over 72}H_6(\nu)\bigg) \nonumber\\
&+& \sigma^3\bigg ({S_5\over 120} H_5(\nu) +
{S_3S_4\over 144}H_7(\nu) +
{S_3^3\over 1296}H_9(\nu)\bigg) + ...\bigg\rbrack d\delta,
\label{eq:pdf}
\end{eqnarray}
where $\nu = \delta/\sigma$ and $H_n(\nu)$ are the Hermite polynomials
\begin{equation}
H_n(\nu) \equiv (-1)^n \exp{(\nu^2/2)}\frac{d^n}{d\nu^n}\exp{(-\nu^2/2)}.
\label{eq:hermite}
\end{equation}
In general, the $\sigma$--dependence of $S_n$ must be taken into account
when applying equation (\ref{eq:pdf})\footnote{In the one dimensional case,
until shell crossing \cite{bk94} $S_3(\sigma) = 6 + 24\sigma^2 + O(\sigma^4)$;
$S_4(\sigma) = 72 + 810\sigma^2 + O(\sigma^4)$; which converge to $S_3 = 6$,
$S_4 = 72$ for infinitesimal values of $\sigma$. In three dimensions the
functional forms for $S_3(\sigma)$, $S_4(\sigma)$ etc. are not known;
results obtained using the Zel'dovich approximation are discussed in
\cite{bk94}.
For nominal values of $\sigma$, good agreement with the PDF obtained from
N-body simulations of the CDM model
is provided by the lognormal model (see \S 3.5) for which \cite{bk94}
$S_3(\sigma) = 3 + \sigma^2$, $S_4(\sigma) = 16 + 15\sigma^2 + 6\sigma^4
+\sigma^6$.}

The quantity in square brackets provides a correction to the initial
Gaussian distribution.
Since the Edgeworth form of the Gram--Charlier series is an asymptotic
expansion in $S_n$, one needs to know all the moments in order to
reconstruct the PDF. In practice however, the first two moments $S_3$
and $S_4$, suffice to determine the maximum of the PDF
to within a reasonable accuracy when $\sigma \le 0.5$ \cite{bk94}.
As a result, measurements of the shape of the PDF near its peak
(which is a statistically
robust quantity), may allow us to determine the lowest moments $S_3$ and
$S_4$ in the weakly non-linear regime when $\sigma \ll 1$. One should
be aware, however, that there are restrictions on the  possible behaviour
of the moments in order that the Edgeworth expansion converges.

Another useful way to employ perturbative techniques is by moving into
the Fourier domain. One can then obtain higher--order corrections to
such quantities as the power--spectrum. The approach we follow
here is that of \cite{jb94}; see also \cite{f84,ggrw86,mss92,f94}.
We begin by taking the Fourier transform of $\delta(\vec{x},\eta)$
as in \S 2;
$\eta$ is the conformal time coordinate $\d\eta=\d t/a(t)$:
\begin{equation}
\hat{\delta}(\vec{k},\eta) = \frac{1}{(2\pi)^{3}} \int \d^{3} \vec{x}
\exp( -i \vec{k} \cdot \vec{x} ) \delta(\vec{x},\eta).
\label{eq:z2_1}
\end{equation}
In linear theory, the velocity field is curl--free and is therefore
completely specified by knowledge of $\theta=\vec{\nabla}\cdot\vec{v}$.
With this definition, equations (\ref{eq:continuity}) and (\ref{eq:Euler})
can be Fourier transformed to yield:
\begin{equation}
\frac{\d\hat{\delta}}{\d\eta} + \hat{\theta} = -\int\d^{3} \vec{k}_1 \int
\d^{3}
\vec{k}_2 \delta_{D} (\vec{k}_1+\vec{k}_2-\vec{k}) \frac{\vec{k}\cdot\vec{k}_1}
{k_1^{2}} \hat{\theta}(\vec{k}_1,\eta) \hat{\delta}(\vec{k}_2,\eta)
\label{eq:z2_2}
\end{equation}
\begin{equation}
\frac{\d\hat{\theta}}{\d\eta} + \frac{\dot{a}}{a} \hat{\theta}
+\frac{6}{\eta^{2}} \hat{\delta} =
-\int \d^{3}\vec{k}_1 \int \d^{3} \vec{k}_2 \delta_D (\vec{k}_1+\vec{k}_2
-\vec{k}) \frac{k^{2} \vec{k}_1 \cdot \vec{k}_2}{2k_1^{2}k_2^{2}}
\hat{\theta}(\vec{k}_1,\eta)\hat{\theta}(\vec{k}_2,\eta),
\label{eq:z2_3}
\end{equation}
where $\delta_D$ is the Dirac delta function. Recall that the correct
definition of the power--spectrum is then:
\begin{equation}
\langle \hat{\delta}(\vec{k}_1,\eta) \hat{\delta} (\vec{k}_2,\eta) \rangle
= P(k_1,\eta) \delta_D (\vec{k}_1+\vec{k}_2).
\label{eq:z2_4}
\end{equation}
We now proceed as we did in real space by expanding $\hat{\delta}$ and
$\hat{\theta}$ as perturbation series:
\begin{equation}
\hat{\delta}(\vec{k},\eta) = \sum_{n=1}^{\infty} a^{n}(\eta) \hat{\delta}_n
(\vec{k}),
\label{eq:z2_5}
\end{equation}
\begin{equation}
\hat{\theta}(\vec{k},\eta) = \sum_{n=1}^{\infty} \dot{a}(\eta) a^{n-1}
\theta_n(\vec{k}).
\label{eq:z2_6}
\end{equation}
The general solution to eqs. (\ref{eq:z2_2}) and (\ref{eq:z2_3}) can
be written:
\begin{eqnarray}
\hat{\delta}_n(\vec{k}) & = & \int \d^{3} \vec{q}_1 \ldots \int \d^{3}
\vec{q}_n
\delta_D (\vec{q}_1 + \ldots \vec{q}_n -\vec{k}) F_n (\vec{q}_1 \ldots
\vec{q}_n)
\prod_i\hat{\delta}_i(\vec{q}_i)\nonumber\\
\hat{\theta}_n(\vec{k}) & = & \int \d^{3} \vec{q}_1 \ldots \int \d^{3}
\vec{q}_n
\delta_D (\vec{q}_1 + \ldots \vec{q}_n -\vec{k}) G_n (\vec{q}_1 \ldots
\vec{q}_n)
\prod_i\hat{\delta}_i(\vec{q}_i).
\label{eq:z2_7}
\end{eqnarray}
The functions $F_n$ and $G_n$ must be obtained from the recursion relations:
\begin{eqnarray}
F_n(\vec{q}_1\ldots\vec{q}_n) & = & \sum_{m=1}^{n-1}
\frac{G_m (\vec{q}_1 \ldots \vec{q}_n)}{(2n+3)(n-1)} \left[ (1+2n)
\frac{\vec{k}\cdot\vec{k}_1}{k_1^{2}} F_{n-m} (\vec{q}_{m+1}\ldots\vec{q}_{n})
\right. \nonumber\\
  & &  \left. + \frac{k^{2} \vec{k}_1\cdot\vec{k}_2}{k_1^{2}k_2^{2}}
  G_{n-m} (\vec{q}_{m+1}\ldots\vec{q}_n) \right]\nonumber\\
G_n(\vec{q}_1\ldots\vec{q}_n) &  =  &\sum_{m=1}^{n-1}
\frac{G_{m}(\vec{q}_1\ldots
\vec{q}_m)}{(2n+3)(n-1)} \left[ 3\frac{\vec{k}\cdot\vec{k}_1}{k_1^{2}}
F_{n-m} (\vec{q}_{m+1}\ldots \vec{q}_{n})\right.\nonumber\\
  & & \left.  + n\frac{k^{2} \vec{k}_1\cdot\vec{k}_2}{k_1^{2}k_2^{2}}
  G_{n-m} (\vec{q}_{m+1} \ldots \vec{q}_n) \right].
  \label{eq:z2_8}
  \end{eqnarray}
where $\vec{k}_1 \equiv \vec{q}_1 + \ldots \vec{q}_{m}$,
$\vec{k}_2 \equiv \vec{q}_{m+1} + \ldots \vec{q}_n$, $\vec{k}=\vec{k}_1+
\vec{k}_2$ and $F_1=G_1=1$. An expansion of this type is also given
by \cite{ggrw86}.

These expressions can be used to construct the higher--order
power spectrum by calculating the required number of terms
in the perturbation expansion for $\hat{\delta}$ and calculating
the required expectation values. For example, define
\begin{equation}
\langle \hat{\delta}_{m}(\vec{k}) \hat{\delta}_{n-m} (\vec{k}') \rangle
=P_{m,n-m}(k) \delta_D (\vec{k}+\vec{k}')
\label{eq:z2_9}
\end{equation}
and the power--spectrum to second order can be shown to be
\begin{equation}
P(k,\eta) = a^{2}(\eta) P_{1,1}(k) + a^{4}(\eta) [P_{2,2}(k)+2P_{1,3}(k)]
+ O(a^{6}).
\label{eq:z2_10}
\end{equation}
Notice that the power--spectrum to second order requires calculation
of the fourth--order terms in $\hat{\delta}(\vec{k})$. The required
integrals need to be performed numerically: for specific
examples see \cite{j81,v83,jsb84,c90,mss92,jb94,f94}.

For an application of perturbative methods to non--Gaussian linear
density fields, see \cite{fs93}. Perturbation series for a spherical
harmonic expansion are discussed by \cite{zh93}. Some aspects
of the perturbation expansion for open Universe models
are covered by \cite{mf91,clmm94}.

These methods are clearly useful (but a bit laborious) and can
extend calculations of statistical quantities such as the
power--spectrum and higher--order moments onto large scales
where they still show a non--linear signal but cannot be
reliably probed by computational methods. Nevertheless, there
are a number of problems which mean that one should look
for alternative methods to complement such perturbative techniques.
First, there is the problem of convergence. There is no proof
that the perturbation series (\ref{eq:het1}),
(\ref{eq:z2_5}) and (\ref{eq:z2_6}) actually
converge, nor do we know what the domain of convergence is if they
do. As soon as the second--order contributions to the power--spectrum
become important, so do the third-- and higher--order contributions
so one is never sure exactly how much to trust a calculation to
a given order. Having said that, this method does seem to give
results compatible with the $N$--body simulations
for a useful part of the parameter space \cite{jb94}.
Moreover, although useful for statistical information, the method
does not provide an insight into the geometrical properties
of the evolving distribution. For these reasons it is useful to
explore other dynamical approximations which might complement
the perturbative method.

\subsection{The Zel'dovich Approximation}
In 1970, Yakov B. Zel'dovich proposed an ingenious extrapolation of linear
gravitational instability into the nonlinear regime \cite{z70},
by following
perturbations in
the particle trajectories, rather than in the density field. In the
following, we will use ``particle'' and ``fluid element''
interchangeably, because we are applying this technique
to the case of a continuous medium rather than a discrete
set of particles. The Zel'dovich approximation
has the remarkably simple form:
\begin{equation}
\vec{r} = a\vec{x} (\vec{q}, t) = a(t) \left[
\vec{q} + D_+(t) \vec{u}(\vec{q} )\right],
\label{eq:z1a}
\end{equation}
where $\vec{x}$  is the final
(i.e. Eulerian) comoving coordinate of a particle initially located
at a Lagrangian co-ordinate $\vec{q}$. In eq. (\ref{eq:z1a}),
$a(t)$ is the  cosmic scale factor, and $D_+(t)$ is the growing mode of
the linear density contrast, defined in \S 2. We note that an
extension of the Zel'dovich
approximation which incorporates the decaying mode has been
discussed in \cite{b87}. The vector field
$\vec{u} (\vec{q} )$
is the initial velocity field of the particle
($\vec{u} (\vec{q} ) = \vec{v}/a\dot{D}_+ \equiv d\vec{x}/dD_+$).
We confine our attention to initially irrotational flows so that
$\vec{u}(\vec{q}) = - \vec{\nabla} \Phi_0(\vec{q})$, where
$\Phi_0(\vec{q})$ is the linear velocity potential, which is
related to the
Newtonian gravitational potential $\phi$ by $\phi = A^{-1}\Phi_0$ where
$A = 2/(3H^2a^3)$;
for a spatially flat matter dominated Universe $A$ is a
constant (see \S 2.2 and \cite{ko89,kof91}).
The Kelvin circulation theorem then ensures that the velocity flow remains
irrotational at least until orbit crossing. After this, there are regions
of multistream flow and, although each stream remains curl--free the net
velocity field is no longer potential.
Any initial rotation present in the velocity field is strongly damped during
expansion in the period before orbit-crossing.
Vorticity can however be generated in multistream flows within
pancakes, as shown in \cite{d73}. Attempts to incorporate vorticity
directly into the Zel'dovich approximation have been discussed
by \cite{buch92,bs93}. A generalisation of the
Zel'dovich approximation based on the principle
of least action is given in \cite{gmmy93}; see also
\cite{sb94}. An application of
the Zel'dovich approximation to cosmic string wakes has been discussed
in \cite{pbs90}. Relativistic generalisations of these approximations
are discussed in \S 4.8.

By reparametrising the time coordinate $\tau = D_+(t)$,
equation (\ref{eq:z1a}) can be
made to look just like the inertial motion of a particle moving with constant
velocity $\vec{u}$:
\begin{equation}
\vec{x}  = \vec{q} + \tau \vec{u}(\vec{q}).
\label{eq:z1}
\end{equation}
An essential feature of inertial motion given random initial conditions, is the
intersection of particle trajectories leading to the formation of
singularities in the density field.
A similar effect is predicted also by the Zel'dovich approximation, and can be
estimated quite simply by applying the continuity condition (mass conservation)
\begin{equation}
dM = \rho_0~ \d^3\vec{q} = \rho(\vec{x},t)  ~ \d^3\vec{x}.
\label{eq:z2}
\end{equation}
As a result we get the following expression for the density in
terms of the Jacobian of the
transformation from $\vec{q}$ to $\vec{x}$:
\begin{eqnarray}
\frac{\d V_e}{\d V_l} & = &
 J\left[ \frac{\partial \vec{x}}{\partial \vec{q}}\right]\nonumber\\
  & = &
\vert \delta_{ij} - D_+(t)
\frac{\partial^{2}\Phi_0}{\partial q_i\partial q_j}\vert\nonumber\\
  & = & \left[1 - D_+(t)\lambda_1(\vec{q})\right]
 \left[1 - D_+(t)\lambda_2(\vec{q})\right]\left[1 - D_+(t)\lambda_3(\vec{q})
 \right], \label{eq:zel3}
 \end{eqnarray}
where $dV_e \equiv \d^3\vec{x}$ and $dV_l \equiv \d^3\vec{q}$,
are the Eulerian and Lagrangian volume elements, respectively, and
$\lambda_1, \lambda_2$ and $\lambda_3$ are the eigenvalues of the
deformation tensor
\begin{equation}
d_{ij} \equiv \frac{\partial^{2} \Phi_0}{\partial q_i \partial
q_j}. \label{eq:z_s}
\end{equation}
The expression for the density distribution is therefore
\begin{eqnarray}
\rho(\vec{r}, t) & = &
\frac{\rho_0}{a^{3}} \left[1 - D_+(t)\lambda_1(\vec{q})\right]^{-1}
\left[1 - D_+(t)\lambda_2(\vec{q})\right]^{-1}
\left[1 - D_+(t)\lambda_3(\vec{q})\right]^{-1}
\nonumber \\
 & = & \frac{\rho_0}{a^{3}}
 (1 - D_+I_1 + D_+^2I_2 - D_+^3I_3)^{-1}, \label{eq:z4}
 \end{eqnarray}
where $ I_1 = \lambda_1 + \lambda_2 + \lambda_3$, $I_2 = \lambda_1\lambda_2 +
\lambda_2\lambda_3 + \lambda_1\lambda_3$ and
$I_3 = \lambda_1\lambda_2\lambda_3$ are the invariants of
$d_{ij}$. The eigenvalues of the deformation tensor define a
unique coordinate system
in which a parallelipiped preserves its original form upon deformation.

In general, the eigenvalues $\lambda_i$ can be either positive or negative.
A very useful analytical formula giving the joint probability distribution
of an ordered set of eigenvalues $\lambda_1 > \lambda_2
> \lambda_3$, corresponding to a Gaussian potential field, was
obtained by Doroshkevich \cite{d70},
who showed that
\begin{equation}
P(\lambda_1, \lambda_2, \lambda_3) =
\frac{5^{3}.27}{8\pi \sigma_{0}^{6}\sqrt{5}}
\exp \left( \frac{-3I_1^{2}+15I_2/2}{\sigma_0^{2}} \right)
(\lambda_1-\lambda_2)(\lambda_2-\lambda_3)(\lambda_1-\lambda_3),
\label{eq:z5}
\end{equation}
where $\sigma_0^{2}$ is the initial dispersion of the overdensity
field; $\sigma(t) = a(t)\sigma_0$ is the corresponding {\em rms}
linear density at time $t$ in an $\Omega_0=1$ Universe.

A consequence of this result for a gaussian random field $\Phi_0$ is that
the probability for one eigenvalue to be
positive and the remaining two to be negative is $42\%$ (the same as that for
two positive and one negative eigenvalue). A volume element
satisfying this criterion will collapse along
one direction, resulting in the formation of two dimensional
pancake-type singularities
in the density field, which arise when $1 - D_+(t)\lambda_1 = 0$;
see Fig. (\ref{fig:4zel}).

We take this opportunity to remark that in this respect the
Zel'dovich approximation is very similar to the equations
describing the propagation of light rays in geometric optics.
The formation of caustics -- alternating bright and dark regions
in the intensity of light -- is a well--known phenomenon in the
latter context; see \cite{zms83,sz89,kof91}. Furthermore,
the relationship between
the singularities occuring in the Zel'dovich approximation and Catastrophe
Theory has been explored in \cite{asz82,bruce86,a86,agv85,sz89}.

\begin{figure}
\vspace{15cm}
\caption{Caustics arising in the Zel'dovich approximation are shown for a
two dimensional realisation of eq. (\protect{\ref{eq:z1a}})
with random (Gaussian)
initial conditions.}
\label{fig:4zel}
\end{figure}

The situation in which two of the eigenvalues are positive and one negative
leads to the formation of one dimensional filaments, which typically form at
the intersection of pancakes. Collapse along all three directions
($\lambda_1, \lambda_2, \lambda_3 > 0$) has a much smaller probability of
occuring ($8\%$), and leads to the formation of zero dimensional singularities
-- clumps, which arise when two nearby filaments intersect.
The reverse situation when all three eigenvalues are negative, corresponds
to expansion along all
three directions and leads to the formation of voids. Properties
of voids in the  framework of the Zel'dovich approximation and in the adhesion
model are discussed in \S 5.

An interesting feature of the pancake model is the percolation of high
density regions which occurs generically in the Zel'dovich approximation.
For Gaussian initial conditions such a situation is rather unlikely
to occur since, in order for regions to form a percolating network it is
essential that their volume (area) exceed $16\%$ ($50\%$) in
3D (2D) \cite{sz89}. On the other hand,
in the Zel'dovich approximation,
we find that pancakes percolate despite the fact that they
occupy a negligible fraction of the total volume. The reason for this is
quite simple \cite{sz89}: pancakes, filaments and clumps
correspond to regions in Lagrangian space (L-space)
where $\lambda_1 > 0$. For a Gaussian
distribution such regions occupy roughly $92\%$ of the total volume in 3D
($79\%$ of the area in 2D). As a result such regions percolate in L-space.
Since the Zel'dovich approximation (\ref{eq:z1a}) is a topology--preserving
homomorphism,
the requirements of continuity guarantee that any such
mapping $\vec{q} \rightarrow \vec{x}$
which percolates in L-space will do so in Eulerian space
(E-space) as well.

The internal consistency of the Zel'dovich approximation can be assessed
by means of the following argument \cite{drs73}.
Differentiating
equation (\ref{eq:z1a})
twice with respect to time allows us to determine
the acceleration of the fluid element.
A dynamical estimate of the density $\rho_{dyn}$
is then provided by the Poisson equation (\ref{eq:Poisson}):
$\vec{\nabla}_r\ddot{\vec{r}} = - \nabla_r^2\tilde\phi = - 4\pi G\rho_{dyn}$.
If the Zel'dovich approximation were exact, the values of
$\rho_{dyn}$ determined dynamically from the Poisson equation and
$\rho$ calculated using the continuity conditions
(\ref{eq:z2}--\ref{eq:z4})
would coincide. The fractional difference
$(\rho_{dyn} - \rho)/\rho \equiv \tilde\delta$ therefore
provides us with a measure of the accuracy of the Zel'dovich approximation.
Performing this calculation we find that
\begin{equation}
\tilde\delta = -~ D_+^2 I_2 + 2 D_+^3 I_3, \label{eq:zel5}
\end{equation}
where $I_2$ and $I_3$ are the invariants of the deformation tensor introduced
earlier.

{}From the above form for the fractional error $\tilde\delta$ it is clear that
$\tilde\delta = 0$ if $\lambda_2 = \lambda_3 = 0$,
i.e. the Zel'dovich approximation is exact in one dimension, and in higher
dimensional situations when the collapse is planar.
The generic nature of pancake formation has been strikingly demonstrated
for a wide variety of initial conditions by N-body simulations, e.g.
\cite{ms89}.
It is worth noting that the error term $\tilde\delta$ remains finite even
during
pancake formation, when $D_+(t) \simeq \lambda_1^{-1}, $ and $\rho \rightarrow
\infty$. It is useful to call
$\rho_{dyn}$ the {\em dynamical} density
and  $\rho$ the {\em  continuity} density.
Momentum conservation leads to $\rho_{dyn}$,
mass conservation to $\rho$.
For laminar flows the Zel'dovich approximation can be suitably inverted and
expressed in Eulerian space as
$\vec{q}(\vec{x}) = \vec{x} - D_+\vec{u}[\vec{q}(\vec{x})]$,
the continuity equation then yields \cite{ndbb91} :
\begin{equation}
\delta = \big\vert\frac{\partial\vec{q}}{\partial \vec{x}}
\big\vert - 1
= - D_+\tilde I_1 + D^2_+\tilde I_2 - D^3_+\tilde I_3,
\label{eq:z6}
\end{equation}
where $\tilde I_1, \tilde I_2, \tilde I_3$ are the invariants of the
deformation tensor
$\partial\vec{u}(\vec{x})/\partial\vec{x}$ evaluated
in Eulerian space. The POTENT procedure was used in \cite{ndbb91}
to determine $\vec{u}(\vec{x})$; then eq. (\ref{eq:z6}) was used
to reconstruct
the form of the density field in the weakly
non-linear regime.

Evaluating the acceleration according to the
Zel'dovich approximation (\ref{eq:z1a}) and substituting its value
in the Euler equation  (\ref{eq:Euler}),
we obtain $\delta_{dyn} = \rho_{dyn}/\bar{\rho} - 1 = - \vec{\nabla}
\vec{v}/H f a$ $(\vec v = a\dot{D}_+\vec u)$
which, surprisingly, is also
just the linear velocity - density relation (\ref{eq:l9}).
It can be shown \cite{ndbb91}
that in general, $\delta_{dyn}$ overestimates the real value
of the density, whereas $\delta$ tends to underestimate it, so that
$(\delta + \delta_{dyn})/2$ often provides a better
estimate of the actual density contrast than either $\delta$ or $\delta_{dyn}$.

At relatively early times when the density contrast
is still linear $D_+(t) < 1$,
the higher order terms in eq. (\ref{eq:z4})
 -- $b^2 I_2$ and $b^3 I_3$ -- can be ignored
relative to $b I_1$, with the result that eq. (\ref{eq:z4})
reduces to
\begin{equation}
\rho(\vec{x}, t) \simeq \frac{\rho_0}{a^{3}}(1 + D_+ I_1)
\end{equation}
or, more usefully,
\begin{equation}
\frac{\delta \rho}{\rho} \simeq ~D_+(t) I_1 = D_+(t)
(\lambda_1 + \lambda_2 + \lambda_3) =  D_+(t)
\nabla^2\Phi_0.
\label{eq:z7}
\end{equation}
Since $\Phi_0 \propto \phi$,
we recover the linearised Poisson equation.  Comparing
eq. (\ref{eq:z7}) with eq. (\ref{eq:z4}), we find that
the predictions of linear theory and the
Zel'dovich approximation (ZA) as to the  formation sites of the first nonlinear
objects are somewhat different.
Whereas the Zel'dovich approximation predicts that
the first pancakes will collapse near high peaks of the largest eigenvalue
$\lambda_1$, linear theory predicts that the first nonlinear objects will
form near the maxima of
$\delta\rho/\rho  \propto (\lambda_1 + \lambda_2 + \lambda_3)$.
For very asymmetric configurations of the primordial density field
($\lambda_1 >> \lambda_2, ~~\lambda_3$), the results of ZA and linear theory
are
expected to correlate, however important differences can also arise especially
in regions where some of the eigenvalues are negative. For instance
regions in which $\lambda_1 > 0$ and $\lambda_2, \lambda_3 \ll 0$ so that
$\sum_i\lambda_i < 0$, will remain forever underdense according to linear
theory, yet would correspond to sites for pancake formation according to ZA.

An interesting feature of one dimensional pancake collapse is that
the surface density of pancake matter remains finite even as the
density within pancakes approaches infinity.
(For planar collapse the density profile in the vicinity of a pancake located
at $\vec{x}_1$ is $\rho \propto \vert \vec{x} - \vec{x_1}\vert ^{-{1\over
2}}$.)
This leads to finite, well defined values for both the gravitational
potential and the gravitational
force field outside pancakes. A discussion
of the physics at the sites of caustic formation
can be found in  \cite{dss78,bcsw84,ss-m84,zn83}.

Using eq. (\ref{eq:z5}) in conjunction with eq. (\ref{eq:z4})
it is possible to derive an
expression for $P(\rho, a)$ -- the probability distribution function
for the density (PDF)
in the Zel'dovich approximation \cite{kof91,kbgnd94,bk94}:
\begin{equation}
P(\tilde\rho, t) = \frac{9\times 5^{3/2}}{4\pi\tilde{\rho}^3 N_s \sigma^4}
\int_{3 \tilde{\rho}^{-1/3}}^{\infty}
\d s e^{-{(s - 3)^2\over 2\sigma^2}} (1 + e^{-{6s\over \sigma^2}})
(e^{-{\beta_1^2\over 2\sigma^2}} + e^{-{\beta_2^2\over 2\sigma^2}} -
e^{-{\beta_3^2\over 2\sigma^2}}),
\label{eq:z8}
\end{equation}
where $\sigma$ is the variance, defined by
\begin{equation}
\sigma^2(R,t) = 4\pi D_+^2(t)\int{P(k)W^2(kR)k^2 dk},
\end{equation}
and
\begin{equation}
\beta_n(s) = \sqrt{5}~s\left(\half + \cos\left[{2\over 3}(n-1)\pi
+ {1\over 3} \arccos(54/\tilde{\rho} s^3 - 1)\right]\right),
\end{equation}
in which $\tilde\rho = \rho/\bar{\rho}$ and
$N_s$ is the mean number of streams in the flow; $N_s = 1$ in the single
stream regime indicating the absence of shell crossing.

For small values of the variance $\sigma  \ll 1$ we recover the linear result
\begin{equation}
P_0(\rho) = {1\over \sqrt {2\pi}\sigma} \exp
\left[ -{(\rho - \bar\rho)^2\over
2\sigma^2}\right].
\label{eq:z9}
\end{equation}
The Gaussian density distribution described by equation (\ref{eq:z9})
is valid  only
at early times when departures from the mean density are small. At late times
$\delta\rho/\rho$
can become as large as we like in regions where $\lambda_1 > 0$.
In regions where $\lambda_1 <0$ (i.e. voids) on the other hand,
the density is always bounded from below by $\rho > 0$,
so that $\delta\rho/\rho  \ge - 1$.
As a result, the initially symmetric distribution (\ref{eq:z9})
gradually develops an asymmetry
and the probability distribution function, equation (\ref{eq:z8})
grows progressively more non--Gaussian with time; see Figure (\ref{fig:4pdf}).
It is interesting to note that for large densities
$\rho >> \bar{\rho}$ eq. (\ref{eq:z8})
has the asymptotic form $P(\rho, t) \propto \rho^{-3}$. Since the density
distribution near a pancake will depend upon the
distance `$x$' to the pancake we get
$P(\rho)d\rho
\propto dx$ which gives $\rho \propto x^{-{1/2}}$,
which correctly reproduces the density profile in the vicinity of pancakes.
A discussion of the density PDF in ZA after final smoothing
can be found in \cite{bk94}; also see \cite{ps93}.

\begin{figure}
\vspace{15cm}
\caption{The probability density function for the lognormal
(solid line) and ZA (dashed line) for three different smoothing
lengths: $R_s = 18, 12, 6h^{-1}$Mpc.,
corresponding to $\sigma = 0.11, 0.26, 0.55$.
Filled regions correspond to the results of $N$--body simulations of a
standard CDM model.
Reproduced, with permission, from \protect{\cite{kbgnd94}}.}
\label{fig:4pdf}
\end{figure}

For small $\vert\rho - \bar{\rho}\vert$, it can be shown
that \cite{kbgnd94}
the probability density function resembles the log-normal distribution
\cite{cj91}.  This is demonstrated in
Fig. (\ref{fig:4pdf}), where
the probability distribution function for the density is shown both for the
lognormal distribution and the Zel'dovich approximation.
When we compare these two approximations with $N$--body dynamical simulations,
we find that for small values of the smoothing length corresponding to
$\sigma \le 0.26$
ZA models the $N$--body results somewhat better than
the lognormal distribution, whereas for larger values of $\sigma$
the reverse is true \cite{kbgnd94}.

In contrast to the density distribution, the one--point velocity distribution
function retains its original Gaussian form to a  large extent,
even in the non-linear regime, until orbit crossing.
{}From the observational viewpoint, the probability distribution functions for
the density and the velocity could be used to statistically discriminate
between initially non-Gaussian distributions (such as those predicted by
string/texture models of galaxy formation), and distributions that evolved
from Gaussian initial conditions. Preliminary results based on the POTENT
reconstruction (\S 8.10) and the IRAS survey (\S 8.1) have been obtained
in \cite{kof91,kbgnd94}; they indicated that models with Gaussian
initial conditions are compatible with present observations.

The Zel'dovich approximation also leads to interesting evolution equations
for the gravitational and velocity potentials in the non-linear regime.
To demonstrate this,
we assume that the only operative mode of gravitational instability is
the growing mode. In linear theory this guarantees that the velocity field
is irrotational. Requiring this to be the case
in the non-linear regime as well
(this is in fact guaranteed by Kelvin's circulation theorem
until shell crossing) and
rewriting the Euler equation and the Bernoulli equations
(\ref{eq:Euler}) and (\ref{eq:Berno}) in terms
of the comoving velocity $\vec{u} = (\vec{v}/\dot{a}a) = \d\vec{x}/\d a$
and the comoving velocity potential $\vec{u} = - \vec{\nabla} \Phi$,
$\Phi = \phi_v/(a^2\dot{a})$, we obtain (for a pressureless medium)
\begin{equation}
{\partial\vec{u}\over \partial a} +
(\vec{u}\vec{\nabla}_x)\vec{u} =
- {3\over 2 a}\left[A\vec{\nabla}_x\phi + \vec{u}\right]
\label{eq:z10a}
\end{equation}
\begin{equation}
{\partial \Phi\over \partial a} - \half (\nabla_x\Phi)^2 =
{3\over 2 a} (A\phi - \Phi),
\label{eq:z10b}
\end{equation}
where $A =  (3H^{2}a^{3}/2)^{-1}$. The quantity
$A$ does not depend upon time if the Universe is flat and matter dominated.
Similarly the continuity and Poisson equations become
\begin{equation}
{\partial \eta\over \partial a} + \vec{\nabla}_x (\vec{u}\eta) = 0
\label{eq:z10c}
\end{equation}
\begin{equation}
\nabla^2\phi = {\Omega \delta\over A~a},\label{eq:z10d}
\end{equation}
where $\eta = a^3\rho$ is the comoving density.
As mentioned earlier, the expression for the dynamical density in the
Zel'dovich approximation reduces to the linear expression:
$\delta_{dyn} = - \vec\nabla\vec v/(H f a)$ \cite{ndbb91}, when combined
with the Poisson equation (\ref{eq:z10d}) this leads to the following
relationship between the velocity and gravitational potentials (in a
spatially flat matter dominated Universe): $\Phi = A\phi$
which is equivalent to saying that the peculiar fields for the
velocity and the acceleration are parallel to one another.
(In the one dimensional case, $\Phi = A\phi$ is exactly
true even in
the nonlinear regime. As a result the Zel'dovich {\em ansatz}
ceases to be
an ``approximation'' and provides an exact treatment of nonlinear gravitational
instability in one dimension, until particle trajectories cross).
Substituting this result into the Euler and Bernoulli equations
(\ref{eq:Euler}) and (\ref{eq:Berno}), we obtain:
\begin{equation}
{\partial\vec{u}\over\partial a} + (\vec{u} \vec{\nabla})\vec{u} = 0;
\label{eq:z11a}
\end{equation}
\begin{equation}
{\partial \Phi\over \partial a} - \half (\nabla_x\Phi)^2 = 0;
\label{eq:z11b}
\end{equation}
\begin{equation}
{\partial \phi\over \partial a} - \half (\nabla_x\phi)^2 = 0.
\label{eq:z11c}
\end{equation}
Equation  (\ref{eq:z11a}) gives $\d u/\d a = 0, ~(\d/d a =
\partial/\partial a + \vec{u} \vec{\nabla}_x)$, implying that the
particle motion is inertial.
The equation for the velocity potential $\Phi$ has the following interesting
analytical solution:
\begin{equation}
\Phi (\vec{x}, a) = \Phi_0(\vec{q}) - {(\vec{x} - \vec{q})^2\over 2a}
\label{eq:z12}
\end{equation}
The gravitational potential evolves according to an identical equation.
Differentiating
eq. (\ref{eq:z12}) alternately with respect to $~\vec{x}$ and $~\vec{q}$, we
finally get
\begin{eqnarray}
\vec{x} &  = & \vec{q}  - a(t)\vec{\nabla}_q\Phi\nonumber\\
\vec{u} &  = & - \vec{\nabla}_x\Phi(\vec{x}, a)  =
- \vec{\nabla}_q\Phi_0(\vec{q}),
\label{eq:z13}
\end{eqnarray}
recovering the Zel'dovich approximation.
The fact that the velocity is always parallel to the acceleration vector
makes particles move along straight lines in the Zel'dovich approximation.

It is interesting that the evolution equation for the potential (\ref{eq:z11b})
and its solution (\ref{eq:z12})
allow one to reconstruct the primordial form of the velocity potential
$\Phi_0$ from its present value $\Phi$. (The same applies also to
the gravitational potential $\phi$.) Attempts to do this using the
POTENT procedure have been discussed by
\cite{nd92}. This involves an integration of the
Zel'dovich--Bernoulli equation (\ref{eq:z11b}).

As demonstrated recently by Gramann \cite{g93b} a somewhat different
evolution equation for the gravitational and velocity potentials would
result if we were to substitute the expression for the
continuity density in the Zel'dovich approximation (\ref{eq:z6}) into the
Eulerian continuity equation (\ref{eq:z10c}). The relationship between
the gravitational and velocity potentials would then become
\begin{equation}
\Phi = \phi - a(t)C_g
\label{eq:g1}
\end{equation}
(we put $A = 1$ for simplicity), where the function $C_g$ is determined from
\begin{equation}
\nabla^2C_g = \sum_{i=1}^2\sum_{j=i+1}^3\left[{\partial^2\phi
\over\partial x_i^2}{\partial^2\phi\over\partial x_j^2} -
({\partial^2\phi\over\partial x_i\partial x_j})^2\right]
\label{eq:g2}
\end{equation}
the corresponding evolution equation for the gravitational potential would then
be modified to
\begin{equation}
{\partial\phi\over\partial a} - {1\over 2}(\nabla_x\phi)^2 = C_g.
\label{eq:g3}
\end{equation}
The Zel'dovich-Continuity equation (\ref{eq:g3}) appears to be an improvement
over the Zel'dovich-Bernoulli equation (\ref{eq:z11b}) when tested against
N-body simulations \cite{g93b}. (One can replace $a(t)$ by $D_+(t)$ for greater
generality in the above equations.)

We would like to conclude this discussion of the Zel'dovich
approximation by
mentioning that,
from the  point of view of an observer, the peculiar velocity flow
associated with overdense regions can give rise to redshift space caustics.
Such caustics arise when overdense regions cease expanding along our line
of sight and begin to contract.
Investigating this effect in the linear regime, Kaiser \cite{k87}
showed that for a
spherical density enhancement, redshift caustics occur just inside  the
turnaround radius.
Some insight into this effect can also be obtained using the Zel'dovich
approximation.
Differentiating eq. (\ref{eq:z1a})  with respect to time we get
\begin{equation}
\vec{V} = {\d\vec{r}\over \d t} = H\vec{r}  - a\dot D_+\vec{\nabla}
\Phi.
\label{eq:z14}
\end{equation}
The second term in the right hand side of eq. (\ref{eq:z14})
indicates the departure from
the pure Hubble flow: $\vec{V}_{H} = H\vec{r}$. Aligning our
coordinate system so
that the radial direction $\vec{r}$ is along the $z$-axis, we obtain the
following expression for the redshift distance to the particle
(or galaxy)
\begin{equation}
\vec{z} = {\vec{V} \over H} = \vec{r}_3 - f~a(t)~D_+(t)~(\vec{\nabla}
\Phi)_3
= a\vec{q}_3 - (1 + f)~a(t)~D_+(t)~(\vec{\nabla} \Phi)_3,
\label{eq:z15}
\end{equation}
where $f$ is the usual growth factor, $f \simeq \Omega_0^{0.6}$ (see \S 2.2
eq. (\ref{eq:9b})).
Following the deformation of a unit volume element from the Lagrangian
coordinate system ($q_1, q_2, q_3$), to the modified Eulerian coordinate
system ($r_1, r_2, z = \vec{V}/H$), and using the continuity equation,
we get \cite{sz87}
\begin{equation}
\rho (\vec{r}, t) = \frac{\rho_0}{a^{3}} \mbox{ det}\,
\left( \begin{array}{ccc}
	 1-D_{+} d_{11} & -D_{+}d_{12} & -D_{+} d_{13}\\
	 -D_{+} d_{12} & 1-D_{+} d_{22} & -D_{+} d_{23}\\
	 -(1+f) D_{+}d_{13} & -(1+f)D_{+}d_{23} & 1-(1+f)
	 D_{+} d_{33} \end{array} \right),
\label{eq:z16}
\end{equation}
where $d_{ij}$ denotes the deformation tensor,
defined by eq. (\ref{eq:z_s}).
In the distinctive case when the eigenvectors corresponding to the eigenvalues
$\lambda_1, \lambda_2, \lambda_3$ are aligned along the line of sight, equation
(\ref{eq:z16}) simplifies to give
\begin{equation}
\rho (\vec{r}, t) =
\frac{\rho_0}{a^{3}} (1 - D_+\lambda_1)^{-1}(1 - D_+\lambda_2)^{-1}
(1 - D_+(1 + f)\lambda_3)^{-1}.
\label{eq:z17}
\end{equation}
{}From eq. (\ref{eq:z17})
it follows that a redshift singularity along the line of sight
will arise when $D \equiv D_z = [(1 + f)\lambda_3]^{-1} \simeq
[(1 + \Omega_{0}^{0.6})\lambda_3]^{-1}$,
since $D_z < D_* = 1/\lambda_3$, a line of sight singularity
will always precede the formation of a real density singularity.
This is a simple consequence of the fact that collapse must be
preceded by ``turnaround'', i.e. the expansion must stop before the
contraction can begin.
The overdensity of matter at the time of formation of the
redshift singularity
can be obtained from eq. (\ref{eq:z4}),
in particular for $\Omega_0=1$, $D_z = 1/2\lambda_3$
and $\rho_z \simeq \rho_0/ a^3 (1 - D_z \lambda_3)
= 2\rho_0/ a^3$, giving
$\delta \simeq 1$, which corresponds to the extrapolated
linear overdensity of a spherical mass fluctuation at turnaround.

Although the Zel'dovich approximation is extremely accurate in predicting the
density field until the formation of pancakes and other caustics, its validity
breaks down soon after shell crossing. Improvements and extensions of the
Zel'dovich approximation can be broadly divided into two main categories:
\begin{itemize}
\item
those that like ZA are formulated within the Lagrangian
formalism;
\item those which have been conceived within the Eulerian framework.
\end{itemize}
To the former belong the truncated Zel'dovich approximation and the
adhesion model \cite{gss85,gss89}
as well as approximations based on
Lagrangian perturbation theory \cite{mabpr91,buch89,buch92,bjcp92}.
To the latter -- the so-called {\em frozen flow} and {\em linear potential}
approximations \cite{mlms92,bsv93,bp94}.

\subsection{The Truncated Zel'dovich Approximation (TZA)}
An important feature of hierarchical clustering is that not all
scales are in the non-linear regime at any given epoch.
The simplest extension of the Zel'dovich approximation -- the truncated
Zel'dovich
approximation (TZA) -- is  based upon this fact, and the observation that
modes contributing to the
breakdown of ZA are those that have gone nonlinear at a given epoch.
By selectively filtering all modes with $k \ge k_{NL}$ (where $k_{NL}$ is
epoch--dependent), we can prevent shell crossing and the
consequent ``smearing out'' of
caustics that occurs in the straightforward ZA \cite{cms93,kpsm92,mps94}.

If one chooses the
criterion ${\delta\rho/\rho} \simeq 1$ to mark the onset
of non-linearity then the non-linear length scale $k_{NL}^{-1}$
can be determined from the equality
\begin{equation}
\sigma(R,t) \equiv \langle ({\delta\rho/\rho})^2\rangle^{1\over 2}
= D_+(t)\left(4\pi\int_{k_H}^{k_{NL}} P(k) k^2 dk\right)^{1\over 2} = 1,
\label{eq:z17a}
\end{equation}
where $k_H = 2\pi a(t)/\lambda_H$ is the comoving horizon
scale, $P(k)$ is the power spectrum. For power law spectra $P(k) \propto k^n$
\begin{equation}
k_{NL}^{-1} \propto D_+^{2/(n+3)}(t).
\end{equation}
Clearly an increase in $D_+(t)$ will be accompanied by an increase in $k_{NL}
^{-1}$, indicating that, as the Universe evolves, successively
longer wavelength modes go nonlinear.

TZA has been compared with ZA and N-body
simulations \cite{cms93,mps94,ssmpm94}. The result
of these studies is that
TZA performs considerably better than ZA,
particularly at late times and for
primordial spectra with more large scale power,
when the evolved matter distribution is compared
point--by--point with a full $N$--body simulation
with the same initial conditions. The form
of filter which optimises TZA seems to be
a Gaussian window: $\exp [- k^2/2k_G^2]$ ($k_{NL} \le
k_G \le 1.5 k_{NL}$) applied to the initial power spectrum $P(k)$
\cite{mps94}. The success of TZA may at least partially, be attributed
to the fact that the evolved gravitational potential resembles the {\em
smoothed} initial potential much more so than it does the unsmoothed initial
potential.
As a result although small scale features are wiped out in TZA,
large scale features
are accurately described by it, since particles are made to move in a
smoothed potential which closely resembles the true {\em evolved} potential
at any given time \cite{pm95,melp95}.
An application of TZA to the theory of the clustering
of Abell clusters is described in \cite{bcm94}.

An interesting application of TZA for determining the typical mass
displacement of objects $d_{rms}$ was suggested by Shandarin \cite{s94}.
{}From ZA (\ref{eq:z1a}) one finds
\begin{eqnarray}
d_{rms} & \equiv & \langle\left(\vec x - \vec q\right)^2\rangle^{1\over
2} =
D_+(t)\vec u_{rms}\nonumber\\
\vec u_{rms} & = & \langle\vec u(\vec q)^2\rangle^{1\over 2}.
\label{eq:z17b}
\end{eqnarray}
Smoothening the original velocity field on non-linear length scales
$k > k_{NL}$ we get
\begin{equation}
\vec u_{rms}^2(k_{NL}) = 4\pi\int_{k_H}^{k_{NL}}P(k) dk.
\label{eq:z17c}
\end{equation}
Combining eqns. (\ref{eq:z17a}), (\ref{eq:z17b}) and (\ref{eq:z17c})
gives
\begin{equation}
d_{rms}^2(k_{NL}) = {\int_{k_H}^{k_{NL}}P(k) dk\over
\int_{k_H}^{k_{NL}}P(k) k^2 dk}.
\label{eq:z17d}
\end{equation}

It is interesting that $d_{rms}$ can sometimes be greater than the
non-linear length scale $k_{NL}^{-1}$. This can be clearly seen for
power-law spectra with sufficient large scale power for which
(\ref{eq:z17d}) reduces to
\begin{equation}
d_{rms}(k_{NL}) = \left(- {n+3\over n+1}\right)^{1\over
2}\left({k_H\over
k_{NL}}\right)^{n+1\over 2} k_{NL}^{-1} \,\,\,\,\, -1 > n > -3.
\label{eq:z40e}
\end{equation}
Thus clumps (with a mass given by (\ref{eq:z39})) can move coherently
travelling a distance $\sim d_{rms} >> k_{NL}^{-1}$ for $n < -1$.
As pointed out by Shandarin \cite{s94} this might provide an explanation
for
the presence of large coherent structures on
scales which are still evolving according to linear theory in
cosmological scenario's such as CDM. It is
interesting to note in this context that the scale at which the phases
of $\delta_k$ can depart significantly from their original values is
close to $d_{rms}$ \cite{brmg91,s94}. Comparatively little is
known about the precise behaviour of the phases of the Fourier
components, however, and this is one of the big gaps in our knowledge
of the clustering process \cite{p80,ss91,sms91,sod92}.

\subsection{Lagrangian Perturbation Theory}
The great success of ZA
is at least partly due to its being formulated
in Lagrangian space rather than in Eulerian space.
This allows one to follow
particle trajectories and the accompanying evolution of overdensities for a
much longer period of time in ZA -- effectively until shell crossing when
$\delta \rightarrow \infty$ -- than was possible in an Eulerian framework
which usually breaks down when $\delta \sim  1$.
Pursuing this logic further, a general Lagrangian perturbation approach
in which, as in ZA, the fundamental
perturbed quantity is taken to be the displacement field
rather than the overdensity might be expected to be more accurate
\cite{mabpr91,bjcp92,buch89,buch92,buch93,buch94,be93,l93,jbc93,g93a,mss94,bchj94,cat94}.

To demonstrate some aspects of this approach let us consider the dynamical
behaviour of particle trajectories in Lagrangian space.
We assume for simplicity
that the Universe is spatially flat and matter dominated. Extensions of our
argument to a curvature dominated Universe may be found in
\cite{bjcp92}.
As in ZA, we assume that the comoving
Eulerian coordinate of a particle $\vec{x}$ and its comoving
Lagrangian coordinate $\vec{q}$ are related by a displacement
field $\vec{\Psi}$:
\begin{equation}
\vec{x} = \vec{q} + \vec{\Psi}(t,\vec{q}).
\label{eq:z19}
\end{equation}
Introducing the displacement matrix
\begin{equation}
M_{ik}(t, \vec{q})={\partial x_i\over \partial q_k}=\delta_{ik}+
{\partial \Psi_i\over \partial q_k}
\label{eq:z20}
\end{equation}
we find that $M_{ik}$ satisfies the equations
\begin{equation}
{\partial \over \partial t}\left(a^2{\partial M_{ik}\over \partial t}
\right) M^{-1}_{ki}+{2a^2\over 3t^2}(J^{-1}-1)=0,
\label{eq:z21a}
\end{equation}
\begin{equation}
\epsilon_{ikl}\dot{M}_{km}M^{-1}_{ml}=0,
\label{eq:z21b}
\end{equation}
where $J=|\mbox{ det}  M_{ik} |$ is the Jacobian of the
transformation
eq. (\ref{eq:z19}), and $\epsilon_{ikl}$ is the Levi-Civita tensor.
Equation (\ref{eq:z21b}) is the condition for
potential motion in Eulerian space \cite{zn83}.
The resulting density and velocity fields can be determined from
\begin{equation}
{\rho\over \rho_0}\equiv \delta + 1 =J^{-1},~~~~\vec{u} =
a\left({\partial \vec{x}\over \partial t}\right)_{\vec{q}}= a\dot
{\vec{\Psi}}.
\label{eq:z22}
\end{equation}
In the weakly nonlinear regime, equations (\ref{eq:z21a}) \& (\ref{eq:z21b})
can be solved by
expanding $\vec{\Psi}$ in a series
$\vec{\Psi} = \vec{\Psi}^{(1)} + \vec{\Psi}^{(2)} + ..$; higher orders
in $\vec{\Psi}^{(n)}$ ($n>1$) being related to lower orders via an iterative
procedure \cite{mabpr91,buch92,l93}.
In a spatially flat matter dominated Universe $\vec{\Psi}^{(n)}$
are separable
at every perturbative order
\begin{equation}
\Psi_i^{(n)}=D_+^n(t)\psi_i^{(n)}(\vec{q}),~~~~
D_+(t) \propto a(t),
\label{eq:z23}
\end{equation}
the lowest order solution being the Zel'dovich approximation
\begin{equation}
x_i = q_i + D_+(t)\psi_i^{(1)}(\vec{q}),~~~~
\psi_i^{(1)}= -{\partial \Phi(\vec{q})\over \partial q_i}.
\label{eq:z24}
\end{equation}
The second order terms $\vec{\Psi} =\vec{\Psi}^{(1)}+\vec{\Psi}^{(2)}$
give rise to what we might call the {\em post-Zel'dovich approximation}
(hereafter PZA) where
\begin{equation}
\psi_{i,i}^{(2)} = -{3\over 14}\left[(\psi_{i,i}^{(1)})^2
- \psi_{i,j}^{(1)}\psi_{j,i}^{(1)}\right],
\label{eq:z25a}
\end{equation}
\begin{equation}
\psi_{i,j}^{(2)} = \psi_{j,i}^{(2)},
\label{eq:z25b}
\end{equation}
Derivatives are defined the with respect to the Lagrangian
coordinate $\vec{q}$; summation
over repeated indices is assumed.
As a result,
\begin{equation}
x_i = q_i + D_+(t)\psi_i^{(1)}(\vec{q}) +
D_+^2(t)\psi_i^{(2)}(\vec{q}).
\label{eq:z26}
\end{equation}
The third--order Lagrangian approximation (or the {\em post-post-Zel'dovich
approximation}, hereafter PPZA) results in
$\vec{\Psi}=\sum_{n=1}^3\vec{\Psi}^{(n)}$ where
\begin{equation}
\psi_{i,i}^{(3)}=  -{5\over 9}(\psi_{i,i}^{(2)} \psi_{j,j}^{(1)}
- \psi_{i,j}^{(2)} \psi_{j,i}^{(1)}) - {1\over 3}\mbox{ det}
\left[\psi_{i,j}^{(1)}
\right],
\label{eq:z27a}
\end{equation}
\begin{equation}
\psi_{i,j}^{(3)} - \psi_{j,i}^{(3)} = {1\over 3}\left(\psi_{i,k}^{(2)}
\psi_{k,j}^{(1)} - \psi_{j,k}^{(2)} \psi_{k,i}^{(1)}\right)
\label{eq:z27b}
\end{equation}
\cite{jbc93,buch94}.
It should be mentioned that, beginning with the third order approximation,
the particle displacement field is no longer potential in
higher order Lagrangian approximations.
(In one dimension $\psi^{(3)} = \psi^{(2)} = 0$, a consequence of the fact
that the Zel'dovich approximation is an exact solution to the equations of
motion (until shell crossing) in this case.)

Many of the results originally derived for the Zel'dovich approximation have
recently been extended to higher order Lagrangian theories and especially to
the post-Zel'dovich approximation (PZA)
\cite{buch93,be93,jbc93,l93,g93a,mss94}.
Spherical collapse in Lagrangian theories has
been investigated in \cite{mss94}; see \S 4.7. A treatment suitable for
non--flat Universes is presented in \cite{cat94}.
The moments of the density distribution have been calculated in
\cite{bjcp92,mss94,hbcj94}; see \S 4.7.
One can also incorporate the PZA into the POTENT reconstruction
{\em ansatz} by deriving an analytical form for the density-velocity
relation in PZA \cite{g93a}.
Higher--order Lagrangian theories have been compared with other
approximations, including the frozen flow (FF)
and linear potential (LP) approximations
\cite{mss94}; the conclusion is that such Lagrangian methods
invariably perform better than either FF or LP in the
weakly non--linear regime (see \S 4.7). The great strength of the
ZA when compared with these other methods is that its simplicity allows
many properties to be calculated analytically.

As was the case with the Zel'dovich approximation, higher order Lagrangian
approximations  break down soon after shell crossing and the formation of the
first pancakes. This problem can be partially remedied if (as in ZA) we
introduce a cutoff scale in the original power spectrum which filters all
scales that are going non-linear at a given epoch, the resulting approximations
are then the truncated versions of PZA, PPZA etc. \cite{bmw94}.

\subsection{The Adhesion Approximation}
As discussed in \S 4.2 the Zel'dovich approximation works very
well until the formation of the first nonlinear structures -- pancakes.
Since particle motion is essentially inertial in ZA, particles passing through
one and the same pancake do not feel each others gravitational
influence.
As a result, the stabilising effect which high density regions (caustics)
have on particle trajectories is absent in
ZA, and particles are allowed to  sail through caustics and are even
pushed away from them after passing through. This leads to
an artificial broadening of pancakes, and results in their ultimate
disappearence. This limitation is shared by
extensions of ZA such as the post Zel'dovich approximation etc.,
which break down order by order, soon after shell crossing.

This tendency is countered in the adhesion model (hereafter AM) -- a
nonperturbative extension of the Zel'dovich approximation -- in which particles
are made to stick together once
they enter into caustics, mimicking the viscous effects of gravity on small
scales.
As a result pancake thickness is stabilised in
the adhesion model, just as it is in N-body simulations.

In order to see how this is accomplished we note that the
combined effect of the terms in the right hand side of the Euler
equation (\ref{eq:z10a})
is like a force whose presence helps to keep the thickness of
pancakes finite.
In the ZA the right hand side in eq. (\ref{eq:z10a}) is set equal to zero so
that the resulting motion of a particle is inertial:
$\d\vec{u}/\d a = 0$.
Solving the coupled set of nonlinear equations (\ref{eq:z10a}--\ref{eq:z10c})
is not mathematically feasible.
However, by replacing the right hand terms in equation (\ref{eq:z10a})
by a mock viscosity term introduced to mimic the adhesive
effects of nonlinear gravity on small scales, we obtain
\begin{equation}
{\partial\vec{u}\over \partial a} + (\vec{u}\vec{\nabla})\vec{u} =
\nu\nabla^2\vec{u}.
\label{eq:z28}
\end{equation}
Equation (\ref{eq:z28})is well known from studies in
turbulence as the three dimensional
generalisation of Burgers' equation \cite{b74,w74}.
For potential motion $\vec{u} = - \vec{\nabla}\Phi$
which we shall be considering,
eq. (\ref{eq:z28})
can be solved analytically by performing the Hopf-Cole substitution
$\Phi (\vec{x}, a) = - 2\nu \log U(\vec{x}, a)$.
As a result eq. (\ref{eq:z28}) translates
into the familiar diffusion equation
\begin{equation}
{\partial U\over \partial a} = \nu\nabla^2 U.
\label{eq:z29}
\end{equation}
Solving eq. (\ref{eq:z29}) (omitting some
intermediate steps) we obtain \cite{gss85,gss89}:
\begin{equation}
\Phi(\vec{x}, a) = - 2\nu~ \log \left[{1\over (4\pi \nu a)^{3\over 2}}~
\int d^3\vec{q}~ \exp \left( - {1\over 2\nu} S(\vec{x},a;\vec{q}\right)
\right],
\label{eq:z30}
\end{equation}
\begin{equation}
\vec{u} (\vec{x}, a) = \frac{\int \d^3\vec{q} \left(\frac{\vec{x}-\vec{q}}{a}
\right)
\exp [- S(\vec{x}, a; \vec{q})/2\nu]}{
\int \d^3q \exp [- S(\vec{x}, a; \vec{q})/2\nu]},
\label{eq:z31}
\end{equation}
where the ``action''
\begin{equation}
S(\vec{x}, a; \vec{q}) = - \Phi_0(\vec{q}) + {(\vec{x} - \vec{q})^2
\over 2 a(t)},
\label{eq:zaction}
\end{equation}
satisfies equation (\ref{eq:z11b}).
The trajectory of a particle can now be determined from
equation (\ref{eq:z31}) by solving
the integral equation \cite{nd90,wg90a}
\begin{equation}
\vec{x}(\vec{q}, a) = \vec{q} + \int_0^a \d a' \vec{u}[\vec{x}(\vec{q},
a'), a'],
\label{eq:z32}
\end{equation}
and the resulting density can be determined from the continuity
equation
\begin{equation}
\rho (\vec{x}, a) = \rho_0(\vec{q}) \vert\partial \vec{x}/ \partial
\vec{q}\vert^{-1},
\label{eq:z33}
\end{equation}
For small (but finite) values of the viscosity parameter $\nu \rightarrow 0$,
the integrals in equations
(\ref{eq:z30}) and (\ref{eq:z31}) can be evaluated using
the method of steepest
descents. (Note that a small $\nu$ implies a large Reynolds number $R_0 =
u_0 l_0/\nu$, $u_0$ and $l_0$ being respectively, the characteristic
amplitude and coherence length of the primordial velocity field
\cite{gsy83}.)
The resulting solutions acquire a simplified
form \cite{gss85,gss89,kps90,kpsm92}:
\begin{equation}
\Phi (\vec{x}, a) = -2\nu \log\,\sum_\alpha j_\alpha \exp \left(
-{1\over 2\nu} S(\vec{x}, a; \vec{q}_\alpha)\right)
\label{eq:z34}
\end{equation}
and
\begin{equation}
\vec{u}(\vec{x}, a) = \frac{\sum_\alpha \left(
{\vec{x} - \vec{q}_{\alpha} \over a}\right)
j_{\alpha} \exp \bigl\lbrack -{1\over 2\nu} S(\vec{x}, a; \vec{q}_{\alpha})
\bigr\rbrack}{
\sum_\alpha j_{\alpha} \exp \lbrack -{1\over 2\nu} S(\vec{x}, a;
\vec{q}_{\alpha})\rbrack },
\label{eq:z35}
\end{equation}
where $\vec{q}_\alpha$ are the Lagrangian coordinates  that
minimise the action (\ref{eq:zaction}), i.e.
\begin{eqnarray}
S_{\alpha} \equiv S(\vec{x}, a; \vec{q}_\alpha) & = &
- \Phi_0(\vec{q}_\alpha) +
{(\vec{x} - \vec{q}_\alpha)^2\over 2 a(t)} = min,\nonumber\\
j_\alpha & = & \Bigl\lbrack \mbox{ det} \bigl (\delta_{ij} -
{\partial^2\Phi_0\over
\partial q_i\partial q_j}\bigr )\Bigr\rbrack ^{-\half}_{q=q_{\alpha}}.
\end{eqnarray}
The points $\vec{q}_\alpha$ that minimise the action
satisfy the Zel'dovich
equation $\vec{x} (\vec{q}_\alpha, a) = \vec{q}_{\alpha} -
a\vec{\nabla}_q\Phi_0(\vec{q}_{\alpha})$;
$S(\vec{x}, a; \vec{q}) \ge S(\vec{x}, a; \vec{q}_\alpha)$
for any $\vec{q} \ne \vec{q}_{\alpha}$.

Equations (\ref{eq:z34}) and (\ref{eq:z35}) have an interesting
geometrical interpretation
according to which
the formation of caustics in E-space can be related to the degree of the
mapping from $\lbrace\vec{q}_\alpha\rbrace$ to $\lbrace\vec{x}
\rbrace$ -- the apex of the paraboloid (Fig. (\ref{fig:4adh})).
At early times this mapping
is one--to--one, since the curvature of the paraboloid
$P(\vec{x},a;\vec{q})$
is greater than that of the primordial potential $\phi_0$ (or equivalently
$\Phi_0$). At such times the Zel'dovich approximation is universally valid.
With passing time, as the curvature of the paraboloid decreases it begins
to touch the potential in more than one point, resulting in a mapping from
$\lbrace\vec{q}_{\alpha}\rbrace$ to $\lbrace\vec{x} \rbrace$ which is
many-to-one.
This indicates the onset of the epoch of pancake formation.
A pancake forms at $\vec{x}$ when the paraboloid
(with apex at $\vec{x}$) touches
the potential in exactly two points (so that the mapping from
$\lbrace\vec{q}_{\alpha}\rbrace$ to $\lbrace\vec{x} \rbrace$
is two-to-one).
Filaments form at the intersections of pancakes; for them the mapping
is three-to-one,
and knots at the intersections of filaments, the mapping being four-to-one.
The density of matter in these structures is directly related to the degree
of the mapping,
so that knots are denser than filaments, which in turn are denser than
pancakes \cite{sss94}.

The masses and velocities of knots can be obtained in this {\em ansatz}
as follows.
In two dimensions the mass of a knot at $\vec{x}$ is given by
$M(\vec{x}) \simeq \bar{\rho}
\Delta (\vec{q}_1, \vec{q}_2, \vec{q}_3)$, where $\Delta (\vec{q}_1,
\vec{q}_2, \vec{q}_3)$
is the area of a triangle with vertices at $\vec{q}_1, \vec{q}_2,
\vec{q}_3$, and
$\bar\rho(t)$ is the mean density of matter in the Universe.
In three dimensions, $M(\vec{x}) \simeq \bar\rho ~D(\vec{q}_1, \vec{q}_2,
\vec{q}_3, \vec{q}_4)$,
where $D(\vec{q}_1, \vec{q}_2, \vec{q}_3, \vec{q}_4)$
is the volume of a pyramid with vertices
at $\vec{q}_1, \vec{q}_2, \vec{q}_3, \vec{q}_4$.
Similarly, the velocity of the knot $U$ in two dimensions
can be determined by solving the
following system  \cite{kpsm92} of algebraic equations:
\begin{eqnarray}
\vec{U}\cdot(\vec{q}_1 - \vec{q}_2) & = & \Phi_0(\vec{q}_2) - \Phi_0(
\vec{q}_1)\nonumber\\
\vec{U}\cdot(\vec{q}_1 - \vec{q}_3)  & = & \Phi_0(\vec{q}_3) - \Phi_0(
\vec{q}_1)
\label{eq:z36}
\end{eqnarray}
(and similarly in 3D).
Equation (\ref{eq:z36}) describes the velocity $\vec{U}$ in terms of a
vector defined in a three dimensional space $(\Phi_0, \vec{q})$.
The magnitude of $\vec{U}$ is proportional to the area of the triangle
constructed by joining the osculating points of the potential
$\Phi_0(\vec{q}_{\alpha})$.
Its direction is given by the projection of the normal to the plane of
this triangle onto the Lagrangian plane defined by the vectors
($\vec{q}_1, \vec{q}_2$).
In one dimension a knot is formed at the apex of a paraboloid
touching the potential in exactly two points.
As a result its mass is given
exactly by $m = (\vert \vec{q}_1 - \vec{q}_2\vert)\times \bar\rho$,
and its velocity by
$\vec{u} = \d\vec{x}/\d a = \vec{n} (
\vert \phi_{q_1} - \phi_{q_2}\vert) /(\vert q_1 - q_2\vert)$,
the direction vector $\vec{n}$ pointing from {\em higher} to {\em lower}
values of the gravitational potential.
The mass function of knots evaluated using the adhesion approximation,
shows good agreement with the Press-Schechter mass function
for a wide range of spectra, in both  one and  two dimensions
\cite{whps91,kpsm92}.

The formation and evolution of large scale structure is described by the
adhesion model as a two stage process. During the first stage matter
falls into pancakes and then moves along them towards filaments and then along
filaments to collect finally in knots. At the end of the first stage
the formation of the skeleton of the large scale structure is complete
and virtually all
of the matter in the Universe is located in one of three structural units:
pancakes, filaments or knots.
This happens at the epoch when $k_{NL}^{-1} \sim R_* = \sqrt{\cal {D}}\sigma_1/
\sigma_2$, $R_*$ characterising the average separation between peaks of the
gravitational potential. (The quantities
$\sigma_0, \sigma_1, \sigma_2$ are the standard deviations in the
 potential and its first and second derivatives respectively;
${\cal {D}}$ is the dimensionality of space.)
The second stage sees the deformation of the large scale structure skeleton
due to the dynamical motion of pancakes, filaments and especially knots
towards each other, caused by their mutual gravitational attraction.
This stage witnesses the merger of knots and the disappearance of small
voids.
illustrated in Fig. (\ref{fig:5adh1d}) for the one dimensional case.
During this stage the paraboloid is extremely broad and can therefore
feel very large features in the primordial gravitational potential.
As a result, the corresponding picture of large scale structure
often shows the
presence of large ``secondary pancakes'' whose size
is related to the characteristic correlation scale of the initial potential
$R_\phi = \sqrt{\cal{D}}\sigma_0 /\sigma_1$ \cite{kpsm92}.
Secondary pancakes should be distinguished from primary pancakes
which form at the onset of gravitational instability.

In addition to providing a simple geometrical means to determine the masses
and momenta of clumps, the adhesion model can also be used to determine the
sizes of voids and the associated void spectrum
for a given cosmological model.
Some properties of voids obtained using the adhesion ansatz are summarised
in \S 5.3 (see also \cite{sss94}).
\begin{figure}
\vspace{15cm}
\caption{The geometrical representation of a paraboloid descending onto
the (two-dimensional) gravitational potential in order to demarcate
sites of structure formation.
The paraboloid is tangent to the potential in {\em Lagrangian} space, its
apex shows the location of caustics (filaments and clumps)
in {\em Eulerian} space. Adapted, with permission,
from \protect{\cite{kps90}}.}
\label{fig:4adh}
\end{figure}

Statistical properties of solutions to Burgers equation
have been studied, both in and out of the cosmological context
\cite{k79,gs81,gss85,gss89,dk90,vdfn94}.
Numerical studies aimed at determining, among other things, the
velocity distribution and mass function of knots have been performed in
one dimension \cite{whps91}
and in two dimensions  \cite{kpsm92}.
In addition, the evolution of voids \cite{sss94}
and filamentary structures \cite{nd90,kpsm92} have been discussed
at length.

An interesting result that  emerges from these studies
is that for power law spectra $P(k) \propto k^{n}$
the typical mass in knots grows as \cite{gss85,gss89,k79,sz89}:
\begin{equation}
 M \propto a(t)^{6/( n + 3)}\,\,\,\,\,\,\,\,\,\,\,\,-1 \le n \le 1
 \label{eq:z39}
 \end{equation}
For steeper spectra $n > 1$
the adhesion model predicts $M \propto a^{3/2}$ \cite{whps91,sz89}.
It is interesting to compare the predictions of the adhesion
model with those arising from semi-analytic models of hierarchical clustering
such as the Press-Schechter approximation \cite{ps74}, discussed in
\S 3.4.
PS suggests that equation (\ref{eq:z39}) is valid over a
somewhat broader range $- 3 \le n
\le 4$. For very steep spectra $n > 4$ PS predicts the limiting rate
$M \propto a(t)^{6/7}$ which is slower than the limiting growth rate
predicted by the adhesion model.\footnote{As in \S 4.2 the preceeding
discussion can be made more general by replacing
$a(t)$ with $D_+(t)$, the growing linear density contrast.}

The adhesion model has been found to work very well in explaining the
large scale features of the structure of
the Universe. Comparison with N-body simulations has shown remarkable
agreement till very late nonlinear times \cite{wg90a,wg90b,s91,kbgnd94,sss94}.
The main drawback of the adhesion model is probably the fact that one cannot
follow particle
trajectories within pancakes and filaments, which remain arbitrarily thin
in models where the coefficient of viscosity $\nu$ is taken to be very small.
Note that $\nu$ has dimensions of (length)$^2$, so that the thickness of
pancakes is proportional to $\sqrt \nu$.

\subsection{The Frozen Flow and Linear Potential Approximations}
In addition to the adhesion approximation, other recent attempts to prevent
the artificial thickening of pancakes present in ZA, have led to the
development of the Frozen Flow (henceforth FF) and Linear Potential (LP)
approximations.
Unlike the Lagrangian space based approximations considered earlier,
both FF and LP are formulated in Eulerian language.
For FF see \cite{mlms92}; LP is discussed in  \cite{bsv93,bp94}.
In FF a particle moving along a fixed trajectory constantly upgrades its
velocity to the local value of the linear velocity field. The equations
of motion in FF can be derived very simply from equation (\ref{eq:z10a})
by neglecting the non-linear term $(\vec u\nabla)\vec u$ and setting
$\vec u = - A\nabla\phi_0$ so that
\begin{equation}
\frac{\partial\vec u}{\partial a} = 0.
\label{eq:ff}
\end{equation}
As a result the comoving velocity field of a particle remains frozen to its
initial value ${\vec u} ({\vec x},t) = {\vec u} ({\vec q} ={\vec x}) $.
Since particles loose memory of previous motion they behave as inertialess
tracers of the primordial velocity field which is a fundamental reversal
of the assumptions made in the Zel'dovich approximation.
Continuity demands that the component of the force field
perpendicular to the plane of the pancake -- $\nabla (\phi_0)_{\perp}$
vanish at sites of pancake formation.
As a result particle velocities directed towards the pancake
get progressively smaller as the site of
pancake formation is approached.
The velocity component parallel to the pancake however remains unaffected
with the result that, as a particle approaches a pancake site it slows down
and modifies its velocity, in such a manner so as to move along the pancake
towards lower values of the gravitational potential,
converging ultimately in a clump at a  local minimum of
$\phi_0$. This asymptotic convergence guarantees that particle trajectories
will never intersect in FF and that the resulting pancake will always
remain thin.
A drawback of FF is that, with the passage of time most particles
collect in minima of the potential where $\vec u = 0$, as a result no
further evolution of large scale structure
is possible within the framework of this model.

The density contrast in FF can be estimated from the continuity equation
(\ref{eq:continuity}) or (\ref{eq:z10c})
which gives
\begin{equation}
\eta(\vec{x},a) = \eta_0(\vec{q}) \exp \left( \int_{a_0}^{a} \d a'
\vec{\nabla} \vec{u} \left[ \vec{x} (\vec{q},a),a' \right] \right).
\label{eq:z43}
\end{equation}
Substituting $\vec{\nabla}\vec{u} = - \nabla_x^2\phi_0(\vec{x}) =
- \delta (\vec{x})/a = $ constant, we finally get
\begin{equation}
\eta(\vec{x},a) = \eta_0(\vec{q}) \exp \int_{a_0}^{a} \d a'
\delta_{+} \left[ \vec{x}(\vec{q},a') \right],
\label{eq:z44}
\end{equation}
where $\delta_+$ is the unevolved value of $\delta (\vec{x})/a$ in the
linear regime, i.e.
the value of the linear density field integrated over the trajectory of
the particle equals the logarithm of the density.
Equation (\ref{eq:z44}) presents a generalisation
of the log-normal distribution \cite{cj91} according to which
the linear and nonlinear density contrasts are related through
$\eta (\vec{x}, t) = \eta_0\exp[\delta (\vec{x}, t)]$,
$\delta (\vec{x}, t)$
being the linear density
contrast at a fixed coordinate value $\vec{x}$.
In fact \cite{mss94} the probability
distribution
function in FF in the weakly nonlinear regime reproduces the log-normal
result (\ref{eq:ln_dis}) for $\sigma\ll 1$.
Both FF and the log-normal
distribution give similar results when applied to particles located at local
maxima or minima of the initial gravitational potential when there is little
change in the particle distribution. Since at early times voids
can be associated with peaks in the gravitational potential,
FF predicts $\delta_{FF} \simeq \exp[- \vert\delta_+ (\vec{x})\vert~a(t)] - 1
\rightarrow - 1$, indicating that voids are evacuated fairly rapidly.
Comparison with the spherical top hat model shows that FF leads to a somewhat
faster evacuation of voids than predicted (see \S 4.7 and
\S 5.1).

The linear potential approximation (LP) is based on the observation that
the gravitational potential usually evolves much more slowly than the
density field. This is clearly indicated by the linearised Poisson equation
$\nabla^2\varphi \propto \delta/a$, ($\delta \ll 1$) ,
which demonstrates that $\phi = \phi_0=$ constant,
if the Universe is spatially flat and matter dominated so that
$\delta \propto a$. The Euler equation in this case simplifies to
\begin{equation}
\frac{\d\vec{u}}{\d a} + \frac{3}{2a} \vec{u} = - \frac{3}{2a} \vec{\nabla}
\phi_0.
\label{eq:z45}
\end{equation}
indicating that particle trajectories follow the
gradient of the primordial potential -- effectively moving along its lines
of force. Unlike FF, particle trajectories in LP can and do intersect at
sites of pancake formation, which correspond to the directional minima of the
primordial gravitational potential.
Particle motion is oscillatory in such regions, which prevents the resulting
pancake from becoming too thick. As in FF, the subsequent motion of particles
within pancakes follows the primordial potential, with virtually all the
particles collecting in clumps, corresponding to local minima of the initial
potential, at late times.  (Note that in both FF and LP particle trajectories
can be curved, which sets them apart from the straight line trajectories
of the Zel'dovich aproximation.)

An essential limitation of both FF and LP is that, since in both approximations
particles move according to the dictat of
the primordial velocity/gravitational
potential, neither is equipped to describe the
merger of clumps, which is the predominant feature of gravitational instability
at late times, especially in hierarchical models such as CDM.
This limitation is not shared by the adhesion model
as demonstrated in \cite{kpsm92,sss94,ssmpm94}; see also \S 4.7.

A common feature shared by all the approximations discussed in this chapter
(i.e. ZA, P$^{(n)}$ZA, AM, FF and LP),
is that they allow a fully parallel treatment, making them ideally suited
for efficient implementation on highly parallel computers.

\subsection{A Comparison of Non--linear Approximation Methods}
The relative accuracy of nonlinear approximation methods can be established
by testing them both against each other as well as against exact theoretical
predictions (where these are available) or alternately,
against the results of N-body simulations.

In the weakly nonlinear regime, the predictions made by higher--order
Eulerian perturbation theory can be used to test
the accuracy of different approximation methods
\cite{gw87,ms94,mss94,bsbc94}. In the strongly nonlinear
regime on the other hand, comparison with N-body simulations can serve to
discriminate between the different approximations
\cite{cms93,ssmpm94,mlmm94,msw94,bmw94,mps94,be94,mbw94}.

In the weakly nonlinear regime ($\delta < 1$) the density field can be
described
by means of a perturbative expansion discussed in \S 4.1.
To compare our approximations with the full perturbative expansion,
we simply need to solve the equation describing spherical top-hat collapse
for the
different approximations and compare the resulting generating functions
term--by--term with the full perturbative result given by Eqn. (\ref{eq:z51a})
and Eqn. (\ref{eq:z51c}). One can also
then proceed to determine the $S_n$ in the quasi--linear hierachy using
Eqn's. (\ref{eq:z49}) \& (\ref{eq:z50}), which we shall do later.

Let us now sumarise the results of this procedure \cite{mss94}.
For the straightforward Zel'dovich approximation we find
\begin{equation}
G_{\delta} = \left( 1-\frac{\tau}{3}\right)^{-3} - 1;
\label{eq:z52}
\end{equation}
for the post--Zel'dovich approximation (PZA, see \S4.4) this
becomes
\begin{equation}
G_{\delta} = \left(1-\frac{\tau}{3}-\frac{\tau^{2}}{21}\right)^{-3} - 1;
\label{eq:z53}
\end{equation}
and for PPZA,
\begin{equation}
G_{\delta} = \left( 1-\frac{\tau}{3} -\frac{\tau^{2}}{21} -\frac{23\tau^{3}}
{1701} \right)^{-3} -1.
\label{eq:z54}
\end{equation}
(Everywhere $G_{\delta} \equiv \delta_{sph}$ the density contrast in the
spherical top-hat model.)

For the approximations formulated in Eulerian space we get:
\begin{equation}
G_{\delta} = \left( \frac{\sqrt{2\tau}}{\sin \sqrt{2\tau}} \right)^{3} -1
\label{eq:z55}
\end{equation}
for the linear potential approximation and
\begin{equation}
G_{\delta}=\exp(\tau) -1
\label{eq:z56}
\end{equation}
for the frozen flow approximation.

Equations (\ref{eq:z52})--(\ref{eq:z56}) can be used in
conjunction with equation (\ref{eq:z49}) and (\ref{eq:z50})
to determine
$\nu_n$ and $S_n$ by expanding $G_{\delta}$ in the vicinity of $\tau = 0$.
The values of $S_1, .. S_6$ obtained for the five approximations
we consider are listed in Table 1.

\begin{table}
\begin{center}
\caption{Comparison of the hierarchical coefficients $S_n$ for the
various approximations discussed in the text.}
\begin{tabular}{lllllll}
    & Exact & ZA & PZA & PPZA & LP & FF\\
$S_{3}$ & 4.857 & 4 & 4.857 & 4.857 & 3.4 & 3\\
$S_{4}$ & 45.89 & 30.22 & 44.92 & 45.89 & 21.22 & 16\\
$S_{5}$ & 656.3 & 342.2 & 624.4 & 654.6 & 196.4 & 125\\
$S_{6}$ & 12653 & 5200 & 11666 & 12568 & 2429 & 1296\\
\end{tabular}
\end{center}
\end{table}
One can see that taking higher and higher order terms in the
Lagrangian theory leads to successively more accurate approximations.
As far as the $S_n$ are concerned, the PPZA is the most accurate
of the approximations we have looked at, followed in decreasing
order of accuracy by PZA, ZA, LP and FF \cite{mss94}. This conclusion is
further borne out by a study of the moments
$T_n = \langle\theta^n\rangle_c/\langle\theta^2\rangle^{n-1}$ of the scalar
velocity variable $\theta \equiv H^{-1}\vec{\nabla}\vec{v}$ \cite{mss94}.

Notice again that the values of $S_n$ depend also on smoothing:
filtering the final density field using a top-hat filter leads
to the following expressions for scale invariant spectra ($P(k) \propto k^n$):
for the ``exact'' perturbative case
\cite{jbc93,b94b}: $S_3 = \frac{34}{7} - (n + 3)$, $S_4 = 45.89 - \frac{62}{3}
(n + 3) + \frac{7}{3} (n + 3)^2$; for the Zel'dovich approximation
\cite{bk94}, $S_3 = 4 - (n + 3)$, $S_4 = 30.22 - \frac{50}{3}
(n + 3) + \frac{7}{3} (n + 3)^2$;  (also see \S 4.1, \S 4.10).
For the quantity $\theta$ we get, after top-hat
smoothing: for the exact
perturbative case, $T_3 = - \Omega^{-0.6}\lbrack 26/7 - (n+3)\rbrack$;
for the Zel'dovich aproximation, $T_3^{(ZA)} = - \Omega^{-0.6}
[2 - (n+3)]$. We see that the Zel'dovich approximation consistently
underestimates the skewness: for $-1 < n < 0.7$ $T_3^{(ZA)}$ and $T_3$
even have differing signs \cite{b94b}.

\begin{figure}
\vspace{15cm}
\caption{The evolution of a spherical density contrast in the
different approximations is plotted against the exact analytical
result. The exact solution is labelled by 1; the post-post-Zel'dovich
approximation by 2; the post-Zel'dovich approximation by 3;
the Zel'dovich approximation by 4; the linear potential approximation by 5;
the frozen flow approximation by 6;
and the linear theory solution by 7.
Reproduced, with permission, from \protect{\cite{mss94}}.}
\label{fig:4tophat}
\end{figure}

A further study of spherical top-hat collapse in the
different approximations shows that
ZA and its extensions are much
closer to the exact solution than either FF or LP right until the
time of recollapse; see Fig. (\ref{fig:4tophat}). Evidently
nonlinear
approximations formulated in Lagrangian space (such as ZA, PZA etc.) are
considerably more accurate in the weakly non--linear regime than those which
are formulated in Eulerian space (LP, FF); see
\cite{gw87,bjcp92,ms94,bsbc94}.

An interesting difference between the Lagrangian and Eulerian approximations
discussed above is that the former are essentially single step mappings whereas
the latter involve following the trajectory of a particle
or fluid-element through
several timesteps. This makes the former more amenable to reconstruction
methods
such as POTENT \cite{nd92}.

In the strongly non--linear regime (after shell crossing) the ZA breaks down
and has to be abandoned in favour of its logical successor, the adhesion
model. In this regime $\delta >> 1$ and one can no longer rely on a
perturbative analysis to discriminate between the different approximation
methods. The logical course to take then appears to be to test the different
non--linear approximations against the results of N-body simulations assuming
identical initial conditions \cite{cms93,ssmpm94,mbw94,bmw94,msw94}.
(For a discussion of $N$--body methods, see \S 7.)
The problem is that
one has to be careful in choosing the correct statistical comparison
between the $N$--body and test density fields. It is possible for
an approximation scheme to produce a pattern of filaments and knots
which has a very similar visual appearance to the full dynamical
simulation, but has the main structures slighly displaced from their
corresponding positions in the full simulation. On the other hand, one
could argue that a true dynamical approximation should be able to
move matter to the right place and that this statistic is therefore
a reasonable test. Finding the right
statistic to perform a comparison which is more specific to pattern
than location is a difficult task which has not yet been fully solved.

A recent comparison of non-linear approximations against the results of
N-body simulations addresses this question in some detail by choosing
different statistical measures (filamentary statistics, void probability
function etc. see \S 8)
to discriminate between different approximation methods \cite{ssmpm94}.
Some of the results of \cite{ssmpm94} are shown in Fig. (\ref{fig:4compar}).
The analysis of \cite{ssmpm94} seems to indicate that local Eulerian
approximations to
gravitational instability such as FF and LP begin to gradually breakdown
as the non-linear length scale $k_{NL}^{-1}$ becomes comparable to
$R_* =  \sqrt{\cal{D}}{\sigma_1 / \sigma_2} $
-- characterising the mean separation between peaks (or troughs) of the
gravitational potential.
The reason for this is clear, particles in FF and LP follow the gradients of
the initial gravitational potential asymptotically approaching the minima of
the potential in FF, and oscillating about them in LP. Neither FF nor LP
takes into account the possible presence of long range modes in the
gravitational potential, a fact that is used to great advantage
in both AM and TZ to move particles coherently over large scales.
As a result the adhesion approximation fares considerably better than FF or LP,
giving results that are reasonably accurate until the non-linear length scale
becomes comparable to the coherence length of the potential,
$k_{NL}^{-1} \sim R_{\phi}$
($R_{\phi} = \sqrt{\cal{D}} {\sigma_0 /\sigma_1}$). For models
with a considerable spread of power such as CDM $R_{\phi} \gg R_*$, and we
expect the adhesion model to perform better than FF or LP at late times
\cite{ssmpm94}.

\begin{figure}
\vspace{15cm}
\caption{A comparison of four non-linear approximations: the adhesion model
(AM), the frozen flow
approximation (FF), the linear potential approximation (LP),
and the truncated Zel'dovich approximation (TZ),
is made against the results of N-body simulations. (The simulations
have been performed using $512^2$ particles in two dimensions.
The initial conditions are scale invariant, with $P(k) \propto k^n$
and $n = 0$:  this corresponds to
$n = - 1$ in three dimensions.)
The results of the adhesion model are shown superimposed on the results of
N-body simulations. The comparison is made for two distinct epochs:
the panels on the left correspond to $k_{NL}^{-1} \sim R_*$ and the panels on
the right to $k_{NL}^{-1} \sim R_{\phi}$.
Reproduced, with permission,
from \protect{\cite{ssmpm94}}.}
\label{fig:4compar}
\end{figure}

\subsection{Relativistic Approaches}
Throughout this discussion we have concentrated upon Newtonian
approaches to the growth of inhomogeneities. This is
generally justified by the fact that one is dealing always
with peculiar motions which are much smaller than the velocity
light and that, in the matter--dominated regime, the relativistic
effects of pressure are presumably negligible.\footnote{ It should
be pointed out that it is essential to treat petrurbations relativistically
if one wishes to follow their evolution on scales larger than the
Hubble radius $\lambda_H \simeq ct$. A gauge-invariant treatment of
density perturbations was developed in \cite{bar80}, subsequent discussions
can be found in \cite{e90,mfb92,ll93b,ss84}.}
Nevertheless, there
are some reasons for considering a relativistic approach. Firstly,
Newtonian theory implicitly possesses the physically disturbing
property that gravitational influence is transmitted instantaneously
(``action at a distance'') and thus violates causality requirements.
Secondly, the biggest problem in a Newtonian treatment is posed by the
need to solve the Poisson equation; this can be avoided in a
relativistic treatment because, if the constraint equations of General
Relativity are satisfied at early times, the evolution equations
guarantee that they will be satisfied at all later times.
Finally, the relativistic approach is conceptually extremely
different from the Newtonian approach and the consequent
difference in formal construction may lead to approximation
schemes that are different from those developed from the Newtonian
approach. We shall confine ourselves to brief remarks on
some of the latest developments in this area.

This final point is especially pertinent at the present time because
of recent studies of the dynamics of self--gravitating fluid
flows. Following a much earlier suggestion of Ellis \cite{ell71},
Matarrese et al. \cite{mps93} suggested that the Einstein equations
be approximated by neglecting the magnetic part of the Weyl tensor.
The truncated equations thus obtained are
strongly local in that they
do not require any information about neighbouring spatial points
to solve for the motion of any point. This approach,
which has become known as the ``silent universe''
approach led to
an exact solution of the Einstein equations and also allows
various properties to be calculated numerically. It was
subsequently shown that the exact solution
obtained in \cite{mps93} corresponds to the well--known
Szekeres solution of Einstein's equations \cite{cpss94},
which is in fact unstable. The properties of this
instability led the authors of
\cite{bj94} to argue that the outcome of a generic
collapse would be a cigar or spindle rather than the
pancake predicted by Newtonian theory. This analysis
was criticised on the grounds of the validity of the
approximation scheme used \cite{kp94}. Further discussions
and developments of this approach can be found in
\cite{mps94a,mps94b,bmp94,ed94,bh94}.
 Although the final word has not been said, it appears that the
approximation of neglecting the magnetic part of
the Weyl tensor is not valid in general, but it
may be a useful approach in situations admitting specific symmetries.

An alternative relativistic approach to structure
uses the powerful and elegant approach of Hamilton--Jacobi
theory \cite{cpss94,ssc94,sals95}.
 Using powerful Hamilton--Jacobi methods, they refined
the analysis of Lifshitz and Khalatnikov \cite{lk64}
who considered a Taylor series expansion of
Einstein's equations. In fact, they consider an
'improved Taylor series' which ensures that
the 3-metric of general relativity remains
positive definite. In this way, they
recover the Zel'dovich approximation and
its higher order generalisations.  A comparison between the relativistic
and Newtonian theory is now being made
at higher orders.\footnote{The authors wish to thank Dave Salopek for
a discussion on these issues.}

Where these approaches will eventually lead is uncertain at the present
time but there is certainly active interest in ideas emerging from
relativistic considerations and there is sure to be a substantial
investment of effort in this area in the near future.

%% file: ssec5.tex
\section{Non--linear Approximations based on Geometry}
Up to now, we have concentrated upon the evolution
of denser--than--average structures as clustering develops.
However, the dominant contribution to the perceived visual
geometry of galaxy clustering is the nature of the
underdense regions, called voids. We shall discuss the
statistical characterisation of voids in more detail
in \S 8.6. In this section we discuss various ways of trying
to understand the evolution of underdensities in the
gravitational instability scenario.

\subsection{Void Expansion}
Ever since the discovery of the spectacular Bo\"{o}tes void in 1981,
successively larger redshift surveys have confirmed that most of the
volume of the Universe is contained in voids -- underdense regions,
virtually devoid of galaxies. Voids are separated from each other by
a network of galaxies, sometimes said to constitute a cellular or
honeycomb structure \cite{ejs80,koss81,dgh86,vgh91}. Although
estimates of typical void diameters vary, sizes ranging from $10$
to $50h^{-1}$ Mpc have been quoted as being typical of voids
in our neighbourhood \cite{kf91,vgh91,dc94}.

The conventional picture of voids has often been to associate them
with underdense regions in the primordial density field, which expand
faster than their surroundings causing matter to be evacuated from
within them and swept up into a shell at their periphery. The crucial role
played by voids in the development of large scale structure
was emphasized more than a decade ago by Zel'dovich and co--workers
\cite{zes82,zs82}. Most of the early work on voids was, however,
carried out under the simplifying assumption that voids are spherically
symmetric, or else belong to a regular symmetric honeycomb structure
\cite{p82,hor83,hsw83,fg84,rg91}. Icke's ``Bubble Theorem'' \cite{i84}
lends support to the idea that isolated voids can be treated as being
spherical,
by showing that aspherical underdense regions tend to become
spherical in time -- the time reversal of the well--known
Lin, Mestel \& Shu instability \cite{lms65}. The bubble theorem
\cite{cm83,f83,iu84,b85,bvg90} has
since been verified using numerical simulations\footnote{It should be
pointed out that voids in an ensemble compete with each other
for space as they expand, and therefore need not remain spherical.
Consequently, the bubble
theorem, which is true for individual isolated voids, does not necessarily
hold for a more realistic set of initial conditions
\protect{\cite{sss94,bv94}}.}.
Although most treatments of void evolution have been within a Newtonian
framework, attempts to model voids within a fully general--relativistic
framework have also been undertaken
\cite{mss82,s82,ms83,sm83,lp85,pl86,pl88,osv83,bc90,c91}; the
study of voids has been extended to include models with a
cosmological constant \cite{mw90}.

Theoretical modelling of individual voids has largely dealt with voids
that are {\em either} compensated by an
overdensity at the void boundary, leading to
a vanishing overall density contrast, {\em or} without
any such compensation. In the latter case, the
overall density contrast is negative. Such uncompensated voids
are often modelled by a spherical top-hat in which the density
is constant within a spherical region. A common feature of both
compensated and uncompensated void models is the gradual appearance
of narrow overdense ridges surrounding voids. Such ridges arise
because matter lying close to a void center tends to expand faster
than that near the edges and thus overtakes denser matter lying
ahead of it, resulting in shell crossing and the piling up
of material at the periphery of voids.

\begin{figure}
\vspace{15cm}
\caption{The initial density profile of a void. The solid line
corresponds to the radial density distribution of an uncompensated
void, and the dashed line to the boundary shell of a compensated
void. Adapted, with permission,  from \protect{\cite{b85}}.}
\label{fig:5voidprofile}
\end{figure}

The initial density distribution in compensated and uncompensated voids
is shown in Figure (\ref{fig:5voidprofile}).
The void has initial radius $R_i$ and
underdensity $\delta_i$ (at a time $t_i$), and is surrounded in the
case of a compensated void by an overdense shell of radius
$\alpha R_i$ ($\alpha >1$).

Examining the dynamical behaviour of both compensated and uncompensated
voids, Bertschinger \cite{b85} found several families of self--similar
solutions in which the expansion of the void at late times could be
described solely in terms of dimensionless quantities:
$\alpha$, $\delta_i$, $r/R_i$ and $t/t_i$. These solutions are similar
in form to the Sedov similarity solutions describing a point explosion
in a uniform medium \cite{s59,soy75,ito83}. In the case of a collisionless
fluid, such as non--baryonic dark matter, the expansion of the
void is accompanied by a transfer of energy from the interior of the
void onto the peripheral shell. This leads to shell--crossing and
collisionless phase--mixing which leads to the development of a
self--similar solution which has no memory of the initial conditions.
On the other hand, if the material is collisional (e.g. baryons),
shell--crossing is prevented by the formation of shock waves. The
subsequent dissipation of kinetic energy into heat leads to the eventual
wiping out of memory of the initial conditions, and the development
of a (different) self--similar solution describing the propagation of the
shock front. In both collisional and collisionless cases the void
boundary expands as $\lambda \propto t^{4/5}$. In the collisional case
the swept--up mass of the compensated shell increases as $t^{2/5}$.

Bertschinger also found similarity solutions for uncompensated voids
embedded in (a) collisionless, (b) collisional and (c) a mixture of
collisional and collisionless media \cite{b85}. In all cases
a dense shell forms around the void whose swept--up mass grows as
$t^{2/3}$ and whose expansion rate is $\lambda \propto t^{8/9}$.
This is somewhat faster than the case of a compensated void;
see also \cite{sm83,b83}.

It is often argued that, since the density of matter within a void
is always bounded by $\vert \delta \vert \le 1$, one can therefore
expect linear theory to be valid for a much longer duration within voids
than is the case for overdense regions. This
is not strictly the case, however. According to linear theory
the density contrast within an underdense region scales as
$\delta \propto -a(t)$ (in a flat Universe), whereas in reality
the void underdensity must be constrained to have a magnitude less
than unity. The result of this is that linear theory greatly
overestimates the expansion rate of voids; see Figure (\ref{fig:5void}).
We have also taken the opportunity to show, in Figure (\ref{fig:5void}),
the behaviour of the underdensity in different non--linear approximations
which we discussed in the previous section. All these approximations
tend to underestimate the density contrast.

\begin{figure}
\vspace{15cm}
\caption{The underdensity in the spherical top-hat model estimated
from different approximations, plotted against the exact solution. The
exact solution is the solid line; the Zel'dovich approximation is ZA;
the linear or {\it frozen}--potential approximation is FP;
the frozen--flow approximation
is FF; linear theory is LIN.}
\label{fig:5void}
\end{figure}

Studies of individual voids, although they clearly provide important
clues about the dynamical behaviour of underdense regions, do suffer
from the fact that they tend to ignore the presence of neighbouring
voids. Such studies need therefore to be supplemented by other
approaches which model voids as part of a larger ensemble of
voids of different shapes and sizes. We shall discuss
two such approaches here: the Voronoi model, which is based on
geometric arguments, and the adhesion model, which we discussed
in \S 4. The statistical characterisation of voids using the
 void probability function is discussed in \S 8.6.

\subsection{The Voronoi Foam}
In the Voronoi model, the large--scale structure of the Universe
(voids, clusters and superclusters), is described by a geometrical
skeleton of walls, filaments and nodes which together constitutes
a Voronoi foam \cite{v08}.

\begin{figure}
\vspace{15cm}
\caption{Stereoscopic pair of three Voronoi cells sharing a common boundary.
Stars indicate the Voronoi nuclei. Reproduced, with permission, from
\protect{\cite{vi89}}.}
\label{fig:5cells}
\end{figure}

The Voronoi foam is constructed as a tesselation, a mathematical
partioning of space into three--dimensional polyhedral units
(polygons in two dimensions). Adjoing polyhedra share common walls,
edges and vertices; see Figure (\ref{fig:5cells}).
The tesselation is contructed around
and contains that part of the space which is nearer to its nucleus
than any other nucleus. The resulting tesselation (which is unique)
is called the Voronoi tesselation.

The physical motivation for the Voronoi model is the observation
that voids will tend to expand more quickly than surrounding
material. As a result, matter tends to flow out of voids and
will generally become trapped in wall--like regions between
adjacent voids. An additional assumption that is usually made
in the case of the Voronoi model is that all voids expand at the same
rate, so that walls separating neighbouring voids will in general
form the perpendicular bisectors of the line joining the centers
of void expansion \cite{iv87}. Intersections of walls define
regions of higher density (filaments)  and intersections
of filaments define zero--dimensional subunits
(nodes). Matter in the Voronoi model is taken to flow out of voids
onto walls, then along walls into filaments and finally into the nodes
which are the vertices of the Voronoi polyhedra.

The mathematical prescription used for constructing a Voronoi foam --
the Voronoi tesselation -- has been applied in fields as diverse as
metallurgy, geology, forestry, molecular physics and astronomy
\cite{m53,k66,m70,skm87}. The first practical implementation of the
idea in studies of large--scale structure was made by van de Weygaert
\& Icke \cite{vi89}, who used the Voronoi model to mimic cluster correlations
by identifying each Voronoi node in their simulations with a
cluster of galaxies; see also \cite{iv87,ms84}.

Their basic prescription to construct the Voronoi skeleton was as follows:
a set of randomly (Poisson) distributed points were chosen to designate
the Voronoi nuclei, the centers of void expansion. The number
density of nuclei is the only free parameter in this model.
Knowing the number--density of nuclei, one can construct the tesselation
using either a kinemetical or geometrical approach. The geometrical
approach is particularly simple to follow in two dimensions,
since the basic structural units are then just filaments
and nodes. A filament always bisects the line joing two neighbouring
Voronoi nuclei. The end--points of a filament are nodes. Every node
marks the termination of exactly three filaments. One can construct
a Voronoi tesselation starting from any given Voronoi nucleus,
and following one of the filaments that surround it in either a clockwise
or an anti--clockwise direction until the filament
terminates in a node. The next filament bordering the same cell is then
chosen and followed until it also ends. This procedure is repeated until
one ends up back at the original node. Repeating this procedure for all
the Voronoi nuclei in a given sample, leads easily to the Voronoi
tesselation \cite{wph91,w92}.

The geometrical method described above is relatively easy to apply in two
dimensions but acquires considerable complexity in higher--dimensional
spaces. The kinematic approach, on the other hand, can easily be
generalised to an arbitrarily large number of spatial dimensions. The
gist of the kinematic approach is that test particles are used to
trace the Voronoi foam. The following procedure is adopted
by van de Weygaert \& Icke \cite{vi89}: a test particle is chosen
so that it is initially located close to a (randomly--chosen)
nucleus, called the {\em parent nucleus} $N_1$. The particle
is then given a velocity which is proportional to its distance from
the nucleus and directed radially away from it, so that
\begin{equation}
\vec{v} =\vec{x}_1-\vec{r}_1.
\end{equation}
The particle is assumed to be located at $\vec{x}_1$ and the nucleus
at $\vec{r}_1$. The motion of the particle is modified by means of an
interative procedure: $\vec{x}_1\rightarrow \vec{x}_1 +
\vec{v}\Delta t$, until the particle becomes closer to another
nucleus
$N_2$ (located at $\vec{r}_2$)
than it is to its parent nucleus. At this point the test particle
would have crossed a cell wall and would now be located just inside the cell
defined by the nucleus $N_2$. When the particle is equidistant from
the two nuclei, it is constrained to move only within the cell wall,
thus staying equidistant from both nuclei, until it becomes closer
to a third nucleus $N_3$ than the previous two, indicating that it has
entered a third cell by crossing a filament which (in three dimensions)
forms at the intersections of three cells defined by the nuclei
$N_1$, $N_2$, $N_3$. The particle is now constrained to move
along the filament by modifying its motion:
\begin{eqnarray}
\vec{v} & \rightarrow & (\vec{v}\hat{f})\hat{f}, ~~~
\hat{f} =\hat{n}_{13} \times \hat{n}_{12}; \nonumber\\
\hat{n}_{13} & = &
{\vec{r}_1-\vec{r}_3\over \vert \vec{r}_1 - \vec{r}_3\vert},~~~~~
\hat{n}_{12}  =
{\vec{r}_1-\vec{r}_2\over \vert \vec{r}_1 - \vec{r}_2\vert};
\end{eqnarray}
$\hat{f}$ is the unit vector along the filament.
(The vectors $\hat{n}_{13}$ and $\hat{n}_{12}$ are normal to
walls defined by the
nuclei $N_1  N_2$ and $N_1 N_3$ respectively.)
The particle moves
along the filament until it is closer to a fourth nucleus
$N_4$. The location of the particle at this juncture is close to
a node, i.e. it is equidistant from four nuclei.
After this, the particle does not move any further because it has
traced out the locations of all the basic structural units
in the vicinity of the parent nucleus $N_1$ \cite{vi89}.
Repeating for all nuclei leads to a Voronoi tesselation.
An excellent recent discussion of the construction of three dimensional
Voronoi tesselation's can be found in \cite{vande94}.

\begin{figure}
\vspace{15cm}
\caption{A two--dimensional Voronoi foam constructed with
clustered nuclei.}
\label{fig:5voronoi}
\end{figure}

As mentioned earlier, the only free parameter in the Voronoi foam model
is the mean number--density of Voronoi nuclei which can, in principle,
be determined by normalising it to astronomical observations. Making
the assumption that the nodes in the Voronoi simulation can be identified
with Abell clusters of richness class ${\cal R}\ge 1$, van de Weygaert
\& Icke \cite{vi89} were able to determine the inter--nuclear separation
to be $\simeq 104h^{-1}$ Mpc\footnote{The number--density of Abell
clusters is assumed to be $6\times 10^{-6}h^{3}$ Mpc$^{-3}$; when compared
with the ratio of nodes to nuclei in a Voronoi tesselation (6.733:1),
this gives the number--density of nuclei.} Interestingly, one can
then find the form of the two--point correlation function of the
clusters (i.e. Voronoi vertices) to be
\begin{equation}
\xi(r) =\left( \frac{r}{r_0}\right)^{-\gamma},
\end{equation}
with $r_0 = 32h^{-1}$ Mpc and $\gamma=1.97$ which is quite close to
the observational results for the cluster--cluster correlation
function (\S 8.2) \cite{bs83,b88}. However, as we discuss in \S 8,
more recent analyses of cluster correlations have indicated that the
clustering scale could be much smaller than the original Bahcall--Soneira
claim: $r_0\simeq 14h^{-1}$ Mpc. This is a potentially serious problem
for the Voronoi model with unclustered nuclei \cite{s88,se91}.

{}From a mathematical viewpoint the Voronoi tesselation around
randomly--distributed nuclei is appealing because of its simplicity
and because many analytical results are known about the statistical
properties of the tesselation in this case \cite{m89}. However, if the
Voronoi model is to provide a useful approximation to physical processes
that resulted in the formation of large--scale structure, the Voronoi
nuclei must be associated with some physically--significant spatial
locations. In models of structure formation based on explosions,
Voronoi nuclei can be associated with detonation sites, provided
all explosions have equal strength and all the explosions occur
at the same time. The arrival of two shock fronts in the same place
will result in the formation of a Voronoi wall. Three shocks will
lead to filaments and four to nodes. Furthermore, since there is
no {\em a priori} prescription exists for the explosion sites
it is quite reasonable in this case to take them as being randomly
distributed \cite{wod89,wwd90,w92}.

On the other hand, in  more conventional models  structure formation
proceeds by gravitational instability, a process which generically
leads to the formation of a cellular structure \cite{ms90}; see \S 4.
In order to apply the Voronoi idea to this case, one might associate
Voronoi nuclei with centers of the primordial underdensities or
peaks of the primordial gravitational potential which are the
progenitors of voids \cite{ks88,sss94}. However, in most scenarios
of structure formation the gravitational potential is spatially
distributed in the manner of a Gaussian random field. Since peaks
of a Gaussian random field are clustered, it seems a proper
application of the Voronoi technology in this case must involve
a clustered distribution of nuclei. This inevitably introduces more
freedom into the model through the introduction of parameters needed
to describe the clustering; see Figure (\ref{fig:5voronoi}).

Although we have assumed for simplicity that expansion rates
in different underdensities are identical, this is also not going
to be strictly correct. A study of voids performed using the adhesion
model demonstrates that void expansion rates generally depend on the
size of the gravitational potential fluctuation inside them
\cite{sss94}. One might therefore seek to modify the Voronoi
idea by assigning different weights to Voronoi nuclei, in
order to incorporate this variation of expansion speeds. The
simplest tesselation which includes such an effect is called
the {\em Johnson--Mehl tesselation} \cite{jm39}. As we shall
see, this closely resembles the geometrical structure obtained
by evolving with the adhesion approximation. A Johnson--Mehl
tesselation will also describe an explosion model in which the
explosions have varying degrees of intensity. In the Johnson--Mehl
tesselation, walls joining adjacent nuclei are not planes and do
not bisect the line joining the nuclei. Although it is more
difficult to construct using simple geometrical arguments, it
can be easily generated using a variation on the kinematical
approach discussed above. It can also be constructed using the
adhesion approximation, as described in the next section.

\subsection{The Adhesion Approximation Revisited}

We saw in \S4.5 how the adhesion approximation provides a simple
geometrical way to determine the masses and momenta of ``knots''.
It also furnishes an elegant prescription for finding the volumes
of void regions. This prescription involves dividing Lagrange space
into {\em stuck} and {\em free} regions by means of the following
procedure. The status of a particle originally located at
$\vec{q}_0$ can be determined by descending a paraboloid
$P(\vec{x},a; \vec{q}) = (\vec{x}-\vec{q})^{2}/2a(t) +h$ onto the
linear gravitational potential $\phi_0$, in such a manner as to
be tangential to the potential at $\vec{q}_0$; see Figure (\ref{fig:5adh1d}).

\begin{figure}
\vspace{15cm}
\caption{The geometrical prescription of the adhesion approximation.
A paraboloid is descended onto the gravitational potential in order
to demarcate {\em stuck} and {\em free} Lagrangian regions. The peaks
of the potential correspond to voids in the Zel'dovich approximation.
The particle having Lagrangian co--ordinate $\vec{q}_0$ is free
in the uppermost figure, and has just entered a caustic in the middle
figure. The apex of the paraboloid, $\vec{x}$, describes the
location of the caustic (clump with mass $m$) in Eulerian space.
Middle and lower panels illustrate the merger of clumps.
Adapted, with permission, from \protect{\cite{sss94}}.}
\label{fig:5adh1d}
\end{figure}

If the paraboloid touches or intersects the potential at a point
other than $\vec{q}_0$ then the particle in questions has already
entered a caustic, otherwise it has not. Using this procedure, we can
divide Lagrangian space at any time into {\em stuck} and {\em free}
regions. Stuck regions correspond to particles already in caustics,
whereas free regions have not yet shell-crossed and
are therefore still expanding via the Zel'dovich
relations, equations (\ref{eq:z1}) \& (\ref{eq:z4}), and
correspond to voids \cite{p89,s89,sss94}.

\begin{figure}
\vspace{15cm}
\caption{The distribution of caustics (dots) is plotted superimposed upon
{\em stuck} (unshaded)
and {\em free} (shaded) Lagrangian regions for
{\it early} times when cellular structure is just beginning
to form. The distribution of caustics in this case was obtained by
moving the border between stuck and free regions using the Zel'dovich
approximation.
Reproduced, with permission, from
\protect{\cite{sss94}}.}
\label{fig:5pancake1}
\end{figure}

\begin{figure}
\vspace{15cm}
\caption{The distribution of caustics (dots) is plotted superimposed upon
{\em stuck} (unshaded)
and {\em free} (shaded) Lagrangian regions for
{\it late} times when most of matter is already in caustics and
cellular structure
is fully formed. Reproduced, with permission,
 from \protect{\cite{sss94}},
where details of the simulations are provided.}
\label{fig:5pancake2}
\end{figure}

The location of caustics in Eulerian space can be found very simply
by moving the border between stuck and free Lagrangian regions
by means of the Zel'dovich approximation \cite{sss94}.
In Figure (\ref{fig:5pancake1}) and Figure (\ref{fig:5pancake2})
the location of caustics (filaments and knots) is
shown superimposed upon pictures of stuck and free regions. At early
times, most particles have not yet fallen into pancakes, so that
the resulting free regions percolate in Lagrangian space, whereas
stuck regions do not; Figure (\ref{fig:5pancake1}).
At late times the situation is reversed since
most of the matter is now in caustics, so that stuck regions percolate
whereas free regions exist as isolated regions; Figure (\ref{fig:5pancake2}).
The percolation
of stuck and free Lagrangian regions is closely related to the issue
of the global topology of large--scale structure, i.e. whether it is
bubble--like or sponge-like; a discussion of this issue can be
found in \S 8.8 and in ref. \cite{sss94b}.

This procedure so far
has given us the volume of voids in Lagrangian space. In order to
obtain their volume in Eulerian space (or real space) we apply
the volume deformation formula obtained from the Zel'dovich
approximation:
\begin{equation}
\d V_e = \d V_l [1-a(t)\lambda_1(\vec{q})][1-a(t)\lambda_2(\vec{q})]
[1-a(t)\lambda_3(\vec{q})],
\label{eq:void_el}
\end{equation}
where $\d V_e$ and $\d V_l$ are the volume elements in Eulerian and
Lagrangian space respectively. Summing over all elementary volume
elements within a given Lagrangian space void now gives us the
volume of the void in Eulerian space:
\begin{equation}
V_e = \sum \d V_e = \sum_{i=1}^{N}
\d V_l(i) [1-a(t)\lambda_1(\vec{q})][1-a(t)\lambda_2(\vec{q})]
[1-a(t)\lambda_3(\vec{q})],
\label{eq:void_el1}
\end{equation}
where $N$ denotes the number of elementary free volume elements in
a given Lagrangian--space void. This quasi--analytic method is very
simple and can be used to obtain the spatial distribution and
sizes of individual voids, as well as the integrated void spectrum
\cite{sss94}.

We would like to mention at this point that voids defined according to
the adhesion model constitute a subclass\footnote{Similarly, voids in the
Zel'dovich
approximation constitute a subclass of voids defined according to
linear theory. This can be seen from the fact that linear theory voids
satisfy the inequality $\sum_i\lambda_i < 0$ (which follows trivially
from the Poisson equation (\ref{eq:Poisson})). Zel'dovich voids, on the
other hand, satisfy a more restrictive class of initial conditions
$\lambda_1, \lambda_2, \lambda_3 < 0$, which follows from the requirement
$dV_E > dV_L$ in equation (\ref{eq:void_el}).} of voids defined in
the Zel'dovich approximation. This follows from the fact that
in the Zel'dovich approximation, a necessary and sufficient condition
for the existence of underdense regions is that all eigenvalues of the
deformation tensor are negative. This is not the case in the adhesion
approximation. The reason for this is that the adhesion approximation
is inherently non--local. One can see in Figure (\ref{fig:5adh1d}) that
a region which is declared a void at an earlier epoch is no longer
one at a later epoch when it has merged with two neighbouring
large voids. Thus not all peaks act as void progenitors at all times.
This example also underlines an important feature of non--linear
gravitational instability. This is that the evolution of small voids
is prone to the tidal influence of surrounding larger voids,
which can cause the void to shrink and even disappear. The disappearance
of small voids in the course of cosmological evolution is
accompanied by a flow of power from small scales to larger scales,
as illustrated in Figure (\ref{fig:5power}) \cite{km92,kpsm92,bv94,sss94}.
For scale invariant spectra $P(k) \propto k^n$, $-1 \le n < 1$,
this results in the
following late time growth law for the mean void diameter:
$D_{void} \propto D_+^{2/(n+3)}(t)$,
where $D_+(t)$ is the growing mode of the linear density contrast
\cite{gss89,sss94}; recall that
$D_+ \propto a(t)$ if the Universe is spatially flat and matter--dominated.

\begin{figure}
\vspace{15cm}
\caption{The void spectrum (number fraction of voids as a function
of diameter) is shown for the truncated power--law spectrum
$P(k) \propto k$ for $k<k_c$;
$P(k)=0$ for $k\geq k_c$ where $k_c=16\times k_f$ and
$k_f$ is the fundamental mode of the box. The figures from top to
bottom correspond to three different expansion epochs described by the
scale factor $a$.
(The scale factor $a$ is also just
the linear density contrast on the grid scale.) Reproduced, with
permission, from \protect{\cite{sss94}}.}
\label{fig:5power}
\end{figure}

\begin{figure}
\vspace{15cm}
\caption{The value of the normalised primordial gravitational potential
$\phi/\phi_{rms}$ evaluated at the centers of voids (in Lagrangian space)
is shown plotted
against the void diameter for the COBE normalised CDM spectrum.
Reproduced, with permission, from \protect{\cite{sss94}}.}
\label{fig:5phi}
\end{figure}

An interesting feature to emerge from studies of voids performed using
the adhesion approximation is that the sizes of voids tend to be
strongly correlated with the height of the primordial gravitational
potential at void centers (defined in Lagrangian space),
with larger voids originating from those
regions where the initial potential is higher; see Figure (\ref{fig:5phi}).
However, voids will not in general have a one-to-one association
with high peaks in the potential simply
because complicated features in the
primordial potential -- ridges, saddle--points, etc. -- can sometimes
lead to the formation of voids in those regions where the potential
does not have a well defined maximum \cite{sss94}.

At late times, voids in Lagrangian space occupy a negligible fraction
of the total volume, and it is straightforward to implement the
adhesion approximation by considering only the maximum value
of the gravitational potential in separate localised regions
(Lagrangian--space voids) as the progenitors of structure. We call
this the ``needle approximation'' since the gravitational
potential is effectively replaced by a set of discrete vertical needles
on which the paraboloid is moved in order to map out the skeleton
of the large--scale structure; see Figure (\ref{fig:5needle}).

\begin{figure}
\vspace{15cm}
\caption{The adhesion approximation at late times -- the ``needle
approximation''.}
\label{fig:5needle}
\end{figure}

The location of the needle is similar in many respects to the location
of the Voronoi nuclei, and indeed the needle approximation gives a
spatial geometry which is in many respects equivalent to a Johnson--Mehl
tesselation.

An important characteristic of the adhesion model, which is not shared
by most approximation schemes, is that it constructs a skeleton that
evolves dynamically with time. This can be expressed in geometrical
language by saying that the paraboloid mapping the large--scale
distribution surveys larger and larger regions of the gravitational
potential as the Universe evolves. It can therefore describe such
dynamical processes as the merger of knots and the disappearance of
small voids, which is not possible in an approach which is
manifestly local in construction such as the Zel'dovich approximation.
It is also clear that if all peaks of the gravitational potential
were taken to have equal heights, the skeleton of large--scale structure
would freeze at late times in the adhesion framework. The adhesion
approximation would in this case be entirely equivalent to a Voronoi
tesselation, performed using the local maxima of the gravitational
potential as sites for the Voronoi nuclei.

Studies of voids
have also shown that they can be populated by substructures, such as
mini--Zel'dovich pancakes and filaments, which run through  voids
bounded by larger pancakes \cite{sss94,vv93}. The likelihood
of a void being populated by substructure increases with its size at
a given time, and decreases with time because voids get emptier
as they evolve. Interestingly, evidence for substructure within
existing large voids has been reported for both the Bo\"{o}tes void
\cite{whkllcow92} and for voids in the CfA survey \cite{pgmk92}.

%% file: ssec6.tex
\section{Statistical Solutions and Scaling Arguments}
In this section we discuss methods for obtaining statistical
information about non--linear clustering in an analytical approach.
In these approaches, one abandons the idea of covering the evolution
of individual fluid elements and concentrates  instead  upon
global statistical properties like the correlation functions.
Even with this simplification, analytical progress is difficult and
one normally has to resort to further approximations.
This normally
involves the assumption of some form of scale--invariance (for which
there is some evidence from the observed properties of the galaxy
distribution; Sec 8).

\subsection{Scaling model for the Correlation Function}
We begin with a simple model for the evolution of the two--point
covariance function of the matter fluctuations which was
first suggested and studied in \cite{hklm91} for the case
where $\Omega=1$. The simplest way of understanding this model is
by reference to the spherical collapse model we discussed in
\S 3.1. Recall that, in the case where $\Omega=1$, a spherical inhomogeneity
virialises at an actual density contrast of order around 100 at a
time when the density contrast predicted by linear theory would be
of order unity. Now the two--point covariance function of the
matter fluctuations is, roughly speaking, the `average' profile
of a spherical clump. Assuming that collapse occurs when the
integrated overdensity reaches some critical value, it seems that
the relevant quantity to deal with should be
the volume--averaged correlation function, defined by
\begin{equation}
\bar{\xi}(r) =\frac{3}{r^{2}} \int_0^x \xi(x) x^2 dx.
\end{equation}
A density enhancement of a factor $1+\delta$ may be thought of
as arising through a spherical collapse by a factor $(1+\delta)^{1/3}$
in radius so one could further expect a given integrated
non--linear correlation on a scale $r_{NL}$
to have arisen from linear fluctuations on a scale $r_L$ in the
initial conditions:
\begin{equation}
r_L=\left[ 1+\bar{\xi}_{NL}(r_{NL}) \right]^{1/3} r_{NL}.
\end{equation}
We must now conjecture that the non--linear integrated correlation
function is a universal function of the linear one, after taking
account the change of scale:
\begin{equation}
\bar{\xi}_{NL}(r_{NL})=f [\bar{\xi}_L(r_L)].\label{eq:hklm_fit}
\end{equation}
The success of this approach depends upon finding the appropriate
function $f$ which should be independent of the form of the initial
power spectrum (or, equivalently, correlation function).
We can see quite easily that, for small $x$, the function $f(x)$
should be $f(x)\simeq x$ to reproduce linear theory. The spherical
collapse model further suggests that $f(x=1)\simeq 100$. If we assume
that at large $x$ the fluctuations enter a period of stable clustering
\cite{p80}, we should get $\xi_{NL}\propto a^{3}$. But the linear
covariance function scales as $a^{2}$. Asymptotically, therefore,
we have $f(x) \propto x^{3/2}$ for large $x$. In between these two
extremes, one can calibrate the observed behaviour of $\bar{\xi}$
against numerical simulations and find a fitting formula for the
intermediate range of scale \cite{hklm91}. Remarkably, it does
appear that a scaling formula of the form (\ref{eq:hklm_fit})
does indeed exist that satisfies the two asymptotic limits:
\begin{equation}
f(x)=\frac{x+0.358x^3+0.0236x^6}{1+0.0134x^3+0.00202x^{9/2}}.
\end{equation}
A generalisation of this formula to non--flat universes and models
with a cosmological constant has been given in \cite{pd94}, who also
showed how to apply similar techniques to the evolution of the
power spectrum rather than the correlation function;
it has also been discussed from a theoretical point of view in
\cite{np94}.

\subsection{Statistical Mechanics: the BBGKY hierarchy}
To obtain a more detailed statistical description of a
non--linear self--gravitating system requires a much more
sophisticated approach than that above, and this leads us
into a treatment based on statistical mechanics.
The pioneering investigation by Davis and Peebles \cite{dp77}
into the hierarchical closure of the BBGKY equations is of such
historical importance in the understanding of non--linear gravitational
clustering that we will discuss it here in some detail. We begin
with some basic statistical mechanics.

Consider a (classical) system of $N$ particles, interacting with each other
via two--body forces described by a potential $\phi(\vec{x}_i, \vec{x}_j)$,
where $\vec{x}_i$ is the position of the $i$-th particle ($i=1, \ldots N$).
We take the particles to be identical, so the Hamiltonian for the system
is symmetric under interchange of particle positions; the potential is
taken to be a function only of the particle separations:
$\phi_{i,j}= \phi(\vec{x}_i, \vec{x}_j) = \phi(|\vec{x}_i - \vec{x}_j|)$.

The phase--space for this system is $6N$--dimensional, denoted by
$\Gamma = \gamma_1 \oplus  \gamma_2 \ldots \oplus \gamma_N$ where $\gamma_i$
is a one--particle phase space. The system is described by  a point in the
phase--space $\Gamma$; let the probability density function (pdf) of this
point be $P_N$. The evolution of $P_N$ is governed by Liouville's
equation:
\begin{equation}
\frac{\partial P_N}{\partial t} + \sum_i \vec{v}_i \cdot \frac{\partial
P_N}{\partial \vec{x}_i} + \sum_{i<j} \vec{a}_i \cdot \frac{\partial P_N}
{\partial \vec{v}_i} = 0.
\label{eq:liou}
\end{equation}
The particle accelerations $\vec{a}_i$ are given by
\begin{equation}
\vec{a}_i = - \sum_{i\neq j} \frac{\partial \phi_{i,j}}
{\partial \vec{x}_j}.
\end{equation}
The fact that this equation involves a $6N$--dimensional function $P_N$
encourages one to seek a lower--dimensional representation. The way
to do this is to marginalise the distribution $P_N$ over part of the
full phase space. For example, to construct a one--point distribution
$P_1(\gamma_1)$, one simply marginalises over all but one of the
phase space elements:
\begin{equation}
P_1(\gamma_1) = \int P_N d\gamma_2 d\gamma_3 \ldots d\gamma_N.
\end{equation}
Marginalising both sides of eq. (\ref{eq:liou}) leads to
\begin{equation}
\frac{\partial P_1}{\partial t} + \vec{v}_1 \cdot \frac{\partial P_1}
{\partial \vec{x}_1} + (N-1)
\int \frac{\partial \phi (|\vec{x}_1 - \vec{x}_2|)}{\partial \vec{x}_2}
\frac{\partial P_2}{\partial \vec{v}_1} d\gamma_2=0. \label{eq:liou2}
\end{equation}
(The factor of $(N-1)$ comes from the sum over $i\neq j$ terms; \cite{s85}).
Equation (\ref{eq:liou2}) makes the problem clear. We cannot solve for
$P_1$ without knowledge of $P_2(\gamma_1, \gamma_2)$ and we will not
be able to get $P_2$ without knowledge of $P_3$ and so on. We will
therefore be led to a hierarchy of $N$ equations and, apparently, no
better off than with the original Liouville equation (\ref{eq:liou}).
The physical reason for this is that the evolution of a phase--space point
depends on the distribution of all the other points, which in turn
depends upon the position of the original point. The fact that $P_1$
depends only on $P_2$ explicitly is by virtue of the two--body nature of
the force between the particles.

Having established the need for an hierarchy, we now derive it
in the standard form called the BBGKY equations, after its inventors
Born, Bogoliubov, Green, Kirkwood and Yvon \cite{b75,s85}. It is
convenient to change the definition of our functions $P_n$ by
including a factor in the marginalising rule:
\begin{equation}
P_n(\gamma_1,\ldots, \gamma_n) = \frac{(N/\rho)^{N}}{(N-n)!}
\int P_N(\gamma_1,\gamma_2 \ldots \gamma_N) d\gamma_{n+1}\ldots d\gamma_N.
\label{eq:marg}
\end{equation}
In this equation, $\rho$ is the mean real--space number--density of
objects (particles).
The factor in front of the integral in eq. (\ref{eq:marg}) takes account
of the number of ways of interchanging labels among the objects
within $P_N$; including this factor in the marginalising rule prevents
the number of particles $N$ from appearing explicitly in the equations.
$P_n(\gamma_1 \ldots \gamma_n)$ is now the probability of finding any
$n$ objects simultaneously in the phase space elements $\gamma_1 \ldots
\gamma_n$. Marginalising both sides of eq. (\ref{eq:liou2}) according
to this rule and integrating gives, fairly straightforwardly,
\begin{equation}
\left[ \frac{\partial}{\partial t} + \sum_{i=1}^{n} \vec{v}_i
\cdot \frac{\partial}{\partial \vec{x}_i} -
\sum_{i}^{n} \frac{\partial \phi_{i,j}}{\partial \vec{x}_i}
\frac{\partial}{\partial \vec{v}_i} \right] P_n =
\sum_{i=1}^{n} \int \frac{\partial \phi_{i,j}}{\partial \vec{x}_i}
\frac{\partial}{\partial \vec{v}_i} P_{n+1} d\gamma_{n+1}.
\label{eq:bbgky}
\end{equation}
Details of the steps involved in this derivation are given in
\cite{s85}; see also \cite{b75}.
The equations (\ref{eq:bbgky}) -- there are $N$ of them -- are called
the BBGKY equations and can be applied to any system evolving according
to equation (\ref{eq:liou}).
The equations (\ref{eq:bbgky}) can also be derived
from the Vlasov equation \cite{rf92}.

To apply the preceding formalism to the case of cosmological
perturbations is not immediately straightforward because the
system is expanding \cite{p80}.
The effective Lagrangian\footnote{This Lagrangian,
in conjunction with the principle
of least action, can be used to trace back galaxy motions
and recover the initial configuration of, for example,
the Local Group; see \protect{\cite{p90,brc94,p94}}.}
for a particle moving
in an expanding universe, in comoving coordinates, is:
\begin{equation}
L = \frac{1}{2} m a^{2} \dot{x}^{2} - m\phi(\vec{x}).
\label{eq:lagran}
\end{equation}
The canonical momentum is $\vec{p}=ma^{2} \dot{\vec{x}}$; if
$\phi=0$ then $\vec{p}$ is conserved so that $\vec{v} =
a(t) \dot{\vec{x}} \propto 1/a(t)$. The potential obeys Poisson's
equation (\ref{eq:Poisson}). We can now write down the BBGKY equations
in an appropriate form:
\begin{equation}
\frac{\partial P_n}{\partial t} + \vec{v}_i \cdot \frac{\partial P_n}{
\partial \vec{x}_i} + \frac{Gm^{2}}{a} \int d\gamma_{n+1}
\frac{\vec{x}_i - \vec{x}_{n+1}}{|\vec{x}_i - \vec{x}_{n+1}|^{3}}
\frac{\partial P_{n+1}}{\partial \vec{v}_i}=0, \label{eq:cbbgky}
\end{equation}
where we implicitly sum over the index $i$, from 1 to $n$.

Now we have to worry about trying to close this hierarchy. The approach
of Davis and Peebles \cite{dp77}, which we shall initially follow,
was to solve for the first two moments of the distribution function
after invoking a separation between velocity and position space
and a hierarchical form for the three--point correlation function
of the particle positions, as follows. First, we construct the
first three moments of the velocity distribution after assuming that
our system is statistically homogeneous, so that the one--point
velocity distribution function, $P_1$ can be taken to be
independent of position:
\begin{equation}
\int d^{3} \vec{v} P_{1} (\vec{v}) = \bar{n} a^{3}
\end{equation}
\begin{equation}
\int d^{3} \vec{v} P_{1} (\vec{v}) \vec{v} = 0
\end{equation}
\begin{equation}
\int d^{3} \vec{v} P_1 (\vec{v}) v^{i}v^{j} = \frac{1}{3}
\bar{n}
a^{3} \langle v_1^{2} \rangle \delta^{ij},
\end{equation}
where $\bar{n}$ is the mean number--density of particles and
$\langle v_1^{2} \rangle$ is the one--particle velocity dispersion;
$v^{i}$ denotes the $i$--th component of the velocity.

We now need integrals over the two-- and three--point distributions,
which we denote by $P_2(1,2)$ and $P_3(1,2,3)$:
\begin{equation}
\int d^{3}\vec{v}_1 d^{3}\vec{v}_2 P_2(1,2) = (\bar{n}a^{3})^{2}(1+\xi_{12})
\end{equation}
\begin{equation}
\int d^{3}\vec{v}_1 d^{3}\vec{v}_2 d^{3} \vec{v}_3 P_3(1,2,3) =
(\bar{n}a^{3})^{3}(1+ \xi_{12}+\xi_{23}+\xi_{13} +\zeta_{123}),
\end{equation}
where $\xi_{12}=\xi(\vec{x}_1-\vec{x}_2)$, etc. Finally, we need
the first two pairwise relative velocity moments:
\begin{equation}
\int d^{3} \vec{v}_1 d^{3} \vec{v}_2 P_2(1,2) v_{12}^{i}
= (\bar{n}a^{3}) \langle v_{12}^{i} \rangle (1+\xi_{12})
\end{equation}
\begin{equation}
\int d^{3} \vec{v}_1 d^{3}\vec{v}_2 v_{12}^{i} v_{12}^{j}
= (\bar{n}a^{3})^{2} \langle v_{12}^{i} v_{12}^{j} \rangle (1+\xi_{12}),
\end{equation}
where $\vec{v}_{12} = \vec{v}_{1} - \vec{v}_{2}$.

The above definitions are substituted into eq. (\ref{eq:cbbgky})
for $n=2$ and we are left with three equations:
\begin{equation}
\frac{\partial \xi_{12}}{\partial t} + \frac{1}{a}
\frac{\partial}{\partial x_{12}^{i}} [ \langle v_{12}^{i} \rangle
(1+\xi_{12}) ] = 0; \label{eq:bbd1}
\end{equation}
\begin{eqnarray}
\frac{1}{a} \frac{\partial}{\partial t} [a \langle v_{12}^{i}\rangle
(1+\xi_{12})] +
\frac{1}{a} \frac{\partial}{\partial x_{12}^{j}}
[\langle v_{12}^{i} v_{12}^{j} \rangle (1+\xi_{12})]
& + & \nonumber\\
G \bar{\rho} a \int d^{3}\vec{x}_3 (\xi_{13}+\xi_{23}+\zeta_{123})
\left( \frac{x_{13}^{i}}{x_{13}^{3}} - \frac{x_{23}^{i}}{x_{23}^{3}}
\right) &  = & 0; \label{eq:bbd2}
\end{eqnarray}
\begin{eqnarray}
\frac{1}{a^{2}} \frac{\partial}{\partial t} \left[ a^{2}
\langle v_{12}^{i} v_{12}^{j} \rangle (1+\xi_{12}) \right]
+ \frac{1}{a} \frac{\partial}{\partial x_{12}^{k}}
\left[ v_{12}^{i} v_{12}^{j} v_{12}^{k} (1+\xi_{12}) \right]
& + & \nonumber \\
G\bar{\rho} a \int d^{3} \vec{x}_3 \left\{
\left[ \frac{1}{2} (1+\xi_{13})v_{13}^{i} - \frac{1}{2} (1+\xi_{23})
v_{23}^{i} + \langle P_3 v_{12}^{i} \rangle \right] \right. & \times &
\nonumber\\
\left. \left( \frac{x_{13}^{j}}{x_{13}^{3}} - \frac{x_{23}^{j}}{x_{23}^{3}}
\right)  (i \Leftrightarrow j) \right\}
& = & 0. \label{eq:bbd3}
\end{eqnarray}
These are dynamical equations governing the evolution of the first
few correlation functions and velocity moments. Notice that
the average $\langle P_3 v_{12}^{i} \rangle$ is over the three--body
phase space distribution: it is a three--body weighted pairwise
velocity.

The hierarchy, or at least the first few levels of it, are now
in a more familiar language to cosmologists but we still have to
introduce some argument in order to close it. We are helped by
the assumption of large--scale homogeneity and isotropy; the
resulting symmetries reduce the number of independent tensor
components. We thus expect the mean pairwise streaming motion
$\langle v_{12}^{i} \rangle=v(x_{12})\hat{x}_{12}^{i}$,
i.e. aligned with the separation; the velocity dispersion
can be split into longitudinal and transverse polarisations:
\begin{equation}
\langle v_{12}^{i} v_{12}^{j} \rangle = \hat{x}^{i} \hat{x}^{j}
\Pi(x_{12}) + (\delta^{ij} - \hat{x}^{i}\hat{x}^{j})\Sigma(x_{12}).
\end{equation}
The notation here follows \cite{f82,rf92};
beware of different definitions of
$\Sigma$ and $\Pi$ \cite{dp77}!
We now need to model the three--body weighted pair velocity,
the three--point correlation function, $\xi_{123}$
and the third velocity moment $\langle v^{i} v^{j} v^{k} \rangle$.

In the limit of weak correlations, one can merely ignore the
higher--order correlations \cite{p80}, or treat them term by
term in a perturbative expansion \cite{f84,ggrw86,b92a,b92b}. This
approach has led to many interesting results in the quasi--linear
regime.

When the correlations are strong, however, three--point correlations
will be even larger in amplitude than two--point correlations, so
we cannot use such a scheme. The Davis--Peebles approach \cite{dp77}
is to close the hierarchy using a phenomenological model. From
the observed behaviour of the three--point correlation function, it
is reasonable to model $\zeta_{123}$ by the hierarchical form:
\begin{equation}
\zeta_{123} = Q(\xi_{12}\xi_{13}+ \xi_{12}\xi_{23} + \xi_{13}\xi_{23})
\end{equation}
(see section 8.3). Indeed such a scaling may also be relevant for the
correlation functions of even higher order, as we discuss later.

To close the velocity moments is rather harder. Here we part company
with \cite{dp77} and follow the steps taken by \cite{rf92}.
The former authors assumed, without much justification, that the
skewness of the velocity distribution should vanish. Less
restrictive considerations lead to the following form for the
required average:
\begin{equation}
\langle v^{i} v^{j} v^{k} \rangle =
A(3v\Pi +v^{3}) \hat{x}^{i} \hat{x}^{j} \hat{x}^{k} + Bv \Sigma
(\hat{x}^{i} \delta^{jk} + \hat{x}^{j} \delta^{ik} + \hat{x}^{k}
\delta^{ij} - 3 \hat{x}^{i} \hat{x}^{j} \hat{x}^{k} ).\label{eq:abdef}
\end{equation}
(The Davis--Peebles form of this expression has $A=B=1$).
The more general form for the three--body weighted pairwise
velocity is
\begin{equation}
\langle P_3 v_{12}^{i} \rangle = v_{12}^{i}(\xi_{13}+\xi_{23})
+v_{13}^{i}\xi_{23} - v_{23}^{i} \xi_{13} + Q^{\ast}
[v_{12}^{i} \xi_{12} (\xi_{13}+\xi_{23})+(v_{13}^{i}-v_{23}^{i})
\xi_{13}\xi_{23}].
\end{equation}
(The analogous Davis--Peebles result has $Q^{\ast}=Q$). Given this
result one can actually determine $A$ and $B$ in terms of $Q$ and
$Q^{\ast}$. Remarkably, it seems that Davis and Peebles got lucky
in that the choice $Q=Q^{\ast}$ requires that $A=B=1$.
They obtained a consistent approximation, when they could quite easily
not have done so.
Substituting  these forms into the dynamical equations
(\ref{eq:bbd1}), (\ref{eq:bbd2}) \& (\ref{eq:bbd3}) leads to
four equations for the
four unknowns $\xi$, $v$, $\Pi$ and $\Sigma$ which we will
not reproduce here; see \cite{rf92}.

In the strong clustering limit ($\xi\gg 1$) one can simplify
the equations considerably by taking a power--law form for the
two--point correlation function as a solution of the
first dynamical equation (\ref{eq:bbd1}), which becomes in this limit:
\begin{equation}
a\frac{d\xi}{da}- \frac{1}{x^{2}} \frac{\partial}{\partial x^{2}}
\left(x^{2}\xi\right) = 0.
\end{equation}
The solution is of the form $\xi \sim a^{\beta} x^{-\gamma}$
with $\beta=3-\gamma$ but $\gamma<2$ for the equations to
converge. From this solution one finds $\Sigma \sim \Pi
\sim a^{2-\gamma}x^{2-\gamma}$. These functional forms are
independent of $A,B, Q,Q^{\ast}$. It turns out that $B=1$
for consistency, but $A$ is determined by
\begin{equation}
A = 1- 4\left(\frac{Q}{Q^{\ast}} -1 \right)
\left( \frac{(\Sigma/\Pi)+\gamma-2}{15-6\gamma} \right),
\end{equation}
so that $A=B$ if and only if $Q=Q^{\ast}$. Notice however
that the power--law solution here exists for any value of
$\gamma$. To obtain the final result, one assumes that
self--similar clustering, which is the ansatz we have
used to close the hierarchy, continues all the way
from the linear regime. In the self--similar solution,
we write $\xi(x,t)=\xi(st)$, where $s=x/t^{\alpha}$
for some power $\alpha$. For an initial power--law power
spectrum of index $n$, linear perturbation theory gives
$\xi \sim t^{4/3} x^{-(3+n)}$. If we assume that this
solution matches onto one of the infinite number of scaling solutions
in the strongly non--linear regime, we are left with an
equation relating $\gamma$ and $n$:
\begin{equation}
\gamma= \left( \frac{9+3n}{5+n} \right). \label{eq:gamman}
\end{equation}

Obviously this treatment rests on a number of simplifying
assumptions whose validity can be questioned\cite{f82,hbpr86,h88,rf92,bhar94}.
These results are nevertheless generally  taken as confirming the
assumption that gravitational
clustering evolves hierarchically in the non--linear regime.
In our view, the issue of whether this is indeed
a correct interpretation of the BBGKY analysis is far from
settled. For a start, the analysis of \cite{s80}, suggests that
$\xi$ should tend asympotically to a slope of $\gamma=2$
in the non--linear limit, independent of $n$. Numerical
experiments using self--similar initial conditions apparently
produce self--similar behaviour, but do not have power--law
correlation functions \cite{efwd88}. Simulations with non--self--similar
initial conditions do not evolve self--similarly and have correlation
functions whose slope varies with time; indeed the slope $\gamma$
seems quite happy to pass through $\gamma=2$ without caring
\cite{defw85}. Hamilton \cite{h88} has further objected that
cluster--cluster correlations and galaxy--galaxy correlations
should be equal if the distribution were truly self--similar and
so these considerations do not apply to the real Universe.
These points seem to confirm the conclusion of
\cite{rf92} that the hierarchical solution of the BBGKY equations
may be unstable to non--self--similar perturbations. Given that
cosmologically--interesting primordial spectra are not usually
of scale--free form one should be very cautious about using this
type of analysis in anything other than qualitative arguments.

As a final remark in this section, we point out that all the
problems associated with the closure ansatz can be avoided
if one deals directly with the Liouville equation for
$P_{N}$, rather than
marginalising to the $P_{n}$. One suggestion as to how to
avoid the hierarchy thus induced is by dealing with the de
Finetti generating function of the distributions, rather than
the distributions themselves. The equation for the time
evolution of this function turns out to be remarkably simple
and may eventually lead to progress in this field, although
it is a novel approach which has only just begun to be
investigated \cite{gg92}.

\subsection{Thermodynamics}
Since the difficulties involved in climbing the BBGKY ladder seem
so extreme, even for only the third rung, one is tempted to
abandon this approach and try for an even more general description
of the properties of self--gravitating material.

In a number of papers, Saslaw and co--workers
\cite{sh84,s85,s86,s89,s92,cs86,scii90,ss93,iis93}
have developed a
thermodynamic approach to galaxy clustering in which they
concentrate upon the distribution function $f(N)$, the
probability\footnote{We use $f(N)$ and $P_N(V)$ interchangeably
to define the same physical quantity; see \S 8.}
of finding $N$ particles in a (randomly--positioned)
spatial volume $V$.
The assumptions involved in this theory
are perhaps even more simplified than in the BBGKY closure problem
but, to coin a phrase, the proof of the pudding is in the eating and
the predictions of the theory do seem to be in good accord
with observations of galaxy counts.

Basically, the theory assumes that a kind of gravitational
quasi--equilibrium operates in galaxy clustering. Now, obviously,
there is no thermodynamical equilibrium overall in a self--gravitating
system; evolution to at least one singularity seems to be the
generic outcome. What one requires is that evolution of a clustering
hierarchy should take place ``slowly'' enough that each successive
phase of the evolution is approximately described by an
equilibrium law. Further objections can be raised, and some
counter--arguments are discussed extensively in \cite{sh84}.
A more careful derivation of the distribution, with slightly
relaxed assumptions is given in \cite{scii90}.

The main result of the theory is that the distribution function
$f(N)$ has the (unique) form:
\begin{equation}
f(N)= \frac{\bar{N}(1-b)}{N!} \left[ \bar{N}(1-b) +Nb \right]^{N-1}
\exp \left( -\bar{N}(1-b)-Nb) \right).\label{eq:therm1}
\end{equation}
This distribution has many interesting statistical properties
\cite{s89}, including the fact that it is invariant under
projection from 3D to 2D like the Poisson distribution.
In eq. (\ref{eq:therm1}), $\bar{N}$ is the mean number of objects
in $V$; $\bar{N}=\bar{n}V$ and $b$ is a parameter which is the
ratio of potential energy to kinetic energy in the system:
$b=-W/2K$. For a Poisson distribution $b=0$ and
\begin{equation}
f(N)= {{\bar N}^N\over {N!}}\exp \left(-\bar {N}\right).
\label{eq:poiss}
\end{equation}
In linear theory $b=3f^{2}/4\Omega$, which reduces
to $b=3/4$ for $\Omega_0=1$. In the fully relaxed
state described by virial equilibrium, $b=1$. Perhaps one
of the reasons for the success of this theory is that the
parameter $b$ only changes slightly during the course of
evolution and, although the appropriate $b$ may well be
different in more--evolved (i.e. denser) regions, the distribution
averaged over the whole system still resembles eq. (\ref{eq:therm1}).
Some more detailed properties of the distribution (\ref{eq:therm1})
are discussed in \S 8.5 where we also discuss the properties
of its moments.

To make more detailed statements about $b$ requires
one to go back to both the BBGKY hierarchy (sec. 6.1) and
use the cosmic energy equation \cite{p80}. On this basis,
one can show that in a cosmological model where the scale
factor grows as $a(t) \propto t^{\alpha}$, the value of
$b$ should be given by
\begin{equation}
b = \frac{2\alpha+1}{2\alpha+2},
\end{equation}
which is in good accord with the observed value $b=0.70\pm 0.05$, taken
from the best fitting form of (\ref{eq:therm1}) to the
counts of galaxies in the Zwicky catalogue \cite{cs86}. More recently,
Saslaw has shown that the time--evolution of $b$ is described by the
relation:
\begin{equation}
a(t) = a_{\ast} b^{1/8} (1-b)^{-7/8}.
\end{equation}

Some authors, e.g. \cite{s92}, have claimed that this is the behaviour
followed by $N$--body numerical simulations of self--gravitating
systems, but this has been questioned \cite{bsd91}.
There is evidence from the latest galaxy redshift surveys
that $f(N)$ is indeed well fitted by the form (\ref{eq:therm1});
there are, however, other distribution functions, such as the
Poisson--lognormal model \cite{cj91}, which fit the observations
just as well \cite{sms93}.

\subsection{Fractals and Scaling}
The self--similar properties that seem to be implied by both
observations and the BBGKY analysis described above, lead one
naturally to a description of the mass distribution in the
language of fractal sets \cite{m77,m82,borg94}. The prevalence of
techniques based on fractal geometry in fields such as condensed
matter physics has led to an interest in applying these methods
to the cosmological situation.

To get a rough idea of the fractal description consider the
mass contained in a sphere of radius $r$ around a given
galaxy. In the case where $\xi(r)\gg 1$ we have
\begin{equation}
M(r) \propto \xi(r) r^{3},
\end{equation}
since $\xi(r)$ has a power--law form with a slope of around
$\gamma=1.8$ then we have that $M(r) \propto r^{1.2}$.
In the language of fractals, this corresponds to a {\em correlation dimension}
of $D=3-\gamma=1.2$. One can interpret this very simply by noting
that if mass is distributed long one--dimensional structures
(filaments) then $M(r)\propto r$; two--dimensional sheets would
have $M\propto r^{2}$ and a space--filling homogeneous distribution
would have $M\propto r^{3}$. A fractional dimension like that observed
indicates a fractal structure like those discussed by Mandelbrot
\cite{m77,m82}.

Many attempts have been made to apply this kind of description
to the observed universe \cite{p74,gr75,efh79,p80,pi87,cps88,cp92}.
A completely fractal universe actually would cause a
lot of problems for us, because it need not tend to
homogeneity on large scales and we would not therefore be able
to do statistical cosmology! Fortunately, it seems that the
real Universe is not described by a simple fractal of this
kind \cite{mj90}; a more sophisticated description is needed,
in terms of {\em multifractals} \cite{pv87,borg94}.
Nevertheless, simple scaling arguments can be useful for
describing the behaviour of simple statistical properties of the
matter distribution; see, for example, \cite{efh79,hklm91,pd94}.

One can understand a multifractal model in terms of the
simple argument given below. Suppose that, around any individual
point, the mass contained within $r$ scales as $M(r) \propto
r^{\alpha}$, but that $\alpha$ varies from point to point:
$\alpha=\alpha({\bf x})$. In a simple fractal -- a {\it monofractal} --
$\alpha$ would be independent of spatial position. If we
make the assumption that the subset of points upon which $\alpha$
takes a given value is itself such a monofractal, then the set as a whole
consists of a set of interwoven simple fractals
of different dimensionality and is therefore termed a multifractal.
This is therefore a generalisation of the concept of a fractal, but
it is not completely general.
The function $f(\alpha)$ defines the dimensionality of
the sets of constant $\alpha$: it must have only one maximum
at $\alpha_0$; $f(\alpha_0)=D_0$ is the {\it capacity dimension}
of the whole set (sometimes called the {\it Hausdorff dimension} or,
more loosely, the {\it fractal dimension}).
The crucial property that a multifractal set possesses is that
the  scaling dimensions for the $q$--th moment of cell
counts, the {\it generalised dimensions}, are not independent of $q$:
\begin{equation}
D_q = \lim_{r\rightarrow 0} \frac{1}{(q-1)} \frac{ \log \sum_{i=1}^{N(r)}
[n_i(r)/N]^{q}}{\log r}
\end{equation}
($q\neq 1$; $n_i(r)$ is the occupation number of the $i$--th cell
of side $r$; $N$ is the number of cells.
An alternative definition is required for
$q=1$, but we shall not discuss this special case here.
For a monofractal, $D_q=D_0$ for all $q$.
We shall discuss the methods for extracting $D_q$ from the
data in Sec. 8.9; it is clear that $D_q \neq D_0$ for both
data and simulations \cite{mj90,mjdv90,mc94} and that the
multifractal description provides a good qualitative model for the
form of the clustering.

It is perhaps helpful to consider a simplified
two--dimensional model which demonstrates
the difference between monofractal and multifractal distributions.
 First, consider a square, divided into four equal square pieces.
To each of the four
sub-squares is assigned a number
$p_i$ ($i=1,2,3,4$), where $\sum p_i=1$.
Each sub-square is then divided in the same way as the original,
and each subdivision is randomly assigned one of the $p_i$ as
above. This process is repeated an arbitrarily large number
of times. If it is repeated $n$ times we have generated a
$2^n \times 2^n$ array of numbers, each one of which is a product
of $n$ of the $p_i$, randomly chosen. It is quite easy to show that
the $D_q$ for this distribution can be written
\begin{equation}
D_q = (1-q)^{-1} \log_2 \left( p_1^{q}+p_2^{q}+p_3^{q}+p_4^q\right),
\end{equation}
for $n\rightarrow \infty$. The proof of this is given by Meakin
\cite{meak87}. Now suppose $p_4=0$ and $p_1=p_2=p_3=1/3$. In this
case $D_q=D_0$ for all $q$ and the model is a monofractal: each
occupied cell has been produced by the same kind of scaling as all its
neighbours. If the $p_i$ are not degenerate, e.g $p_1=0.4$,
$p_2=0.3$, $p_3=0.2$ and $p_4=0.1$ then the $D_q$ are a function
of $q$ and the model must be a multifractal. One can see that
each occupied cell at the lowest level is produced by a different kind
of scaling, which depends on the particular assignment of a
$p$ at each stage of the hierarchy.

Although multifractal models can give a good insight into the
scaling properties of the galaxy distribution, there is as yet
no connection between this description and the dynamical origin
of clustering. For the time being such models must be regarded
as purely phenomenological. Examples of multifractal models
based on hiearchical fragmentation have been discussed
\cite{mjdv90,jcm92}
but the models involved are quite crude. One interesting point
can be raised in connection with some of the previous discussion
in this paper. As we shall see, the observed distribution of
scaling indices corresponds roughly to a quadratic $f(\alpha)$
spectrum, at least around the maximum. Now $f(\alpha)$ is related
to $D_q$ by a Legendre transform:
\begin{equation}
\tau(q)=\alpha q -f(\alpha), \,\,\,\,\,\,\,\,\, \alpha(q) =\frac{\d\tau}
{\d q}, \,\,\,\,\,\,\,\,\,\, D_q = \frac{\tau(q)}{(q-1)}. \label{eq:legendre}
\end{equation}
A parabolic $f(\alpha)$ is therefore equivalent to a linear dependence
of $D_q$ upon $q$; if the $f(\alpha)$ degenerates to a $\delta$--function
then $D_q$ is constant, as discussed before. It can be shown that
the former situation leads to a log--normal distribution of cell--counts
\cite{jcm92}; this shows the consistency of the lognormal
approximation \cite{cj91}. More generally, multifractal sets
can be described as spatially intermittent, being characterised
by large--void regions and dense clusters phenomenologically like
the observed distribution. It has been suggested that the word
{\em heterotopic} should be used to describe clustering of this
kind \cite{jcm92}.

%% file: ssec7.tex
\section{Numerical Methods}
We  have demonstrated that many aspects of the structure
formation process
can be understood analytically, with the aid of the various
approximation schemes discussed in previous sections. Nevertheless,
many situations remain outside the reach of analytical methods and one
is compelled to adopt a numerical approach to solve the equations of
motion of the gravitating matter. In general, numerical simulations are
also used to test the applicability of analytical approximations
under the assumption that they give the closest possible representation
of the `true' non-linear density field. Although this review is
intended to cover the analytical schemes in most detail, we feel it
is appropriate to give a brief description of the various
numerical methods on offer. Part of the purpose of doing this is to
remind the reader that these methods are themselves approximate, and
one should not assume that numerical simulation techniques are therefore
necessarily the best way to answer every possible question about
the non--linear clustering pattern.

Any numerical simulation will have limitations in terms of size
and resolution. These limitations stem from unavoidable constraints
on memory and computational time. The size constraint means that
one can only handle the evolution of a relatively small, but
hopefully `representative', piece of the initial hypersurface upon
which the primordial density perturbations are laid down.
One must therefore at least run an ensemble of simulations to
assess the `cosmic variance' in any statistic extracted from the
final simulation configuration. It is not possible in a simulation to
assess easily the effect of varying the initial power spectrum
or fluctuation statistics: one has to run the simulation again
from scratch for each choice. Despite these limitations, numerical
simulations have led to considerable increases of our understanding
of the clustering process over the past two decades.

The literature already contains many excellent reviews of numerical
techniques in cosmology, so we shall not go into great detail here
about optimal parameter choices, resolution effects and numerical
artefacts. What we shall do is run briefly through the basic
techniques and point the reader towards more detailed references
where appropriate.

\subsection{Direct $N$--body Simulation}
The simplest way to compute the non--linear evolution of a
cosmological fluid is to represent it as a discrete set of
particles, and then sum the (pairwise) interactions between particles
directly to calculate the Newtonian force on each one.
With the adoption of a (small) timestep, one can use
the resulting acceleration to update the particle velocity
and then its position. The new positions can then be
used to re-calculate the interparticle forces, and so on.
One should note at the outset that  particle techniques are not
actually intended to represent the motion of a discrete set of particles.
The particle configuration is itself an approximation to a fluid.
There is also a numerical problem with  summation of interparticle
forces:  the Newtonian gravitational force
between two particles increases as the particles approach each other
and it is therefore necessary to choose an extremely small timestep
to resolve the large velocity changes this induces. A very small timestep
would require the consumption of enormous amounts of CPU time, so
one usually avoids the problem by treating each particle not as
a point mass, but as an extended body. The practical upshot of this
is that one modifies the Newtonian force law between particles:
\begin{equation}
\vec{F}_{ij} = \frac{Gm^{2}(\vec{x}_j-\vec{x}_i)}{
\left( \epsilon^{2} + | \vec{x}_i-\vec{x}_j|^{2} \right)^{3/2}},
\label{eq:ns_1}
\end{equation}
where the particles are at positions $\vec{x}_i$ and they all have the
same mass $m$; the form of this equation avoids infinite forces
at zero separations.
The parameter $\epsilon$ in eq. (\ref{eq:ns_1})
is usually called the {\em softening length} and it acts to
suppress two--body forces on small scales. This is equivalent
to replacing point masses by extended bodies with a size of
order $\epsilon$. Since we are not supposed to be dealing with
the behaviour of a set of point masses anyway, the introduction
of a softening length is quite reasonable but it means one cannot
trust the distribution of matter on scales of order $\epsilon$
or less.

If we suppose our simulation contains $N$ particles, then the
direct summation of all the $N$ interactions to compute the
acceleration of each particle requires a total of
$N(N-1)/2$ evaluations of (\ref{eq:ns_1}) at each timestep.
This is the crucial limitation of these methods: they tend
to be very slow, with the timescale scaling roughly as $N^{2}$.
The maximum number of particles for which it is practical to
use direct summation is of order $10000$, which is not
sufficient for realistic simulations of large--scale structure
formation. This method was, however, used in the early days
of structure formation studies \cite{agt79,a84,bt87};
see also \cite{dknpss80}. A standard
direct $N$--body code, written by Sverre Aarseth, is published
in \cite{bt87}.

\subsection{Particle--Mesh Techniques}
The usual method for improving upon direct $N$--body summation
for computing inter--particle forces is some form of `particle--mesh'
($PM$) scheme. In this scheme the forces are solved by assiging mass points
to a regular grid and then solving Poisson's equation on the grid.
The use of a regular grid allows one to use Fourier transform methods
to recover the potential which leads to a considerable increase in
speed. A comprehensive account of $PM$ methods is given in the
standard textbook on particle methods \cite{he81} and cosmological
applications are described by \cite{ee81} and \cite{edfw85}. We
shall merely give an outline here, following \cite{edfw85}.
The basic steps in a $PM$ calculation are as follows.

In the following, $\vec{n}$ is a vector representing a grid position
(the three components of $\vec{n}$ are integers); $\vec{x}_i$ is the
location of the $i$--th particle in the simulation volume; for simplicity
we adopt a notation such that $G\equiv 1$, the length of the side of
the simulation cube is unity and the total mass is also unity;
$M$ will be the number of mesh--cells along one side of the simulation
cube the total number of cells being $N$; the vector $\vec{q}=\vec{n}/M$.
First we assign mass points to the grid:
\begin{equation}
\rho(\vec{q})= \frac{M^{3}}{N} \sum_{i=1}^{N} W(\vec{x}_i-\vec{q}),
\label{eq:ns_2}
\end{equation}
where $W$ defines a weighting scheme designed to assign mass to the
mesh scheme. We then calculate the potential by summing over
the mesh
\begin{equation}
\phi(\vec{q})=\frac{1}{M^{3}} \sum_{\vec{q}'} {\cal G} \left(
\vec{q}-\vec{q}' \right) \rho(\vec{q}');
\label{eq:ns_3}
\end{equation}
compute the resulting forces at the grid points,
\begin{equation}
\vec{F}(\vec{q}) = -\frac{1}{N} \vec{D}\phi;
\label{eq:ns_4}
\end{equation}
and then interpolate to find the forces on each particle,
\begin{equation}
\vec{F}(\vec{x}_i) = \sum_{\vec{q}} W(\vec{x}_i-\vec{q}) \vec{F} (\vec{q}).
\label{eq:ns_5}
\end{equation}
In equation (\ref{eq:ns_4}), $\vec{D}$ is a finite differencing scheme
used to derive the forces from the potential. We shall not go into
the various possible choices of weighting function $W$ in this
brief treatment: possibilities include `nearest gridpoint' (NGP),
`cloud--in--cell'' (CIC) and `triangular shaped clouds' (TSC).
Each has pros and cons which are described by \cite{edfw85}, who
also discuss possible choices of finite--differencing scheme
for eq. (\ref{eq:ns_4}). For a general text on particle methods,
see \cite{he81}.

The most important advantage of the $PM$ method is that it allows
a fast summation of the forces. The potential on the grid can be
written
\begin{equation}
\phi(l,m,n) = \sum_{p,q,r} \hat{\cal G} (p,q,r) \hat{\rho}
(p,q,r) \exp \left[ \frac{2\pi i}{M}(pl+qm+rn) \right],
\label{eq:ns_6}
\end{equation}
where the `hats' denote Fourier transforms of the relevant
mesh quantities. There are different possibilities for the
Green's function $\hat{\cal G}$, the most straightforward
being simply
\begin{equation}
\hat{\cal G}(p,q,r) = \frac{-1}{\pi\left(p^2+q^2+r^2\right)},
\label{eq:ns_7}
\end{equation}
unless $p=q=r=0$, in which case $\hat{\cal G}=0$;
alternative, more complicated, forms for the Green's function
are possible \cite{edfw85}.
Equation (\ref{eq:ns_6}) represents a sum, rather than the
convolution in equation (\ref{eq:ns_3}), and it's evaluation
can therefore be performed much more quickly.
A Fast--Fourier Transform is basically a processs which is
of order $N\log N$ in the number of grid points and this
represents a substantial improvement for large $N$
over the direct summation technique. The price to be paid
for this is that the usual Fourier summation method implicitly
requires that the simulation box have periodic boundary
conditions. This is probably the most reasonable choice
of boundary conditions for simulating a `representative'
part of the Universe. Methods do exist, however, for handling other
types of boundary, such as isolated ones, and one can add
an exterior potential by linear superposition.

The potential weakness with this method is the comparatively
poor force resolution on small scales because of the finite
spatial size of the mesh. A substantial increase in spatial
resolution can be achieved by using instead the
`particle--particle--particle mesh' method ($P^{3}M$), also
described in detail by \cite{ee81,he81,ks83,edfw85,bg91}.
Here, the short--range resolution of the algorithm is
improved by  adding a correction to the mesh force. This
contribution is obtained by summing directly all the force
contributions from neighbours within some fixed distance
$r_s$ of each particle. A typical choice for $r_s$ will be
around 3 grid units.
Alternatively, one can use a modified
force law on these small scales to assign any particular density
profile to the particles, similar to the softening procedure
demonstrated in equation (\ref{eq:ns_1}). This part of the
force calculation may well be quite slow, so it is advantageous
merely to calculate the short--range force at the start for
a large number of points spaced linearly in radius, and then
find the actual force by simple interpolation. The long--range
part of the force calculation is done by a variant of the
$PM$ method described earlier.

One of the problems with straightforward $P^{3}M$ is that
when clustering is highly developed, many particles in dense
regions have separations $<r_S$. This means that the algorithm
spends much of its time calculating particle--particle forces
directly and the saving introduced by using the mesh potential
can be lost. An ingenious resolution of this problem was
suggested by Couchman \cite{cou91}. This idea involves using
sub--grids to speed up the direct summation of the short range forces.
Essentially, one zooms in on a small part of the original simulation
grid and uses a finer mesh to apply a $PM$ calculation within
this part. The algorithm can speed up a large $N$--body simulation
by a factor of 100 or more \cite{bg91}; it also allows
a fully parallel implementation. Using these methods it
is feasible to run simulations\footnote{Notice that
the FFT algorithm is optimal if the number of mesh points
is a power of two, so most simulations involve
parameters obeying this choice.} of $256^{3}$ particles on
a $512^{3}$ mesh within a reasonable time, if one has access
to a supercomputer. Even a fairly ordinary desktop workstation
can easily cope with $128^{3}$ particle simulations.

Variants of the $PM$ and $P^{3}M$ technique are now the standard
workhorses for cosmological clustering studies. Different workers
have slightly different interpolation schemes and choices of
softening length, etc. Whether one should use $PM$ or $P^{3}M$
in general depends upon the degree of clustering one wishes
to probe. Strongly non--linear clustering in dense environments
probably requires the force resolution of $P^{3}M$. For larger
scale structure analyses where one does not attempt to probe
the inner structure of highly condensed objects, $PM$ is
probably good enough. One should, however, recognise that the
short--range forces are not computed exactly, even in $P^{3}M$,
so the apparent extra resolution may not necessarily be saying
anything physical. Moreover, both $P^3M$ and the tree codes we
discuss below handle fewer particles than $PM$ for the same
investment of CPU time, leading to poorer mass resolution and
greater discreteness effects. As Melott has pointed out,
``More resolution isn't necessarily better resolution''
\cite{m90b}. Nevertheless, in situations where one needs
to follow the evolution of perturbations into the strongly
non--linear regime, for instance when one is looking at
the formation and evolution of individual galaxies or clusters,
the $P^{3}M$ algorithm may represent a substantial improvement
over plain $PM$.

\subsection{Tree Codes}
An alternative procedure for enhancing the force resolution
of a particle code whilst keeping the necessary CPU demands within
reasonable limits is to adopt an hierarchical subdivision procedure.
The generic name given to this kind of technique is `tree codes'.
The basic idea is to treat distant clumps of particles as single
massive pseudo--particles. The first application of this kind of
method to cosmology was suggested by Barnes \& Hut \cite{bh86}
whose algorithm involves a mesh which is divided into cells
hierarchically such that every cell which contains more than one
particle is divided into $2^{3}$ sub--cells. If any of the resulting
sub--cells contains more than one particle, that cell is sub--divided
again. There are some subtleties involved with communicating
particle positions up and down the resulting `tree', but it is
basically quite straightforward to treat the distant forces
using the coarsely--grained distribution contained in the high
level of the tree, while short--range forces use the finer
grid. The problem with such codes is that, as well as running
slower than $P^3M$ for the same resolution,
they do require substantially greater
memory resources. Their use in cosmological contexts has so
far therefore been quite limited, as memory is often the
limiting factor for simulation work, rather than CPU speed.
More details of the procedure, as well as various tests
and a comparison with particle--mesh techniques are given
by \cite{bh86,bh88,sii90,sw94}.

There has been a growing effort in recent years to parallelise N-body
codes in order to take advantage of the massively parallel machines such
as the CM-5 and the  CRAY T3D. These methods distribute the computational
load evenly amongst N-processors (where N can be as large as 1024 for the
CM5). A discussion of parallel tree codes can be found in
\cite{jks91,wqsz92}.

\subsection{Hydrodynamical Methods}
In most of this section, and indeed most of this article, we
have concentrated upon schemes for evolving the purely gravitational
interactions of the distribution of matter. This may be reasonable
in structure formation scenarios where the dominant contribution is
from collisionless dark matter, but in strongly non--linear
structures one expects pressure forces due to the baryonic
component to become dynamically important, perhaps resulting
in shock formation and dissipation. Moreover, one ideally wants
to know about cooling and star formation in collapsing structures
to have some idea of how to relate the distribution of material
with the distribution of luminosity revealed in galaxy catalogues.
The field of cosmological hydrodynamics is very much in its
infancy, and it is fair to say that there are no analytic
approximations that can be implemented with great confidence in this
kind of analysis, though some phenomenological models of galaxy
formation have been suggested based on analytical
considerations \cite{wf91,cole91,ls91,lgrs93,kw93,kwg93,cafnz94,kgw94}.
For the most part, however, hydrodynamics requires
a  numerical approach; we shall briefly review the two main
methodologies in this section.

\subsubsection{Smoothed Particle Hydrodynamics}
In smoothed--particle hydrodynamics (SPH) one typically
represents the fluid as a set of particles in the same way
as in the $N$--body gravitational simulations described
in \S 7.1--7.3. Densities and gas forces at particle locations
are thus calculated by summing pairwise forces between particles.
Since pressure forces are expected to fall off rapidly with separation,
above some smoothing scale $h$ (see below) it is reasonable to
insert the gas dynamics into the part of a particle code that
details the short--range forces, i.e. the particle--particle
part of a $P^{3}M$ code. This makes it natural
to insert SPH dynamics into a $P^{3}M$ code, as suggested
by Evrard \cite{e88} in the following manner.

One technique used to determine local densities and pressure gradients
is known as kernel estimation. This is essentially equivalent to
convolving a field $f(\vec{x})$ with a filter function $W$ to produce
a smoothed version of the field:
\begin{equation}
f_s(\vec{r}) = \int \d^{3} \vec{x} f(\vec{x}) W(\vec{x}-\vec{r}),
\label{eq:ns_8}
\end{equation}
where $W$ contains some implicit smoothing scale $h$; one possible
choice is a Gaussian. If $f(\vec{x})$ is just the density
field arising from the discrete distribution of particles then
it can be represented simply as the sum of $\delta$--function
contributions at each particle location $\vec{x}_i$ and one
recovers equation (\ref{eq:ns_2}). We need to represent the
pressure forces in the Euler equation (\ref{eq:Euler}): this
is done by specifying the equation of state of the fluid
$P=(\gamma-1)\epsilon\rho$, where $\epsilon$ is the thermal energy
and $\rho$ the local density and $P$ the pressure.
Now one can write the pressure
force term in equation (\ref{eq:Euler}) as
\begin{equation}
-\left(\frac{\vec{\nabla}P}{\rho} \right)=
-\vec{\nabla} \left(\frac{P}{\rho}\right) - \left(\frac{P}{\rho^{2}}\right)
\vec{\nabla} \rho. \label{eq:ns_9}
\end{equation}
Now the gradient of the smoothed function $f_s$ can be
written
\begin{equation}
\vec{\nabla} f_s (\vec{r}) = \int \d^{3} \vec{x}
f(\vec{x}) \vec{\nabla} W(\vec{x}-\vec{r}),
\label{eq:ns_10}
\end{equation}
so that the gas forces can be obtained in the form
\begin{equation}
\vec{F}^{\rm gas}_i = - \left( \frac{\vec{\nabla}P}{\rho} \right)_i
\propto - \sum_j \left( \frac{P_i}{\rho_i^{2}} + \frac{P_j}{\rho_j^{2}}
\right) \vec{\nabla} W(\vec{r}_{ij}).
\label{eq:ns_11}
\end{equation}

The form of equation (\ref{eq:ns_11}) guarantees conservation of linear
and angular momentum when a spherically symmetric kernel $W$ is used.
(The smoothing scale $h$ need not be constant \cite{e88},
but care must be taken to include derivatives of $h$
in a self--consistent manner to be sure the equations
are conservative \cite{np93}.) The
adiabatic change in the internal energy of the gas can similarly be
calculated:
\begin{equation}
\frac{\d \epsilon_i}{\d t} \propto \left(\frac{P_i}{\rho_i^{2}}\right)
\sum_j \vec{\nabla} W(\vec{r}_{ij}) \cdot \vec{v}_{ij},
\label{eq:ns_12}
\end{equation}
where $\vec{v}_{ij}$ is the relative velocity between particles.
For collisions at a high Mach number, thermal pressure will not
prevent the particles from streaming freely, but in real gases
there is molecular viscosity which prevents penetration of gas
clouds. This is modelled in the simulations by introducing a
numerical viscosity, the optimal form of which depends upon the nature
of the simulation being attempted; see the discussion in \cite{e88}.

It is also possible to implement SPH on the back of a gravitational
tree code, as described in \S 7.3. A substantial study by
Hernquist \& Katz \cite{hk89} has recently been devoted to such methods
so we shall not describe them in more detail here.
Cosmological applications of SPH methods and their variants can be
found in \cite{kg91,nb91,khw92,nw93,kzw93,esd94}.
Note also that
tree--SPH methods have been used extensively in stellar collapse
problems, e.g. \cite{np93}.

\subsubsection{Eulerian Hydrodynamics}
The advantage of particle--based methods is that they are
Lagrangian and consequently follow the motion of the fluid. In
practical terms, this means that most of the computing effort
is directed towards places where most of the particles are, and
therefore, where most resolution is required. As mentioned
above, particle methods are the standard numerical tool for
cosmological simulations.

Classical fluid dynamics, on the other hand, has usually
followed an Eulerian approach where one uses a fixed
(or perhaps adaptive) mesh. Codes have been developed which
conserve flux and which integrate the Eulerian equations
of motion rapidly and accurately using various finite--difference
approximation schemes. It has even proved possible to
introduce methods for tracking the behaviour of shocks
accurately -- something which particle codes ofted struggle
to achieve. Two examples of Eulerian codes used for
cosmology are detailed in \cite{cen92} and \cite{rokc93}.
Typically, these codes can treat many more cells than an
SPH code can treat particles, but the resolution is usually
not so good in some regions because the cells will usually be
equally spaced rather than being concentrated in the interesting
high--density regions. We refer the reader to the above
references for details of these codes; we have no space to
describe them here.

An extensive comparison between Eulerian and Lagrangian
hydrodynamical methods has recently been performed
in \cite{hocrh93}, which we
recommend to anyone thinking of applying these techniques in
a cosmological context. Each has its advantages and disadvantages.
Density resolution is usually better in state--of--the--art
Lagrangian codes, except in low--density regions where the number
of particles is so small that statistical effects dominate.
On the other hand,  Eulerian codes generally resolve the
temperature much better, except in very high density regions
where the numerical viscosity is artificially large.

\subsection{Initial Conditions and Boundary Effects}
To complete this section, we make a few brief remarks about
starting conditions for $N$--body simulations, and the effect
of boundaries and resolution on the final results.

Firstly, one needs to be able to set up the initial conditions
for a numerical simulation in a manner appropriate to the
cosmological scenario under consideration. For most models
this means making a random--phase realisation\footnote{Non--Gaussian
initial conditions are discussed by,
for example, \protect{\cite{mmlm91,wc92}}.}
of the power--spectrum -- see \S 2.4.
This is usually achieved by setting up
particles initially exactly on the grid positions, then
using the Zel'dovich approximation, equation (\ref{eq:z1a}),
to move them such as to create a density field with the
required spectrum and statistics. The initial velocity field
is likewise obtained from the primordial gravitational
potential. This procedure is
quite straightforward and is discussed, for example, in
\cite{dknpss80,edfw85}.
One should beware, however, the effects of the
poor $\vec{k}$--space resolution at long wavelengths.
The assignment of $\vec{k}$--space amplitudes requires
a random amplitude for each wave vector contained in the
reciprocal--space version of the initial grid. As the wave
number decreases, the discrete nature of the grid becomes
apparent. For example, there are only three (orthogonal)
wave vectors associated with the fundamental mode of the box.
When amplitudes are assigned via some random--number generator,
one must take care that the statistically poor sampling
of $\vec{k}$--space does not lead to spurious features
in the initial conditions.
One should use a simulation
box which is rather larger than the maximum scale at which
there is significant power in the initial power--spectrum.
At the opposite end of the resolution
question is the
finite spacing of the grid; this means that one should not
trust structure on scales corresponding to greater than
the Nyquist frequency of the initial grid, as these
cannot be resolved by the mesh. Indeed, the resolution will be
determined by whichever is the higher of the particle and mesh
Nyquist frequencies.

Related to this warning is the role of boundary effects on the
simulation. Recall that the particle--mesh methods require the
imposition of periodic boundary conditions, which may also
distort the structure on scales of order the simulation
box. An ingenuous test of the effect of periodic boundary
conditions was recently devised by Farrar \& Melott \cite{fm90}
who imposed the topology of a Klein bottle upon the simulation
box to see how much this altered the results with respect to the
usual situation of a cube with opposite faces identified.
The results suggest that boundary effects are minimal for the
kind of simulation parameters used in most cosmological
studies.

%% file: ssec8.tex
\section{Statistical Description of Galaxy Clustering}
To put our theoretical considerations into context we shall,
in this section, give a concise review of the observed
properties of the galaxy and cluster distributions. We also
give a fairly complete discussion of the various statistical
descriptors used to characterise the essential features of the
clustering pattern. Of course, different approximations will
reproduce properties of the fully non--linear density field
with varying degrees of accuracy. As we have seen before, in \S 4.2
and \S 4.7, techniques
like the Zel'dovich approximation should reproduce well the large--scale
geometry and topology of the distribution, but might be less accurate
at matching `hard' statistical properties like the correlation functions
and cell--counts. Although we have been able in this review to explain
how some approximations lead to estimates of low--order moments
and correlation functions, their behaviour in terms of more
sophisticated statistical properties has only recently begun to be investigated
\cite{ssmpm94}.
We review these methods here to give some idea of the work still
to be done.

Let us, however, lodge an important reservation about all the
statistical methods that utilise only positional (and not velocity)
information about galaxies. As we mentioned in \S2.4, we do not
know how to relate galaxy counts to mass fluctuations. So when
we extract statistical parameters from a galaxy survey there is
no unique way to relate that to parameters which we can calculate
theoretically which are sensitive to the total mass. There
could well exist a complex bias \cite{k84,dr87,k91,bcfw93,c93}
which intervenes in such
a way that galaxy statistics are a gross distortion of the
underlying mass distribution. In \S 8.10 we discuss recent attempts
to use the other half of phase space, peculiar velocities,
to try to probe the full mass distribution, but as we shall see,
attempts to do this are in their infancy and the data are meagre.

We begin by briefly reviewing the kinds of cosmological
clustering data that are available at the present.

\subsection{Galaxy Surveys}
Although the clustering of galaxies was observed empirically
in the days of John Herschel, it is only in the last few
decades that it has been possible to construct complete and
well--controlled samples of galaxies with well--defined
selection criteria. It is only more recently still that we
have been able to build such surveys in three dimensions,
using breakthroughs in instrumentation. Here we give a
brief historical approach to the most recent developments
in this field.

\subsubsection{Selection Functions}
Let us first define some simple terms. Since galaxies do not
have the same intrinsic brightness, different objects can be
seen at different distances from the observer with a detection
method of a fixed flux (apparent luminosity) sensitivity. If
$n(L)$ is the number of galaxies per unit luminosity, then
$n(L)\d L$ is the number in the luminosity range $(L, L+\d L)$.
If the sample is {\em complete} over some volume, then the
Luminosity Function (LF) is defined by the number of galaxies
per unit luminosity per unit volume: $\Phi(L)\d L=n(L)\d L/V_s$, where
$V_s$ is the sample volume. If the luminosity function is
universal, then $\Phi(L)$ is not a function of position
and will be the same  functional form for any sample selection criteria.
This indeed appears to be the case and the correct
form is given by:
\begin{equation}
\Phi(L)\d L = \Phi^{\ast} \left( \frac{L}{L^{\ast}} \right)^{\alpha}
\exp \left(-\frac{L}{L^{\ast}}\right) d\left(\frac{L}{L^{\ast}}\right),
\label{eq:Schechter}
\end{equation}
which is called the Schechter luminosity function \cite{s76}. Typical values
for the parameters are $\alpha\simeq -1.25$ and $L^{\ast}\simeq
10^{10} L_{\odot}$; $\Phi^{\ast}$ is an overall normalisation
constant, giving the number--density of galaxies in the sample.

In general, of course, we cannot construct a sample consisting
of every galaxy within a given volume. The usual method is
to select all galaxies above some fixed flux limit. If this
apparent luminosity limit is $l$, then one can easily calculate
the intrinsic luminosity $L(l,r)$ which a galaxy would have to
have at distance $r$ to be just at the threshold of inclusion
in the sample. Let us put this in dimensionless form
by defining $x(r)$ to be this limiting magnitude,
$x(r)=L(r)/L^{\ast}$ and $x_{\ast}$ to be the dimensionless
luminosity of the faintest source included in the survey at
any distance. The selection function $\phi(r)$ is then
\begin{equation}
\phi(r) = \frac{\int_{x(r)}^{\infty} x^{\alpha} e^{-x} \d x}
{\int_{x_{\ast}}^{\infty} x^{\alpha} e^{-x} \d x}.
\label{eq:selection}
\end{equation}
This equation (\ref{eq:selection}) is
appropriate for optical galaxies; IRAS galaxies, for example, have
a different luminosity function, with
a much shallower high--luminosity cutoff than this and so
can be used to sample comparatively larger scales
\cite{lea94}. Because some galaxies are missed if one deals with
a magnitude limited survey, some statistical correction must be
applied  to galaxies at large radial distances from the observer. This
generally involves weighting galaxies at large $r$ by
\begin{equation}
f(r) = \frac{1}{\phi(r)},
\end{equation}
but different weightings are appropriate for different statistical
treatments, such as when one wishes to construct a minimum variance
estimator for some statistic.

\subsubsection{Projected Catalogues}
Before the availability of redshifts for large samples of galaxies,
the main source of clustering was two--dimensional information
about galaxy positions on the celestial sphere. Historically
important catalogues of this kind were the Zwicky and Jagellonian
sky surveys. A dramatic improvement in the quality and quantity
of data was achieved with the Lick galaxy catalogue, in the form
of counts of galaxies in $10 \times 10$ arcminute cells compiled
by Shane \& Wirtanen \cite{shwi67}. Although over 25 years old, this catalogue,
which contains approximately $10^{6}$ galaxies, is still yielding
new and important constraints on models of galaxy formation and
clustering, e.g. \cite{cp91}.

Doubts have been raised about the Shane \& Wirtanen and earlier catalogues,
on the basis of the
relatively crude visual methods used to count the galaxies
\cite{cp91,p80,bb87}. More
recent catalogues of this kind are compiled using automated
plate--scanning devices such as APM
\cite{mesl90} and COSMOS \cite{cnl92}.
The enormous
numbers of galaxies that can be counted above the flux limit
used in these surveys to some extent compensates for the
lack of redshift information, and these two--dimensional
catalogues are still yielding important constraints on
structure formation scenarios.

\subsubsection{Redshift Surveys}
Given the comparatively long time needed to obtain the spectrum
of a faint galaxy, one needs to plan observational strategies
carefully in order to obtain a well--controlled sample of galaxy
redshifts. For a general review of the art of the redshift survey,
see \cite{hay91}.

One method is to examine only a small area of the sky, but with
a very low limiting flux density. According to this approach, one
obtains a narrow cone or pencil--beam shaped survey volume.
Prominent examples of the use of this kind of survey
are the void searches of Kirschner et al. \cite{koss83}, and
the north--south galactic pole pencil beam analysis of
\cite{beks90}. Here, of course, the information one obtains
is effectively one--dimensional.

Another possible strategy is to look at a larger angle,
but to a somewhat higher limiting flux density. This can
be achieved by looking at a narrow angular strip as
favoured by the CfA survey program \cite{dgh86,dgh89,wtkgh90}.
These surveys produce data which is effectively almost
two--dimensional, usually represented in the form of a
wedge plot. Surveys covering a larger area of the sky,
and thus generating truly three--dimensional samples include
the Perseus--Pisces sample \cite{hg88}, the local
supercluster survey \cite{tf87}, the Southern Sky
Redshift Survey (SSRS) \cite{dea88} and the large--area
CfA survey \cite{dhlt82}.
The combined updated SSRS and CfA surveys include over 15000 galaxies
(to magnitude limit $m_{B(0)} = 15.5$) and cover over 3 steradians of the sky
\cite{dc94}. Looking ahead, the upcoming Sloan survey is underway
which should provide redshifts of around a million galaxies with
$m_B>19.5$ by the end of the century.

Of course, with a terrestrial telescope, one is always constrained
to cover that part of the sky accessible from a given site.
Not so with a satellite. The IRAS satellite detected a great
number of extragalactic sources, yielding a parent all-sky
catalogue containing approximately 12000 galaxies. Follow--up
redshift surveys, include a complete survey down to a limiting
flux of 2 Jansky at 60 micron \cite{sdyh92}, and a randomly--selected
one-in-six subset of these galaxies down to 0.6 Jansky
\cite{lea94}. The
latter of these, named QDOT by the collaboration, is one
of the richest sources of cosmological data. The shallow
fall--off of the luminosity function of IRAS galaxies
means that the the survey probes scales $> 100h^{-1}$ Mpc.

\subsubsection{Rich Clusters}
One of the most striking features of the distribution of galaxies,
obvious even without detailed statistical analysis, is the
existence of extremely dense objects like the Coma cluster.
Such objects contain many hundreds of galaxies in a volume
where, on average, one would expect only a few if galaxies
were not clustered. The most natural assumption to make is
that these ``rich'' clusters of galaxies are structures
which are fully (or almost fully-) collapsed objects. Since
the typical mass of these objects is of order $10^{15} M_{\odot}$,
they represent a much larger mass and length scale than
galaxy samples can probe.

The first systematic survey of the properties of rich clusters
was performed by Abell \cite{a58}. He made a catalogue of the
clusters he could see in projection on the sky by counting objects
inside an ``Abell'' radius (approximately $1.5$ Mpc). The clusters
he detected were ranked into `richness classes' according to
how many bright galaxies were within the Abell radius; he also
assigned clusters into `distance classes' according to an estimated
distance. The original catalogue contained 2712 clusters
in the redshift range $0.02 \le z \le 0.2$
in the northern hemisphere; an extension to the southern
hemisphere was published after Abell's death
\cite{aco89}.
The spatial density of Abell clusters decreases rapidly with increasing
richness,
the mean density of $R \ge 1$ clusters is $\sim 10^{-5}$ h$^3$ Mpc$^{-3}$.
\footnote{A less restrictive selection criteria for clusters was employed
by Zwicky in his catalogue of $\sim 9,700$ clusters. Zwicky's clusters are
often of lower density than Abell's and sometimes contain two or more
subclumps within themselves \cite{zhwkk61}; for a review see \cite{b88}.}

As we shall see later, clusters are themselves clustered, so that
Abell clusters might be expected to trace structure on very large
scales\footnote{Clusters of clusters are usually called
`superclusters', but superclusters are much less well--defined
objects than Abell clusters, because they correspond to a much
smaller overdensity; $\delta \sim 1$ for superclusters,
whereas $\delta > 100$ for clusters.}. The problem is, however, that rich
clusters
are defined in projection on the sky and, since the Abell radius
is quite large, it is possible that many chance superpositions
of galaxies might occur, leading to contamination of the catalogue
especially in the poorer clusters. This, together, with the fact that
Abell's catalogue was compiled by a relatively primitive eyeball
technique, has led many workers to be suspicious that the
structures traced by the rich clusters are merely an artifact of the
selection procedure. A fuller list
of references about the possible effect of contamination
on the statistics of superclustering is given in
\S 8.2. More recently, automated plate--scanning devices
such as APM and COSMOS have been used to define cluster samples using
similar criteria to those that Abell used, but with a smaller
effective `Abell' radius to reduce the possibility of projection
contamination \cite{dems92,edsm92,ncgl92}. This catalogue covers, however,
a much smaller part of the sky than the combined Abell/ACO sample.

To construct a three--dimensional cluster catalogue, one must
proceed from a two--dimensional parent catalogue in a manner similar
to the procedure for a galaxy redshift survey. The main difference
is that one really needs too many galaxy redshifts in each cluster
because cluster galaxies have large peculiar motions and, as
discussed above, some of the galaxies may be line--of--sight interlopers
at a different distance from the bulk of the cluster. The large number
of redshifts required means that relatively few clusters have
published redshifts. The largest compilation of Abell cluster
redshifts is published in \cite{sr91}, but there are several
erroneous redshifts and some are based on only one cluster galaxy.
It is also possible to estimate redshifts using the known
correlation between $m_{10}$ (the apparent magnitude
of the tenth--brightest
cluster galaxy) and distance. The available data sources are listed in
\cite{pw92}, which also contains a discussion of the procedure for
estimating redshifts. To avoid projection contamination altogether,
one might use clusters selected in X--ray surveys, because rich clusters
contain hot gas whose temperature should furnish an estimate of the
virial mass of the cluster \cite{bw92a}. The use of X-ray selected clusters
for cosmological studies is, however, in its infancy and we shall
not discuss it further in this paper.

\subsubsection{Peculiar Motions}
As we remarked earlier, galaxy positions probe only half the relevant
phase--space needed to fully specify the dynamical behaviour
of the mass distribution. Peculiar velocities not only provide
an insight into the other half, but are sensitive to the total
amount of mass present. One might expect them therefore to
provide some information about the amount of bias present.
The amount of information available is, however, quite small and
the errors are difficult to control. One first needs a sample of galaxies
with measured redshifts $z$. One then needs to measure the distance
to the galaxies by some means which is independent of the
redshift. This typically means the Tully--Fisher relationship
for spiral galaxies and the Faber--Jackson relationship for
elliptical galaxies \cite{fb88,lyea88,burs90}. The peculiar motion
of the galaxy in the radial direction from the observer is
then
\begin{equation}
v_{p}(\vec{r})= cz-H_0 r.
\end{equation}
Galaxies are just assumed to be tracers of the velocity flow
in this analysis so one does not need to worry too much
about the possible presence of a bias.

The problem with these studies is managing to find distances
to a sufficient accuracy. While redshifts are quite easy to
measure to very high accuracy, distances can be obtained directly
to no better than 10--20\%. What is worse is that one uses
the continuity equation to relate velocities to mass (at this
level one does have to worry about the bias), and any systematic
dependence of the performance of the distance indicator upon the
local density will generate spurious peculiar velocities in dense
environments and thus throw the whole picture into confusion.
There is indeed some evidence of this kind of effect
\cite{gl93} so one should interpret velocity field studies very carefully.
We discuss some of the methods of analysis in \S 8.10.

\subsection{The two--point correlation function}
For many years the standard tool for quantifying the clustering of
extragalactic objects has been the two--point correlation function,
denoted $\xi(r)$ \cite{tk69,p80,dp83,shfm89}. This is defined by
\begin{equation}
\d P_2 = \bar{n}^{2} \left[ 1+ \xi(r) \right] \d V_1 \d V_2,
\label{eq:def_2pt}
\end{equation}
where $\d P_2$ is the (infinitesimal) probability of finding two
objects in the (infinitesimal) volumes $\d V_1$ and $\d V_2$;
$\bar{n}$ is the mean number--density of objects. We have implicitly
assumed in (\ref{eq:def_2pt}) that the correlation function depends
only upon the separation of the two volume elements, i.e. that the
distribution is statistically homogeneous and isotropic. It is
obvious from the definition that $\xi(r)$ represents the excess
probability of finding an object at a distance $r$ from a given
object, compared to a random (Poisson) distribution. This
definition differs by a factor $\bar{n}$ from the quantities used
in condensed matter physics \cite{z79,pi87}. Notice that this
statistic need not be applied only to galaxies, one could also
calculate the correlation function of clusters, quasars, radio--galaxies
or anything that can be represented by a point process.

Of course, this definition is useful only if we have three--dimensional
information about the positions of the galaxies. Fortunately, the
analogous definition of the two--point correlation function
in projection on the sky also turns out to be useful:
\begin{equation}
\d P_2 = \bar{n}_{\Omega}^{2} \left[ 1+w(\theta) \right] \d\Omega_1
\d\Omega_2. \label{eq:def_2pt_2}
\end{equation}
In this case we are looking at the excess probability of finding
a galaxy at an angle $\theta$ from a given galaxy, in an obvious
notation. The usefulness of this description is that, if the
luminosity function of galaxies does not depend upon their position,
then one can use the Limber equations \cite{l54} to deproject
the distribution and obtain $\xi(r)$ from $w(\theta)$. This means
that one can make use of the much larger projected catalogues
to determine the real--space correlation function.

The usual method for estimating $\xi(r)$ uses a random Poisson
point process generated with the same sample boundary and selection
function as the real data; one can then estimate $\xi$ according to
\begin{equation}
1+\hat{\xi}(r) \simeq \frac{DD(r)}{RR(r)},
\end{equation}
or
\begin{equation}
1+\hat{\xi}(r)  \simeq \frac{DD(r)}{DR(r)},
\end{equation}
where $DD(r)$, $RR(r)$ and $DR(r)$ are the number of pairs with
separation $r$ in the actual data catalogue, in the random
catalogue and with one member in the data and one in the
random catalogue, respectively. We have assumed
for simplicity that the real and random catalogues have the
same number of points (which they need not). The second
of these estimators is more robust to boundary effects (e.g. if
a cluster lies near the edge of the survey region), but they
both give the same result for large samples. A thorough discussion
of these and other estimators is given in \cite{ham93}.

There has been some difficulty in the past in knowing how to
assign errors to the estimation of $\xi(r)$ from a finite sample;
a recent innovation has been the use of the bootstrap technique
to estimate more realistic errors \cite{bbs84,lfb86} than the
usual quasi--Poisson approach \cite{p80}, but this is still not
perfect.

These definitions apply to a discrete set of points distributed in either
two or three dimensions. In the previous discussion of density
fields we have been dealing with a continuous density field and
this description is not appropriate. For a continuous density field
the analogous quantity is the autocovariance function:
\begin{equation}
\langle \delta (\vec{x}_1) \delta(\vec{x}_2) \rangle
= \xi(|\vec{x}_1-\vec{x}_2|)=\xi(r),
\end{equation}
assuming statistical homogeneity and isotropy. In the case
where galaxies occur randomly with a probability per
unit volume proportional to $(1+\delta)$ then the
form of $\xi$ is the same for both continuous and discrete
definitions. If there is a strong bias, however,
then this need not be the case \cite{cf91,c93}.
Observationally, it is well--established that the two--point
correlation function of optical galaxies has the
power--law form\footnote{Galaxies selected from the IRAS sample
have a somewhat smaller
correlation length and gentler slope
$r_{0,g}\simeq 3.79h^{-1} Mpc, \gamma \simeq 1.57$ \cite{sdyh92,fdsyh93}.}:
\begin{equation}
\xi_{gg}(r) = \left(\frac{r}{r_{0,g}}\right)^{-\gamma}
\,\,\,\,\,\,\, r_{0,g}\simeq 5h^{-1}\, \mbox{Mpc}, \label{eq:res_2pt}
\end{equation}
with $\gamma=1.8$ for $r < 10h^{-1}$ Mpc \cite{dp83,shfm89}.
On larger scales the
real--space correlation function is hard to measure from redshift
surveys because of the limited size, so one can use the angular
correlation function estimate (\ref{eq:def_2pt_2}) and the Limber
equation. The results for $w(\theta)$ show a power--law
slope on small scales (with a slope $1-\gamma$, expected from the
properties of the Limber equation). On larger angular scales,
an analysis of the Lick map suggested a break in the power--law
and comparatively little large--scale clustering \cite{p80}.
More recently, however, a similar analysis of the APM and COSMOS
maps have suggested a much larger amplitude of $\xi(r)$ on large
scales, incompatible, for example, with the predictions of the
standard CDM model \cite{mesl90,cnl92,ll93b}. A direct determination
of the large--scale correlation function of IRAS galaxies also
indicates that the slope is different from that quoted for
bright optical galaxies on similar scales \cite{mc94}.

The correlation function measured in redshift space $\xi_r$ can
differ significantly from its value in real space $\xi$ because
of the effect of peculiar velocities of galaxies \cite{k87}. In the
estimates of the two--point correlation function on small scales,
many of the pairs with small $r$ belong to rich clusters
which have large intrinsic velocity dispersions. Redshift
and distance are therefore not equivalent for these galaxies.
In terms of the linear growth factor $f$, defined by eq. (\ref{eq:9b}),
the mapping between redshift space and real space introduces
a multiplicative error in the radial direction
in the linear regime:
$\xi_r(\vec{r}) \simeq C(\Omega_0)\xi(\vec{r})$,
where $C(\Omega_0)=1+(2f/3)+(f^{2}/5)$, and $f$ is the
usual growth factor $f(\Omega_0) \simeq \Omega_0^{0.6}$.
It is possible to use the observed anisotropy in redshift
space to estimate
$f$ and thus $\Omega_0$ \cite{ham92,h93},
but this linear analysis requires
one to be looking at very large scales.

It is worth also mentioning the properties of the rich cluster
distribution. Rich clusters seem to have a correlation function
which is of the same form as
equation (\ref{eq:res_2pt}), but with
a different correlation length: $r_{0,c}\simeq 12-25 h^{-1}$ Mpc.
The exact value of this is controversial, because of arguments
about possible projection contamination in the cluster catalogues
\cite{bs83,pgh86,s88,b88,dbpo89,odbps90,se91}, but it certainly has
a higher amplitude than the galaxy correlation function. Indeed,
recent surveys of objects of varying richness such as galaxy groups,
Abell clusters, X-ray clusters, superclusters and so on
indicate that richer (and consequently
rarer) objects exhibit stronger clustering (see figure (\ref{fig:clust}).
It has been
suggested that a good fit to observations is
provided by the two--point correlation function $\xi (r) = (0.4 d_i/r)^{1.8}$
where $d_i$ is the mean separation of objects belonging to a given
richness class \cite{b88,bw92}. Furthermore,
a cross-correlation analysis between clusters and galaxies in the
Abell and Lick catalogues respectively, gives
$\xi_{cg} = (r_{cg}/r)^2$ with correlation length $r_{cg} = 15 \pm
3 h^{-1}$ Mpc \cite{sp77,le88}.

The increased amplitude of cluster clustering compared with galaxy
clustering, and Kaiser's interpretation of this fact \cite{k84},
led to the work on the peaks model discussed in \S 3.

\begin{figure}
\vspace{15cm}
\caption{The amplitude of the two-point correlation function (scaled to 1 Mpc.)
between objects belonging to different ``richness classes''
is shown as a function of their mean separation $d_i$.
Reproduced, with permission, from {\protect{\cite{bw92}}}.}
\label{fig:clust}
\end{figure}

\subsection{Higher--order correlations}
The two--point correlation function is a useful statistic, but
it only affords a complete description of the cosmological density
field if galaxies constitute a Poisson sample of a Gaussian
random field. To obtain a more complete statistical description
we need to include probabilities of finding three objects
in three disjoint volumes, and so on. For a set of $n$ objects
we need to know the $m$--point correlation functions for
$2 \leq m \leq n$ in order to have a full description.

The definition of the higher--order correlation functions is
a straightforward generalisation of equation (\ref{eq:def_2pt}):
\begin{equation}
\d P_m =\bar{n}^{m} \left[ 1+\xi^{(m)}(r) \right]
\d V_{1} \ldots \d V_{m}, \label{eq:def_mpt}
\end{equation}
The function $\xi^{(m)}(r)$, however, contains contributions from
correlations of orders less than $m$. For example, the number
of triplets is larger than a random distribution  partly because there
are more pairs than in a random distribution:
\begin{equation}
\d P_3 = \bar{n}^{3} \left[ 1+\xi_{12} +\xi_{23} + \xi_{31}
+\zeta_{123} \right].\label{eq:def_3pt}
\end{equation}
The part of $\xi^{(3)}$ which does not depend on $\xi_{ij}$,
usually written $\zeta_{123}$, is called the {\em irreducible}
or {\em connected} three--point function. The four--point
correlation function $\xi^{(4)}$ will contain terms in
$\zeta_{ijk}$, $\xi_{ij}\xi_{kl}$ and $\xi_{ij}$ which must
be subtracted to give the connected function $\eta_{1234}$.
An equivalent term to describe the connected moment of order
$m$ is the {\em cumulant}, $\kappa_m$. Since one is generally
most interested in the connected $n$--point functions,
the cumulants are the most natural functions
to use in many applications.

To construct higher--order cumulants of a {\em continuous} field
one makes judicious use of the {\em cluster expansion}
\cite{b92}. The cumulants $\kappa_n$ are defined
as the part of the expectation value $\delta_1 \ldots \delta_n$
which cannot be expressed in terms of expectation values of
lower order; $\delta_1\equiv \delta(\vec{x}_1)$, etc.
To determine the cumulant in terms of
$\langle\delta_1\delta_2\ldots\delta_n\rangle$
for any order one simply expresses the required expectation
value as a sum over partitions of the set $\{1,\ldots,n\}$;
the cumulant is just the part corresponding to the
unpartitioned set \cite{b92}. For example, the possible partitions
of the set $\{1,2,3\}$ are ($\{1\},\{2,3\}$), ($\{2\},\{1,3\}$),
($\{3\},\{1,2\}$), ($\{1\},\{2\},\{3\}$) and the unpartitioned
set ($\{1,2,3\}$). This means that the expectation value
can be written
\begin{equation}
\langle \delta_1\delta_2\delta_3\rangle
= \langle\delta_1\rangle_c\langle\delta_2\delta_3\rangle_c
+\langle\delta_2\rangle_c\langle\delta_1\delta_3\rangle_c
+\langle\delta_3\rangle_c \langle\delta_1\delta_2\rangle _c+
\langle \delta_1 \rangle_c\langle \delta_2 \rangle_c\langle\delta_3
\rangle_c + \langle \delta_1\delta_2\delta_3\rangle_c
\end{equation}
The cumulant $\kappa_3\equiv\langle \delta_1\delta_2\delta_3
\rangle_c$, $\kappa_2=\langle\delta_1\delta_2\rangle_c$, etc.
Since $\langle \delta \rangle=0$ by construction,
$\kappa_1=\langle \delta_1\rangle_c = \langle \delta_1\rangle=0$
\cite{f84}.
Moreover, $\kappa_2=\langle \delta_1\delta_2 \rangle_c = \langle
\delta_1\delta_2 \rangle$. This definition can be extended to higher
order quite straightforwardly. All the cumulants for
$n>2$ are zero for a Gaussian random field; for such a field
the odd expectation values are all zero, and the even ones
can be expressed as combinations of $\langle\delta_i\delta_j\rangle$
in such a way that the connected part is zero.

One can define the cumulants corresponding to a {\em discrete} set
of points in an analogous manner. Equation (\ref{eq:def_mpt})
gives the  the probability of finding $m$ galaxies in the $m$ (disjoint)
volumes $\delta V_i$ in terms of the total $m$--point
correlation function $\xi^{(m)}$. This function, however,
also contains contributions from correlations of lower order than
$m$ and a more useful statistic is the {\it reduced} or {\it connected
correlation function} which is simply that part of $\xi^{(m)}$ which
does not depend on correlations of lower--order; we shall use
$\kappa_m$ for the connected part of $\xi^{(m)}$. One can
illustrate the way to extract the reduced correlation function
simply using the three--point function as an example.
Using the cluster expansion in the form given in the preceding paragraph
but, this time, assigning the
 the single partitions $\langle \delta_i \rangle$ the value of unity for
point distributions rather than the zero
value one uses in the case for continuous fields \cite{f85},
we find
\begin{equation}
\delta^3 P_3=n^3_V [1+\xi(r_{12})+\xi(r_{23})+\xi(r_{31})+\zeta
(r_{12}, r_{23}, r_{31})]
\delta V_1 \delta V_2 \delta V_3,
\end{equation}

\noindent where $\zeta\equiv \kappa_3$ is
the reduced three--point function or, equivalently, the discrete
cumulant. The
terms  $\xi(r_{ij})$ represent the excess number of triplets
one gets compared with a random distribution (described by the `1')
just by virtue of having more pairs than in a random distribution;
the term $\zeta$ is the number of triplets above that expected
for a distribution with a given two--point correlation function.
{}From now on we shall drop the term ``connected'' or ``reduced'';
whenever we use an $m$--point correlation function, it will
be assumed to be the reduced one.

As with the two--point function, the higher--order covariances
of a continuous density field have the same form
as the higher--order correlations of a spatial
point process if the point process is generated from the
continuous field according to
the Poisson model. In other cases this need not be so.
The estimation methods and error analysis can be generalised
to these functions from the two--point case \cite{p80}.

Observationally, it seems that the three--point cumulant is
fairly well described by the hierarchical form \cite{pg75,gp77}:
\begin{equation}
\kappa_{3} = Q(\xi_{12}\xi_{23} + \xi_{23}\xi_{31} + \xi_{31}\xi_{12}).
\label{eq:hier_3pt}
\end{equation}
with $Q = 1.0 \pm 0.1$ \cite{f84a,sbl84,mssz92,ssb92,p93}.
Objections to this claim have also been lodged \cite{s79,blm87,biggh93}.
This observation is the motivation for the Davis--Peebles closure
{\em ansatz} for the BBGKY hierarchy; see \S 6.1.

The generalisation of eq. (\ref{eq:hier_3pt}) to $n>3$ involves
a bit of combinatorial analysis:
\begin{equation}
\kappa_n = \sum_{g} Q_{n,g} \sum_{l} \prod_{(n-1)} \xi_{ij}.
\label{eq:hier_npt}
\end{equation}
The notation here means a product over the $n-1$ edges linking $n$
objects, summed over all relabellings of the objects ($l$)
and summed again over all distinct $n$--tree graphs with a
given topology $g$ weighted by a coefficient $Q_{n,g}$. The
four point term must therefore include two coefficients, one
for `snake' connections and the other for `star' graphs. This is illustrated
in Figure (\ref{fig:8ho}).

\begin{figure}
\vspace{15cm}
\caption{In the hierarchical model, different topologies of graphs
connecting the $n$--points must be assigned a different weight.
For $n=2$ and $n=3$, the different graphs connecting the points
are topologically equivalent, but for $n=4$ there are two
distinct topologies. The topological difference can be seen by
considering the result of cutting one edge in the graph.
The first `snake' topology is such that
connections can be cut to leave either two pairs, or one pair and
a triplet. The second cannot be cut in such a way as to leave
two pairs; this is a `star' topology. There are 12 possible
relabellings of the snake and 4 of the star.
For the $n=5$ function, there are three distinct
topologies, illustrated in the figure with 5, 60 and 60
relabellings. We leave it as an exercise for the reader to show
that $n=6$ has 6 different topologies, and a total of 1296
different relabellings.}
\label{fig:8ho}
\end{figure}

Attempts have been made to extract the
hierachical coefficients for all $n\leq 9$
\cite{g94}, but this is certainly a laborious way to describe
the galaxy distribution. Rather than look at the behaviour of the
moments separately, it might be better in general to construct
the full distribution function of cell--counts and compare that
with theory. Furthermore,
because of the finite number of galaxies
in any sample, it becomes very difficult to measure the correlation
functions on large scales as $n$ increases.

Although the hierarchical model appears successful at fitting
the observed correlations, one should be cautious about whether
this does actually mean that the Davis--Peebles form of
self--similar evolution is at work. For a start, such a hierarchy
also developes to leading order in perturbation theory \cite{b92,b92b}.
Also, one must bear in mind that other models for $\kappa_3$
might be difficult to distingish on large scales from eq. (\ref{eq:hier_3pt}).
For example, the lognormal model produces three--point correlations of the
Kirkwood \cite{k35} form:
\begin{equation}
\kappa_3 = (\xi_{12}\xi_{23} + \xi_{23}\xi_{31} + \xi_{31}\xi_{12})
+ \xi_{12}\xi_{23}\xi_{31}, \label{eq:kirkwood}
\end{equation}
so that the hierarchical coefficient is $Q$, but there is an additional
triplet term which does not fit the hierachical form. Notice, however,
that this triplet term will be less than the hierarchical terms for
small $\xi$ and the model therefore looks hierachical on large scales.
A discussion of this problem in the context of
higher--order cluster correlations
is given in \cite{dc93b}.

\subsection{Power Spectra}
There are many advantages, particularly on large scales, in not
measuring the two--point correlation function directly, but
through its Fourier transform. The Wiener--Khintchin theorem states
that, for a statistically homogeneous random field, the
two--point correlation function is the Fourier transform of the
{\em power--spectrum} (\S 2.4):
\begin{eqnarray}
\xi(x) & = & \int \d^{3} \vec{k} \exp (i \vec{k} \cdot \vec{x})
P(k)\nonumber\\
P(k) & = & \frac{1}{(2\pi)^{3}} \int \d^{3} \vec{x} \exp (-i \vec{k}
\cdot \vec{x} ) \xi(x).
\label{eq:wkthm}
\end{eqnarray}
The power--spectrum is, roughly speaking,
proportional to the mean square amplitude of Fourier modes of the
distribution. For power law primordial spectra with a short range cut-off:
$P(k) \propto k^n$ $k < k_c$, $P(k) = 0$ $k \ge k_c$,
$\xi(x) \propto  x^{-(3+n)} ~(n > -3)$, $x >> k_c^{-1}$;
which can be used to
deduce the power spectrum from a knowledge of $\xi$ in regions where
$\xi $ can be represented as a power law.
Being actually a spectral density function, however,
$P(k)$ must have units of volume \cite{b92}.
This simple fact seems to have
caused much confusion in the literature, and many authors insist
on inserting unnecessary factors of the sample volume $V$
into equation (\ref{eq:wkthm}) to account for this \cite{p80,p92}.
To avoid this possible dependence upon the sample volume it is
more useful to deal with a dimensionless power--spectrum
$\Delta^{2}(k) \equiv 4\pi k^{3} P(k)$. (Still, one must beware
of different conventions with the factors of $2\pi$ in the definition
of the Fourier transform.)

Studies based on the clustering of optical and IRAS galaxies seem to indicate
that the observations can be described by a power spectrum which follows an
approximate power-law form with a slope which ranges from $n \simeq -2$ on
scales
$\lambda \le 30 $h$^{-1}$ Mpc to $n \simeq - 1$ on scales $30 < \lambda <
50 - 100$ h$^{-1}$ Mpc \cite{bf91,k91,vpgh92,fdsyh93,fkp93,pvgh94,dvghp94}.
On very large scales $\lambda > 700$ Mpc. the shape of the power spectrum
is determined by the angular fluctuations in the Microwave Background Radiation
which indicate $n = 1.1 \pm 0.32$ \cite{ghbbw94,wss94}.
The observed power spectrum is compared with
the standard cold dark matter power spectrum in figure
(\ref{fig:power}).

\begin{figure}
\vspace{15cm}
\caption{The power spectrum of matter reconstructed from observations of
the Cosmic Microwave Background and from galaxy clustering data. The boxes on
the
left refer to MBR measurements conducted with (from left to right)
COBE, FIRS, Tenerife, SP91-13pt, Saskatoon, Python, ARGO, MSAM2, MAX-GUM and
MAX-MuP, MSAM3.
The data points correspond to the following large scale surveys:
stars (APM), crosses (CfA), triangles (IRAS-1.2Jy), squares (IRAS-QDOT).
The solid line corresponds to the CDM spectrum normalised to COBE.
The MBR power spectrum (for $\Omega_B = 0.06$) is shown at the bottom.
Reproduced, with permission, from {\protect{\cite{wss94}}}.}
\label{fig:power}
\end{figure}

Peacock and collaborators have explored
the power--spectrum of galaxy clustering for a number of different
samples \cite{pn91,p91,p92,pd94} and the results are reasonably
well fitted by the functional form:
\begin{equation}
\Delta^{2}(k) = \frac{(k/k_0)^{\alpha}}{1+(k/k_c)^{\alpha-\beta}}~,
\label{eq:ps_obs}
\end{equation}
which describes a break between two power laws.
The best--fitting value for the
parameters are $\alpha = 1.50 \pm 0.03$, $\beta = 4.0 \pm 0.5$,
$k_{c} \simeq 0.039 \pm 0.002 h$ Mpc$^{-1}$, $k_0
\simeq 0.29 \pm 0.01 h$ Mpc$^{-1}$, but $k_0$ depends quite sensitively
upon the accuracy of the various selection functions.
(The value of $\beta$ is related to the primordial spectral index on large
scales, $\beta = 4$ corresponds to the Harrison-Zel'dovich
scale-invariant spectrum.)
The form (\ref{eq:ps_obs}) on large scales,
is similar to a low--density CDM
spectrum or a mixed CHDM spectrum \cite{ebw92,pn92}.
The effect of non--linear
evolution on small scales (high $k$) is discussed by
Peacock \& Dodds \cite{pd94} using the {\em ansatz} of
Hamilton {\em et al} \cite{hklm91}, which is discussed in \S 6.
The power spectrum of Abell cluster correlations has also
been computed \cite{pw92}; the results are consistent with
a rather large value for the correlation length, $r_0
\simeq 21 h^{-1}$ Mpc and indicate that the clustering strength does
depend on the cluster richness.
Estimates of the value of the spectral index on cluster scales indicate
$-1 \le n \le -2$, \cite{ha91,fdsyh93}.
The redshift--space anisotropy of $P(k)$ can also be used to
estimate $\Omega_0$ by analogy with the two--point correlation
function \cite{kea91,gcb93,fdsyh93,cfw94,ht95}.

Since the power spectrum is the Fourier transform of the two--point
correlation function, it would seem likely that similar transforms
of the $n$--point correlation functions would also prove to be useful
descriptors of galaxy clustering. This is indeed the case. For example,
the Fourier transform of the three--point correlation function is
known as the {\em bispectrum} \cite{p80,fm85,bf91}.
The use of higher--order spectra is not widespread, but they
may turn out to be a very effective way to detect non--Gaussian
fluctuation statistics on very large scales.

\subsection{Cell--counts}
A simple but useful way of measuring the correlations of galaxies
on large scales which does not suffer from the problems of the
correlation functions is by looking at the distribution of galaxy
counts in cells, $P_{N}(V)$ (defined as the probability  of finding exactly
$N$
objects in a randomly placed volume $V$), or the low--order moments of this
distribution such as the skewness $\gamma$ and variance $\sigma^{2}$.
Indeed some of the earliest quantitative analyses of galaxy
clustering adopted this approach \cite{h34,z57,shwi67}.
More recently, observations of the spatial distribution of IRAS galaxies
indicate that $\sigma  = 0.5 \pm 0.1$ in a box of size ($30 h^{-1}Mpc)^3$
\cite{sea91}.

The counts in cells probability $P_{N}(V)$ is the discrete analogue of
the probability distribution function (PDF) $P(\delta)$ whose moments are
defined as
\begin{equation}
<\delta^n> = \int_{-1}^{\infty}{\delta}^n P(\delta)d\delta
\end{equation}
For a discrete distribution of point sources this generalises to
\begin{equation}
\mu_m(V) \equiv \langle\left({\Delta N\over \bar{N}}\right)^m \rangle
= \sum_{N=0}^{\infty} \left({N - \bar{N}\over \bar{N}}\right)^m P_N (V)
\end{equation}
where $\mu_2 \equiv \sigma^2, \mu_3 \equiv \gamma$;
$\bar{N}$ is the mean number of galaxies in a cell of volume $V$.
Models such as the lognormal (\ref{eq:ln_dis})
and the gravitational quasi--equilibrium
distribution (\ref{eq:therm1}) allow one to test the whole distribution
of cell counts against the theory. Both the Poisson--sampled lognormal
and the thermodynamic distribution seem to fit the observations
fairly well.

Using only the moments of the cell-count distribution does result in
a loss of information, but the advantage is a simple relationship
between the moments and the $\kappa_n$, e.g.:
\begin{equation}
\sigma^{2}(V)\equiv \langle \left(\frac{\Delta N}{\bar{N}} \right)^{2}\rangle
= \frac{1}{\bar{N}} + \bar{\kappa}_2
\label{eq:dis_var}
\end{equation}
where
\begin{equation}
\bar {\kappa}_n = \frac{1}{V^{n}} \int \ldots \int \d V_1 \ldots \d V_n
\kappa_n(\vec{r}_1\ldots \vec{r}_n).
\end{equation}
Apart from the Poisson term -- which is a discreteness effect --
the second order moment is simply an integral over the two--point
correlation function. The same is true for higher order moments, but the
discreteness terms are more complicated and the integrals must be
taken over the cumulants. For example, the skewness $\gamma$ can be
written
\begin{equation}
\gamma(V) \equiv \langle \left( \frac{\Delta N}{\bar{N}} \right)^{3} \rangle
= -\frac{2}{\bar{N}^{2}} + \frac{3\sigma^2}{\bar{N}} + \bar{\kappa}_3=
\frac{1}{\bar{N}^{2}}+ \frac{3\bar{\kappa}_2}{\bar{N}} + \bar{\kappa}_3.
\label{eq:dis_skw}
\end{equation}
In terms of $\gamma$, our hierarchical parameter $S_3$ is just
$\gamma/\sigma^{4}$ ($S_3 \simeq \bar{\kappa}_3/\bar{\kappa}_2^2$ for
large $\bar{N}$) \cite{cf91}. Equation (\ref{eq:dis_var})
provides a good
way to measure the two--point correlation function on large scales
and its use is discussed in \cite{eea90,sea91}. Use of the skewness
as a descriptor is discussed in \cite{cf91,bjcp92,jbc93,bks93}.

The usual formulation is to write the ratio of the $n$--th order
moment to the $(n-1)$--th power of the variance as $S_n$.
A moment's thought will then show that for a hierarchical model,
the $S_n$ should be constant, independent of cell volume.
For the simply hierachical distribution (\ref{eq:hier_3pt})
we have that $S_{3} = 3Q$, which seems to be in reasonable agreement with
measured skewnesses \cite{cf91}. Of course, there should be
some scale dependence of clustering properties if the initial
power spectrum is not completely scale--free, so one would not expect
$S_3$ to be accurately constant on all scales in, for example,
the CDM model. It is, however, a very slowly--varying quantity.
It should be stressed that the higher order $\bar{\kappa}_n$
become increasingly sensitive to
the presence of rare peaks of high density in a distribution and
one should expect substantial errors in determinations of them
from samples of relatively small size.
Applications of the count probability approach to galaxy
catalogues are dicussed in \cite{eea90,lepm92,gaz92,bsdfyh93,gf94};
the use of count probabilities in the analysis
of N-body simulations may be found in
\cite{cf91,bh92,wc92,liis93,fms93,lmmm94}.
\footnote{The role of the skewness in analysing the Cosmic Microwave
Background is discussed in \protect{\cite{ls93,luo94,mss95}}.}

As noted in \S 4.1, higher order cumulants are related to lower order
cumulants in the weakly non-linear regime, by the correlation hierarchy
$\bar{\kappa}_n = S_n\bar{\kappa}_2^{n-1}, n > 1$
\cite{b92b,p80,f84,ggrw86}.
As discussed above and in \S 8.3, such a scaling law can also be predicted
by the BBGKY equations in the fully nonlinear regime.
It is interesting that the
observed values of $S_3 \simeq \bar{\kappa}_3/\bar{\kappa}_2^2$ and
$S_4 \simeq \bar{\kappa}_4/\bar{\kappa}_2^3$ do appear to be
approximately constant over a wide range of scales.
(Studies of the IRAS sample indicate $S_3^{IRAS} \simeq 1.5 \pm 0.5$ on scales
ranging from 1 to 25 h$^{-1}$ Mpc \cite{bsdfyh93}).
Thus within the considerable errors, there seems to be a roughly
hierarchical behaviour of the clustering data \cite{cf91,bjcp92,jbc93,bsdfyh93}
consistent  with most gravitational instability models of structure
formation.

One might be tempted to use the observed value of
$S_3^{OBS}$ in conjunction with the quasi-linear theory prediction:
$S_3 = \langle\delta^3\rangle_c/\langle\delta^2\rangle^2 = 34/7 - (n+3)$
to determine the value of the spectral index $n$.
However the fact that light need not trace mass makes this
difficult in practice: the observed value $S_3^{OBS}$ and $S_3$ are
related by $S_3^{OBS} = S_3/b$, where $b = \delta^{OBS}/\delta$ is the
linear biasing factor. Thus, we need an independent estimate of $n$ in order to
determine $b$ and
vice-versa. A bias--independent means of determining the spectral
index might be provided  by the
skewness and kurtosis of velocity fields which are discussed in \S 8.10.

It is interesting to note that many different approaches to the
clustering problem agree with each other as far as the relatively
low--order moments of the cell--count distribution. As we saw in
\S 4, different dynamical approximations agree with perturbation
theory in predicting hierarchical behaviour in the skewness.
Moreover, the lognormal model (\S 3.5) also produces, in the
limit of large scales, a value of $S_3 = 3$. Although this is
actually a Kirkwood model it does behave on large scales very
much like an hierarchical model. A completely different approach
using the gravitational quasi--equilibrium distribution
of \S 6.2, also gives a constant $S_3=(1+2b)$. For values of $b$
around $0.65-0.75$, this also agrees roughly with the observations.
Although it is encouraging that these different approximations
do agree with each other to a reasonable degree and also seem
to behave in roughly the same way as the data, it is advisable
to be cautious here. The skewness is a relatively crude statistical
descriptor and many different non--Gaussian distributions have the
same skewness, but very different higher--order moments. One
could proceed by measuring higher and higher--order moments from
the data, but we do not feel that this is a very efficient way to
proceed. It is perhaps better to focus instead upon the distribution
function of cell--counts, $P_N(V)$, rather than its moments. The
problem is that, except for a few special cases, it is not possible
to derive the distribution function analytically even in the
limit of large $V$. The observed form of $P_N(V)$ may be
fairly well matched by the gravitational quasi--equilibrium distribution
(though see \cite{bsd91}),
and also by the Poisson--Lognormal model \cite{cj91,mss94}; see
\S 3.5.

The distribution function of galaxy counts
can be related to the moments (or integrals over the correlation
functions) by the {\em moment generating function}; see \cite{f85}.
It should be pointed out, however, that the moment generating
function does not necessarily exist for all statistical distributions:
the necessary series does not converge, for example, for the lognormal
distribution \cite{cj91}. For hierarchical cumulants,
equation (\ref{eq:hier_npt}),
the series does converge; scaling properties of the observed $P_{N}(V)$
do seem to be in accord with the hierarchical model
\cite{ml87,ml91,fghms89,bs89a,vgh91}, but scaling seems to be
broken on large scales.

\subsection{Voids}
The distribution function of galaxy counts leads naturally on to the
{\em Void Probablity Function} (VPF) -- the probability that a randomly
selected volume $V$ is completely empty.
Properties of voids are also appealing for intuitive reasons:
these are the features that stand out most strikingly
in the visual appearance of the galaxy distribution.
As demonstrated in \cite{w79,bs89a} the generating function of the count
probabilities
${\cal P}(\lambda) = \sum_{N=0}^{\infty} \lambda^N P_N(V)$ can be shown to be a
sum over the {\em mean value} of cumulants of all orders
\begin{equation}
\log {\cal P}(\lambda) = \sum_{n=1}^{\infty} {(\lambda - 1)^n\over {n!}}
(\bar{N})^{n} \bar {\kappa}_n
\label{eq:vpf1}
\end{equation}
Setting $\lambda = 0$ in eq. (\ref{eq:vpf1}) we obtain
\begin{equation}
\log P_0 (V) = \sum_{n=1}^{\infty} \frac{(-\bar{N})^{n}}{n!} \bar {\kappa}_n,
\label{eq:vpf}
\end{equation}
again as long as this
sum converges \cite{cj91}.
In an unclustered Universe, $\bar{\kappa}_1=1$
(because this is a discrete distribution: \S 8.3) and
$\bar {\kappa}_n = 0$ for $n \ge 2$ and
\begin{equation}
P_0 (V) = \exp(-\bar{N}), ~~~~P_N (V) =
{(\bar{N})^{N}\over {N!}} \exp (-\bar{N})
\end{equation}
which is just the Poisson distribution.
In the weakly nonlinear regime equation (\ref{eq:vpf1}) can be rewritten as
\cite{b92b}
\begin{equation}
\log {\cal P}(\lambda)
= -\frac{\varphi\lbrack(1-\lambda)\bar{N}\sigma^2\rbrack}
{\sigma^2}
\end{equation}
where $\varphi(y)$ is the generating function for the moments $S_p$
\begin{equation}
\varphi(y) = \sum_{p=1}^{\infty} S_p\frac{(-1)^{p-1}}{p!}y^p
\end{equation}
the resulting VPF has the form
\begin{equation}
P_0 (V) = \exp\left[-\bar{N}\frac{\varphi(y)}{y}\right]
\end{equation}
where $y = \bar{N}\sigma^2$.

The VPF is quite easy to extract from simulations
and real data and turns out to depend strongly upon correlations of
all orders and is therefore a potentially useful diagnostic of the
clustering: many authors have investigated its properties
\cite{f86,ml87,ml91,oppw86,p85,pp86,fghms89,vgh91,eegs91,ssmpm94,lw94}.
These studies
again seem to support the view that clustering on scales immediately
accessible to observations is roughly hierachical in form.

Although the VPF is unquestionably a  useful statistic, it pays
no attention to the geometry of the voids, or their topology.
Typically one uses a spherical test volume so a flat or filamentary
void will will not register in the VPF with a $V$ corresponding
to its real volume. Moreover,
because the voids which seem most obvious to the eye are not actually
completely empty, these do not get counted in the VPF statistics.
It would be good to have an alternative statistic for measuring
void properties. One possibility is the void spectrum discussed in
\cite{kf91,km92,sss94} and \S5.3, which seems to be a good statistical
discriminator between models of structure formation.

\subsection{Geometrical Descriptors}
The possibility that galaxies lie preferentially along filamentary
or sheet--like structures separating large voids has aroused considerable
interest during the last decade \cite{jet78,dgh86,dgh89,wtkgh90}.
Because of the
possible connection between these structures and the classical Zel'dovich
pancake scenario, strenuous attempts have been made to find objective
methods for quantifying the geometrical properties of galaxy clustering.
(Bearing in mind the `Canals of Mars' fiasco, one should also be aware of
the problem that one's eye is easily
misled into constructing filaments and such structures from images
where they do not really exist \cite{bb87}.)

One of the first methods suggested was based on {\em percolation theory}
and was borrowed from condensed matter physics \cite{z79,zel82,sh83,zes82}.
To illustrate
the method, suppose we have $n$ galaxies in a sample cube of side $L$;
the mean separation of galaxies is defined to be $l=Ln^{-1/3}$. Around
each galaxy we construct a sphere of diameter $d=bl$, where $b$ is a
dimensionless quantity. If $b$ is small, virtually all the spheres
will be disjoint; as it increases, neighbouring spheres will overlap
in clusters forming {\em percolated regions} while isolated spheres
will still be disjoint. At some critical value $b^{\ast}$, the
largest percolated region traverses the whole volume: $L_{p}(b)
=L$ and $b=b^{\ast}$, where $L_p$ is the size of the largest percolated
region. The value of $b$ depends on the local geometry of the
distribution. For a regular lattice, $b=1$ and for a random (Poisson)
distribution the most probable $b^{\ast}$ is $b^{\ast}=b_{p}^{\ast}=0.86$.
A distribution containing filaments and sheets would percolate more
easily than this and would have $b<0.86$; a system
characterised by isolated dense clumps would have $b>0.86$.
For discussion of this method, including sampling problems and
more detailed percolation statistics, see
\cite{zes82,sh83,bb83,dw85,i92,ks93,sz89}.

There is some degree of consensus in the literature that the
percolation method is not very successful at discriminating between
the kind of data and theoretical models involved in cosmology. For
example, HDM and CDM models which both have $b^{\ast}\simeq 0.65$.
To some extent, however, this conclusion refers not so much to the
percolation method itself, but to the attempt to encode all the
properties of the percolation process in one single parameter
$b^{\ast}$. Since the critical percolation threshold is determined
by the size of the largest structures in the simulation box, one
might have expected this not to be a very sensitive discriminant.
A more detailed analysis might well prove more successful. For example,
consider what happens as the percolation radius is increased. In the
beginning, one has a large number of isolated clusters. As the radius
is increased, these tend to merge together producing a smaller number
of larger clusters some of which may be highly elongated. Eventually,
when percolation occurs, one cluster is the size of the simulation box,
but there may be other clusters remaining which are smaller than this
biggest one. In a refinement of the percolation method
Klypin \& Shandarin \cite{ks93} have shown that two
parameters are very sensitive diagnostics of the way percolation proceeds:
the size of the largest cluster (as a function of $b$) and
the mean cluster size (excluding the largest one). Instead of using
one number to characterise the onset of percolation, one therefore
uses two functions which describe different aspects of the process.
It remains to be seen whether this can be a useful technique in
practice.

Various statistics have been suggested to search for specifically
filamentary signatures in the galaxy distribution. Kuhn and
Uson \cite{ku82} suggested one should use the mean misalignment
angle between consecutive steps of a self--avoiding random--walk
through the distribution. It seems that the effect of Poisson
noise and sample errors is to easily de-rail the walk from
a filament so the method is of limited usefulness. A similar
problem besets the ridge--finding algorithm \cite{mtg83}:
just one spurious galaxy position can alter the results
markedly, so the algorithm is insufficiently robust for realistic
data sets. A more promising approach is to
use the quadrupole moment of the galaxy distributions \cite{f85,v86},
or structure functions derived from the moment of inertia
tensor of the distribution \cite{bs92} which are
described below.

Consider a set of $N_k$ particles located within a radius $R_k$ of a
randomly chosen $k$~th particle.
Let $x_j^\alpha$ ($\alpha = 1,2,3)$ denote the position vector of the
$j$~th particle relative to the $k$~th particle in this ensemble.
One can construct the first and second ``moments'' of the distribution of
particles located within a distance $R$ of the $k$~th particle:
\begin{equation}
M_k^{\alpha} (R) = {1\over N_k} \sum_{j=1}^{N_k} x_j^\alpha,
\label {moment1} \end{equation}
\begin{equation}
M_k^{\alpha\beta}(R) = {1\over N_k} \sum_{j=1}^{N_k} x_j^\alpha x_j^\beta.
\label {moment2} \end{equation}
One can then define the local moment of inertia tensor about the centre
of mass of each particle by
\begin{equation}
I_k^{\alpha\beta}(R) = M_k^{\alpha\beta}(R) - M_k^{\alpha}(R)M_k^{\beta}(R)
\label{inertia}
\end{equation}
and its average
\begin{equation}
I^{\alpha\beta}(R) = {1\over N} \sum_{k=1}^N I_k^{\alpha\beta}(R),
\label{inertia2}
\end{equation}
where $N$ denotes the number of centers used in carrying out
the average.

Diagonalising $I^{\alpha\beta}(R)$ we determine its eigenvalues $I_1, I_2,
I_3$ (arranged in decreasing order of magnitude) and the ratio's
\begin{equation}
\nu = \left({I_2\over I_1}\right)^{1\over 2};
{}~~~~~\mu = \left({I_3\over I_1}\right)^{1\over 2}.
\end{equation}
The structure functions $S_1, S_2, S_3$ are then defined as follows
\cite{bs92}
\begin{equation}
S_1 = \sin \left[ \frac{\pi}{2} (1-\nu)^p \right]\label{eq:s1}
\end{equation}
\begin{equation}
S_2 = \sin \left[ \frac{\pi}{2} a(\mu,\nu) \right]\label{eq:s2}
\end{equation}
\begin{equation}
S_3 = \sin \left(\frac{\pi\mu}{2}\right).\label{eq:s3}
\end{equation}
where $p=\log 3/\log 1.5$, $a(\mu,\nu)$ is given by solving the equation
\begin{equation}
\frac{\nu^{2}}{a^{2}} - \frac{\mu^{2}}{a^{2}(1-\alpha a^{1/3} +\beta a^{2/3})}
=1,
\end{equation}
the parameters $\alpha$ and $\beta$ are
\begin{equation}
\alpha=1.9;
\end{equation}
\begin{equation}
\beta=-\left(\frac{7}{8}\right)9^{1/3}+\alpha 3^{1/3}.
\end{equation}
These somewhat involved definitions ensure that
$0\leq S_i\leq 1$ for all $i$. Thus $(S_1,S_2,S_3)=(0,0,1)$ for a sphere;
$(S_1,S_2,S_3)=(0,1,0)$ for a flat sheet; $(S_1,S_2,S_3)=(1,0,0)$ for a
filament. Intermediate values of $S_i$ indicate the departure of a
structure from its {\it eikonal} shape: sphere, pancake or filament.
For details see \cite{bs92}.

A related measure, also constructed out of the moment of inertia tensor
is the Vishniac quadrupole statistic \cite{f85,v86}
\begin{equation}
Q_k(R) =
{2 \sum_{\alpha, \beta} M_k^{\alpha\beta} I_k^{\alpha\beta} -
 \sum_{\alpha}  M_k^{\alpha\alpha}
  \sum_{\beta} I_k^{\beta\beta}
  \over
  \left ( \sum_{\gamma} M_k^{\gamma\gamma} \right )^2},
\label {fs} \end{equation}
and
\begin{equation}
Q(R) = {1\over N} \sum_{k=1}^N Q_k(R),
\end{equation}
where $N$ is the number of centres chosen in carrying out the average.
The statistic $Q(R)$ always lies in the range $[0,1]$. It attains its
maximum value $Q = 1$ for a straight line passing through the centre,
and is zero for a uniform distribution of points.
A comparison of $Q$ and the structure functions $S_1, S_2, S_3$ is
discussed in \cite{pc94}. Applications of $Q$ to study the growth of
filamentarity during gravitational clustering are
discussed in \cite{nd90,ssmpm94}.

These techniques, although robust, can still be confused at low signal-to-noise
levels \cite{nd90,ssmpm94}. One way to improve their performance might be
to enhance the filamentary pattern before trying
to quantify its existence with some numerical statistic.
This might be accomplished by
determining $S_i$ and $Q$ for regions of a specified overdensity \cite{sssk95}.
Another way would be to preprocess the data by constructing
the {\em minimal spanning tree} (MST) of the set of points \cite{pc94}.
The MST is a geometrical construction which owes its origins to graph
theory. Consider a distribution of points (say, galaxies) in three-dimensional
space. One can define a graph on this set to be a collection of nodes
(galaxies) and edges (lines joining galaxies). The degree of a node is the
number of edges which emerge from it; a sequence of edges joining nodes
is called a path; a closed path is a circuit; if any two nodes within a
graph have at least one path connecting them then such a graph is said to be
{\it connected}. A connected graph which has no circuits is called a
{\it tree}; if a tree contains all the nodes within a graph then it is
a {\it spanning tree}. The {\em minimal} spanning tree of a data set is
the tree whose total edge length is the smallest. (A tree is said to be unique
if no two edge lengths within it are equal.)

It is sometimes convenient to modify the MST of a data set in order to
eliminate accidental linkages which can be quite common if the data set is
noisy. One way of doing this is by the operations of {\it pruning} and {\it
separating} which are illustrated in Figure (\ref{fig:8mst})
and are carried out as follows: we define a {k branch} to be a path of
$k$ edges connecting a node of degree 1 to a node of degree exceeding
2, in such a way that all intervening nodes are of degree 2.
A MST is said to be {\it pruned to a level $p$} if all $k$-branches with
$k \le p$ have been removed. Pruning serves to remove extraneous `foliage'
on the MST which is not directly related to the main structure of the pattern.
One can also {\it separate} a tree by removing any edges whose length
exceeds some cut-off scale $l_0$ usually defined in terms of the mean length
of edges in the MST $<l>$. Separation tends to minimise `accidental'
linkages, the tree after separation consists of several disjoint pieces
which can be analysed further.

An advantage of the MST is that it emphasises filaments embedded in background
noise thus enhancing the signal-to-noise ratio in any quantitative statistic
such as the structure functions $S_1, S_2, S_3$ and $Q$ \cite{pc94}.
A simple algorithm of constructing the MST, the so--called
{\it Greedy Algorithm}, is described in \cite{bbs85}.

\begin{figure}
\vspace{15cm}
\caption{The Minimal Spanning Tree used to enhance structure in
a clustered point set. The diagram (a) shows a diagrammatic
set of points in two dimensions; (b) shows the MST of the set (a);
(c) shows the result of pruning --  all nodes of degree one
connected to nodes with degree exceeding two have been removed;
(d) a separated, pruned tree -- all edges of (c) exceeding a
certain critical length have been removed. Adapted, with permission,
from \protect{\cite{bbs85}}.}
\label{fig:8mst}
\end{figure}

\subsection{Topological Descriptors}
Interesting though the geometry of the galaxy distribution may
be, such studies do not tell us about the {\em topology}
of clustering. Or, in other words, its connectivity. One is
typically interested in the question of how the individual
filaments, sheets and voids join up and intersect to form the
global pattern. Is the pattern cellular, having isolated
voids surrounded by high--density sheets, or is it
more like a sponge in which under-- and over--dense regions
interlock?

Looking at `slice' surveys gives the strong visual impression
that we are dealing with bubbles; pencil beams perhaps
re-inforce this impression by suggesting that a line--of--sight
intersects at more--or--less regular intervals with walls
of a cellular pattern. One must be careful of such impressions,
however, because of elementary topology. Any closed curve
in two dimensions must have an inside and an outside, so that
a slice through a sponge--like distribution will appear to
exhibit isolated voids just like a slice through a cellular
pattern. It is important therefore that we quantify this
kind of property using well--defined topological descriptors.

The problem of quantifying the topology of large--scale structure
has been discussed in the literature for many years \cite{d70,sz83}.
More recently,
in an influential series of papers, Gott and collaborators
\cite{gmd86,hgw86,gea89,m90a} have developed a particular
method for doing just this. We shall not discuss this approach
in much
detail, because of the existence of an extensive review of
this work by Melott \cite{m90a}.

Briefly, the method makes use of a topological invariant
known as the {\em genus}, related to the {\em Euler--Poincar\'{e}}
characteristic, of the iso--density surfaces of the distribution.
To extract this from a sample, one must first smooth the galaxy
distribution with a filter (usually a Gaussian is used) to remove
the discrete nature of the distribution and produce a continuous
density field. By defining a threshold level on the continuous
field, one can construct excursion sets (sets where the field
exceeds the threshold level) for various density levels. An
excursion set will typically consist of a number of regions, some
of which will be simply connected and others of which will be
multiply connected. If the density threshold is labelled by
$\nu$, the number of standard deviations of the density away
from the mean, then one can construct a graph of the genus
of the excursion sets at $\nu$ as a function of $\nu$: we
call this function $G(\nu)$. The genus can be formally
expressed as an integral over the intrinsic curvature $K$ of the
excursion set surfaces, $S_{\nu}$ by means of the Gauss - Bonnet
theorem:
\begin{equation}
4\pi\left[ 1- G(\nu) \right] = \int_{S_{\nu}} K \d A,
\label{eq:ga_bo}
\end{equation}
where the integral is taken over each compact two--dimensional
surface in the excursion set. Roughly speaking, the genus for
a single surface is the number of handles the surface posesses;
a sphere has no handles and has zero genus, a torus has one and
therefore has a genus of one. For technical reasons to do
with the effect of boundaries, it has become conventional
not to use $G$ but $G_{S}=G-1$ \cite{gea89}. In terms
of this definition, multiply connected surfaces have
$G_S\ge 0$ and simply connected surfaces have $G_{S}<0$.

To compute topological properties of discrete data sets,
it is usually  convenient
to work with the discrete analogue of equation (\ref{eq:ga_bo})
which is easy to derive
once we note that the curvature of a discrete surface is localised
in each of its $N$ vertices so that
\begin{equation}
K(x) \simeq D_i \delta (x - x_i)
\label{eq:ga_bo1}
\end{equation}
where $D_i$ is the deficit angle evaluated at the $i^{th}$ vertex of a
polyhedral surface.
Substituting equation (\ref{eq:ga_bo1}) in equation (\ref{eq:ga_bo}) we
obtain
\begin{equation}
4\pi\left[ 1- G(\nu) \right] = \sum_i^N D_i
\label{eq:ga_bo2}
\end{equation}
which is the required analog of equation (\ref{eq:ga_bo}).

\begin{figure}
\vspace{15cm}
\caption{A polyhedral torus. This figure is adapted from
\protect{\cite{m90a}}.}
\label{fig:8torus}
\end{figure}

To see how equation (\ref{eq:ga_bo2}) can be used to determine the genus
of a compact polyhedral surface,
let us consider the simple but illustrative example of a torus shown in
figure (\ref{fig:8torus}).
All the curvature of the polyhedral torus is concentrated in each
of its 16 vertices, the deficit angle at each vertex being
$D_i = 2\pi - \sum_i V_i$, where $\sum_i V_i$ is the sum of the vertex
angles
around a given vertex. Thus, each of the outer eight corners of the
torus
contributes a positive deficit angle:
$D_i = 2\pi - {3\pi/ 2} = {\pi/ 2}$, whereas each of the eight inner
corners of the torus will contribute a negative
deficit angle $D_i = 2\pi - {5\pi/ 2} = -{\pi/ 2}$. The sum total of
all the deficit angles is therefore $\sum_{i=1}^{16}D_i = 8\times{\pi/
2} - 8\times{\pi/ 2} = 0$, implying $G = 1$ for a torus. (A numerical
algorithm for determining the genus of a two or three-dimensional
discrete manifold is given in \cite{m90a}.)

One of the great advantages of
using the genus measure to study large scale structure, aside
from its robustness to errors in the sample, is that all Gaussian
density fields have the same form \cite{d70,a81,hgw86}  of $G_S(\nu)$:
\begin{equation}
G_{S}(\nu) = A \left(1-\nu^{2}\right) \exp \left( - \frac{\nu^{2}}{2}
\right). \label{eq:gen_gau}
\end{equation}
This means that if one smooths the field enough to remove the
effect of non--linear displacements of galaxy positions,
the genus curve should look Gaussian for any model evolved
from Gaussian initial conditions, regardless of the form of
the initial power--spectrum, which only enters through the
normalisation factor $A$ in Eqn. (\ref{eq:gen_gau}).
This makes it a potentially powerful test of non--Gaussian
initial fluctuations, or models which involve non--gravitational
physics on large  scales to form large--scale structure.
The observations support the interpretation of Gaussian initial
conditions, although the distribution looks non--Gaussian
on smaller scales. The nomenclature for the non--Gaussian
distortion one sees is a `meatball shift': non--linear
clustering tends to produce an excess of high--density
simply--connected regions, compared with the Gaussian curve.
What one would expect to see in the standard picture of gravitational
instability from Gaussian initial conditions, is therefore
a `meatball' topology when the smoothing scale is small,
changing to a sponge as the smoothing scale is increased. This
is indeed what seems to be seen in the observations; an example
is shown in Figure (\ref{fig:8genus}).

\begin{figure}
\vspace{15cm}
\caption{The genus curve is shown for galaxies in the CfA, a small shift
towards meatball-like topology is discernable.
Reproduced, with permission, from \protect{\cite{gea89}}.}
\label{fig:8genus}
\end{figure}

The smoothing required also poses a problem, however, because
present redshift surveys sample space only rather sparsely and
one needs to smooth rather heavily to construct a continuous
field in the first place. A smoothing on scales much larger
than the scale at which correlations are significant will tend
to produce a Gaussian distribution by virtue of the central
limit theorem. The power of this method is therefore limited
by the scale of smoothing required which, in turn, depends
on the space--density of galaxies. This is a particular
problem in a topological analysis of the QDOT redshift
survey \cite{mea92}.

Topological information can also be obtained from two--dimensional
data sets, whether these are simply projected galaxy positions
on the sky (such as the Lick map, or the APM survey) or `slices',
such as the various CfA compilations \cite{dgh86}. There are some
subtleties, however. Firstly, as discussed above,
two--dimensional information topology
does not distinguish between `sponge' and `swiss--cheese' alternatives.
Nevertheless, it is possible to
assess whether, for example, the mean density level ($\nu=0$)
is dominated by underdense or overdense regions so that one can
distinguish swiss--cheese and meatball alternatives to some extent.
The topological quantity usually discussed in two--dimensional
work is the {\em Euler--Poincar\'{e}} characteristic $\Gamma$
of the excursion sets, which is roughly speaking the number of disjoint
regions minus the number of holes in such regions. This is
analogous to the genus, but has the interesting property that it
is an {\em odd} function of $\nu$ for a two--dimensional
Gaussian random field, unlike $G(\nu)$ which is
even. In fact the mean value of $\Gamma$ per unit area
takes the form
\begin{equation}
\langle \Gamma (\nu)  \rangle = B \nu \exp \left( -\nu^{2}/2 \right),
\label{eq:ep_Gau}
\end{equation}
where $B$ is a constant which depends only on the (two--dimensional)
power spectrum of the random field \cite{a81,c88}. Notice that
$\langle\Gamma\rangle<0$ for $\nu<0$ and $\langle \Gamma\rangle>0$
for $\nu>0$. A curve shifted to the left with respect to this
would be a meatball topology, and to the right would be a
swiss--cheese (in the restricted sense described above).
The most obviously useful application of this method is to
look at projected catalogues, the main problem being that
if the catalogue is very deep,
each line of sight contains a superposition of many
three--dimensional structures. This projection acts to suppress
departures from Gaussian statistics by virtue of the central
limit theorem \cite{dc93a}.
Nevertheless, useful information
is obtainable from projected data because of the sheer size of
the data sets available \cite{mchgw89,gmpl92,cp91,pvc92,dc93a,cmplmm93};
as is
the case with three--dimensional studies, the analysis reveals a
clear meatball shift which is what one expects in the gravitational
instability picture. An analysis of the genus curve for topological
defect toy models was given in  \cite{bks93,gpj90}.
An application of the two dimensional genus
characteristic to analysing anisotropy in the
Cosmic Microwave Background is discussed in \cite{c88,gpj90,stb93,luo94}.
More sophisticated generalised
 morphological measures of large scale structure topology
and connectedness
based on Minkowski functionals have been studied in \cite{mbw94b}.

\subsection{Dimensions and Scaling}
The obvious way to test the idea of multifractal scaling using
observations is to attempt to extract estimates of the $f(\alpha)$
spectrum or, equivalently, the generalised dimensions $D_{q}$
from the catalogue. Care must be taken when doing this, however,
because the formal defintions given in Sec. 6.3 require one
to take the limit of very small box--size. Obviously, when dealing
with a point process one cannot do this because in this
limit all boxes will be empty.  One therefore looks for scaling
over a finite range of box sizes.

The standard way to extract information from catalogues is via
partition functions. For a distribution which has multifractal
scaling properties (see \S 6.3), the function $Z(q,r)$, defined by
\begin{equation}
Z(q,r) = \frac{1}{N} \sum_{i=1}^{N} n_{i}^{q-1} \propto
r^{\tau(q)}, \label{eq:zpart}
\end{equation}
where $r$ is the size of the cell. By graphing $Z$ against
$r$ one can extract $\tau(q)$ for any $q$ and this is
related to $D_{q}$ by eq. (\ref{eq:legendre}). The $Z$--partition
function does not behave well, however, for $q<2$ because the
low values of $q$ are dominated by low--density regions
whose statistics are severely distorted by discreteness effects,
as discussed above. For $q<2$ one therefore uses an inverse
function -- the $W$--partition function, defined by
\begin{equation}
W(\tau, n) = \frac{1}{N} \sum_{i=1}^{N} r_{i} (n)^{-\tau}
\propto n^{1-q}. \label{eq:wpart}
\end{equation}
Here $r_{i}(n)$ is the radius of the smallest sphere centred
upon a point $i$ which encloses $n$ neighbours.

Given these two estimators one can extract all the $D_q$
from a galaxy catalogue \cite{jmse88,mj90,mjdv90,mc94}.
Although this technique is relatively new, the results
are encouraging. The CfA survey shows clear multifractal
scaling properties having, for example, $D_2=1.3$ and
$D_0=2.1$; a more recent analysis of the QDOT survey,
which predominantly probes the linear regime, has
also good scaling properties but quite different
to those of CfA: $D_2=2.77$ and $D_0=2.90$.
Notice that in the monofractal models discussed
by, for example, Pietronero \cite{pi87,cps88,cp92} all the
$D_q$ would be identical. It remains
to be seen whether these analyses will lead to an
understanding of the dynamical origin of the multifractal
scaling observed.

\subsection{Velocity Statistics}
Because peculiar velocity data is known only for a relatively
small number of galaxies, statistical studies of galaxy motion
have been restricted to quite simple descriptors, such as the
net flow in a given volume \cite{lyea88}.
For Gaussian primordial density fields one can
construct useful probability distributions, such
as \cite{sv90}
\begin{equation}
P(v) = \sqrt{\frac{54}{\pi}} \int_{v_1}^{v_2}
\left(\frac{v}{\sigma_v}\right)^{2} \exp
\left[ -\frac{3}{2}\left(\frac{v}{\sigma_v}\right)^{2} \right]
\frac{\d v}{\sigma_v},
\end{equation}
which is the probability that velocity is observed
between $v_1$ and $v_2$ on a smoothing scale $R$
in a randomly--selected volume:  $\sigma_v^{2}$ is given by\footnote{
Comparison of $\sigma_v$ with the mean square density fluctuation
$\sigma$ [see (\ref{eq:lad7})],
shows that large
scales are weighed more heavily in the former. For power
law spectra $P(k) \propto k^n$, $\sigma_v^2(R) \propto R^{-(1+n)}$,
the observed value of $\sigma_v(R)$ can therefore be used to place constraints
on the
value of the spectral index $n$.}
\begin{equation}
\sigma_v^{2}(R) = 4\pi (H_0f)^{2}\int P(k)W^2(kR) \d k.
\label{eq:sigma_v}
\end{equation}
where $f \simeq \Omega^{0.6}$.
More information is contained in two--point correlation
statistics of $\vec{v}(\vec{x})$. Firstly, the scalar
velocity correlation function:
\begin{equation}
\xi_v(r)=\langle \vec{v}(\vec{x})\cdot \vec{v}(\vec{x}') \rangle
=4\pi (H_0 f)^{2} \int_{0}^{\infty} P(k) j_0(kr)dk~,
\end{equation}
where $j_0(x)$ is a spherical Bessel function of order zero and
$r=|\vec{x}-\vec{x}'|$ \cite{p87,k89,gdswy89,gjo89}.

Attempts to move further to consider the properties
of the velocity correlation tensor, which takes
into account the direction of $\vec{v}$ and
not just its magnitude,
were developed in detail by
Gorski \cite{g88,gdswy89}. There are a number of subtleties
involved in velocity statistics that one does not encounter
with spatial distributions so we shall point a couple out
here. Let us assume that we have a continuous density
field and its related velocity field $\vec{v}(\vec{x})$;
that is to say, we know the velocity at every point $\vec{x}$
and the density is also defined at $\vec{x}$. The
relevant autocovariance function of $\vec{v}$ must be
a tensor:
\begin{equation}
\langle v^{i}(\vec{x}_1) v^{j}(\vec{x}_2) \rangle = \Psi^{ij}
(\vec{x}_1,\vec{x}_2).
\end{equation}
Using statistical homogeneity and isotropy, we can decompose
the tensor $\Psi$ into transverse and longitudinal parts
in terms of scalar functions $\psi_{\perp}$ and $\psi_{\parallel}$:
\begin{equation}
\Psi^{ij}(\vec{x}_1,\vec{x}_2) = \psi_{\parallel} (x_{12})
n^{i}n^{j} + \psi_{\perp} (x_{12}) \left( \delta^{ij} - n^{i}n^{j}
\right),
\end{equation}
where
\begin{equation}
\vec{n} = \left(\vec{x}_1 - \vec{x}_2\right)/x_{12}.
\end{equation}
If $\vec{u}$ is any vector satisfying $\vec{u} \cdot \vec{n}=0$
then
\begin{eqnarray}
\psi_{\parallel} & = & \langle (\vec{n}\cdot\vec{v}_1) (\vec{n}\cdot\vec{v}_2)
\rangle \nonumber\\
\psi_{\perp} & = & \langle (\vec{u}\cdot\vec{v}_1) (\vec{u}\cdot\vec{v}_2)
\rangle.
\end{eqnarray}
It can be shown that, in the linear regime, $\vec{\nabla} \times
\vec{v} = 0$ and there is a consequent relationship between the
longitudinal and transverse functions:
\begin{equation}
\psi_{\parallel} = \frac{\d}{\d r} \left[ r \psi_{\perp}(r) \right]
\end{equation}
One can test whether the actual velocity field obeys this relationship
using the appropriate covariance tensor. Of course, one needs
good peculiar motion data and the statistics available at present
are not particularly good \cite{gdswy89}.
One also only measures a peculiar motion in the radial
direction from the observer. There is, however,
another problem which is spelt out at length by Bertschinger
\cite{b92}: when one measures galaxy velocities one is not
extracting a continuous velocity field. This has a complex and
poorly--understood effect upon the final statistics.

A more sophisticated approach to the use of velocity information
is provided by the POTENT analysis method \cite{bd89,dbf90,dbysdh93};
for a recent extensive review of this approach see \cite{dek94}.
The technique makes use of the fact that
in the linear theory of gravitational instability the velocity field
is curl--free and can therefore be expressed as the gradient
of a potential. We saw in \S 2.2 Eqn. (\ref{eq:l15})
that this velocity potential turns out to be simply
proportional to the linear theory value of the gravitational potential.
Because the velocity field is the gradient of a potential $\Phi$,
one can use the purely radial motions revealed by redshift and
distance information to map  $\Phi$ in three dimensions:
\begin{equation}
\Delta\Phi(r,\theta,\phi) = -\int_{0}^{r} v_r (r^{'},
\theta,\phi) \d r. \label{eq:potent}
\end{equation}
Once the potential has been mapped, one can recover the remaining components of
$v^{\alpha}$ by differentiating $\Phi$ along transverse directions.
One can now solve for the density
field using the linearised continuity equation
\begin{equation}
\delta = -{1\over af H}\vec{\nabla}\cdot\vec{v}
\end{equation}
or its quasi--linear generalisation \cite{b92a}
\begin{equation}
\delta = \left[1 - \frac{2}{3afH}\vec{\nabla}\cdot\vec{v}\right]
^{3/2} - 1
\end{equation}
Since the velocity field probes the underlying distribution of both dark
and luminous matter, the POTENT approximation provides a potentially powerful
means for determining $\Omega$.
An extension
of this techique using the Zel'dovich approximation instead of linear
theory is discussed in \cite{ndbb91} (see \S 4.2). Great things are claimed
of the POTENT analysis, but one should be cautious of these
until larger and more reliable peculiar velocity data sets are
available.

{}From the continuity equation, eq. (\ref{eq:continuityb}) or (\ref{eq:l9})
one sees that the velocity field is more sensitive to the presence of longer
wavelengths in the primordial spectrum than the density field.
We therefore expect the velocity field to show smaller departures from
linear behaviour
than the density field when both are smoothed over identical scales.
It is therefore useful to extend the quasilinear analysis of \S 4.1 and
\S 4.7 to velocity fields. We define for this purpose a scalar
velocity variable $\theta(\vec x) \equiv (1/H) \nabla \vec{v}(\vec x)$,
in terms of which the skewness and the kurtosis of the velocity field
have the form
\begin{eqnarray}
T_3 & = & {\langle\theta^3\rangle\over \langle\theta^2\rangle^2}\nonumber\\
T_4 & = & {\langle\theta^4\rangle - 3\langle\theta^2\rangle^2\over
\langle\theta^2\rangle^3}
\label{eq:t3t4}
\end{eqnarray}
(the angular brackets denote an ensemble average).
For Gaussian fields, $T_3 = T_4 = 0$. In the weakly nonlinear
regime, gravity induces a departure from the initial Gaussianity which
results in the following expressions for $T_3$, $T_4$, for scale free
initial spectra $P(k) \propto k^n$ with top-hat final smoothing
\cite{jbc93,b94b,mss94}
\begin{eqnarray}
T_3 & = &- \frac{1}{\Omega^{0.6}}\,\left[\frac{26}{7} - (n+3)\right]
\nonumber\\
T_4 & = & {1\over \Omega^{1.2}}\,\left[ \frac{12088}{441} - \frac{338}{21}(n+3)
+\frac{7}{3}(n+3)^2\right]
\label{eq:t3t4b}
\end{eqnarray}
If $n$ is known then equation's (\ref{eq:t3t4b}) can be used to determine
the value of $\Omega$.
On the other hand, the ratio $T_4/T_3^2$ is virtually independent of
$\Omega$ :
\begin{equation}
{T_4\over T_3^2} = {(12088/441) - (338/21)(n+3) + (7/3)(n+3)^2\over
\big\lbrack(26/7) - (n+3)\big\rbrack^2},
\label{eq:t3t4c}
\end{equation}
and can therefore be used to test the gravitational instability hypothesis if
the value of $n$ is known independently \cite{b94b}.
For instance galaxy survey's seem to indicate
$-2 \le n \le -1$ on cluster scales (see \S 8.4), which leads to the
prediction: $1.55 \le T_4/T_3^2 \le 1.85$, for data
smoothed on such scales.

These above methods of course require one to know peculiar
motions for a sample of galaxies. As we have mentioned
previously, one can still use the existence of peculiar
velocities in a statistical sense even if one cannot measure
a velocity for each individual source. Correlations and
power--spectra in particular show a characteristic distortion
when they are viewed in redshift space rather than real space.
The two--point correlation function will, for example, be
elongated in the radial direction. Given the paucity of
available peculiar velocity data, it seems to us that this
type of analysis is the most promising approach to
the use of cosmological velocity information at present
\cite{kea91,ham92,gcb93,h93,fdsyh93,fea94,cfw94,ht95}.

%% file: ssec9.tex
\section{Discussion and Conclusions}
In this review we have attempted to give a comprehensive account of the many
facets of gravitational instability, focussing particularly upon
its dynamical and statistical aspects.

It is well established that gravitational instability proceeds
along several distinct epochs over which its characteristic features change
continuously. It is perhaps helpful, by way of a summary, to delineate
these epochs and how they can be described by the various
techniques we have reviewed in this paper.

During the first epoch density fluctuations are much smaller
than unity and consequently an adequate description is provided by linear
theory. During this epoch fluctuations,
as long as the thermal velocity of matter is small,
evolve self-similarly
$\delta(\vec x, t) \propto \Delta_+(t)\delta(\vec x)$, as a result of which
the fluctuation spectrum retains many of its original features,
particularly the presence or absence of phase correlations.

Linear theory begins to break down when $\delta \sim 1$, marking
the onset of the quasi-linear, or weakly non-linear, epoch. During this epoch,
mode--mode coupling between the different Fourier components of the density
contrast becomes important, resulting in the build--up of phase correlations.
Non--Gaussian features in the density field then become important, even
when the initial fluctuations are Gaussian.

Our understanding of the strongly non--linear regime
has undergone a radical change
during the last two decades. Two decades ago the {\em top--down} model of
galaxy formation suggested by Zeldovich and co-workers in connection with
adiabatic models of structure formation and the {\em bottom--up},
or hierarchical, model suggested by Peebles in relation to isothermal
models of galaxy formation were regarded as being mutually exclusive.
The former produced correlated structures on very large scales whereas
clustering in the latter was essentially featureless and self--similar.
The first hints that this rigid demarcation of the two scenarios
was not, in fact, correct appeared in the early
eighties in studies involving non-baryonic models (such as CDM).
These models develope structures from small to large scales and
so might be called {\em hierarchical} and yet displayed many
large--scale features usually associated with pancake models.
These results received further confirmation from high resolution
N-body simulations and non-linear approximations such as the adhesion model.
It is now widely believed that pancakes form generically in models of
hierarchical clustering, the precise size of the pancake being determined by
the form of the primordial fluctuation spectrum.

It is interesting that pancakes, though prominent at the time of appearance,
are actually best regarded as intermediate asymptotes which fade out and
ultimately disappear as matter
within them moves towards filaments (which form at the intersections of
pancakes) and then along filaments towards clumps (which form where
filaments intersect). The period during which pancakes, filaments and clumps
coexist may be called the {\em cellular epoch}. Once most of the
matter has collected in clumps, gravitational instability proceeds
 hierarchically as neighboring clumps attract one another, finally
merging to form more and more massive  aggregates of
gravitationally bound matter. This later epoch of gravitational instability
is characterised by the gradual flow of power from smaller to larger scales
accompanied by the elimination of small voids and also, occasionally, by the
emergence of large coherent structures -- {\it secondary pancakes} -- in models
whose primordial spectra have signficant large scale power (including CDM).

Clearly the physics of the different epochs characterising gravitational
instability will vary continuously as will the statistical properties of
the Universe during these separate epochs. It is here that non-linear
approximations to gravitational dynamics accompanied by statistical methods
play
a vital role in the development of our understanding of these processes,
and how they apply to the real Universe.

While some features of gravitational instability, particularly its
dynamical aspects, are reasonably well understood, others, including a
comprehensive statistical treatment (such as the one based upon the BBGKY
hierarchy) are still in the developmental stage. Consequently,
some quasi-analytical treatments which can adequately address
certain dynamical aspects of clustering may also can fail to come up with
satisfactory answers to statistical questions and vice versa.
An illustration of this is the Press-Schechter approximation which,
while accounting fairly accurately for
the mass (or multiplicity) function of bound objects such as galaxies,
fails to say anything at all about the large scale clustering properties of
the distribution.

It is becoming widely acknowledged that large scale features
in the galaxy distribution such as superclusters and voids are real physical
entities and can no longer be regarded solely as products of a visual
system sensitive to detecting filamentarity.
As a result there is a growing need for statistical measures that can provide
a qualitative description of such large scale features as exist in the galaxy
distribution. It is also realised that whereas statistical descriptors such as
the two point correlation function  are very useful in characterising
the clustering properties of the galaxy distribution on small scales,
they say next to nothing about the over-all clustering pattern of galaxies
in the Universe (whether {\em bubble-like} or {\em sponge-like}) and
must therefore be complemented by other statistical measures some of which
are discussed in \S 8. The development of analytical techniques to follow
the evolution of structure must therefore be accompanied by a comprehensive
understanding of the behaviour of the various pattern diagnostics
when applied to different degrees of evolution from different
primordial power spectra.

We have concentrated upon analytical approaches in this article, but
there are of course many situations were an analytic treatment is
not practical, particularly if one seeks a detailed comparison
of evolved distributions with observational properties. We
therefore fully acknowledge the enormous value of computational
methods in this kind of study. Without the $N$--body simulation,
progress in this area during the last two decades would have been limited
indeed. Nevertheless, simulating the process of structure
formation is not the same as understanding it and we feel most
strongly that analytical studies should be the focus of much
more theoretical effort in the forthcoming years. We have shown
that much can be explained and understood using quite simple
mathematical techiques, and that this is an extremely active area.
As further observational data are obtained, about the clustering
of galaxies and the primordial fluctuations (from the cosmic
microwave background), to complement these theoretical analyses,
there is good reason to be optimistic that we will, before long,
be able to understand the large--scale structure of the
Universe.

%% file: spaper.bbl
\begin{thebibliography}{999}
\bibitem{a84}
S.J. Aarseth, in: J. Brackbill and B.I. Cohen (eds), {\em Methods
of Computational Physics} (Academic Press, New York, 1984) p. 1.

\bibitem{agt79}
S.J. Aarseth, J.R. Gott and E.L. Turner, {\em Astrophys. J.},
{\bf 236} (1979) 43.

\bibitem{aw84}
L.F. Abbott and M.B. Wise, {\em Nucl. Phys. B.},
{\bf 135} (1984) 279.

\bibitem{a58}
G.O. Abell, {\em Astrophys. J. Suppl.}, {\bf 3} (1958) 211.

\bibitem{aco89}
G.O. Abell, H.G. Corwin and R.P. Olowin, {\em Astrophys. J. Suppl.},
{\bf 70} (1989) 1.

\bibitem{a81}
R.J. Adler, {\em The Geometry of Random Fields} (Wiley, New York, 1981).

\bibitem{ab69}
J. Aitchison and J.A.C. Brown, {\em The Lognormal Distribution
with special reference to its uses in Economics}
(Cambridge University Press, Cambridge, 1969).

\bibitem{aj90}
L. Appel and B.J.T. Jones, {\em Mon. Not. R. astr. Soc.},
{\bf 245} (1990) 522.

\bibitem{asz82}
V.I. Arnol'd, S.F. Shandarin and Ya.B. Zel'dovich,
{\em Geophys. Astrophys. Fluid. Dyn.}, {\bf 20} (1982) 111.

\bibitem{a86}
V.I. Arnol'd, {\em Catastrophe Theory} (Springer, Berlin, 1976).

\bibitem{agv85}
V.I. Arnol'd, S.M. Gusein-Sade, A.N. Varchenko
{\em Singularities of Differentiable Maps, vol. 1}
(Birkhauser, Boston, 1985).

\bibitem{bab90}
A. Babul, {\em Astrophys. J.}, {\bf 349} (1990) 429.

\bibitem{bs92}
A. Babul and G.D. Starkman, {\em Astrophys. J.}, {\bf 401} (1992) 28.

\bibitem{bv94}
A. Babul and R. van de Weygaert, {\em CITA preprint},
{\em Nature} submitted (1994).

\bibitem{bp94}
J.S. Bagla and T. Padmanabhan, {\em Mon. Not. R. astr. Soc.},
{\bf 266} (1994) 227.

\bibitem{b88}
N.A. Bahcall, {\em Ann. Rev. Astron. Astrophys.} {\bf 26} (1988) 631.

\bibitem{bs83}
N.A. Bahcall and R.M. Soneira, {\em Astrophys. J.}, {\bf 270} (1983)
20.

\bibitem{bw92}
N.A. Bahcall and M.J. West, {\em Astrophys. J.}, {\bf 392} (1992) 419.

\bibitem{b75}
R. Balescu, {\em Equilibrium and Non--equilibrium Statistical Mechanics}
(Wiley, New York, 1975)

\bibitem{bs89a}
R. Balian and R. Schaeffer, {\em Astr. Astrophys.},
{\bf 220} (1989) 1.

\bibitem{bs89b}
R. Balian and R. Schaeffer, {\em Astr. Astrophys.},
{\bf 226} (1989) 373.

\bibitem{bar80}
J.M. Bardeen, {\em Phys. Rev. D}, {\bf 22} (1980) 1882.

\bibitem{bbks86}
J.M. Bardeen, J.R. Bond, N. Kaiser and A.S. Szalay,
{\em Astrophys. J.}, {\bf 304} (1986) 15.

\bibitem{bh86}
J. Barnes and P. Hut, {\em Nature}, {\bf 324} (1986) 446.

\bibitem{bb87}
J.D. Barrow and S.P. Bhavsar, {\em Quart. J. R. astr. Soc.},
{\bf 28} (1987) 109.

\bibitem{bbs84}
J.D. Barrow, S.P. Bhavsar and D.H. Sonoda, {\em Mon. Not. R. astr. Soc.},
{\bf 210} (1984) 19P.

\bibitem{bbs85}
J.D. Barrow, S.P. Bhavsar and D.H. Sonoda, {\em Mon. Not. R. astr. Soc.},
{\bf 216} (1985) 17.

\bibitem{bs93}
J.D. Barrow and P. Saich, {\em Class. Quant. Grav.}, {\bf 10} (1993)
79.

\bibitem{be53}
H. Bateman and A. Erdelyi, {\em Higher Transcendental Functions},
Vol. 1 (McGraw--Hill, New York, 1953).

\bibitem{be93}
C.M. Baugh and G. Efstathiou,  {\em Mon. Not. R. astr. Soc.}, {\bf 265}
(1993) 165.

\bibitem{be94}
C.M. Baugh and G. Efstathiou,  {\em Mon. Not. R. astr. Soc.}, {\bf 267}
(1994) 323.

\bibitem{bf91}
D.J. Baumgart and J.N. Fry, {\em Astrophys. J.}, {\bf 375} (1991) 25.

\bibitem{b92a}
F. Bernardeau, {\em Astrophys. J.}, {\bf 390} (1992) L61.

\bibitem{b92b}
F. Bernardeau, {\em Astrophys. J.}, {\bf 392} (1992) 1.

\bibitem{b94}
F. Bernardeau, {\em Astrophys. J.}, {\bf 427} (1994) 51.

\bibitem{b94b}
F. Bernardeau, {\em Astrophys. J.}, {\bf 433} (1994) 1.

\bibitem{bk94}
F. Bernardeau and L.A. Kofman, {\em Astrophys. J.}, in press (1994).

\bibitem{bsbc94}
F. Bernardeau, T.P. Singh, B. Banerjee and S.M. Chitre,
{\em Mon. Not. R. astr. Soc.}, {\bf 269} (1994) 947.

\bibitem{b83}
E. Bertschinger, {\em Astrophys. J.}, {\bf 268} (1983) 17.

\bibitem{b85}
E. Bertschinger, {\em Astrophys. J. Suppl.}, {\bf 58} (1985) 1.

\bibitem{b85b}
E. Bertschinger, {\em Astrophys. J. Suppl.}, {\bf 58} (1985) 39.

\bibitem{b87}
E. Bertschinger, {\em Astrophys. J.}, {\bf 316} (1987) 489.

\bibitem{b92}
E. Bertschinger, in: V.J. Martinez, M. Portilla and D. Saez (eds),
{\em New Insights into the Universe. Proceedings, Valencia, Spain
1991} (Springer, Berlin, 1992) p. 65.

\bibitem{bd89}
E. Bertschinger and A. Dekel, {\em Astrophys. J.},
{\bf 336} (1989) L5.

\bibitem{bdfdb90}
E. Bertschinger, A. Dekel, S.M. Faber, A. Dressler and
D. Burstein, {\em Astrophys. J.}, {\bf 364} (1990) 370.

\bibitem{bg91}
E. Bertschinger and J.M. Gelb, {\em Computers in Physics},
{\bf 5} (1991) 164.

\bibitem{bj94}
E. Bertschinger and B. Jain, {\em Astrophys. J.}, {\bf 431} (1994)
486.

\bibitem{bh94}
E. Bertschinger and A.J.S. Hamilton, {\em Astrophys. J.},
{\bf 435} (1994) 1.

\bibitem{bhar94}
S. Bharadwaj, {\em Astrophys. J.}, {\bf 428} (1994) 419.

\bibitem{bb83}
S.P. Bhavsar and J.D. Barrow,
{\em Mon. Not. R. astr. Soc.}, {\bf 205} (1983) 61P.

\bibitem{bl88}
S.P. Bhavsar and E.N. Ling, {\em Astrophys. J.},
{\bf 331} (1988) L63.

\bibitem{bbk92}
S. Bildhauer, T. Buchert and M. Kasai, {\em Astron. Astrophys.}, {\bf 263}
(1992) 23.

\bibitem{bt87}
J. Binney and S. Tremaine, {\em Galactic Dynamics} (Princeton University
Press, Princeton, 1987).

\bibitem{bvg90}
O. Blaes, J.V. Villumsen and P. Goldreich, {\em Astrophys. J.}, {\bf 361}
(1990) 331.

\bibitem{bfpr84}
G.R. Blumenthal, S.M. Faber, J.R. Primack and M.J. Rees,
{\em Nature}, {\bf 311} (1984) 517.

\bibitem{bcglp92}
G.R. Blumenthal, L.N. da Costa, D.S. Goldwirth, M. Lecar and T. Piran,
{\em Astrophys. J.},
{\bf 388} (1992) 234.

\bibitem{bw92a}
H. B\"{o}hringer and G. Wiedenmann, in:
V.J. Martinez, M. Portilla and D. Saez (eds),
{\em New Insights into the Universe. Proceedings, Valencia, Spain
1991} (Springer, Berlin, 1992) p. 127.

\bibitem{bond94}
J.R. Bond, in: M. Sasaki (ed), {\em Relativistic Cosmology},
(Academic Press 1994), p. 23.

\bibitem{bcek91}
J.R. Bond, S. Cole, G. Efstathiou and N. Kaiser, {\em Astrophys. J.},
{\bf 379} (1991) 440.

\bibitem{bc88}
J.R. Bond and H.M.P. Couchman, in: A.A. Coley, C.C. Dyer and B.O.J.
Tupper (eds), {\em Proceedings of the Second Canadian Conference
on General Relativity and Relativistic Astrophysics}
(World Scientific, Singapore, 1988), p. 385.

\bibitem{bcsw84}
J.R. Bond, J. Centrella, A.S. Szalay and J.R. Wilson,
{\em Mon. Not. R. astr. Soc.}, {\bf 210} (1984) 515.

\bibitem{be84}
J.R. Bond and G. Efstathiou, {\em Astrophys. J.}, {\bf 285}
(1984) L45.

\bibitem{bonds83}
J.R. Bond and A.S. Szalay, {\em Astrophys. J.}, {\bf 277}
(1983) 443.

\bibitem{bss88}
J.R. Bond, A.S. Szalay and J. Silk, {\em Astrophys. J.},
{\bf 324} (1988) 627.

\bibitem{bcdes94}
J.R. Bond, R. Crittenden, R.L. Davis, G. Efstathiou and P.J. Steinhardt,
{\em Phys. Rev. Lett.} {\bf 72} (1994) 13.

\bibitem{bm94a}
J.R. Bond and S.T. Myers, CITA preprint (1994).

\bibitem{bm94b}
J.R. Bond and S.T. Myers, CITA preprint (1994).

\bibitem{bm94c}
J.R. Bond and S.T. Myers, CITA preprint (1994).

\bibitem{bm94d}
J.R. Bond and S.T. Myers, CITA preprint (1994).

\bibitem{b47}
H. Bondi, {\em Mon. Not. R. astr. Soc.}, {\bf 107} (1947) 410.

\bibitem{bc90}
W.B. Bonnor and A. Chamorro, {\em Astrophys. J.}, {\bf 361} (1990) 21.

\bibitem{biggh93}
S.A. Bonometto, A. Iovino, L. Guzzo, R. Giovanelli and M.P.. Haynes,
{\em Astrophys. J.}, {\bf 419} (1993), 451.

\bibitem{blm87}
S.A. Bonometto, F. Lucchin and S. Matarrese, {\em Astrophys. J.},
{\bf 323} (1987) 19.

\bibitem{borg94}
S. Borgani, {\em Physics Reports}, to be published (1994).

\bibitem{bcm94}
S. Borgani, P. Coles and L. Moscardini, {\em Mon. Not. R. astr. Soc.},
{\bf 271} (1994) 223.

\bibitem{bh88}
F.R. Bouchet and L. Hernquist, {\em Astrophys. J. Suppl.},
{\bf 68} (1988) 521.

\bibitem{bh92}
F.R. Bouchet and L. Hernquist,
{\em Astrophys. J.}, {\bf 400} (1992), 25.

\bibitem{bjcp92}
F.R. Bouchet, R. Juszkiewicz, S. Colombi and R. Pellat,
{\em Astrophys. J.}, {\bf 394} (1992) L5.

\bibitem{bsd91}
F.R. Bouchet, R. Schaeffer and M. Davis,
{\em Astrophys. J.}, {\bf 383} (1991) 19.

\bibitem{bsdfyh93}
F.R. Bouchet, M.A. Strauss, M. Davis, K.B. Fisher, A. Yahil
and J.P. Huchra, {\em Astrophys. J.}, {\bf 417} (1993) 36.

\bibitem{bchj94}
F.R. Bouchet, S. Colombi, E. Hivon and R. Juszkiewicz,
{\em Astr. Astrophys.} submitted (1994).

\bibitem{bow91}
R.G. Bower, {\em Mon. Not. R. astr. Soc.},
{\bf 248} (1991) 332.

\bibitem{bcfw93}
R.G. Bower, P. Coles, C.S. Frenk and S.D.M. White,
{\em Astrophys. J.}, {\bf 405} (1993) 403.

\bibitem{bsv93}
T.G. Brainerd, R.J. Scherrer and J. V. Villumsen,
{\em Astrophys. J.} {\bf 418} (1993) 570.

\bibitem{brc94}
E. Branchini and R.G. Carlberg, {\em Astrophys. J.}, {\bf 434} (1994) 37.

\bibitem{bks93}
R. Brandenberger, D. Kaplan and S. Ramsey, Brown University
Preprint, Brown-HET-922 (1993)

\bibitem{beks90}
T.J. Broadhurst, R.S. Ellis, D.C. Koo and A.S. Szalay,
{\em Nature}, {\bf 343} (1990) 726.

\bibitem{bruce86}
J.W. Bruce, {\em J. Lond. Math. Soc. (2)}, {\bf 33} (1986) 375.

\bibitem{bde92}
M. Bruni, P.K.S. Dunsby and G.F.R. Ellis,
{\em Astrophys. J.}, {\bf 395} (1992) 34.

\bibitem{bmp94}
M. Bruni, S. Matarrese and O. Pantano,
{\em Astrophys. J.}, submitted (1994).

\bibitem{buch89}
T. Buchert, {\em Astron. Astrophys.}, {\bf 223} (1989) 9.

\bibitem{buch92}
T. Buchert, {\em Mon. Not. R. astr. Soc.}, {\bf 254} (1992) 729.

\bibitem{buch93}
T. Buchert, {\em Astron. Astrophys.}, {\bf 267} (1993) L51.

\bibitem{buch94}
T. Buchert, {\em Mon. Not. R. astr. Soc.}, {\bf 267} (1994) 811.

\bibitem{be93}
T. Buchert and J. Ehlers, {\em Mon. Not. R. astr. Soc.}, {\bf 264} (1993)
375.

\bibitem{bmw94}
T. Buchert, A.L. Melott, and A.G. Weiss
{\em Astron. Astrophys.}, {\bf 288} (1994) 349.

\bibitem{b74}
J.M. Burgers, {\em The Non--linear Diffusion Equation}
(Reidel, Dordrecht, 1974).

\bibitem{burs90}
D. Burstein, {\em Rep. Prog. Phys.}, {\bf 53} (1990) 421.

\bibitem{cc89}
R.G. Carlberg and H.M.P. Couchman, {\em Astrophys. J.} {\bf 340} (1989) 47.

\bibitem{cl56}
D.E. Cartwright and M.S. Longuet--Higgins, {\em Proc. Roy. Soc.
Lond. Ser. A.}, {\bf 237} (1956) 212.

\bibitem{cat94}
P. Catelan, {\em Mon. Not. R. astr. Soc.}, submitted (1994).

\bibitem{ccmm94}
P. Catelan, P.Coles, S. Matarrese and L. Moscardini, {\em Mon. Not.
R. astr. Soc.}, {\bf 268} (1994) 966.

\bibitem{clmm94}
P. Catelan, F. Lucchin, S. Matarrese and L. Moscardini,
{\em Mon. Not. R. astr. Soc.}, submitted (1994)

\bibitem{cm94}
P. Catelan and L. Moscardini, {\em Astrophys. J.} {\bf 426} (1994) 14.

\bibitem{cm94b}
A. Cavaliere and N. Menci,
{\em Astrophys. J.}, {\bf 435} (1994) 528.

\bibitem{cen92}
R. Cen, {\em Astrophys. J. Suppl.}, {\bf 78} (1992) 341.

\bibitem{cm83}
J.M. Centrella and A.L. Melott {\em Nature}, {\bf 305} (1983) 196.

\bibitem{c91}
A. Chamorro, {\em Astrophys. J.} {\bf 383} (1991) 51.

\bibitem{cprw87}
J.M. Cline, H.D. Politzer, S.-J. Rey and M.B. Wise,
{\em Comm. Math. Phys.}, {\bf 112} (1987) 217.

\bibitem{cole91}
S. Cole, {\em Astrophys. J.}, {\bf 367} (1991) 45.

\bibitem{cafnz94}
S. Cole, A. Arag\'{o}n-Salamanca, C.S. Frenk, J.F. Navarro and S.E. Zepf,
{\em Mon. Not. R. astr. Soc.}, {\bf 271} (1994) 781.

\bibitem{cfw94}
S. Cole, K.B. Fisher and D.H. Weinberg, {\em Mon. Not. R. astr. Soc.},
{\bf 267} (1994) 785.

\bibitem{ck88}
S. Cole and N. Kaiser, {\em Mon. Not. R. astr. Soc.}, {\bf 233} (1988) 637.

\bibitem{ck89}
S. Cole and N. Kaiser, {\em Mon. Not. R. astr. Soc.}, {\bf 237} (1989) 1127.

\bibitem{cps88}
P.H. Coleman, L. Pietronero and R.H. Sanders,
{\em Astr. Astrophys.}, {\bf 200}, L32.

\bibitem{cp92}
P.H. Coleman and L. Pietronero, {\em Phys. Rep.}, {\bf 213} (1992) 311.

\bibitem{c86}
P. Coles, {\em Mon. Not. R. astr. Soc.}, {\bf 222} (1986) 9P.

\bibitem{c88}
P. Coles, {\em Mon. Not. R. astr. Soc.},
{\bf 234} (1988) 509.

\bibitem{c89}
P. Coles, {\em Mon. Not. R. astr. Soc.},
{\bf 238} (1989) 319.

\bibitem{c90}
P. Coles, {\em Mon. Not. R. astr. Soc.}, {\bf 243} (1990) 171.

\bibitem{c92}
P. Coles, in: E.D. Feigelson and G.J. Babu,
{\em Statistical Challenges in Modern Astronomy}
(Springer, New York, 1992) p. 57.

\bibitem{c93}
P. Coles, {\em Mon. Not. R. astr. Soc.},
{\bf 262} (1993) 1065.

\bibitem{cb87}
P. Coles and J.D. Barrow, {\em Mon. Not. R. astr. Soc.},
{\bf 228} (1987) 407.

\bibitem{cd93}
P. Coles and A.G. Davies, {\em Mon. Not. R. astr. Soc.},
{\bf 264} (1993) 261.

\bibitem{ce94}
P. Coles and G.F.R. Ellis,
{\em Nature}, {\bf 370} (1994) 609.

\bibitem{cf91}
P. Coles and C.S. Frenk, {\em Mon. Not. R. astr. Soc.},
{\bf 253} (1991) 727.

\bibitem{cj91}
P. Coles and B.J.T. Jones, {\em Mon. Not. R. astr. Soc.},
{\bf 248} (1991) 1.

\bibitem{cms93}
P. Coles, A.L. Melott and S.F. Shandarin,
{\em Mon. Not. R. astr. Soc.}, {\bf 260} (1993) 765.

\bibitem{cmplmm93}
P. Coles, L. Moscardini, M. Plionis, F. Lucchin, S. Matarrese
and A. Messina,
{\em Mon. Not. R. astr. Soc.}, {\bf 260} (1993) 572.

\bibitem{cmlmm93}
P. Coles, L. Moscardini, F. Lucchin, S. Matarrese and A. Messina,
{\em Mon. Not. R. astr. Soc.}, {\bf 264} (1993) 749.

\bibitem{cp91}
P. Coles and M. Plionis, {\em Mon. Not. R. astr. Soc.},
{\bf 250} (1991) 75.

\bibitem{cnl92}
C.A. Collins, R.C. Nichol and S.L. Lumsden,
{\em Mon. Not. R. astr. Soc.}, {\bf 254} (1992) 295.

\bibitem{col94}
S. Colombi, {\em Astrophys. J.}, {\bf 435} (1994) 536.

\bibitem{c87a}
H.M.P. Couchman, {\em Mon. Not. R. astr. Soc.}, {\bf 225} (1987) 777.

\bibitem{c87b}
H.M.P. Couchman, {\em Mon. Not. R. astr. Soc.}, {\bf 225} (1987) 795.

\bibitem{cou91}
H.M.P. Couchman, {\em Astrophys. J.}, {\bf 368} (1991) L23.

\bibitem{cs86}
P. Crane and W.C. Saslaw, {\em Astrophys. J.}, {\bf 301} (1986) 1.

\bibitem{cbdes93}
R. Crittenden, J.R. Bond, R.L. Davis, G. Efstathiou and
P.J. Steinhardt, {\em Phys. Rev. Lett.}, {\bf 71} (1993)
324.

\bibitem{cpss94}
K.M. Croudace, J. Parry, D.S. Salopek and J.M. Stewart, {\em Astrophys. J},
{\bf 423} (1994) 22

\bibitem{cs88}
E.L. Crow and K. Shimizu (eds), {\em Lognormal Distributions:
Theory and Applications} (Marcel Dekker, New York, 1988).

\bibitem{dea94}
L.N. da Costa, M.J. Geller, P.S. Pellegrini, D.W. Latham, A.P. Fairall,
R.D. Marzhe, C.N.A. Miller, J.P. Huchra, J.H. Calderon, M. Ramella
and M.I. Kurtz, {\em Astrophys. J.}, {\bf 424} (1994) L1.

\bibitem{dea88}
L.N. da Costa, P.S. Pellegrini, W.L.W. Sargent, J. Tonry, M. Davis,
A. Meiksin, D.W. Latham, J.W. Menzies and L.A. Coulson,
{\em Astrophys. J.}, {\bf 337} (1988) 544.

\bibitem{dc94}
L.N. da Costa, in: {\it Cosmic Velocity Fields}, Proceedings
of the ninth IAP Astrophysics meeting; F.R. Bouchet and
M. Lachieze-Rey (eds) (Editions Frontieres 1994) p. 475.

\bibitem{dvghp94}
L.N. da Costa, M.S. Vogeley, M.J. Geller, J.P. Huchra and C. Park,
{\em Astrophys. J. Lett.} in press (1994).

\bibitem{dems92}
G.B. Dalton, G. Efstathiou, S.J. Maddox and W.J. Sutherland,
{\em Astrophys. J.}, {\bf 390} (1992) L1.

\bibitem{dc93a}
A.G. Davies and P. Coles, {\em Mon. Not. R. astr. Soc.},
{\bf 260} (1993) 553.

\bibitem{dc93b}
A.G. Davies and P. Coles, {\em Mon. Not. R. astr. Soc.},
{\bf 262} (1993) 591.

\bibitem{defw85}
M. Davis, G. Efstathiou, C.S. Frenk and S.D.M. White,
{\bf 292} (1985) 371.

\bibitem{dhlt82}
M. Davis, J.P. Huchra, D.W. Latham and J.R. Tonry,
{\em Astrophys. J.}, {\bf 253} (1982) 423.

\bibitem{dp77}
M. Davis and P.J.E. Peebles, {\em Astrophys. J. Suppl.}, {\bf 34} (1977)
425.

\bibitem{dp83}
M. Davis and P.J.E. Peebles, {\em Astrophys. J.}, {\bf 267} (1983)
465.

\bibitem{dhsst92}
R.L. Davis, H.M. Hodges, G.F. Smoot, P.J. Steinhardt and
M.S. Turner, {\em Phys. Rev. Lett.}, {\bf 69} (1992) 1856.

\bibitem{dek94}
A. Dekel, {\em Ann. Rev. Astron. Astrophys.}, {\bf 32} (1994) 371.

\bibitem{dbf90}
A. Dekel, E. Bertschinger and S.M. Faber,
{\em Astrophys. J.}, {\bf 364} (1990) 349.

\bibitem{dbysdh93}
A. Dekel, E. Bertschinger, A. Yahil, M.A. Strauss, M. Davis and
J.P. Huchra, {\em Astrophys. J.}, {\bf 412} (1993) 1.

\bibitem{dbpo89}
A. Dekel, G.R. Blumenthal, J.R. Primack and S. Olivier,
{\em Astrophys. J.}, {\bf 338} (1989) L5.

\bibitem{dr87}
A. Dekel and M.J. Rees, {\em Nature}, {\bf 326} (1987) 455.

\bibitem{dw85}
A. Dekel and M.J. West, {\em Astrophys. J.},
{\bf 288} (1985) 411.

\bibitem{dgh86}
V. de Lapparent, M.J. Geller and J.P. Huchra,
{\em Astrophys. J.}, {\bf 302} (1986) L1.

\bibitem{dgh89}
V. de Lapparent, M.J. Geller and J.P. Huchra,
{\em Astrophys. J.}, {\bf 343} (1989) 1.

\bibitem{d70}
A.G. Doroshkevich, {\em Astrofizica}, {\bf 6} (1970) 320.

\bibitem{d73}
A.G. Doroshkevich, {\em Astrophys. Lett.}, {\bf 14} (1973) 11.

\bibitem{dk90}
A.G. Doroshkevich and E.V. Kotok,
{\em Mon. Not. R. astr. Soc.}, {\bf 246} (1990) 10.

\bibitem{drs73}
A.G. Doroshkevich, V.S. Ryabenkii and S.F. Shandarin,
{\em Astrofizika}, {\bf 9} (1973) 144.

\bibitem{dss78}
A.G. Doroshkevich, S.F. Shandarin and E. Saar,
{\em Mon. Not. R. astr. Soc.}, {\bf 246} (1978) 10.


\bibitem{dknpss80}
A.G. Doroshkevich, E.V. Kotok, I.D. Novikov, A.N. Polyudov, S.F. Shandarin
and Yu.S. Sigov, {\em Mon. Not. R. astr. Soc.}, {\bf 192} (1980) 321.

\bibitem{dkov90}
R. Durrer and I. Kovner, {\em Astrophys. J.}, {\bf 356} (1990)
49.

\bibitem{ds90}
R. Durrer and N. Straumann, {\em Mon. Not. R. astr. Soc.}, {\bf 242}
(1990) 221.

\bibitem{e90}
G. Efstathiou, in: J.A. Peacock, A.F. Heavens and A.T. Davies
(eds), {\em Physics of the Early Universe. Proceedings of the
36th Scottish Universities Summer School in Physics}
(Adam Hilger, Bristol, 1990) p. 361.

\bibitem{eb86}
G. Efstathiou and J.R. Bond, {\em Mon. Not. R. astr. Soc.},
{\bf 218} (1986) 103.

\bibitem{ebw92}
G. Efstathiou, J.R. Bond and S.D.M. White, {\em Mon. Not. R.
astr. Soc.}, {\bf 258} (1992) 1P.

\bibitem{edsm92}
G. Efstathiou, G.B. Dalton, W. Sutherland and S.J. Maddox,
{\em Mon. Not. R. astr. Soc.}, {\bf 257} (1992) 125.

\bibitem{edfw85}
G. Efstathiou, M. Davis, C.S. Frenk and S.D.M. White,
{\em Astrophys. J. Suppl.}, {\bf 57} (1985) 241.

\bibitem{ee81}
G. Efstathiou and J.W. Eastwood, {\em Mon. Not. R. astr. Soc.},
{\bf 194} (1981) 503.

\bibitem{efh79}
G. Efstathiou, S.M. Fall and C. Hogan, {\em Mon. Not. R. astr. Soc.},
{\bf 189} (1979) 203.

\bibitem{efwd88}
G. Efstathiou, C.S. Frenk, S.D.M. White and M. Davis, {\em
Mon. Not. R. astr. Soc.}, {\bf 235} (1988) 715.

\bibitem{eea90}
G. Efstathiou, N. Kaiser, W. Saunders, A. Lawrence,
M. Rowan--Robinson and C.S. Frenk, {\em Mon. Not. R. astr. Soc.},
{\bf 247} (1990) 10P.

\bibitem{er88}
G. Efstathiou and M.J. Rees, {\em Mon. Not. R. astr. Soc.}, {\bf 230} (1988)
5P.

\bibitem{ejs80}
J. Einasto, M. Joeveer and E. Saar, {\em Mon. Not. R. astr. Soc.}, {\bf 193}
(1980) 353.

\bibitem{eegs91}
J. Einasto, J. Einasto, M. Gramann and E. Saar,
{\em Mon. Not. R. astr. Soc.}, {\bf 248} (1991) 593.

\bibitem{ell71}
G.F.R. Ellis, in: R.K. Sachs (ed), {\em General Relativity and
Cosmology} (Academic Press, New York, 1971), p. 104.

\bibitem{eb89}
G.F.R. Ellis and M. Bruni, {\em Phys. Rev. D}, {\bf 40} (1989) 1804.

\bibitem{ed94}
G.F.R. Ellis and P.K.S. Dunsby, Cape Town preprint, {\em Astrophys. J.}
submitted (1994).

\bibitem{es87}
G.F.R. Ellis and W.J. Stoeger, {\em Class. Qu. Grav.}, {\bf 4} (1987)
1679.

\bibitem{e88}
A.E. Evrard, {\em Mon. Not. R. astr. Soc.}, {\bf 235} (1988) 911.

\bibitem{esd94}
A.E. Evrard, F. Summers and M. Davis, {\em Astrophys. J.}, {\bf 422}
(1994) 11.

\bibitem{fb88}
S.M. Faber and D. Burstein, in:
V.C. Rubin and G.V. Coyne (eds), {\em Large--scale Motions in the
Universe} (Princeton University Press, Princeton, 1988) p. 115.

\bibitem{f83}
D. Fargion, {\em Nuovo Cimento B.}, {\bf 77} (1983) 111.

\bibitem{fm90}
K. Farrar and A.L. Melott, {\em Computers in Physics}, {\bf 4} (1990)
185.

\bibitem{fe93}
H.A. Feldman and A.E. Evrard, {\em Int. J. Mod. Phys. D.},
{\bf 2} (1993) 113.

\bibitem{fkp93}
H.A. Feldman, N. Kaiser and J.A. Peacock, {\em Astrophys. J.},
{\bf 426} (1994) 23.

\bibitem{fg84b}
J.A. Fillmore and P. Goldreich, {\em Astrophys. J.}, {\bf 281} (1984)
1.

\bibitem{fg84}
J.A. Fillmore and P. Goldreich, {\em Astrophys. J.}, {\bf 281} (1984)
9.

\bibitem{fdsyh93}
K.B. Fisher, M. Davis, M.A. Strauss, A. Yahil and J.P. Huchra,
{\em Astrophys. J.}, {\bf 402} (1993) 42.

\bibitem{fea94}
K.B. Fisher, M. Davis, M.A. Strauss, A. Yahil and J.P. Huchra,
{\em Mon. Not. R. astr. Soc.}, {\bf 267} (1994) 927.

\bibitem{f82}
J.N. Fry, {\em Astrophys. J.}, {\bf 262} (1982) 424.

\bibitem{f84a}
J.N. Fry, {\em Astrophys. J.}, {\bf 277} (1984) L5.

\bibitem{f84}
J.N. Fry, {\em Astrophys. J.}, {\bf 279} (1984) 499.

\bibitem{f84b}
J.N. Fry, {\em Astrophys. J.}, {\bf 279} (1984) 729.

\bibitem{f85}
J.N. Fry, {\em Astrophys. J.}, {\bf 289} (1985) 10.

\bibitem{f85b}
J.N. Fry, {\em Phys. Lett.}, {\bf 158B} (1985) 211.

\bibitem{f86}
J.N. Fry, {\em Astrophys. J.}, {\bf 306} (1986) 358.

\bibitem{f94}
J.N. Fry, {\em Astrophys. J.}, {\bf 421} (1994) 21.

\bibitem{fghms89}
J.N. Fry, R. Giovanelli, M.P. Haynes, A.L. Melott and R.J.
Scherrer, {\em Astrophys. J.}, {\bf 340} (1989) 11.

\bibitem{fm85}
J.N. Fry and A.L. Melott, {\em Astrophys. J.}, {\bf 292} (1985) 395.

\bibitem{fs93}
J.N. Fry and R.J. Scherrer, {\em Astrophys. J.},
to be published (1993)

\bibitem{fms93}
J.N. Fry, A.L. Melott and S.F. Shandarin,
{\em Astrophys. J.}, {\bf 412} (1993) 504.

\bibitem{gaz92}
E. Gaztanaga, {\em Astrophys. J.}, {\bf 398} (1992) L17.

\bibitem{g94}
E. Gaztanaga, {\em Mon. Not. R. astr. Soc.}, {\bf 268} (1994) 913.

\bibitem{gf94}
E. Gaztanaga and J.A. Frieman, Fermilab preprint 207-A (1994)

\bibitem{gmmy93}
M. Giavalisco, B. Mancinelli, P.J. Mancinelli and A. Yahil,
{\em Astrophys. J.}, {\bf 411} (1993) 9.

\bibitem{gh91}
R. Giovanelli and M.P. Haynes, {\em Ann. Rev. Astr. Astrophys.},
{\bf 29} (1991) 499.

\bibitem{ggrw86}
M.H. Goroff, B. Grinstein, S.--J. Rey and M.B. Wise,
{\em Astrophys. J.}, {\bf 311} (1986) 6.

\bibitem{g88}
K. Gorski, {\em Astrophys. J.}, {\bf 332} (1988) L7.

\bibitem{gdswy89}
K. Gorski, M. Davis, M.A. Strauss, S.D.M. White and A. Yahil,
{\em Astrophys. J.}, {\bf 344} (1989) 1.

\bibitem{gsv92}
K. Gorski, J. Silk and N. Vittorio,
{\em Phys. Rev. Lett.} {\bf 68} (1992) 733.

\bibitem{ghbbw94}
K. Gorski, G. Hindshaw, A.J. Banday, C.L. Bennett and E.L. Wright,
{\em Astrophys. J. Lett.} (1994) submitted.

\bibitem{gott75}
J.R. Gott, {\em Astrophys. J.}, {\bf 201} (1975) 296.

\bibitem{gr75}
J.R. Gott and M.J. Rees, {\em Astr. Astrophys.}, {\bf 45} (1975), 365.

\bibitem{gmd86}
J.R. Gott, A.L. Melott and M. Dickinson,
{\em Astrophys. J.}, {\bf 306} (1986) 341.

\bibitem{gmpl92}
J.R. Gott, S. Mao, C. Park, O. Lahav, {\em Astrophys. J.}, {\bf 385} (1992)
26.

\bibitem{gea89}
J.R. Gott, J. Miller, T.X. Thuan, S.E. Schneider, D.H. Weinberg,
C. Garrie, K. Polk, M. Vogeley, S. Jeffrey, S.P. Bhavsar,
A.L. Melott, R. Giovanelli, M.P. Haynes, R.B. Tully and A.J.S.
Hamilton, {\em Astrophys. J.}, {\bf 340} (1989) 625.

\bibitem{gpj90}
J.R. Gott, C. Park, R. Juszkiewicz, W.E. Bies, D.P.Bennett,
F.R. Bouchet and A. Stebbins, {\em Astrophys. J.}, {\bf 352} (1990) 1.

\bibitem{g93a}
M. Gramann, {\em Astrophys. J.}, {\bf 405} (1993) L47.

\bibitem{g93b}
M. Gramann, {\em Astrophys. J.}, {\bf 405} (1993) 449.

\bibitem{gcb93}
M. Gramann, R. Cen and N.A. Bahcall,
{\em Astrophys. J.}, {\bf 419} (1993) 440.

\bibitem{gw87}
B. Grinstein and M.B. Wise, {\em Astrophys. J.},
{\bf 320} (1987) 448.

\bibitem{gz81}
L.P. Grischuk and Ya.B. Zel'dovich,
{\em Sov. Astr.}, {\bf 25} (1981) 267.

\bibitem{gjo89}
E.J. Groth, R. Juszkiewicz and J.P. Ostriker,
{\em Astrophys. J.}, {\bf 346} (1989) 558.

\bibitem{gp75}
E.J. Groth and P.J.E. Peebles,
{\em Astr. Astrophys.}, {\bf 41} (1975) 143.

\bibitem{gp76}
E.J. Groth and P.J.E. Peebles,
{\em Astr. Astrophys.}, {\bf 53} (1976) 131.

\bibitem{gp77}
E.J. Groth and P.J.E. Peebles,
{\em Astrophys. J.}, {\bf 217} (1977) 385.

\bibitem{gg92}
S.F. Gull and A.J.M. Garrett, in: G. J. Erickson (ed),
{\em Maximum Entropy and Bayesian Methods, Seattle 1991}
(Kluwer, Dordrecht, 1992).

\bibitem{gunn77}
J.E. Gunn, {\em Astrophys. J.}, {\bf 218} (1977) 592.

\bibitem{gg72}
J.E. Gunn and J.R. Gott, {\em Astrophys. J.}, {\bf 176} (1972) 1.

\bibitem{gs81}
S.N. Gurbatov and A.I. Saichev, {\em Sov. Phys. JETP.},
{\bf 53} (1981) 347.

\bibitem{gss85}
S.N. Gurbatov, A.I. Saichev and S.F. Shandarin,
{\em Sov. Phys. Dokl.}, {\bf 30} (1985) 921.

\bibitem{gss89}
S.N. Gurbatov, A.I. Saichev and S.F. Shandarin,
{\em Mon. Not. R. astr. Soc.}, {\bf 236} (1989) 385.

\bibitem{gsy83}
S.N. Gurbatov, A.I. Saichev and I.G. Yakushkin,
{\em Sov. Phys. Usp.}, {\bf 26} (1983) 857.

\bibitem{gz70}
M. Guyot and Ya.B. Zel'dovich, {\em Astr. Astrophys.},
{\bf 9} (1970) 227.

\bibitem{gl93}
R. Guzman and J.R. Lucey,
{\em Mon. Not. R. astr. Soc.}, {\bf 263} (1993) L47.

\bibitem{h88}
A.J.S. Hamilton, {\em Astrophys. J.}, {\bf 332} (1988) 67.

\bibitem{ham92}
A.J.S. Hamilton, {\em Astrophys. J.}, {\bf 385} (1992) L5.

\bibitem{h93}
A.J.S. Hamilton, {\em Astrophys. J.}, {\bf 406} (1993) L47.

\bibitem{ham93}
A.J.S. Hamilton, {\em Astrophys. J.}, {\bf 417} (1993) 19.

\bibitem{hgw86}
A.J.S. Hamilton, J.R. Gott and D. Weinberg,
{\em Astrophys. J.}, {\bf 309} (1986) 1.

\bibitem{hklm91}
A.J.S. Hamilton, P. Kumar, E. Lu and A. Mathews,
{\em Astrophys. J.}, {\bf 374} (1991) L1.


\bibitem{hbpr86}
D. Hansel, F.R. Bouchet, R. Pellat and A. Romani, {\em Astrophys. J.},
{\bf 310} (1986), 23.

\bibitem{h70}
E. Harrison, {\em Phys. Rev. D.}, {\bf 1} (1970) 2726.

\bibitem{hor83}
M.A. Hausman, D.W. Olson and B.D. Roth, {\em Astrophys. J.}, {\bf 270}
(1983) 351.

\bibitem{h66}
S.W. Hawking, {\em Astrophys. J.}, {\bf 145} (1966) 544.

\bibitem{hay91}
M.P. Haynes, {\em Ann. Rev. Astr. Astrophys.}, {\bf 29} (1991) 499.

\bibitem{hg88}
M.P. Haynes and R. Giovanelli, in: V.C. Rubin and G.V. Coyne (eds),
{\em Large--scale Motions in the Universe} (Princetion University
Press, Princeton, 1988) p. 31.

\bibitem{h77}
D.J. Heath, {\em Mon. Not. R. astr. Soc.}, {\bf 179} (1977) 351.

\bibitem{ht95}
A.F. Heavens and A.N. Taylor, {\em Mon. Not. R. astr. Soc.},
in press (1995).

\bibitem{ha91}
J.P. Henry and K.A. Arnoud, {\em Astrophys. J.}, {\bf 372}
(1991) 410.

\bibitem{hk89}
L. Hernquist and N. Katz, {\em Astrophys. J. Suppl.}, {\bf 70}
(1989) 419.

\bibitem{hbcj94}
E. Hivon, F.R. Bouchet, S. Colombi and R. Juszkiewicz,
{\em Astr. Astrophys.} Sunmitted (1994).

\bibitem{he81}
R.W. Hockney and J.W. Eastwood, {\em Computer Simulation
Using Particles} (McGraw--Hill, New York, 1981).

\bibitem{hsw83}
G.L. Hoffman, E.E. Salpeter and I. Wasserman, {\em Astrophys. J.},
{\bf 268} (1983) 527.

\bibitem{hof86}
Y. Hoffman, {\em Astrophys. J.}, {\bf 308} (1986) 493.

\bibitem{hof89}
Y. Hoffman, {\em Astrophys. J.}, {\bf 340} (1989) 69.

\bibitem{hb84}
Y. Hoffman and S. Bludman, {\em Phys. Rev. Lett.} {\bf 52} (1984) 1087.

\bibitem{h89}
J. Holtzman, {\em Astrophys. J. Suppl.}, {\bf 71} (1989) 1.

\bibitem{h34}
E. Hubble, {\em Astrophys. J.}, {\bf 79} (1934) 8.

\bibitem{hv90}
J.--C. Hwang and E. T. Vishniac, {\em Astrophys. J.},
{\bf 353} (1990) 1.

\bibitem{i73}
V. Icke, {\em Astr. Astrophys.}, {\bf 27} (1973) 1.

\bibitem{i84}
V. Icke, {\em Mon. Not. R. astr. Soc.}, {\bf 206} (1984) 1.

\bibitem{iv87}
V. Icke and R. van de Weygaert, {\em Astr. Astrophys.}, {\bf 184}
(1987) 16.

\bibitem{ito83}
S. Ikeuchi, K. Tomisaka and J.P. Ostriker, {\em Astrophys. J.},
{\bf 265} (1983) 583.

\bibitem{iu84}
S. Ikeuchi and M. Umemura, {\em Prog. Theor. Phys.}, {\bf 72} (1984) 216.

\bibitem{i92}
M.B. Isichenko, {\em Rev. Mod. Phys.}, {\bf 64} (1992) 961.

\bibitem{iis93}
M. Itoh, S. Inagaki and W.C. Saslaw, {\em Astrophys. J.}, {\bf 403} (1993)
476.

\bibitem{jb94}
B. Jain and E. Bertschinger, {\em Astrophys. J.}, {\bf 431} (1994)
495.

\bibitem{jd83}
K. Janes and P. Demarque, {\em Astrophys. J.}, {\bf 264} (1983)
206.

\bibitem{js86}
L.G. Jensen and A.S. Szalay, {\em Astrophys. J.}, {\bf 305} (1986)
L5.

\bibitem{jet78}
M. Joeveer, J. Einasto and E. Tago, {\em Mon. Not. R. astr. Soc.},
{\bf 185} (1978) 357.

\bibitem{jm39}
W.A. Johnson and R.F. Mehl, {\em Trans. Am. Inst. Min. Metal. Eng.},
{\bf 135} (1939) 416.

\bibitem{jcm92}
B.J.T. Jones, P. Coles and V.J. Mart\'{\i}nez, {\em Mon. Not. R. astr.
Soc.}, {\bf 259} (1992) 146.

\bibitem{jmse88}
B.J.T. Jones, V. J. Mart\'{\i}nez, E. Saar and J. Einasto,
{\em Astrophys. J.}, {\bf 332} (1988) L1.

\bibitem{j81}
R. Juszkiewicz, {\em Mon. Not. R. astr. Soc.},
{\bf 197} (1981) 931.

\bibitem{jb91}
R. Juszkiewicz and F.R. Bouchet, in: {\em Proceedings of the Second DAEC
Meeting, Meudon, France} eds. G. Mamon and D. Gerbal, (1991).

\bibitem{jbc93}
R. Juszkiewicz, F.R. Bouchet and S. Colombi,
{\em Astrophys. J.}, {\bf 412} (1993) L9.

\bibitem{jwacb93}
R. Juszkiewicz, D.H. Weinberg, P. Amsterdamski, M. Chodorowski and
F.R. Bouchet, preprint IASSNS-AST 93/50 (1993).

\bibitem{jsb84}
R. Juszkiewicz, D.H. Sonoda and J.D. Barrow,
{\em Mon. Not. R. astr. Soc.}, {\bf 209} (1984) 139.

\bibitem{k84}
N. Kaiser, {\em Astrophys. J.}, {\bf 284} (1984) L9.

\bibitem{k87}
N. Kaiser, {\em Mon. Not. R. astr. Soc.},
{\bf 227} (1987) 1.

\bibitem{k88}
N. Kaiser, {\em Mon. Not. R. astr. Soc.},
{\bf 231} (1988) 149.

\bibitem{k89}
N. Kaiser, in: M. Mezetti, G. Giuricin,
F. Mardirossian and M. Ramella (eds),
{\em Large--scale Structure and Motions in the
Universe} (Kluwer, Dordrecht, 1989) p. 197.

\bibitem{k91}
N. Kaiser, in: D.W. Latham and L.N. da Costa (eds)
{\em Proceedings of the Texas/ESO--CERN Symposium on Relativistic
Astrophysics, Cosmology and Fundamental Physics},
Annals of the New York Academy of Sciences,
{\bf 647} (1991) p. 295.

\bibitem{kea91}
N. Kaiser, G. Efstathiou, R.S. Ellis, C.S. Frenk,
A. Lawrence, M. Rowan--Robinson and W. Saunders,
{\em Mon. Not. R. astr. Soc.}, {\bf 252} (1991) 1.

\bibitem{hocrh93}
H. Kang, J.P. Ostriker, R. Cen, D. Ryu, L. Hernquist, A.E. Evrard,
G.L. Bryan and M.L. Norman, {\em Astrophys. J.}, {\bf 430}
(1994) 83.

\bibitem{ky87}
A. Kashlinsky, {\em Astrophys. J.}, {\bf 317} (1987) 19.

\bibitem{ky91}
A. Kashlinsky, in: D. W. Latham  and L.N. da Costa (eds),
{\em Large--scale Structures and Peculiar Motions
in the Universe},  A. S. P. Conference Series {\bf 15} (1991)
p. 103

\bibitem{kg91}
N. Katz and J.E. Gunn, {\em Astrophys. J.}, {\bf 377} (1991) 365.

\bibitem{khw92}
N. Katz, L. Hernquist and D.H. Weinberg, {\em Astrophys. J.},
{\bf 399} (1992) 109.

\bibitem{kqg93}
N. Katz, T. Quinn and J.M. Gelb, {\em Mon. Not. R. astr. Soc.},
{\bf 265} (1993) 689.

\bibitem{kzw93}
N. Katz and S.D.M. White, {\em Astrophys. J.}, {\bf 412} (1993) 455.

\bibitem{kf91}
G. Kauffman and A.P. Fairall, {\em Mon. Not. R. astr. Soc.},
{\bf 248} (1991) 313.

\bibitem{kgw94}
G. Kauffman, B. Guiderdoni and S.D.M. White, {\em Mon. Not. R. astr. Soc.},
{\bf 267} (1994) 981.

\bibitem{km92}
G. Kauffman and A.L. Melott, {\em Astrophys. J.}, {\bf 393} (1992) 415.

\bibitem{kw93}
G. Kauffman and S.D.M. White, {\em Mon. Not. R. astr. Soc.}, {\bf 261}
(1993) 921.

\bibitem{kwg93}
G. Kauffman, S.D.M. White and B. Guiderdoni, {\em Mon. Not. R. astr. Soc.},
{\bf 264} (1993) 201.

\bibitem{ks77}
M. Kendall and A. Stuart, {\em The Advanced Theory of Statistics, Volume 1},
(Griffin \& Co., London, 1977).

\bibitem{k66}
T. Kiang, {\em Zeitschr. f. Astrophys.,} {\bf 64} (1966) 433.

\bibitem{k79}
S. Kida, {\em J. Fluid Mech.}, {\bf 93} (1979) 337.

\bibitem{k35}
J.C. Kirkwood, {J. Chem. Phys.}, {\bf 3} (1935) 300.

\bibitem{koss81}
R.P. Kirshner, A. Oemler, P.L. Schechter and S.A.
Schechtman, {\em Astrophys. J.}, {\bf 248} (1981) L57.

\bibitem{koss83}
R.P. Kirshner, A. Oemler, P.L. Schechter and S.A.
Schechtman, {\em Astr. J.}, {\bf 88} (1983) 1285.

\bibitem{khpr93}
A.A. Klypin, J. Holtzman, J.R. Primack and E. Regos,
{\em Astrophys. J.}, {\bf 416} (1993) 1.

\bibitem{ks83}
A.A. Klypin and S.F. Shandarin, {\em Mon. Not. R. astr. Soc.}, {\bf 204}
(1983) 891.

\bibitem{ks93}
A.A. Klypin and S.F. Shandarin, {\em Astrophys. J.},
{\bf 413} (1993) 48.

\bibitem{ks84}
H. Kodama and M. Sasaki,
{\em Prog. Theor. Phys. Suppl.}, {\bf 78} (1984) 1.

\bibitem{ks87}
H. Kodama and M. Sasaki,
{\em Int. J. Mod. Phys. A.}, {\bf 2} (1987) 491.


\bibitem{ko89}
L.A. Kofman, in: P. Flin and H.W. Duerbeck (eds),
{\em Morphological Cosmology. Proceedings of 11th Krakow
Cosmological School} (Springer, Berlin, 1989) p. 354.

\bibitem{kof91}
L.A. Kofman, in: K. Sato (ed), {\em Proc. IUAP Proceedings
on Nucleosynthesis in the Universe} (Kluwer, Dordrecht, 1991) 495.

\bibitem{kbgnd94}
L.A. Kofman, E. Bertschinger, J.M. Gelb, A. Nusser and A. Dekel,
{\em Astrophys. J.}, {\bf 420} (1994) 44.

\bibitem{kgb93}
L.A. Kofman, N.Y. Gnedin and N.A. Bahcall,
{\em Astrophys. J.}, {\bf 413} (1993) 1.

\bibitem{kp94}
L.A. Kofman and D.Yu. Pogosyan, SISSA preprint: astroph/9403029 (1994).

\bibitem{kps90}
L.A. Kofman, D.Yu. Pogosyan and S.F. Shandarin,
{\em Mon. Not. R. astr. Soc.}, {\bf 242} (1990) 200.

\bibitem{kpsm92}
L.A. Kofman, D.Yu. Pogosyan, S.F. Shandarin and A.L. Melott,
{\em Astrophys. J.}, {\bf 393} (1992) 437.

\bibitem{ks88}
L.A. Kofman and S. F. Shandarin,
{\em Nature}, {\bf 334} (1988) 129.

%

\bibitem{kstar85}
L.A. Kofman and A.A. Starobinsky,
{\em Sov. Astron. Lett.} {\bf 11} (1985) 271.

\bibitem{kt90}
E.W. Kolb and M.S. Turner,
{\em The Early Universe} (Addison--Wesley, Chicago, 1990).

\bibitem{ku82}
J.R. Kuhn and J.M. Uson, {\em Astrophys. J.}, {\bf 263} (1982)
L47.

\bibitem{lc93}
C.G. Lacey and S. Cole,
{\em Mon. Not. R. astr. Soc.}, {\bf 262} (1993) 627.

\bibitem{lgrs93}
C.G. Lacey, B. Guiderdoni, B. Rocca-Volmerange, and J. Silk,
{\em Astrophys. J.}, {\bf 402} (1993) 15.

\bibitem{ls91}
C.G. Lacey and J. Silk, {\em Astrophys. J.}, {\bf 381} (1991) 14.

\bibitem{l93}
M. Lachieze--Rey, {\em Astrophys. J.}, {\bf 408} (1993) 403.

\bibitem{llpr91}
O. Lahav, P.B. Lilje, J.R. Primack and M.J. Rees,
{\em Mon. Not. R. astr. Soc.}, {\bf 251} (1991) 128.

\bibitem{liis93}
O. Lahav, M. Itoh, S. Inagaki and Y. Suto,
{\em Astrophys. J.}, {\bf 402} (1993) 387.

\bibitem{lp85}
K. Lake and R. Pim, {\em Astrophys. J.}, {\bf 298} (1985) 439.

\bibitem{lea94}
A. Lawrence, M. Rowan--Robinson, J. Crawford, I. Parry,
X.-Y. Xia, R.S. Ellis, C.S. Frenk, W. Saunders,
G. Efstathiou and N. Kaiser,
{\em Mon. Not. R. astr. Soc.}, to be published (1995).

\bibitem{l31}
A. Lemaitre, {\em Mon. Not. R. astr. Soc.}, {\bf 91} (1931)
483.

\bibitem{ll92}
A.R. Liddle and D.H. Lyth, {\em Phys. Lett. B.}, {\bf 291} (1992)
391.

\bibitem{ll93a}
A.R. Liddle and D.H. Lyth, {\em Mon. Not. R. astr. Soc.},
{\bf 265} (1993) 379.

\bibitem{ll93b}
A.R. Liddle and D.H. Lyth, {\em Phys. Reports}, {\bf 231} (1993)
1.

\bibitem{lc92}
J.E. Lidsey and P. Coles, {\em Mon. Not. R. astr. Soc.},
{\bf 258} (1992)  57P.

\bibitem{l46}
E.M. Lifshitz, {\em Zh. Eksp. Teoret. Fiz.} {\bf 16} (1946) 587.

\bibitem{lk64}
E.M. Lifshitz and I.M. Khalatnikov,
{\em Sov. Phys. Usp.}, {\bf 6} (1964) 495.

\bibitem{le88}
P.B. Lilje and G. Efstathiou, {\em Mon. Not. R. astr. Soc.},
{\bf 231} (1988) 635.

\bibitem{l54}
D.N. Limber, {\em Astrophys. J.}, {\bf 119} (1954) 665.

\bibitem{lms65}
C.C. Lin, L. Mestel and F.H. Shu, {\em Astrophys. J.}, {\bf 142} (1965)
1431.

\bibitem{lfb86}
E.N. Ling, C.S . Frenk and J.D. Barrow, {\em Mon. Not. R. astr. Soc.},
{\bf 223} (1986) 21P.

\bibitem{lw94}
B. Little and D.H. Weinberg, {\em Mon. Not. R. astr. Soc.},
{\bf 267} (1994) 605.

\bibitem{ljwb94}
E.L. Lokas, R. Juszkiewicz, D.H. Weinberg and F.R. Bouchet,
{\em Mon. Not. R. astr. Soc.}, submitted (1994).

\bibitem{lepm92}
J. Loveday, G. Efstathiou, B.A. Paterson and S.J. Maddox,
{\em Astrophys. J.}, {\bf 400} (1992) L43.

\bibitem{lmm92}
F. Lucchin, S. Matarrese and S. Mollerach,
{\em Astrophys. J.}, {\bf 401} (1992) L49.

\bibitem{lmmm94}
F. Lucchin, S. Matarrese, A.L. Melott and L. Moscardini,
{\em Astrophys. J.}, {\bf 422} (1994) 430.

\bibitem{lhp89}
S. L. Lumsden, A. F. Heavens and J. A. Peacock, {\em Mon. Not. R.
astr. Soc.}, {\bf 238} (1989) 293.

\bibitem{luo94}
X. Luo, {\em Phys. Rev. D}, {\bf 49} (1994) 3810.

\bibitem{ls93}
X. Luo and D.N. Schramm, {\em Phys. Rev. Lett.}, {\bf 71} (1993)
1124.

\bibitem{l64}
D. Lynden--Bell, {\em Astrophys. J.}, {\bf 139} (1964) 1195.

\bibitem{lyea88}
D. Lynden--Bell, S. M. Faber, D. Burstein, R. L. Davies,
A. Dressler, R. J. Terlevich and G. Wegner,
{\em Astrophys. J.}, {\bf 326} (1988) 19.

\bibitem{llb89}
D. Lynden--Bell, O. Lahav and D. Burstein, {\em Mon. Not. R. astr. Soc.},
{\bf 241} (1989) 325.

\bibitem{mesl90}
S. J. Maddox, G. Efstathiou, W. J. Sutherland and J. Loveday,
{\em Mon. Not. R. astr. Soc.}, {\bf 242} (1990) 43P.

\bibitem{ms83}
K. Maeda and H. Sato, {\em Prog. Theor. Phys.}, {\bf 70} (1983) 772.

\bibitem{mss82}
K. Maeda, M. Sasaki and H. Sato, {\em Prog. Theor. Phys.}, {\bf 69}
(1982) 89.

\bibitem{mss92}
N. Makino, M. Sasaki and Y. Suto, {\em Phys. Rev. D.},
{\bf 46} (1992) 585.

\bibitem{m77}
B. B. Mandelbrot, {\em Fractals: Form, Chance, Dimension} (Freeman,
San Francisco, 1977).

\bibitem{m82}
B. B. Mandelbrot, {\em The Fractal Geometry of Nature} (Freeman, New York,
1982).

\bibitem{mhp93}
R. G. Mann, A. F. Heavens and J. A. Peacock, {\em Mon. Not. R. astr. Soc.},
{\bf 263} (1993) 798.

\bibitem{mk94}
S. Mao and C.S. Kochanek, {\em Mon. Not. R. astr. Soc.}, {\bf 268} (1994)
569.

\bibitem{mr93}
D. Maoz and H.--W. Rix, {\em Astrophys. J.}, {\bf 416} (1993) 425.

\bibitem{m91}
H. Martel, {\em Astrophys. J.}, {\bf 377} (1991) 7.

\bibitem{mf91}
H. Martel and W. Freudling, {\em Astrophys. J.},
{\bf 371} (1991) 1.

\bibitem{mw90}
H. Martel and I. Wasserman, {\em Astrophys. J.}, {\bf 348} (1990) 1.

\bibitem{mc94}
V. J. Mart\'{\i}nez and P. Coles, {\em Astrophys. J.}, {\bf 437}
(1994) 550.

\bibitem{mj90}
V. J. Mart\'{\i}nez and B. J. T. Jones, {\em Mon. Not. R. astr. Soc.},
{\bf 242} (1990) 517.

\bibitem{mjdv90}
V. J. Mart\'{\i}nez, B. J. T. Jones, R. Dominguez--Tenreiro and R. van de
Weygaert, {\em Astrophys. J.}, {\bf 357} (1990) 50.

\bibitem{mlms92}
S. Matarrese, F. Lucchin, L. Moscardini and D. Saez,
{\em Mon. Not. R. astr. Soc.}, {\bf 259} (1992) 437.

\bibitem{mps93}
S. Matarrese, O. Pantano and D. Saez, {\em Phys. Rev. D}, {\bf 47}
(1993) 1311.

\bibitem{mps94a}
S. Matarrese, O. Pantano and D. Saez, {\em Phys. Rev. Lett.} {\bf 72}
(1994) 320.

\bibitem{mps94b}
S. Matarrese, O. Pantano and D. Saez,
{\em Mon. Not. R. astr. Soc.}, {\bf 271} (1994) 513.

\bibitem{math90}
J. C. Mather, E. S. Cheng, R. E. Eplee, R. B. Isaacman, S. S. Meyer,
R. A. Shafer, R. Weiss, E. L. Wright, C. L. Bennett, N. W. Boggess,
E. Dwek, S. Gulkis, M. G. Hauser, M. Janssen, T. Kelsall, P. M. Lubin,
S. H. Moseley Jr, T. L. Murdock, R. F. Silverberg, G. F. Smoot and
D. T. Wilkinson, {\em Astrophys. J.}, {\bf 354} (1990)
L37.

\bibitem{math94}
J. C. Mather et al. {\em Astrophys. J.}, {\bf 420} (1994) 439.

\bibitem{ms84}
T. Matsuda and E. Shima, {\em  Prog. Theor. Phys.}, {\bf 71} (1984) 855.

\bibitem{ml87}
S. Maurogordato and M. Lachieze-Rey, {\em Astrophys. J.}, {\bf
320} (1987) 13.

\bibitem{ml91}
S. Maurogordato and M. Lachieze-Rey, {\em Astrophys. J.}, {\bf 369}
(1991) 30.

\bibitem{meak87}
P. Meakin, {\em Phys. Rev. A.}, {\bf 36} (1987) 2833.

\bibitem{mbw94b}
K. R. Mecke, T. Buchert and H. Wagner,
{\em Astr. Astrophys.}, {\bf 288} (1994) 697.

\bibitem{mssz92}
A. Meiksin, I. Szapudi and A. S. Szalay,
{\em Astrophys. J.}, {\bf 384} (1992) 87.

\bibitem{m90a}
A. L. Melott, {\em Physics Reports}, {\bf 193} (1990) 1.

\bibitem{m90b}
A. L. Melott, {\em Comments on Astrophysics}, {\bf 15} (1990) 1.

\bibitem{melp95}
A. L. Melott and J.L. Pauls, {\em Phys. Rev. Lett.} (1994) submitted.

\bibitem{mps94}
A. L. Melott, T. Pellman and S. F. Shandarin,
{\em Mon. Not. R. astr. Soc.}, {\bf 269} (1994) 626.

\bibitem{ms89}
A. L. Melott and S. F. Shandarin,
{\em Astrophys. J.}, {\bf 342} (1989) 26.

\bibitem{ms90}
A.L. Melott and S.F. Shandarin, {\em Nature} {\bf 346} (1990) 633.

\bibitem{mlmm94}
A.L. Melott, F. Lucchin, S. Matarrese, and L. Moscardini,
{\em Mon. Not. R. astr. Soc.}, {\bf 268} (1994) 69.

\bibitem{mbw94}
A. L. Melott, T. Buchert and A. G. Weiss,
{\em Astr. Astrophys.} In Press.

\bibitem{mchgw89}
A.L. Melott, A.P. Cohen, A.J.S. Hamilton, J.R. Gott and D.H. Weinberg,
{\em Astrophys. J.}, {\bf 345} (1989) 618.

\bibitem{msw94}
A.L. Melott, S.F. Shandarin and D.H. Weinberg,
{\em Astrophys. J.} {\bf 428} (1994) 28.

\bibitem{m74}
P. Meszaros, {\em Astr. Astrophys.}, {\bf 37} (1974) 225.

\bibitem{m53}
J. L. Meyerling, {\em Philips Res. Rept.} {\bf 8} (1953) 270.

\bibitem{m70}
R. E. Miles, {\em Math. Biosci.}, {\bf 6} (1970) 270.

\bibitem{m89}
J. Moller, {\em Adv. Appl. Prob. }, {\bf 21} (1989) 37.

\bibitem{mtg83}
J. E. Moody, E. L. Turner and J. R. Gott, {\em Astrophys. J.},
{\bf 273} (1983) 16.

\bibitem{mea92}
B. Moore, C. S. Frenk, D. H. Weinberg, W. Saunders, A. Lawrence,
M. Rowan--Robinson, N. Kaiser, G. Efstathiou and R. S. Ellis,
{\em Mon. Not. R. astr. Soc.}, {\bf 256} (1992) 477.

\bibitem{mhp86}
J.G. More, A.F. Heavens and J.A. Peacock, {\em Mon. Not. R. astr. Soc.},
{\bf 220} (1986) 189.

\bibitem{mmlm91}
L. Moscardini, S. Matarrese, F. Lucchin, and A. Messina,
{\em Mon. Not. R. astr. Soc.}, {\bf 248} (1991) 424.

\bibitem{mabpr91}
F. Moutarde, J. M. Alimi, F. R. Bouchet, R. Pellat and A. Ramani,
{\em Astrophys. J.}, {\bf 382} (1991) 377.

\bibitem{mfb92}
V.F. Mukhanov, H.A. Feldman and R.H. Brandenberger,
{\em Phys. Rep.}, {\bf 215} (1992) 203.

\bibitem{mss94}
D. Munshi, V. Sahni and A. A. Starobinsky,
{\em Astrophys. J.}, {\bf 436} (1994) 517.

\bibitem{mss95}
D. Munshi, T. Souradeep and A. A. Starobinsky,
{\em Astrophys. J.}, submitted (1995).

\bibitem{ms94}
D. Munshi and A. A. Starobinsky, {\em Astrophys. J.},
{\bf 428} (1994) 433.

\bibitem{nf72}
H. Nariai and M. Fujimoto, {\em Prog. Theor. Phys.}, {\bf 47} (1972) 105.

\bibitem{nb91}
J.F. Navarro and W. Benz, {\em Astrophys. J.}, {\bf 380} (1991) 320.

\bibitem{nw93}
J.F. Navarro and S.D.M. White, {\em Mon. Not. R. astr. Soc.},
{\bf 265} (1993) 271.

\bibitem{np93}
R. P. Nelson and J. C. B. Papaloizou, {\em Mon. Not. R. astr. Soc.},
{\bf 265} (1993) 905.

\bibitem{ncgl92}
R.C. Nichol, C.A. Collins, L. Guzzo and S.L. Lumsden,
{\em Mon. Not. R. astr. Soc.}, {\bf 255} (1992) 21P.

\bibitem{np94}
R. Nityananda and T. Padmanabhan,
{\em Mon. Not. R. astr. Soc.}, {\bf 271} (1994) 976.

\bibitem{nd90}
A. Nusser and A. Dekel, {\em Astrophys. J.}, {\bf 362} (1990) 14.

\bibitem{nd92}
A. Nusser and A. Dekel, {\em Astrophys. J.}, {\bf 391} (1992) 443.

\bibitem{nd93}
A. Nusser and A. Dekel, {\em Astrophys. J.}, {\bf 405} (1993) 437.

\bibitem{ndbb91}
A. Nusser, A. Dekel, E. Bertschinger and G. R. Blumenthal,
{\em Astrophys. J.}, {\bf 376} (1991) 6.

\bibitem{osv83}
F. Occhionero, P. Santangelo and N. Vittorio, {\em Astron. Astrophys.}
{\bf 117} (1983) 365.

\bibitem{osw91}
K. A. Olive, G. Steigman and T. P. Walker, {\em Astrophys. J.},
{\bf 380} (1991) L1.

\bibitem{odbps90}
S. Olivier, G. R. Blumenthal, A. Dekel, J. R. Primack and D. Stanhill,
{\em Astrophys. J.}, {\bf 356} (1990) 1.

\bibitem{oppw86}
S. Otto, H. D. Politzer, J. Preskill and M. B. Wise,
{\em Astrophys. J.}, {\bf 304} (1986) 62.

\bibitem{pn92}
T. Padmanabhan and D. Narasimha, {\em Mon. Not. R. astr. Soc.}, {\bf 259}
(1992) 41p.

\bibitem{ps93}
T. Padmanabhan and K. Subramanian, {\em Astrophys. J.},
{\bf 410} (1993) 482.

\bibitem{pad93}
T. Padmanabhan, {\em Structure formation in the Universe}
(Cambridge University Press, Cambridge, 1993).

\bibitem{pv87}
G. Paladin and A. Vulpiani, {\em Physics Reports}, {\bf 156} (1987)
147.

\bibitem{p85}
M. Panek, {\em Mon. Not. R. astr. Soc.}, {\bf 216} (1985) 85.

\bibitem{pgmk92}
C. Park, J. R. Gott, A. L. Melott and I. D. Kharachentsev,
{\em Astrophys. J.}, {\bf 387} (1992) 1.

\bibitem{pvgh94}
C. Park, M.S. Vogeley, M.J. Geller and J.P. Huchra,
{\em Astrophys. J.}, {\bf 431} (1994) 569.

\bibitem{pp67}
R.B. Partridge and P.J.E. Peebles, {\em Astrophys. J.}, {\bf 148} (1967) 377.

\bibitem{pm95}
J. L. Pauls and A.L. Melott,
{\em Mon. Not. R. astr. Soc.} in press (1995).

\bibitem{p91}
J.A. Peacock, {\em Mon. Not. R. astr. Soc.}, {\bf 253} (1991)
1P.

\bibitem{p92}
J.A. Peacock, in: V.J. Martinez, M.Portilla and D. Saez (eds),
{\em New Insights into the Universe. Proceedings, Valencia, Spain
1991} (Springer, Berlin, 1992) p. 1.

\bibitem{pd94}
J.A. Peacock and S. Dodds, {\em Mon. Not. R. astr. Soc.},
{\bf 267} (1994) 1020.

\bibitem{ph85}
J.A. Peacock and A.F. Heavens, {\em Mon. Not. R. astr. Soc.},
{\bf 217} (1985) 805.

\bibitem{ph90}
J.A. Peacock and A.F. Heavens, {\em Mon. Not. R. astr. Soc.},
{\bf 243} (1990) 133.

\bibitem{pn91}
J.A. Peacock and D. Nicholson, {\em Mon. Not. R. astr. Soc.},
{\bf 253} (1991) 307.

\bibitem{pw92}
J.A. Peacock and M.J. West, {\em Mon. Not. R. astr. Soc.},
{\bf 259} (1992) 494.

\bibitem{pc94}
R.C. Pearson and P. Coles, {\em Mon. Not. R. astr. Soc.},
{\bf 272} (1995) 231.

\bibitem{p67}
P.J.E. Peebles, {\em Astrophys. J.}, {\bf 147} (1967) 859.

\bibitem{p74}
P.J.E. Peebles, {\em Astrophys. J.}, {\bf 189} (1974) L51.

\bibitem{p74b}
P.J.E. Peebles, {\em Astr. Astrophys.}, {\bf 32} (1974) 197.

\bibitem{p80}
P.J.E. Peebles, {\em The Large--scale Structure of the Universe}
(Princeton University Press, Princeton, 1980).

\bibitem{p82a}
P.J.E. Peebles, {\em Astrophys. J.}, {\bf 257} (1982) 438.

\bibitem{p82}
P.J.E. Peebles, {\em Astrophys. J.}, {\bf 258} (1982) 415.

\bibitem{p82b}
P.J.E. Peebles, {\em Astrophys. J.}, {\bf 263} (1982) L1.

\bibitem{p84}
P.J.E. Peebles, {\em Astrophys. J.}, {\bf 284} (1984) 439.

\bibitem{p87}
P.J.E. Peebles, {\em Astrophys. J.}, {\bf 315} (1987) L73.

\bibitem{p90}
P.J.E. Peebles, {\em Astrophys. J.}, {\bf 362} (1990) 1.

\bibitem{p93}
P.J.E. Peebles, {\em Principles of Physical Cosmology}
(Princeton University Press, Princeton, 1993).

\bibitem{p94}
P.J.E. Peebles, {\em Astrophys. J.}, {\bf 429} (1994) 43.

\bibitem{pg75}
P.J.E. Peebles and E.J. Groth, {\em Astrophys. J.}, {\bf 196} (1975)
1.

\bibitem{pbs90}
L. Perivolaropoulos, R. Brandenberger and A. Stebbins,
{\em Phys. Rev. D}, {\bf 41} (1990) 1764.

\bibitem{pss67}
V. Petrosian, E.E. Salpeter and P. Szekeres,
{\em Astrophys. J.}, {\bf 147} (1967) 1222.

\bibitem{pi87}
L. Pietronero, {\em Physica}, {\bf 144A} (1987) 257.

\bibitem{pl86}
R. Pim and K. Lake, {\em Astrophys. J.}, {\bf 304} (1986) 75.

\bibitem{pl88}
R. Pim and K. Lake, {\em Astrophys. J.}, {\bf 330} (1988) 625.

\bibitem{pvc92}
M. Plionis, R. Valdarnini and P. Coles, {\em Mon. Not. R. astr.
Soc.}, {\bf 258} (1992) 114.

\bibitem{p89}
D. Yu Pogosyan, {\em Tartu Preprint, Estonian Acad. Sci.} (1989).

\bibitem{pstar93}
D. Yu Pogosyan and A.A. Starobinsky,
{\em Mon. Not. R. astr. Soc.}, {\bf 265} (1993) 507.

\bibitem{pp86}
H.D. Politzer and J. Preskill, {\em Phys. Rev. Lett.},
{\bf 56} (1986) 99.

\bibitem{pw84}
H.D. Politzer and M.B. Wise, {\em Astrophys. J.},
{\bf 285} (1984) L1.

\bibitem{pgh86}
M. Postman, M.J. Geller and J.P. Huchra,
{\em Astr. J.}, {\bf 91} (1986) 1267.

\bibitem{ps74}
W.H. Press and P.L. Schechter,
{\em Astrophys. J.}, {\bf 187} (1974) 425.

\bibitem{pb84}
J.R. Primack and G.R. Blumenthal, in:
F. Mardirossian, G. Giuricin and M. Mezzetti (eds),
{\em Clusters and Groups of Galaxies},
(Reidel, Dordrecht, 1984) p. 435.

\bibitem{phkc94}
J.R. Primack, J. Holtzman, A. Klypin and D.O. Caldwell, {\em Phys. Rev. Lett.},
in press (1995).

\bibitem{rg91}
E. Regos and M.J. Geller, {\em Astrophys. J.}, {\bf 377} (1991) 14.

\bibitem{r45a}
S.O. Rice, {\em Bell Systems Tech. Journal}, {\bf 24} (1945) 46.

\bibitem{r45b}
S.O. Rice, in: N. Wax (ed), {\em Selected Papers on Noise and
Stochastic Processes} (Dover, New York, 1954) p. 133.


\bibitem{rf92}
L. Ruamsuwan and J.N. Fry, {\em Astrophys. J.}, {\bf 396} (1992) 416.

\bibitem{rsv82}
V.A. Rubakov, M.V. Sazhin and A.V. Veryaskin,
{\em Phys. Lett. B.}, {\bf 115} (1982) 189.

\bibitem{brmg91}
B. Ryden and M. Gramann, {\em Astrophys. J.}, {\bf 383} (1991) L33.

\bibitem{rokc93}
D. Ryu, J.P. Ostriker, H. Kang and R. Cen, {\em Astrophys. J.},
{\bf 414} (1993) 1.

\bibitem{sw67}
R.K. Sachs and A.M. Wolfe, {\em Astrophys. J.},
{\bf 147} (1967) 73.

\bibitem{sv80}
V. Sahni, {\em Pramana}, {\bf 15} (1980) 423.

\bibitem{s84}
V. Sahni, {\em One--loop Quantum Gravitational Effects and the
Growth of Density Perturbations in Cosmological Models},
(PhD Thesis, Moscow State University, 1984)

\bibitem{s91}
V. Sahni, in: {\em Proceedings of Texas/ESO--CERN Symposium on
Relativistic Astrophysics, Cosmology and Fundamental Physics,
Annals of the New York Academy of Sciences}, {\bf 647} (1991) 749.

\bibitem{sfs92}
V. Sahni, H. Feldman and A. Stebbins, {\em Astrophys. J.},
{\bf 385} (1992) 1.

\bibitem{sss94}
V. Sahni, B.S. Sathyaprakash and S.F. Shandarin,
{\em Astrophys. J.}, {\bf 431} (1994) 20.

\bibitem{ssp94}
V. Sahni, B.S. Sathyaprakash and D. Pogosyan,
In preparation (1995).

\bibitem{sss94b}
V. Sahni, B.S. Sathyaprakash and S.F. Shandarin, In preparation
(1995).

\bibitem{ss92a}
V. Sahni and T. Souradeep, in:
M. C. Bento, O. Bertolami, J.M. Mourao
and R. F. Picken (eds), {\em Classical and Quantum Gravity:
Proceedings of the First Iberian
Meeting on Gravity, Evora, Portugal 21 - 26 September 1992}
(World Scientific, Singapore, 1993) p. 256.

\bibitem{jks91}
J.K. Salmon, {\em ``Parallel Hierarchical N-body Methods''},
PhD thesis, Caltech (1991) Unpublished.

\bibitem{sw94}
J.K. Salmon and M.S. Warren, {\em J. Comp. Phys.} ,
in press (1994).

\bibitem{sk92}
D.S. Salopek, {\em Phys. Rev. Lett.}, {\bf 69} (1992) 3602.

\bibitem{ssc94}
D.S. Salopek, J.M. Stewart and K.M. Croudace, {\em Mon. Not. R. astr. Soc.},
{\bf 271} (1994) 1005.

\bibitem{sals95}
D.S. Salopek and J.M. Stewart, {\em Phys. Rev. D}, {\bf 51} (1995) 517.

\bibitem{s80}
W.C. Saslaw, {\em Astrophys. J.}, {\bf 235} (1980) 299.

\bibitem{s85}
W.C. Saslaw, {\em Gravitational Physics of Stellar and Galactic Systems}
(Cambridge University Press, Cambridge, 1985).

\bibitem{s86}
W. C. Saslaw, {\em Astrophys. J.}, {\bf 304} (1986) 11.

\bibitem{s89}
W.C. Saslaw, {\em Astrophys. J.}, {\bf 341} (1989) 588.

\bibitem{s92}
W.C. Saslaw, {\em Astrophys. J.}, {\bf 391} (1991) 423.

\bibitem{scii90}
W.C. Saslaw, S.M. Chitre, M. Itoh and S. Inagaki, {\em Astrophys. J.},
{\bf 365} (1990) 419.

\bibitem{sh84}
W.C. Saslaw and A.J.S. Hamilton, {\em Astrophys. J.}, {\bf 276} (1984)
13.

\bibitem{ss93}
W.C. Saslaw and R.K. Sheth, {\em Astrophys. J.}, {\bf 409},
504.

\bibitem{ssmpm94}
B. S. Sathyaprakash, V. Sahni, D. Munshi, D. Pogosyan and A. L. Melott,
{\em Mon. Not. R. ast. Soc.} (1995) In press.

\bibitem{sssk95}
B. S. Sathyaprakash, V. Sahni, S.F. Shandarin and K.B. Fisher,
In preparation 1995.

\bibitem{s82}
H. Sato, {\em Prog. Theor. Phys.}, {\bf 68} (1982) 236.

\bibitem{sm83}
H. Sato and K. Maeda, {\em Prog. Theor. Phys.}, {\bf 70} (1983) 119.

\bibitem{sea91}
W. Saunders, C.S. Frenk, M. Rowan--Robinson, G. Efstathiou, A. Lawrence,
N. Kaiser, R.S. Ellis, J. Crawford, X.--Y. Xia and I. Parry,
{\em Nature}, {\bf 349} (1991) 32.

\bibitem{ss85}
R. Schaeffer and J. Silk, {\em Astrophys. J.}, {\bf 292} (1985) 319.

\bibitem{ss88}
R. Schaeffer and J. Silk, {\em Astrophys. J.}, {\bf 322} (1988), 1.

\bibitem{s76}
P.L. Schechter, {\em Astrophys. J.}, {\bf 203} (1976) 247.

\bibitem{sms91}
R.J. Scherrer, A.L. Melott and S.F. Shandarin,
{\em Astrophys. J.}, {\bf 377} (1991) 29.

\bibitem{soy75}
J. Schwarz, J. P. Ostriker and A. Yahil, {\em Astrophys. J.},
{\bf 202} (1975) 1.

\bibitem{s59}
L.I. Sedov, {\em Similarity and Dimensional Methods in Mechanics}
(Academic Press, New York, 1959)

\bibitem{sp77}
M. Seldner and P.J.E. Peebles,
{\em Astrophys. J.}, {\bf 215} (1977) 703.

\bibitem{sh83}
S.F. Shandarin, {\em  Sov. Astron. Lett.}, {\bf 9} (1983) 104.

\bibitem{s94}
S.F. Shandarin, Kansas University preprint (1994);
To appear in: {\em Proceedings of the Conference ``Cosmic Velocity
Fields'', Paris, July 1993}

\bibitem{sdz83}
S.F. Shandarin, A.G. Doroshkevich and Ya.B. Zel'dovich,
{\em Sov. Phys. Usp.}, {\bf 26} (1983) 46.

\bibitem{sz83}
S.F. Shandarin and Ya.B. Zel'dovich,
{\em Comments on Astrophysics}, {\bf 10} (1983) 33.

\bibitem{sz89}
S.F. Shandarin and Ya.B. Zel'dovich,
{\em Rev. Mod. Phys.}, {\bf 61} (1989) 185.

\bibitem{shwi67}
C.D. Shane and C.A. Wirtanen, {\em Publ. Lick. Obs.},
{\bf 22} (1967), part 1.

\bibitem{s79}
T. Shanks, {\em Mon. Not. R. astr. Soc.}, {\bf 186} (1979) 583.

\bibitem{shfm89}
T. Shanks, D. Hale-Sutton, R. Fong and N. Metcalfe, {\em Mon. Not.
R. astr. Soc.}, {\bf 237} (1989), 589.

\bibitem{ss-m84}
P. Shapiro and K. Struck--Marcell, {\em Astrophys. J. Suppl.},
{\bf 57} (1984) 205.

\bibitem{sbl84}
N.A. Sharp, S.A. Bonometto and F. Lucchin, {\em Astr. Astrophys.},
{\bf 130} (1984) 79.

\bibitem{sms93}
R.K. Sheth, H.J. Mo and W.C. Saslaw, {\em Astrophys. J.}, {\bf 427}
(1994) 562.

\bibitem{sv90}
J. Silk and N. Vittorio, in: J. Audouze and F. Melchiorri (eds),
{\em Confrontation between Theories and Observations in Cosmology:
Present Status and Future Programs. Proceedings of the International
School of Physics ``Enrico Fermi''},
(North--Holland, Amsterdam, 1990) p. 137.

\bibitem{swaga94}
V. Silviera and I. Waga, {\em Phys. Rev. D}, {\bf 50} (1994) 4890.

\bibitem{sea92}
G.F. Smoot, C.L. Bennett, A. Kogut, E.L. Wright, J. Aymon,
N.W. Boggess, E.S. Cheng, G. de Amici, S. Gulkis, M.G. Hauser,
G. Hinshaw, P.D. Jackson, M. Janssen, E. Kaita, T. Kelsall,
P. Keegstra, C. Lineweaver, K. Loewenstein, P. Lubin, J. Mather,
S.S. Meyer, S.H. Moseley, T. Murdock, L. Rokke, R.F. Silverberg,
L. Tenorio, R. Weiss and D.T. Wilkinson, {\em Astrophys. J.},
{\bf 396} (1992) L1.

\bibitem{stb93}
G.F. Smoot, L. Tenorio, A.J. Banday, A. Kogut, E.L. Wright, G. Hinshaw
and C.L. Bennett, {\em Astrophys. J.}, submitted (1993).

\bibitem{sod92}
J. Soda and Y. Suto, {\em Astrophys. J.}, {\bf 396} (1992) 379.

\bibitem{sn85}
L.V. Solov'eva and I.S. Nurgaliev, {\em Sov. Astr.},
{\bf 29} (1985) 267.

\bibitem{sstar85}
L.V. Solov'eva and A.A. Starobinsky, {\em Sov. Astr.},
{\bf 29} (1985) 367.

\bibitem{ss92b}
T. Souradeep and V. Sahni,
{\em Mod. Phys. Lett.}, {\bf A7} (1992) 3541.

\bibitem{sy85}
A.A. Starobinsky, {\em Sov. Astr. Lett.}, {\bf 11} (1985) 133.

\bibitem{ss84}
A.A. Starobinsky and V. Sahni, in:
{\em Modern Theoretical and Experimental Problems of General
Relativity} (MGI Press, Moscow, 1984) p. 77.

\bibitem{skm87}
D. Stoyan, W.S. Kendall and J. Mecke, {\em Stochastic geometry and its
applications}, (Akademie--Verlag, Berlin, 1987)

\bibitem{sd88}
M.A. Strauss and M. Davis, in: V. Rubin and G. V. Coyne (eds),
{\em Large--scale Motions in the Universe} (Princeton University
Press, Princeton, 1988) p. 256.

\bibitem{sdyh92}
M.A. Strauss, M. Davis, A. Yahil and J.P. Huchra,
{\em Astrophys. J.}, {\bf 385} (1992) 421.

\bibitem{sr91}
M.F. Struble and H.J. Rood, {\em Astrophys. J. Suppl.}, {\bf 77}
(1991) 263.

\bibitem{sb94}
M.P. Susperregi and J.J. Binney, {\em Mon. Not. R. astr. Soc.},
{\bf 271} (1994) 719.

\bibitem{s88}
W.J. Sutherland, {\em Mon. Not. R. astr. Soc.}, {\bf 234} (1988) 159.

\bibitem{se91}
W.J. Sutherland and G. Efstathiou, {\em Mon. Not. R. astr. Soc.},
{\bf 248} (1991) 159.

\bibitem{sii90}
Y. Suto, M. Itoh and S. Inagaki, {\em Astrophys. J.}, {\bf 350}
(1990) 492.

\bibitem{ss91}
Y. Suto and M. Sasaki, {\em Phys. Rev. Letters}, {\bf 66} (1991) 264.

\bibitem{sz87}
A.S. Szalay in: {\em Large Scale Structures in the Universe},
Seventeenth advanced course of the Swiss society of Astronomy and Astrophysics,
Saas-Fee (Geneva Observatory, Switzerland, 1987).

\bibitem{szsz93}
I. Szapudi and A.S. Szalay, {\em Astrophys. J.}, {\bf 403} (1993)
43.

\bibitem{ssb92}
I. Szapudi, A.S. Szalay and P. Boschan,
{\em Astrophys. J.}, {\bf 390} (1992) 350.

\bibitem{t34}
R.C. Tolman, {\em Proc. Nat. Acad. Sci.} {\bf 20} (1934) 169.

\bibitem{tk69}
M. Totsuji and T. Kihara, {\em Pub. Astr. Soc. Jap.}, {\bf 21} (1969)
221.

\bibitem{tf87}
R.B. Tully and J.R. Fisher, {\em Nearby Galaxy Atlas} (Cambridge
University Press, Cambridge, 1987).

\bibitem{t90}
E.L. Turner, {\em Astrophys. J.}, {\bf 365} (1990) L43.

\bibitem{vi89}
R. van de Weygaert and V. Icke, {\em Astron. Astrophys.} {\bf 213} (1989)
1.

\bibitem{vdb94}
R. van de Weygaert and A. Babul,
{\em Astrophys. J. Lett.} {\bf 425} (1994) L59.

\bibitem{vv93}
R. van de Weygaert and E. van Kampen, {\em Mon. Not. R. astr. Soc.},
{\bf 263} (1993) 481.

\bibitem{vande94}
R. van de Weygaert, {\em Astron. Astrophys.} {\bf 283} (1994) 361.

\bibitem{vdfn94}
M. Vergassola, B. Dubrulle, U. Frisch and A. Noullez,
{\em Astron. Astrophys.} {\bf 289} (1994) 325.

\bibitem{v83}
E.T. Vishniac, {\em Mon. Not. R. astr. Soc.},
{\bf 203} (1983) 345.

\bibitem{v86}
E.T. Vishniac, in: E.W. Kolb, M.S. Turner, D. Lindley, K. Olive and
D. Seckel (eds),
{\em Inner Space/ Outer Space} (University of Chicago Press, Chicago,
1986) p. 190.

\bibitem{vt87}
N. Vittorio and M.S. Turner, {\em Astrophys. J.}, {\bf 316} (1987)
475.

\bibitem{vgh91}
S.V. Vogeley, M.J. Geller and J.P. Huchra, {\em Astrophys. J.},
{\bf 382} (1991) 44.

\bibitem{vpgh92}
S.V. Vogeley, C. Park, M.J. Geller and J.P. Huchra,
{\em Astrophys. J.}, {\bf 391} (1992) L5.

\bibitem{v08}
G. Voronoi, {\em J. reine angew. Math.} {\bf 134} (1908) 198.

\bibitem{wqsz92}
M.S. Warren, P.J. Quinn, J.K. Salmon and W.H. Zurek,
{\em Astrophys. J.}, {\bf 399} (1992) 405.

\bibitem{w81}
I. Wasserman, {\em Astrophys. J.}, {\bf 248} (1981) 1.

\bibitem{wtkgh90}
G. Wegner, J.R. Thorstensen, M. J. Kurtz, M.J. Geller and J.P. Huchra,
{\em Astr. J.}, {\bf 100} (1990) 1405.

\bibitem{w87}
S. Weinberg, {\em Phys. Rev. Lett.}, {\bf 59} (1987) 2607.

\bibitem{wc92}
D.H. Weinberg and S. Cole, {\em Mon. Not. R. astr. Soc.}, {\bf 259} (1992) 652.

\bibitem{wod89}
D.H. Weinberg, J.P. Ostriker and A. Dekel, {\em Astrophys. J.}, {\bf 336}
(1989) 9.

\bibitem{wg90a}
D.H. Weinberg and J.E. Gunn,
{\em Mon. Not. R. astr. Soc.}, {\bf 247} (1990) 260.

\bibitem{wg90b}
D H. Weinberg and J.E. Gunn,
{\em Astrophys. J.}, {\bf 352} (1990) L25.

\bibitem{whkllcow92}
D. Weistrop, P. Hintzen, R.C. Kennicutt, C. Liu, J. Lowenthal,
K.P. Cheng, R. Oliverson, and B. Woodgate,
{\em Astrophys. J.}, {\bf 396} (1992) L23.

\bibitem{wwd90}
M.J. West, D.H. Weinberg and A. Dekel, {\em Astrophys. J.}, {\bf 353}
(1990) 329.

\bibitem{wss94}
M. White, D. Scott and J. Silk,
{\em Ann. Rev. Astron. Astrophys.}, {\bf 32} (1994) 319.

\bibitem{w79}
S.D.M. White, {\em Mon. Not. R. astr. Soc.}, {\bf 186} (1979) 145.

\bibitem{wf91}
S.D.M.White and C.S. Frenk, {\em Astrophys. J.}, {\bf 379} (1991), 52.

\bibitem{w74}
G.B. Whitham, {\em Linear and Non--linear Waves},
(Wiley, New York, 1974).

\bibitem{w92}
B.G. Williams, {\em Ph.D. Thesis, University of Edinburgh}, (1992).

\bibitem{wph91}
B.G. Williams, J.A. Peacock and A.F. Heavens,
{\em Mon. Not. R. astr. Soc.}, {\bf 252} (1991) 43.

\bibitem{whps91}
B.G. Williams, A.F. Heavens, J.A. Peacock and S.F. Shandarin,
{\em Mon. Not. R. astr. Soc.}, {\bf 250} (1991) 458.

\bibitem{wr94}
E.L. Wright, et al. {\em Astrophys. J.}, {\bf 420} (1994) 1.

\bibitem{y88}
A. Yahil, in: V. Rubin and G.V. Coyne (eds),
{\em Large--scale Motions in the Universe}
(Princeton University Press, Princeton, 1988) p. 219.

\bibitem{zh93}
S. Zaroubi and Y. Hoffman, {\em Astrophys. J.}, {\bf 414} (1993)
20.

\bibitem{z64}
Ya.B. Zel'dovich, {\em Astron. Zh.}, {\bf 41} (1964) 873.

\bibitem{z70}
Ya.B. Zel'dovich, {\em Astr. Astrophys.}, {\bf 4} (1970)
84.

\bibitem{z72}
Ya.B. Zel'dovich, {\em Mon. Not. R. astr. Soc.},
{\bf 160} (1972) 1P.

\bibitem{zel82}
Ya.B. Zel'dovich, {\em Sov. Astron. Lett.}, {\bf 8} (1982) 102.

\bibitem{zes82}
Ya B. Zel'dovich, J. Einasto and S.F. Shandarin,
{\em Nature}, {\bf 300} (1982) 407.

\bibitem{zms83}
Ya.B. Zel'dovich, A.V. Mamaev and S.F. Shandarin,
{\em Sov. Phys. Usp.}, {\bf 26} (1983) 77.

\bibitem{zn83}
Ya.B. Zel'dovich and I.D. Novikov,
{\em The Structure and Evolution of the Universe},
(University of Chicago Press, Chicago, 1983).

\bibitem{zs82}
Ya.B. Zel'dovich and S.F. Shandarin, {\em Sov. Astron. Lett.}, {\bf 8}
(1982) 67.

\bibitem{z79}
J.M. Ziman, {\em Models of Disorder} (Cambridge University Press,
Cambridge, 1979).

\bibitem{z57}
F. Zwicky, {\em Morphological Astronomy} (Springer, Berlin, 1957).

\bibitem{zhwkk61}
F. Zwicky, E. Herzog, P. Wild, M. Karpowicz and C.T. Kowal,
{\em Catalog of Galaxies and Clusters of Galaxies} Vols. 1
- 6, (Pasadena: Calif. Inst. Technol. 1961 - 1968).

\end{thebibliography}
